\documentclass[11pt,a4paper]{article}

\pdfoutput=1

\usepackage[margin=27mm]{geometry}

\usepackage{amssymb,amsmath,amsfonts,amsthm, amscd, mathrsfs, helvet, mathtools}
\usepackage{arydshln}
\usepackage{bm, stmaryrd}
\usepackage{braket}
\usepackage{setspace}
\usepackage{caption}
\usepackage{graphicx,verbatim}
\usepackage{enumitem}
\usepackage[usenames, dvipsnames]{color}
\usepackage[T1]{fontenc}
\usepackage{scalerel}
\usepackage{subfiles}

\usepackage{tikz}    
\usetikzlibrary{arrows,matrix}
\usetikzlibrary{arrows.meta}
\usetikzlibrary{calc} 
\usetikzlibrary{cd}
\usetikzlibrary{matrix,calc}
\usetikzlibrary{decorations.pathreplacing,positioning}
\usetikzlibrary{decorations.markings,shapes.geometric,shapes.misc}

\usepackage{pgfplots}
\usepackage{tikz-3dplot}
 
\tdplotsetmaincoords{60}{115}
\pgfplotsset{compat=newest}

\theoremstyle{plain}
\newtheorem{theorem}{Theorem}[section]
\newtheorem*{theorem*}{Theorem}

\newtheorem*{proposition*}{Proposition}

\newtheorem{def-lem}[theorem]{Definition/Lemma}
\newtheorem{def-prop}[theorem]{Definition/Proposition}

\theoremstyle{remark}

\newtheorem{examplex}[theorem]{Example}

\newenvironment{example}
  {\pushQED{\qed}\examplex}
  {\popQED\endexamplex}

\DeclareFontEncoding{LS1}{}{}
\DeclareFontSubstitution{LS1}{stix}{m}{n}
\DeclareSymbolFont{symbols2}{LS1}{stixfrak}{m}{n}
\DeclareMathSymbol{\typecolon}{\mathbin}{symbols2}{"25}

\definecolor{myPurple}{rgb}{.4,0.2,0.8}

\definecolor{brightRed}{rgb}{1,0,0}

\definecolor{myBlue}{rgb}{0,0,1}
\newcommand{\blue}[0]{\color{myBlue}}
\definecolor{myGreen}{rgb}{0,0.7,0}

\usepackage{hyperref}
\hypersetup{
    colorlinks=true, 
    linktoc=all,     
    linkcolor=myPurple,  
    urlcolor=myPurple,
    citecolor=myGreen,
    menucolor=black,
    linktocpage=true,
}
            
\DeclareFontEncoding{LS1}{}{}
\DeclareFontSubstitution{LS1}{stix}{m}{n}
\DeclareSymbolFont{stixsymbols}{LS1}{stixscr}{m}{n}
\SetSymbolFont{stixsymbols}{bold}{LS1}{stixscr}{b}{n}
\DeclareMathSymbol{\scrb}{\mathalpha}{stixsymbols}{"62}
\DeclareMathSymbol{\scrr}{\mathalpha}{stixsymbols}{"72}
\DeclareMathSymbol{\lax}{\mathalpha}{stixsymbols}{"6C}

\newcommand{\tensor}[1]{{\mathfrak{#1}}}

\newcommand{\nord}[1]{\mathopen{:}#1\mathclose{:}}
\newcommand{\cnord}[1]{\mathopen{\typecolon}#1\mathclose{\typecolon}}

\DeclareMathOperator{\id}{id}

\DeclareMathOperator{\ad}{ad}

\DeclareMathOperator{\End}{End}

\DeclareMathOperator{\rk}{rk}

\DeclareSymbolFont{widetriangleaccent}{OMX}{yhex}{m}{n}
\DeclareMathAccent{\widetriangle}{\mathord}{widetriangleaccent}{"E6}

\def\P{\mathrm{P}}

\def\bsl{\smallsetminus}

\def\sl{\mathfrak{sl}}
\def\su{\mathfrak{su}}

\def\a{\mathsf{a}}

\def\b{\mathsf{b}}

\def\dd{\mathrm{d}}

\def\g{\mathfrak{g}}
\def\x{\mathsf{x}}

\def\H{\mathsf H}

\def\hg{\widehat{\mathfrak{g}}}
\def\hbg{\widehat{\bar{\mathfrak{g}}}}

\def\h{\mathfrak{h}}

\def\q{\mathsf{q}}

\def\Heis{\mathfrak{H}}

\def\Fock{{\mathfrak{F}}}
\def\hFock{{\widehat{\mathfrak{F}}}}
\def\Bos{{\mathfrak B}}
\def\Shi{{\mathsf s}}

\def\L{\mathsf{L}}

\newcommand{\bb}[1]{[\kern-.1em[ #1 ]\kern-.1em]}
\newcommand{\lau}[1]{(\kern-.2em( #1 )\kern-.2em)}

\newcommand{\lbf}[1]{\langle\kern-.2em\langle #1 \rangle\kern-.2em\rangle}
\newcommand{\biglbf}[1]{\big\langle \kern-.25em \big\langle #1 \big\rangle \kern-.25em \big\rangle}

\newcommand{\rbf}[1]{(\kern-.2em( #1 )\kern-.2em)}
\newcommand{\bigrbf}[1]{\big( \kern-.25em \big( #1 \big) \kern-.25em \big)}

\def\CC{\mathbb{C}}
\def\RR{\mathbb{R}}

\def\ZZ{\mathbb{Z}}

\def\O{\mathsf{O}}
\def\R{\mathcal{R}}

\def\L{\mathsf{L}}

\def\Vop{\mathsf V}
\def\Wop{\mathsf W}
\def\X{\mathsf{X}}

\def\ii{\mathsf{i}}

\def\short{\epsilon'\bsl\epsilon}

\def\1{\tensor{1}}
\def\2{\tensor{2}}
\def\3{\tensor{3}}
\def\4{\tensor{4}}

\newlength{\dhatheight}

\tikzset{
  ctrlpoint/.style={%
    draw=gray,
    circle,
    inner sep=0,
    minimum width=1ex,
  }
}

\numberwithin{equation}{section}

\begin{document}

\title{\textbf{Effective Hamiltonians and\\
Wilson--Polchinski renormalisation}}
\author{Ricky Li$^{\chi}$\ and\ Beno\^{\i}t Vicedo$^{\bar\chi}$\vspace{4mm}\\
{\small Department of Mathematics, University of York,}\\
{\small Heslington, York YO10 5GH, United Kingdom.}\vspace{4mm}\\
{\small 
Email: ${}^{\chi}$~\href{mailto:ricky.li@york.ac.uk}{\texttt{ricky.li@york.ac.uk}},
${}^{\bar\chi}$~\href{mailto:benoit.vicedo@gmail.com}{\texttt{benoit.vicedo@gmail.com}}
\vspace{2mm}
}
}
\date{February 2026}

\maketitle

\begin{abstract}
\noindent We develop a novel approach to the Wilsonian renormalisation of Hamiltonians for $2$-dimensional quantum field theories on the cylinder described in the UV by marginally relevant deformations of conformal field theories. To introduce a Wilsonian short-distance cutoff we make essential use of free field realisations of the full vertex operator algebra in the UV. Our method is intrinsically non-perturbative; we derive a Hamiltonian analogue of Polchinski’s equation describing the flows of all couplings.

As a primary example of our general method, we apply it to the marginal anisotropic deformation of the $\su_2$ Wess--Zumino--Witten model at level $1$, which is equivalent to the sine-Gordon model on the cylinder. In particular, we reproduce the standard renormalisation group flow of the sine-Gordon model near the Kosterlitz--Thouless point to second order in the couplings, a result usually derived using Lagrangian/path-integral methods.
\end{abstract}

\setcounter{tocdepth}{3}
\tableofcontents

\section{Introduction and overview}

Renormalisation is one of the cornerstones of quantum field theory. The subject has evolved from the perturbative ``subtraction of infinities'' in Feynman diagram computations in early quantum electrodynamics to the Wilsonian paradigm \cite{KogutWilson}, which formulates quantum field theory in terms of scale-dependent effective descriptions. From this modern perspective, the renormalisation group formalises the relationship between effective quantum field theories defined at different energy scales (inverse-length scales), with high-energy degrees of freedom systematically and continuously integrated out to yield low-energy descriptions.

\subsection{Approaches to renormalisation}

\subsubsection{Rigorous perturbative renormalisation}

The rigorous mathematical underpinning of the perturbative ``subtraction of infinities'' was the culmination of a substantial body of work spanning several decades. It began with the combinatorial study of nested divergences in Feynman graphs, formalised in the Bogolyubov--Parasiuk--Hepp--Zimmermann (BPHZ) theorem \cite{BogoPara, Hepp, Zimmermann:1969jj}. This established that local counterterms can be recursively constructed so as to render $n$-point Green's functions finite to all orders in perturbation theory (see, for instance, \cite{Collins}). The BPHZ renormalisation scheme finds its rigorous justification in the causal Epstein--Glaser axiomatic framework \cite{EpsteinGlaser}, in which renormalisation arises from the non-unique extension of time-ordered operator-valued distributions to coincident points rather than from the subtraction of divergent integrals. Finally, Connes and Kreimer showed \cite{Connes:1999yr, Connes:2000fe} that the BPHZ recursion is equivalent to the Birkhoff factorisation of regularised Feynman rules viewed as characters of the Hopf algebra of Feynman graphs \cite{Kreimer:1997dp}, thereby recasting perturbative renormalisation as a Riemann--Hilbert problem.

\subsubsection{Non-perturbative renormalisation: path-integrals}

The development of the exact, functional, or non-perturbative renormalisation group in the physics literature ran parallel to these mathematical advances in perturbative quantum field theory. Building on Kadanoff’s coarse-graining intuition \cite{Kadanoff:1966wm}, Wilson cast quantum field theory as a flow of effective actions obtained by successively integrating out high-energy modes \cite{KogutWilson}.

This physical picture was first encoded into exact functional differential equations by Wegner and Houghton \cite{Wegner:1972ih} using a sharp momentum cutoff, which Polchinski then significantly improved upon by working instead with a smooth cutoff regulator \cite{Polchinski}. This flow equation was subsequently reformulated by Wetterich \cite{Wetterich:1992yh} in terms of the so-called effective average action. The resulting Wetterich equation is mathematically equivalent to Polchinski's via a functional Legendre transform, but provides a formulation in terms of macroscopic observables that is often more suitable for non-perturbative approximation schemes and numerical analysis; see e.g. \cite{Berges:2002, Delamotte, Dupuis:2020fhh} for extensive reviews.

It was only relatively recently that Costello established a rigorous mathematical foundation for this Wilsonian approach in the context of Euclidean quantum field theory \cite{CostelloRGbook}, see also \cite{CGBook2}, by combining the concept of effective actions with the Batalin--Vilkovisky formalism. Finally, the incorporation of these functional flow methods into the Lorentzian framework of perturbative algebraic quantum field theory was recently achieved in \cite{DAngelo:2022vsh, DAngelo:2023tis}, thereby reconciling the Wilsonian intuition with the rigorous axioms of local covariance.

\subsubsection{Non-perturbative renormalisation: Hamiltonians}

Most formulations of Wilsonian renormalisation are rooted in the path-integral formalism or in perturbative Lagrangian frameworks, where the construction of the S-matrix typically relies on the time-ordering intuition of the Dyson series.

Somewhat surprisingly, comparatively less attention has been devoted to the Hamiltonian counterpart, despite the fact that Wilson’s original intuition \cite{KogutWilson, Wilson1975} was inherently based on the diagonalisation of the Hamiltonian matrix in a basis of truncated states. This operator-theoretic approach was explicitly formalised by Głazek and Wilson \cite{GlazekWilson, Glazek:1994qc} and Wegner \cite{Wegner:1994fdg} through the method of continuous unitary similarity transformations to decouple high-energy modes, a method known as the `Similarity Renormalisation Group', while related schemes were developed in \cite{Minic:1994ff, GubankovaWeg, Walhout1998, Alexanian:1998gz, Alexanian:1998wu}.

In the Truncated Conformal Space Approach (TCSA) developed in \cite{Yurov:1989yu}, a marginally relevant perturbation $\Vop$ of a conformal field theory with Hamiltonian $\H_0 = \L_0 + \bar\L_0$ is studied numerically by truncating the Hilbert space of the conformal field theory to the finite-dimensional subspace of states with energy bounded by some fixed $E_{\rm max}$. In this setting, Wilsonian renormalisation techniques have been numerically used to reduce the dependence of the results on the cutoff $E_{\rm max}$ coming from the truncation, see for instance \cite{Feverati:2006ni, Giokas:2011ix, Rychkov:2014eea}.

More recently, an extensive Hamiltonian renormalisation programme has been developed by Thiemann and collaborators, see for instance \cite{Lang:2017beo, Lang:2017yxi, Thiemann:2022ulb, Zarate:2025qlg}, which provides a rigorous implementation of Kadanoff's coarse-graining idea in the Hamiltonian formalism. Specifically, in this approach, operator valued distributions are smeared using finite-dimensional spaces of test functions labelled by integer resolution scales $M \in \ZZ_{\geq 1}$ and the coarse-graining map is defined from a finer graining $M'$ to a coarser one $M < M'$.

\subsection{The Hamiltonian Polchinski equation} \label{sec: Pol eq intro}

The main goal of this paper is to derive a Hamiltonian analogue of Polchinski's equation for effective Hamiltonians in the Wilsonian framework. We formulate our approach for quantum field theories on the cylinder $S^1 \times \RR$, with spatial compactification $x \sim x + L$, described in the UV by relevant or marginally relevant deformations of $2$-dimensional conformal field theories. We summarise below the content of \S\ref{sec: Renormalisation} where this framework is developed.

\subsubsection{Free field realisations and smooth regularisations}

Standard applications of the Wilsonian renormalisation group typically deal with scalar fields $\phi(x)$ on $\RR^d$, where the separation of scales is implemented by a sharp momentum cutoff on the Fourier modes $\hat{\phi}(k)$. We are interested in perturbations of $2$-dimensional conformal field theories in the UV, which may not intrinsically be described in terms of free fields. To adapt the standard Wilsonian philosophy, our starting point is therefore to make a choice of a \textbf{free field realisation} of the underlying full vertex operator algebra. This is to be contrasted with Costello's approach \cite{CostelloRGbook} which employs the BV formalism and homological methods to treat general interacting theories without necessarily mapping them to free fields.

However, the compact spatial geometry of the cylinder $S^1 \times \RR$ introduces another technical subtlety: the Fourier modes of Hamiltonian fields on $S^1$ are discrete. In this setting, a standard sharp cutoff on the mode numbers $n \in \mathbb{Z}$ would necessarily be an integer, making it impossible to derive continuous renormalisation group flow equations in the form of differential equations with respect to the scale. To circumvent this, we must eschew sharp truncations in favour of \textbf{smooth regularisations}. So instead of truncating the algebra of Fourier modes, we introduce a continuous regulator directly into the canonical commutation relations of the free fields, as detailed in \S\ref{sec: regularisation}. For concreteness, consider a chiral free boson $\chi(x)$ with modes $\b_n$ for $n \in \ZZ$ satisfying the standard Heisenberg Lie algebra relations $[\b_n, \b_m] = n \delta_{n+m,0}$. In the regularised theory with short-distance cutoff $\epsilon > 0$ we replace it by a smoothly regularised chiral free boson $\chi^\epsilon(x)$ whose modes $\b_n^\epsilon$ for $n \in \ZZ$ satisfy the modified relations
\begin{equation} \label{reg chiral algebra intro}
[\b_n^\epsilon, \b_m^\epsilon] = n \, \eta\left(\frac{2\pi |n| \epsilon}{L}\right) \delta_{n+m,0},
\end{equation}
where the regulator $\eta : \mathbb{R}_{\geq 0} \to \mathbb{R}$ is a fixed but arbitrary smooth function such that $\eta(0)=1$ and $\eta(x)$ decays faster than any power as $x \to \infty$.

This smooth regularisation cures the ultraviolet divergences inherent in the short-distance singularities of products of chiral bosons, which typically manifest as divergent series $\sum_{n=1}^\infty n^s$. Indeed, in the regularised theory these are replaced by finite, albeit $\epsilon$-dependent, series of the form $\sum_{n=1}^\infty n^s \eta\left(\frac{2\pi n \epsilon}{L}\right)$. More importantly, when the same regularisation is applied also to the anti-chiral sector, operators which induce deformations away from the conformal field theory, given by zero-modes of products of chiral and anti-chiral operators, have a well-defined action on the Fock space of the free field realisation. A prototypical example is the radius-changing operator $\int_0^L dx \, \partial_x \chi(x) \partial_x \bar\chi(x)$ in the free compactified boson conformal field theory.

\subsubsection{Effective Hamiltonian and Polchinski's equation}

Since the short-distance cutoff $\epsilon > 0$ used in \eqref{reg chiral algebra intro} was arbitrary, physical observables in the theory should not depend on it. Indeed, the core philosophy of Wilsonian renormalisation (see, for instance, \cite{HollowoodBook}) is that there is a notion of `separation of scale' in the sense that the details of the physics at length scales below the one we are concerned about should not be relevant. In particular, physics at length scales longer than the cutoff $\epsilon > 0$ should be independent of $\epsilon$.

To describe how we implement this idea in our setup, we focus here again for concreteness on a theory with field content described by a chiral and anti-chiral boson $\chi(x)$ and $\bar\chi(x)$. Given any two scales $\epsilon' > \epsilon > 0$, we introduce a `shell' chiral boson $\chi^{\epsilon'\bsl\epsilon}(x)$ with modes $\b_n^{\epsilon'\bsl\epsilon}$ for every $n \in \ZZ$ subject to certain commutation relations such that the modes of $\chi^{\epsilon'}(x) + \chi^{\epsilon'\bsl\epsilon}(x)$ satisfy the same commutation relations as those of $\chi^\epsilon(x)$ given in \eqref{reg chiral algebra intro}; and likewise for the anti-chiral sector. Then the Hamiltonian analogue of the procedure of `integrating out' degrees of freedom between the cutoffs $\epsilon$ and $\epsilon'$ amounts to the problem of constructing an \textbf{effective Hamiltonian} $\H_{\rm eff}^{\epsilon'}$ in the theory with larger cutoff $\epsilon'$, i.e. built from the chiral and anti-chiral fields $\chi^{\epsilon'}(x)$ and $\bar\chi^{\epsilon'}(x)$, which describes the same physics as the original Hamiltonian $\H^\epsilon$ in the theory with smaller cutoff $\epsilon$, i.e. built from the chiral and anti-chiral fields $\chi^\epsilon(x)$ and $\bar\chi^\epsilon(x)$.

Our approach to defining the effective Hamiltonian $\H_{\rm eff}^{\epsilon'}$, spelled out in detail in \S\ref{sec: effective Hamiltonians}, is to require that its imaginary-time evolution over an arbitrary but fixed time $T$ matches the imaginary-time evolution of the original Hamiltonian $\H^\epsilon$ at the lower cutoff $\epsilon$ projected down to the long-distance subspace, namely
\begin{equation} \label{Heff def intro}
e^{-T \H^{\epsilon'}_{\text{eff}}} = \P_l e^{-T \H^\epsilon} \P_l.
\end{equation}
Here $\P_l$ denotes the projection onto the long-distance subspace, which consists of states where none of the `shell modes' $\b_n^{\epsilon'\bsl\epsilon}$ are excited, i.e. the subspace where the shell sector is in its vacuum state. In other words, the operator on the right hand side of \eqref{Heff def intro} describes the full imaginary-time evolution of all states in the long-distance subspace which end up back in this subspace after an imaginary-time $T$, including all virtual excursions into the short-distance sector. The effective Hamiltonian $\H_{\rm eff}^{\epsilon'}$ on the left hand side of \eqref{Heff def intro} encodes the same imaginary-time evolution over a time $T$ but without ever leaving the long-distance subspace, incorporating the effective dynamics of the shell degrees of freedom in the fields $\chi^{\epsilon'\bsl\epsilon}(x)$ and $\bar\chi^{\epsilon'\bsl\epsilon}(x)$ which have been eliminated or `integrated out'.

\medskip

This matching of evolution operators is similar in spirit to the path-integral formulation of Wilsonian renormalisation in the presence of a high-energy cutoff. There the effective action $S_{\text{eff}}^{\Lambda'}[\phi^{\Lambda'}]$ for the low-energy field $\phi^{\Lambda'}$ in the theory with cutoff $\Lambda'$ is defined by integrating out shell degrees of freedom $\phi^{\Lambda\bsl\Lambda'}$ between the cutoffs $\Lambda' < \Lambda$ in the path-integral, namely
\begin{equation} \label{Seff def intro}
e^{-S_{\text{eff}}^\Lambda[\phi^{\Lambda'}]} = \int \mathcal{D} \phi^{\Lambda\bsl\Lambda'} \, e^{-S^\Lambda[\phi^{\Lambda'} + \phi^{\Lambda\bsl\Lambda'}]},
\end{equation}
where $S^\Lambda[\phi^\Lambda]$ denotes the action for the high-energy field $\phi^\Lambda = \phi^{\Lambda'} + \phi^{\Lambda\bsl\Lambda'}$ in the theory with cutoff $\Lambda$.
Our definition \eqref{Heff def intro} of the effective Hamiltonian $\H_{\rm eff}^{\epsilon'}$ can be viewed as a Hamiltonian analogue of this standard definition \eqref{Seff def intro} of the effective action. Indeed, one of the main results of the paper, see \S\ref{sec: RG flow}, is a Hamiltonian analogue of \textbf{Polchinski's equation} \cite{Polchinski} derived using \eqref{Heff def intro} in the case when the length scale cutoff varies infinitesimally so that $\epsilon' = \epsilon + \delta \epsilon$ for small $\delta\epsilon$. By working to leading order in $\delta\epsilon$ we derive the beta function for all the couplings.

\medskip

Our definition of the effective Hamiltonian in \eqref{Heff def intro} differs conceptually from the one used in the Similarity Renormalisation Group \cite{GlazekWilson, Glazek:1994qc, Wegner:1994fdg, Minic:1994ff, GubankovaWeg, Walhout1998, Alexanian:1998gz, Alexanian:1998wu}. The key idea behind this approach is to seek a unitary similarity transformation of the Hamiltonian $\H^\Lambda$ in the theory with cutoff $\Lambda$ which decouples the shell degrees of freedom between the cutoffs $\Lambda' < \Lambda$ from the low-energy ones below the cutoff $\Lambda'$. In other words, one is after a unitary operator $\mathsf U_{\Lambda, \Lambda'}$ such that the transformed Hamiltonian $\mathsf U_{\Lambda,\Lambda'} \H^\Lambda \mathsf U_{\Lambda,\Lambda'}^\dag$ is block-diagonal with respect to the low- and high-energy subspaces, so that it sends, in particular, the low-energy subspace to itself. The effective Hamiltonian at the cutoff $\Lambda'$ is then given simply by restricting this transformed Hamiltonian to the low-energy subspace, i.e.
\begin{equation} \label{Heff AM}
\H^{\Lambda'}_{\rm eff} = \P_l \mathsf U_{\Lambda,\Lambda'} \H^\Lambda \mathsf U_{\Lambda,\Lambda'}^\dag \P_l
\end{equation}
where $\P_l$ is the projection onto the low-energy subspace. In practice, however, the Hamiltonian $\H^\Lambda$ can only be block-diagonalised perturbatively to the first few orders in both the couplings and the ratio of energy scales $\Lambda'/\Lambda$, see for instance \cite{Alexanian:1998wu} for a nice application of these ideas to the $4$-dimensional $\varphi^4$ theory. In this approach, the virtual excursions into the shell sector are captured by the unitary transformation which can effectively be seen as mapping the free vacuum state of the shell sector in the regularised theory to the interacting one.

There are also similarities between our definition \eqref{Heff def intro} and the construction of the effective Hamiltonian in the recently proposed Hamiltonian Truncation Effective Theory \cite{Cohen:2021erm, Demiray:2025zqh}. In the method of Hamiltonian Truncation first introduced in \cite{BrooksFrautschi}, or the Truncated Conformal Space Approach of \cite{Yurov:1989yu}, the truncated Hamiltonian is $\P_l (\H_0 + \Vop) \P_l$, where $\P_l$ here denotes the projection onto the finite-dimensional subspace of states with total energy less than some cutoff $E_{\rm max}$, which can be diagonalised numerically. Hamiltonian Truncation Effective Theory \cite{Cohen:2021erm, Demiray:2025zqh}, see also the very recent paper \cite{Maestri:2026hqb}, offers a systematic way of improving on the naive truncation $\P_l (\H_0 + \Vop) \P_l$ of the Hamiltonian by using methods from effective field theory to construct an effective Hamiltonian which takes into account the effects from states above the cutoff $E_{\rm max}$. However, rather than use a similarity transformation as in \eqref{Heff AM}, the effective Hamiltonian is obtained by matching its transition amplitudes between low-energy states to the transition amplitudes of the full Hamiltonian between the same states.

The approaches of the Similarity Renormalisation Group and of Hamiltonian Truncation Effective Theory described above are both influenced by well-known methods to calculate effective Hamiltonians, such as Rayleigh--Schrödinger perturbation theory or Schrieffer--Wolff transformations \cite{SchriefferWolff, BDL}. Yet it is important to observe that these methods rely on a sharp separation of energy scales in the spectrum of the unperturbed theory since they typically involve denominators $(E_h - E_l)^{-1}$ with differences between energy eigenvalues $E_h$ and $E_l$ in the high-energy and low-energy sectors. But in a smooth regularisation scheme like the one we are using \eqref{reg chiral algebra intro}, following Polchinski's original approach to Wilsonian renormalisation using smooth high-energy cutoffs \cite{Polchinski}, the `shell' modes to be integrated out in the construction of the effective Hamiltonian inevitably include a tail with arbitrarily low energies, causing these energy denominators to diverge. Our definition \eqref{Heff def intro} of the effective Hamiltonian, which closely mimics the path-integral definition of the effective action used in \cite{Polchinski}, does not rely on the existence of an energy gap between short- and long-distance degrees of freedom.

Finally, let us note that the use of smooth regularisation in perturbative renormalisation has independently been explored in the recent works \cite{PadillaSmith1, PadillaSmith2}. Their motivation for using smooth regularisations was also inspired by \cite{Tao-blog}, see \S\ref{sec: regularisation}, and so it would be interesting to understand if the framework developed in these papers is related to ours.

\subsection{Main example and future directions}

The main example to which we apply the methods developed in \S\ref{sec: Renormalisation}, briefly outlined above in \S\ref{sec: Pol eq intro}, is the sine-Gordon model. We summarise below the content of \S\ref{sec: sine-Gordon} where this example is studied in detail.

The quantum sine-Gordon model is a tremendously well studied quantum field theory in $2$ dimensions, with a vast literature initiated by Coleman's seminal work \cite{Coleman} on its equivalence to the massive Thirring model. This theory is simultaneously rich in non-perturbative phenomena, in particular through its strong/weak coupling duality with the massive Thirring model, see e.g. \cite{Mandelstam:1975hb}, yet sufficiently simple to be tractable, owing in part to its integrability \cite{Kulish:1976xi, Zamolodchikov:1978xm, Sklyanin:1979pfu, Zamolodchikov:1995xk, Niccoli:2009jq} (see also \cite{Torrielli:2024bpa} for a recent review). As such, it provides an ideal testing ground for exploring new and existing frameworks in $2$-dimensional quantum field theory, which explains why the literature on the sine-Gordon model continues to rapidly expand even to this day. In particular, recent progress was made on describing the sine-Gordon model from the perspective of perturbative algebraic quantum field theory \cite{BahnsRejzner, Zanello}.

A remarkable feature of the quantum sine-Gordon model is that its renormalisation group flow exhibits the famous Berezinskii--Kosterlitz--Thouless (BKT) transition \cite{Berezinskii:1971, Berezinskii:1972, KosterlitzThouless}. This phase transition, characterised by a separatrix dividing the massless and massive phases, has been derived in many different formalisms. For instance, it was originally derived using perturbative Wilsonian renormalisation techniques on the XY-model \cite{Kosterlitz:1974sm}, which lies in the same universality class as the sine-Gordon model, and later using standard field-theoretic counterterm renormalization around the Kosterlitz--Thouless point \cite{Amit:1979ab,Balog:2000qr}, using string theoretic methods \cite{Lovelace:1986kr}, using Polchinski's equation \cite{Oak:2017trw} or using conformal perturbation theory, see e.g. \cite[\S4.6]{FradkinBook}. We refer also to \cite{Frohlich:1981yn, Marchetti1900} for rigorous derivations and to \cite{Dimock:1991ig, Dimock:1993qw, Dimock:1999jt} where rigorous Euclidean path-integral methods are used, relating to the framework of construction quantum field theory \cite{GlimmJaffe}.
While the BKT transition is seen already at second order in the couplings around the Kosterlitz--Thouless point within these perturbative frameworks, the renormalisation group flow of the sine-Gordon model has also been extensively studied non-perturbatively using the functional renormalisation group, namely via the Wetterich equation, see for instance \cite{FRG-sG, FRG-sG2, DavietDupuis} and also \cite{Hariharakrishnan:2024iba} for a recent comparison between the non-perturbative and perturbative analyses.

\medskip

Importantly for us, the sine-Gordon model on the cylinder can be described as a marginally relevant perturbation of a 2-dimensional conformal field theory in the UV, namely the free boson compactified at the self-dual radius. Explicitly, the sine-Gordon model is equivalent to an anisotropic deformation of the $\su_2$ WZW model at level $1$ \cite[\S3g]{Bernard:1990ys}, with Hamiltonian
\begin{equation} \label{H sG intro}
\H_{\rm sG} = \frac{2 \pi}{L} (\L_0 + \bar \L_0) + \frac{g_1}{4\pi}\int_0^L \dd x \big( J^+(x) \bar J^-(x) + J^-(x) \bar J^+(x) \big) + \frac{g_2}{8 \pi} \int_0^L \dd x \, J^3(x) \bar J^3(x).
\end{equation}
The first term on the right hand side is the Hamiltonian of the $\su_2$ WZW model and the two $J \bar J$-deformations are written in terms of the chiral and anti-chiral $\su_2$-currents of the WZW model. The relationship with the sine-Gordon model, which is described by a single boson field $\Phi(x)$ and its conjugate momentum $\Pi(x)$, is seen directly by recalling that the chiral and anti-chiral $\su_2$-currents at level $1$ admit a free field realisation in terms of chiral and anti-chiral free bosons $\chi(x)$ and $\bar\chi(x)$, respectively; see \S\ref{sec: Free field realisations} and \S\ref{sec: quantum sG as def WZW} for details. Under this free field realisation the kinetic term of the sine-Gordon Hamiltonian arises as the sum of the first and last terms in \eqref{H sG intro}, with the chiral decompositions $\Phi(x) = \phi(x) + \bar\phi(x)$ and $\Pi(x) = \partial_x \phi(x) - \partial_x \bar\phi(x)$ related to the chiral and anti-chiral fields $\chi(x)$ and $\bar\chi(x)$ by a Bogolyubov transformation; see \S\ref{sec: changing radius} and \S\ref{sec: usual sG Ham} for details. In particular, the last term in \eqref{H sG intro} is the marginal perturbation inducing a radius change in the compactified free boson from the self-dual radius to a generic radius $R = 2/\beta$ determined by $g_2$. The second term in \eqref{H sG intro} then becomes proportional to the standard cosine potential $\int_0^L \dd x \, \cnord{\cos(\beta \Phi(x))}_\beta$ which is relevant for $\beta^2 < 8 \pi$; see \S\ref{sec: usual sG Ham}.

As a non-trivial check of our approach to Wilsonian renormalisation in the Hamiltonian formalism using a smooth regularisation \`a la Polchinski, outlined in \S\ref{sec: Pol eq intro}, the main goal of \S\ref{sec: sine-Gordon} is to rederive the renormalisation group flow equations to second order of the marginal couplings $g_1$ and $g_2$ in \eqref{H sG intro}, with respect to the continuous short-distance cutoff scale $\epsilon > 0$. With our normalisation conventions we recover the well-known renormalisation group flow
\begin{equation} \label{KT flow intro}
\epsilon \partial_\epsilon g_1 = g_1 g_2, \qquad
\epsilon \partial_\epsilon g_2 = g_1^2
\end{equation}
which exhibits the BKT transition; see Figure \ref{fig:anisotropic flow} in the main text. The renormalisation group flow \eqref{KT flow intro} is universal, i.e. scheme independent, and was given in \cite{Zamolodchikov:1995xk}.
In fact, the flow derived in \cite{Zamolodchikov:1995xk} is an all-loop exact expression in a particular renormalisation scheme, obtained using the Poincar\'e--Dulac theorem to fix the renormalisation scheme ambiguity of the subleading coefficients. Note that a different all-loop extension of the universal leading-order flow \eqref{KT flow intro} was also given in \cite[\S III]{BernardLeClair}. The latter is based on a conjectured exact expression for the beta function of current-current deformations of $2$-dimensional conformal field theories with Kac--Moody current-algebra symmetry \cite{GLM}, obtained using Ward identity considerations. However, a subsequent $4$-loop computation of this renormalisation group flow using conformal perturbation theory in \cite{Ludwig:2002fu} was found to be incompatible with the flow of \cite[\S III]{BernardLeClair} in an arbitrary renormalisation scheme. It would be interesting to push the computations in the present Hamiltonian framework to higher order in perturbation theory to check these all-loop exact extensions of the universal renormalisation group flow \eqref{KT flow intro}.

\medskip

Finally, it would be interesting to apply the Hamiltonian framework developed in this paper to other $2$-dimensional quantum field theories on the cylinder $S^1 \times \RR$, which can be described by marginally relevant deformations of conformal field theories in the UV. For concreteness we have focused on the $\su_2$ WZW model at level $1$ because the associated (anti-)chiral $\su_2$-current algebra admits a simple realisation in terms of a single free (anti-)chiral boson. However, the method is expected to be applicable much more generally to any quantum field theory whose conformal field theory in the UV can be realised in terms of arbitrarily many free fields such as (anti-)chiral bosons, fermions, $\beta\gamma$-systems or $bc$-systems.

An obvious first step would be to explore the Wilsonian renormalisation flow of the massive Thirring model, recently derived using functional renormalisation group methods in \cite{RG-Thirring}, by realising the (anti-)chiral $\su_2$-current algebra at level $1$ in terms of free (anti-)chiral fermions instead of bosons, see \cite{IFrenkel}. Another natural next step is to apply our approach to affine Toda field theories associated with higher rank Lie algebras $\mathfrak{g}$, for which the renormalisation group equations have been derived using standard perturbative methods and which are known to exhibit generalised BKT transitions \cite{Grisaru:1990gf, Grisaru:1990kh, Bernard:1990ys}. Notably, the construction presented here for $\su_2$ should admit a direct generalisation to higher rank in the case of simply-laced $\g$ where the associated untwisted affine Kac--Moody algebra also has a basic representation in terms of $\text{rk}\, \g$ bosons via the Frenkel--Kac construction; see \cite{FrenkelKac} and \cite[\S5.6]{KacVA}. A particularly interesting direction for future research would be the application of these methods to integrable $\sigma$-models such as the Klim\v{c}\'{\i}k model \cite{Klimcik:2008eq}. Indeed, the study of its UV limit in \cite{Kotousov:2022azm} served as one of the primary motivations for the present work, which seeks to provide a systematic framework for constructing $2$-dimensional quantum field theories that are described in the UV as perturbations of conformal field theories with free field realisations.

\subsection*{Acknowledgements}

We would like to thank Edoardo D'Angelo, Ben Hoare, Sylvain Lacroix, Nat Levine, Enrique Moreno, Sandor Nagy, Istvan Nandori, Stefano Negro, Antonio Padilla, Robert Smith and Alessandro Torrielli for useful discussions and correspondences in relation to various aspects of this work. The authors gratefully acknowledge the support of the Leverhulme Trust through a Leverhulme Research Project Grant (RPG-2021-154). B.V.\ also gratefully acknowledges the support of the Engineering and Physical Sciences Research Council (UKRI1723).

\section{Background and motivation} \label{sec: background}

We are interested in studying the renormalisation group flows of $2$-dimensional quantum field theories whose behaviour in the ultraviolet is described by marginally relevant perturbations of $2$-dimensional conformal field theories. For ease of presentation, to motivate our approach we focus on a particular class of $2$-dimensional conformal field theories, the Wess--Zumino--Witten (WZW) models, which exhibit left and right affine Kac--Moody algebra symmetries generated by chiral and anti-chiral currents. In this setting, the marginally relevant operator inducing the renormalisation group flow is a bilinear combination of these chiral and anti-chiral currents.

Throughout the rest of the paper we shall, in fact, further specialise to a particular instance of this general setting which corresponds to the sine-Gordon model, to be discussed in \S\ref{sec: sine-Gordon}. In this section we will therefore present the background relevant for this particular case, but we expect the techniques outlined in the paper to be applicable much more generally.

\subsection{Current algebras} \label{sec: current_algebras}

We recall the definition of the untwisted affine Kac--Moody algebra in terms of generators and relations in \S\ref{sec: currents}. Two commuting copies of this algebra, encoded in the chiral and anti-chiral currents, generate the full affine vertex algebra which underpins the conformal structure of the Wess--Zumino--Witten model \cite[\S 15]{CFTbook}. Since we will be primarily interested in the example of the Lie algebra $\sl_2$, we discuss the specifics of this case alongside the general story.
In \S\ref{sec: Fourier decomp} we describe the usual change of variable from the plane to the cylinder \cite{CFTbook}, giving rise to Fourier mode decompositions of the chiral and anti-chiral currents on $S^1$. Importantly, there is a subtle difference in sign between the currents which leads to a whole class of local operators, coupling the two chiralities together, whose careful treatment will require renormalisation. In \S\ref{sec: EM tensor} we recall the definition of the chiral and anti-chiral Virasoro generators in terms of the currents and express the Hamiltonian of the WZW model in terms of the zero-modes of the stress-energy tensor.

\subsubsection{Chiral and anti-chiral currents} \label{sec: currents}

Affine Kac--Moody algebras can be obtained as central extensions of loop algebras associated with any finite-dimensional semisimple complex Lie algebra $\g$.

Given an arbitrary basis $J^a$ for $a = 1, \ldots, \dim \g$ of the Lie algebra $\g$, we let $f^{ab}{}_c$ denote the associated structure constants, so that $[J^a, J^b] = \sum_c f^{ab}{}_c J^c$. When the reality conditions imposed are such that $(J^a)^\dag = J^a$, it is customary to include an extra factor of $i$ on the right hand side so that $f^{ab}{}_c$ is real; see e.g. \cite[(13.2)]{CFTbook}. We will be interested in different reality conditions so we do not follow this convention. Let $\kappa : \g \times \g \to \CC$ be the normalised Killing form on $\g$ defined for any $\mathsf x, \mathsf y \in \g$ by $\kappa(\mathsf x, \mathsf y) \coloneqq \frac{1}{2 h^\vee} \text{tr}\big( \text{ad}(\mathsf x) \text{ad}(\mathsf y) \big)$, where $h^\vee$ is the dual Coxeter number of $\g$, and denote its components by $\kappa^{ab} = \kappa(J^a, J^b)$. The components define an invertible matrix, since $\kappa$ is non-degenerate, and we denote the components of its inverse by $\kappa_{ab}$. In the case $\g = \sl_2$ of interest in this paper, we fix a standard basis of $J^3, J^+$ and $J^-$ with relations
\begin{equation} \label{sl2 algebra}
[J^+, J^-] = J^3, \qquad [J^3, J^\pm] = \pm 2 J^\pm.
\end{equation}
The dual Coxeter number of $\sl_2$ is $h^\vee = 2$ so that $\kappa(J^3, J^3) = 2$ and $\kappa(J^+, J^-) = 1$.

The untwisted affine Kac--Moody algebra associated with $\g$ is an infinite dimensional Lie algebra, with generators $J^a_n$ for $a=1,\ldots, \dim \g$ and $n \in \ZZ$, and Lie bracket given by
\begin{subequations} \label{Current mode commutation relations}
\begin{equation} \label{Current mode commutation relations 1}
    [J^a_n, J^b_m] = \sum_c f^{ab}{}_c J^{c}_{n + m} + k \kappa^{ab} \, n \delta_{n+m, 0},
\end{equation}
where $k \in \CC$ is called the level. In the chosen normalisation of the Killing form, the so-called \emph{critical} level is $- h^\vee$ and we shall henceforth assume that $k \neq -h^\vee$. Strictly speaking, the level $k$ in the above algebra \eqref{Current mode commutation relations 1} should be replaced by a central element ${\bf 1}$. However, when one is only interested in considering representations of the untwisted affine Kac--Moody algebra $\widehat{\g}$ on which this central element ${\bf 1}$ acts as multiplication by the number $k$, it is standard to work directly with the Lie algebra relations \eqref{Current mode commutation relations 1} and refer to this as the untwisted affine Kac-Moody algebra $\widehat{\g}$ at level $k$, often denoted in the literature simply by $\widehat{\g}_k$.

We introduce a second copy of the untwisted affine Kac--Moody algebra whose generators we denote by $\bar J^a_n$ for $a=1,\ldots, \dim \g$ and $n \in \ZZ$, satisfying the same Lie algebra relations as in \eqref{Current mode commutation relations 1}, namely
\begin{equation} \label{Current mode commutation relations 2}
    [\bar J^a_n, \bar J^b_m] = \sum_c f^{ab}{}_c \bar J^{c}_{n + m} + k \kappa^{ab} \, n \delta_{n+m, 0}.
\end{equation}
\end{subequations}
In principle the level $k$ here could be different from the one appearing in \eqref{Current mode commutation relations 1}, but in the example we shall consider both levels coincide so we shall only consider this case.
To distinguish this second copy of the untwisted affine Kac--Moody algebra from $\widehat{\g}_k$ defined above, we shall denote it as $\widehat{\bar \g}_k$. We further take these two affine Kac--Moody algebras to commute, i.e. we require $[J^a_n, \bar J^b_m] = 0$, which amounts to working with the direct sum Lie algebra $\widehat{\g}_k \oplus \widehat{\bar \g}_k$.

The infinitely many generators of these two affine Kac--Moody algebras can be conveniently organised into the chiral and anti-chiral currents as follows. To each basis element $J^a$ (or more generally to any element of $\mathfrak{g}$) we can associate a pair of holomorphic and anti-holomorphic currents in a complex variable $z$ defined as
\begin{equation} \label{Mode expansion}
J^a[z] \coloneqq \sum_{n \in \mathbb{Z}} J^a_n z^{-n-1}, \qquad
\bar J^a[\bar z] \coloneqq \sum_{n \in \mathbb{Z}} \bar J^a_n \bar z^{-n-1}.
\end{equation}
We use square bracket notation for the arguments $z$ and $\bar z$ since we reserve the more standard bracket notation for the Fourier mode decomposition of the same currents, see \S\ref{sec: Fourier decomp} below.

It is important to stress here that while $\bar z$ denotes the complex conjugate of $z$, the modes $\bar J^a_n$ and $J^a_n$ are independent, and in particular they are \emph{not} related by hermitian conjugation.
Instead, hermitian conjugation is defined using a choice of anti-linear involution $\tau : \g \to \g$ on the Lie algebra $\g$, by letting $\X_n^\dag \coloneqq - (\tau \X)_{-n}$ and
$\bar \X_n^\dag \coloneqq - (\overline{\tau \X})_{-n}$ for every $\X \in \g$ and $n \in \ZZ$. Note that the overall minus sign in these definitions is to ensure that $(\cdot)^\dag$ defines an anti-linear \emph{anti}-involution on the Fourier modes.
When $\g = \sl_2$ we take $\tau : \sl_2 \to \sl_2$ to be the anti-linear involution defining the compact real form $\su_2$ of $\sl_2$, given by $\tau J^3 = - J^3$ and $\tau J^\pm = - J^\mp$. We then have
\begin{equation} \label{reality modes sl2}
(J^3_n)^\dag = J^3_{-n}, \qquad
(J^\pm_n)^\dag = J^\mp_{-n}, \qquad
(\bar J^3_n)^\dag = \bar J^3_{-n}, \qquad
(\bar J^\pm_n)^\dag = \bar J^\mp_{-n}.
\end{equation}

In terms of the currents \eqref{Mode expansion}, the pair of untwisted affine Kac--Moody algebra relations \eqref{Current mode commutation relations} are then equivalently encoded in the singular part of the operator product expansions
\begin{subequations} \label{Current OPE}
\begin{align}
J^a[z]J^b[w] &\sim \frac{k\kappa^{ab}}{(z - w)^2} + \sum_c f^{ab}{}_c \frac{J^c[w]}{z - w}, \\
\bar J^a[\bar z] \bar J^b[\bar w] &\sim \frac{k\kappa^{ab}}{(\bar z - \bar w)^2} + \sum_c f^{ab}{}_c \frac{\bar J^c[\bar w]}{\bar z - \bar w}.
\end{align}
\end{subequations}
We also have the regular operator product expansion $J^a[z] \bar J^b[\bar w] \sim 0$ which encodes the fact that the two copies of the affine Kac--Moody algebra mutually commute.

More general composite operators $\mathcal O[z, \bar z]$, which may depend on both $z$ and $\bar z$, can then be constructed as finite linear combinations of normal ordered products of derivatives of the chiral and anti-chiral currents \eqref{Mode expansion}, namely
\begin{equation} \label{composite operator}
\nord{\frac{1}{m_1!} \partial_z^{m_1} J^{a_1}[z] \ldots \frac{1}{m_r!} \partial_z^{m_r} J^{a_r}[z]} \, \nord{\frac{1}{n_1!} \partial_{\bar z}^{n_1} \bar J^{b_1}[\bar z] \ldots \frac{1}{n_s!} \partial_{\bar z}^{n_s} \bar J^{b_s}[\bar z]}
\end{equation}
with $m_1, \ldots, m_r, n_1, \ldots, n_s \in \ZZ_{\geq 0}$ and $a_1, \ldots, a_r, b_1, \ldots, b_s \in \{ 1, \ldots, \dim \g \}$ for any $r, s \in \ZZ_{\geq 0}$.
Under the state-field correspondence, such composite operators in the $2$-dimensional conformal field theory correspond to particular states in the full affine vertex algebra $\mathbb{F}^k(\g)$. The latter is defined as the module over $\widehat{\g}_k \oplus \widehat{\bar \g}_k$ induced from the trivial representation of the positive part of the Kac--Moody algebra, namely the $1$-dimensional vector space $\CC \ket{0}$ spanned by the vacuum state $\ket{0}$ with defining properties $J^a_n \ket{0} = 0$ and $\bar J^a_n \ket{0} = 0$ for all $n \geq 0$ and $a = 1, \ldots, \dim \g$. Specifically, the state-field correspondence associates the operator \eqref{composite operator} to the state
\begin{equation} \label{composite operator state}
J^{a_1}_{-m_1-1} \ldots J^{a_r}_{-m_r-1} \bar J^{b_1}_{-n_1-1} \ldots \bar J^{b_s}_{-n_s-1} \ket{0} \in \mathbb{F}^k(\g).
\end{equation}
We refer for instance to \cite{Vicedo:2025vql} for a detailed description of the full affine vertex algebra $\mathbb{F}^k(\g)$.

\subsubsection{Fourier mode decompositions} \label{sec: Fourier decomp}

We are interested in $2$-dimensional field theories on the cylinder $S^1 \times \RR$, where $S^1$ represents the compact spatial direction and $\RR$ is the time direction. In the Hamiltonian formalism, such a theory is described by a collection of fields on a constant-time Cauchy surface, i.e. a copy of $S^1$, which therefore admit Fourier mode decompositions. In the context of a $2$-dimensional conformal field theory, this set-up can be obtained by means of the so-called radial quantisation using a coordinate transformation from the plane to the cylinder \cite[\S6.1]{CFTbook}.

\paragraph{Currents on the cylinder.}

Explicitly, let $x \in [0, L]$ be a coordinate along the circle $S^1$ where the length scale $L \in \RR_{>0}$ represents the circumference of the cylinder, and let $t \in \RR$ be the time coordinate along the vertical direction on the cylinder. By introducing the complex coordinate $u = x + i t$ on the cylinder, with the periodic identification $u \sim u + L$, we can map the cylinder to the plane using the conformal transformation $u \mapsto z = e^{2 \pi i u / L}$. Under the inverse transformation, the chiral and anti-chiral currents \eqref{Mode expansion} then get mapped to
\begin{subequations} \label{Fourier mode expansion}
\begin{align}
J^a(u) &= \frac{2 \pi}{L} z J^a[z] = \frac{2 \pi}{L} \sum_{n \in \mathbb{Z}} J^a_n e^{-2 \pi i n u / L}, \\
\bar J^a(\bar u) &= \frac{2 \pi}{L}  \bar z \bar J^a[\bar z] = \frac{2 \pi}{L} \sum_{n \in \mathbb{Z}} \bar J^a_n e^{2 \pi i n \bar u / L},
\end{align}
\end{subequations}
see, e.g., \cite[\S 4.1]{Vicedo:2025vql}. In the latter, the coordinate transformation $z = e^{iu} \mapsto iu$ is considered but using instead the present coordinate transformation $z = e^{2 \pi i u / L} \mapsto iu$ leads to the additional factors of $\frac{2 \pi}{L}$. Note, in particular, that \eqref{Fourier mode expansion} are really the expressions for the $\sl_2$-currents in the coordinate $iu$, rather than $u$. On the left hand sides of \eqref{Fourier mode expansion} we introduced the bracket notation for the arguments $u$ and $\bar u$ of the currents on the cylinder, by contrast with the square bracket notation used in \eqref{Mode expansion} for the currents on the $z$-plane. These are sometimes denoted instead as $J^a_{\text{cyl.}}(u)$ and $J^a_{\text{pl.}}(z)$, respectively, and similarly for the anti-chiral currents, see e.g. \cite{CFTbook}, but since we will only be interested in the cylinder we omit the subscript `cyl.' and instead reserve the bracket notation for the arguments of a current on the cylinder.

Restricting \eqref{Fourier mode expansion} to the constant time slice $u = x \in [0, L]$ leads to the desired Fourier mode decompositions
\begin{equation} \label{Fourier mode expansion real}
J^a(x) = \frac{2 \pi}{L} \sum_{n \in \mathbb{Z}} J^a_n e^{-2 \pi i n x / L}, \qquad
\bar J^a(x) = \frac{2 \pi}{L} \sum_{n \in \mathbb{Z}} \bar J^a_n e^{2 \pi i n x / L}.
\end{equation}
It follows using \eqref{reality modes sl2} that the currents \eqref{Fourier mode expansion} in the $u$ coordinate on the cylinder satisfy the simple reality conditions 
\begin{equation} \label{reality currents sl2}
J^3(u)^\dag = J^3(\bar u), \qquad
J^\pm(u)^\dag = J^\mp(\bar u), \qquad
\bar J^3(u)^\dag = \bar J^3(\bar u), \qquad
\bar J^\pm(u)^\dag = \bar J^\mp(\bar u).
\end{equation}

\paragraph{Local operators.}

We pause here to make a trivial but crucial observation about the Fourier mode expansions \eqref{Fourier mode expansion real}. Notice that the chiral modes $J^a_n$ in the expansion of the current $J^a(x)$ come multiplied by the exponential $e^{-2 \pi i n x / L}$ while the anti-chiral modes $\bar J^a_n$ in the expansion of the current $\bar J^a(x)$ come instead multiplied by the inverse exponential $e^{2 \pi i n x / L}$. This property stems from the fact that the chiral and anti-chiral currents defined in \eqref{Mode expansion} were respectively holomorphic and anti-holomorphic in the variable $z = e^{2 \pi i u /L}$.

And although trivial, this observation has important implications on which kind of local operators, i.e. integrals over $x \in S^1$ of composite operators of the form \eqref{composite operator}, are well defined.
Indeed, integrals of purely chiral (or anti-chiral) composite operators lead to infinite sums of normal ordered monomials of the form $\nord{J^{a_1}_{n_1} \ldots J^{a_r}_{n_r}}$ with $n_1 + \ldots + n_r = 0$. For instance, when $\g = \sl_2$ we may consider the local operator
\begin{equation} \label{chiral local op example}
\frac{1}{2 \pi} \int_0^L \dd x \, \nord{J^3(x) J^3(x)} = \frac{2 \pi}{L} \sum_{n > 0} J^3_{-n} J^3_n.
\end{equation}
From such a normal ordered infinite sum, only finitely many terms can act non-trivially on any given state in a smooth representation of the chiral affine Kac--Moody algebra \eqref{Current mode commutation relations 1}.

In stark contrast, the integrals of composite operators \eqref{composite operator} comprising both chiral and anti-chiral pieces lead to infinite sums of products of chiral creation operators ($J^a_n$ with $n<0$) with anti-chiral creation operators ($\bar J^a_n$ with $n<0$). For instance, when $\g = \sl_2$ an example of a local operator we will be interested in is
\begin{equation} \label{non-chiral local op example}
\frac{1}{2 \pi} \int_0^L \dd x \, J^3(x) \bar J^3(x) = \frac{2 \pi}{L} \sum_{n \in \ZZ} J^3_n \bar J^3_n.
\end{equation}
The construction of $\widehat{\g}_k \oplus \widehat{\bar \g}_k$-representations on which such local operators have a well-defined action requires some care. We will come back to this issue in \S\ref{sec: Renormalisation} below after describing the kind of representation of interest for the sine-Gordon model in \S\ref{sec: Free field realisations}.

\subsubsection{Energy-momentum tensor} \label{sec: EM tensor}

Two important composite operators are the holomorphic and anti-holomorphic components of the energy-momentum tensor, which generate commuting chiral and anti-chiral copies of the Virasoro algebra, given by the Sugawara construction
\begin{subequations} \label{T on plane}
\begin{align}
T[z] &\coloneqq \frac{1}{2 (k+h^\vee)} \sum_{a,b} \kappa_{ab} \, \nord{J^a[z] J^b[z]},\\
\bar T[\bar z] &\coloneqq \frac{1}{2 (k+h^\vee)} \sum_{a,b} \kappa_{ab} \, \nord{\bar J^a[\bar z] \bar J^b[\bar z]}.
\end{align}
\end{subequations}
These operators correspond, under the state-field correspondence of $\mathbb{F}^k(\g)$, to the chiral state $\Omega \coloneqq \frac{1}{2} (k+h^\vee)^{-1} \sum_{a,b} \kappa_{ab} J^a_{-1} J^b_{-1} \ket{0}$ and anti-chiral state $\bar\Omega \coloneqq \frac{1}{2} (k+h^\vee)^{-1} \sum_{a,b} \kappa_{ab} \bar J^a_{-1} \bar J^b_{-1} \ket{0}$, respectively.
They expand as $T[z] = \sum_{n \in \ZZ} \L_n z^{-n-2}$ and $\bar T[\bar z] = \sum_{n \in \ZZ} \bar \L_n \bar z^{-n-2}$ where the chiral and anti-chiral Virasoro modes $\L_n$ and $\bar \L_n$ are given explicitly by
\begin{subequations}
\begin{align}
\L_n &\coloneqq \frac{1}{2 (k+h^\vee)} \sum_{m \in \ZZ} \sum_{a,b} \kappa_{ab} \, \nord{J^a_m J^b_{n-m}},\\
\bar \L_n &\coloneqq \frac{1}{2 (k+h^\vee)} \sum_{m \in \ZZ} \sum_{a,b} \kappa_{ab} \, \nord{\bar J^a_m \bar J^b_{n-m}}.
\end{align}
\end{subequations}
These generate two commuting copies of the Virasoro algebra with central charge $c = \frac{k \dim \g}{k+h^\vee}$.
It is well known, see for instance \cite[(5.138)]{CFTbook} or also \cite[\S 4.1.2]{Vicedo:2025vql}, that under the inverse of the conformal transformation $u \mapsto z = e^{2 \pi i u / L}$, the holomorphic and anti-holomorphic components of the energy-momentum tensor \eqref{T on plane} get mapped to
\begin{equation}\label{T in circular coordinates}
T(u) = \bigg( \frac{2\pi}{L}\bigg)^2 \bigg( z^2 T[z] - \frac{c}{24} \bigg), \qquad
\bar T(\bar u) = \bigg( \frac{2\pi}{L}\bigg)^2 \bigg( \bar z^2 \bar T[\bar z] - \frac{c}{24} \bigg).
\end{equation}
We note here the usual shifts by the constant $- \frac{c}{24}$ which originate from the Schwarzian derivative terms in the transformation properties of the (anti-)holomorphic components of the energy momentum tensor under conformal transformations \cite[\S5.4.1]{CFTbook}.

The Hamiltonian of the WZW model on the circle is given in terms of \eqref{T in circular coordinates} as
\begin{equation} \label{WZW Hamiltonian}
\H_0 \coloneqq \frac{1}{2 \pi} \int_0^L \dd x \, T(x) + \frac{1}{2 \pi} \int_0^L \dd x \, \bar T(x) = \frac{2 \pi}{L} \bigg( \L_0 + \bar \L_0 - \frac{c}{12} \bigg).
\end{equation}

\subsection{Free field realisations}\label{sec: Free field realisations}

In view of studying the renormalisation group flow of the WZW model perturbed by marginally relevant operators using a Wilsonian-type approach, we first need to introduce a Wilsonian cutoff in the WZW model. Naturally, it is tempting to regularise the theory by truncating the (anti-)chiral Kac--Moody currents at some energy cutoff $\Lambda \in \ZZ_{>0}$, keeping only those Fourier modes $J^a_n$ and $\bar J^a_n$ whose mode number $n$ satisfy the bound $|n| \leq \Lambda$. Specifically, this truncation is obtained by quotienting the direct sum of chiral and anti-chiral Kac--Moody algebras with defining relations \eqref{Current mode commutation relations} by the ideal $\langle J^a_n, \bar J^a_n \rangle_{|n| > \Lambda}$, generated by the basis elements we want to discard. However, we see from the first term on the right hand side of \eqref{Current mode commutation relations} that this ideal is in fact equal to the whole algebra. We will later introduce smooth regularisation in \S\ref{sec: smooth regularisations} for which the same problem persists.

If the underlying Lie algebra $\mathfrak{g}$ were abelian, however, then only the second term on the right hand side of the relation \eqref{Current mode commutation relations} would be present, i.e. the central extension term, and there would be no such problem. In this abelian setting, the untwisted affine Kac--Moody algebras \eqref{Current mode commutation relations 1} and \eqref{Current mode commutation relations 2} correspond to chiral and anti-chiral Heisenberg Lie algebras, respectively, and describe the algebras of Fourier modes of derivatives of chiral and anti-chiral \emph{free} bosons; see \S\ref{sec: chiral free boson} below. The adjective `free' here refers to the fact that the Lie bracket of any two modes is central (we have not yet specified any Hamiltonian). Another important example of free bosonic fields is given by the $\beta\gamma$-system, i.e. an infinite-dimensional Weyl algebra.

As we will explain shortly in \S\ref{sec: regularisation}, the algebras of Fourier modes of \emph{free} bosonic fields are easily regularised. In order to regularise the pair of affine Kac--Moody algebras \eqref{Current mode commutation relations} we shall exploit the fact that these always admit free field realisations, such as free fermion representations or the Wakimoto realisation which utilises a collection of both free chiral bosons and $\beta\gamma$-systems. For simplicity, in this paper we shall focus on the simplest type of free field realisation of the untwisted affine Kac--Moody algebra. Namely, when $\g$ is simply laced and the level is $k=1$ we have access to the basic, or vertex, representation of \eqref{Current mode commutation relations 1} in terms of $\rk \g$ chiral free bosons, see e.g. \cite[Theorem 14.8]{KacBook} or \cite[\S15.6.3]{CFTbook}, and similarly for \eqref{Current mode commutation relations 2} in terms of $\rk \g$ anti-chiral free bosons. In fact, for the purpose of describing the sine-Gordon model in \S\ref{sec: sine-Gordon} later, we shall be interested in the simplest such free field realisation in the case when $\g = \sl_2$, for which only a single (anti-)chiral free boson is needed \cite[\S15.6.1]{CFTbook}.

Since it will play a crucial role in our description of the sine-Gordon model, and to set out our conventions, we begin in \S\ref{sec: chiral free boson} by recalling the details of the basic representations of the chiral and anti-chiral affine Kac--Moody algebras for $\sl_2$ at level $1$. We then relate these to the compactified free boson in \S\ref{sec: compactified boson}.

\subsubsection{Chiral and anti-chiral free bosons} \label{sec: chiral free boson}

We begin by introducing the infinite-dimensional Lie algebra $\Heis_{\rm log}$ of the modes $\x_0, \bar \x_0, \b_n$ and $\bar \b_n$ for $n \in \mathbb{Z}$ of the chiral and anti-chiral free bosons subject to the relations
\begin{subequations} \label{Compactified boson commutation relations}
\begin{alignat}{2}
\label{Compactified boson commutation relations a} [\b_n, \b_m] &= n\delta_{n+m,0}, &\qquad [\bar\b_n, \bar\b_m] &= n\delta_{n+m,0}, \\
\label{Compactified boson commutation relations b} [\x_0, \b_n] &= \frac{i}{\sqrt{4 \pi}} \delta_{n,0}, &\qquad [\bar \x_0, \bar \b_n] &= \frac{i}{\sqrt{4 \pi}} \delta_{n,0},
\end{alignat}
\end{subequations}
for all $m,n \in \ZZ$, with all other Lie brackets between generators being zero. We define hermitian conjugation $(\cdot)^\dag : \Heis_{\rm log} \to \Heis_{\rm log}$ on the generators as
\begin{equation} \label{reality boson modes}
\x_0^\dagger = \x_0, \qquad \bar\x_0^\dagger = \bar\x_0, \qquad \b_n^\dagger = \b_{-n}, \qquad \bar \b_n^\dagger = \bar \b_{-n}.    
\end{equation}
We let $\Heis \subset \Heis_{\rm log}$ denote the Lie subalgebra spanned by $\b_n$ and $\bar\b_n$ for all $n \in \ZZ$ subject to the relations \eqref{Compactified boson commutation relations a}. We also let $\Heis_+$, respectively $\Heis_-$, denote the subalgebra spanned by $\b_n$ and $\bar\b_n$ for $n \geq 0$, respectively $n < 0$. Let $\Bos \coloneqq U(\Heis)$ denote the universal enveloping algebra of $\Heis$. The reason for excluding the zero-modes $\x_0$ and $\bar\x_0$ is that these will play a separate role later, entering the definition of intertwining operators between $\Bos$-modules.

\paragraph{Basic representation on the plane.}

The chiral and anti-chiral free bosons are defined by their mode decompositions
\begin{equation} \label{Chiral compact bosons decomposition}
\chi[z] = \chi_+[z] + \chi_-[z], \qquad
\bar \chi[\bar z] = \bar \chi_+[\bar z] + \bar \chi_-[\bar z],
\end{equation}
which we separate into creation and annihilation parts (including, respectively, the zero modes $\x_0$, $\bar \x_0$ and $\b_0$, $\bar\b_0$) as
\begin{subequations} \label{Chiral compact bosons definitions}
\begin{alignat}{3}
\chi_+[z] &\coloneqq \x_0 + \frac{i}{\sqrt{4\pi}} \sum_{n < 0} \b_n \frac{z^{-n}}{n}, &\qquad
\chi_-[z] &\coloneqq - \frac{i}{\sqrt{4\pi}}\b_0\log z + \frac{i}{\sqrt{4\pi}} \sum_{n > 0} \b_n \frac{z^{-n}}{n},\\
\bar \chi_+[\bar z] &\coloneqq \bar\x_0 + \frac{i}{\sqrt{4\pi}} \sum_{n < 0} \bar \b_n \frac{\bar z^{-n}}{n}, &\qquad
\bar \chi_-[\bar z] &\coloneqq - \frac{i}{\sqrt{4\pi}}\bar\b_0\log \bar z + \frac{i}{\sqrt{4\pi}} \sum_{n > 0} \bar \b_n \frac{\bar z^{-n}}{n}.
\end{alignat}
\end{subequations}
Here we implicitly make a choice of branch cut for the logarithm so that the annihilation parts $\chi_-[z]$ and $\bar\chi_-[\bar z]$ of the chiral and anti-chiral free bosons \eqref{Chiral compact bosons decomposition} are only defined on the complex $z$-plane with a cut from the origin to infinity. We will come back to this point shortly.

The commutation relations \eqref{Compactified boson commutation relations} of the modes of the chiral and anti-chiral free bosons \eqref{Chiral compact bosons decomposition} are equivalently encoded in the singular parts of the operator product expansions
\begin{equation} \label{Boson OPE}
\chi[z]\chi[w] \sim - \frac{1}{4 \pi} \log(z - w), \qquad
\bar\chi[\bar z]\bar\chi[\bar w] \sim - \frac{1}{4 \pi} \log(\bar z - \bar w).
\end{equation}
The normalisation factor of $\frac{1}{4\pi}$ appearing in front of $-\log(z-w)$ here is directly related to the factors of $\frac{1}{\sqrt{4\pi}}$ which we chose to include in \eqref{Compactified boson commutation relations} and \eqref{Chiral compact bosons definitions}. Other factors are sometimes used in the literature, a common choice being $1$ as in \cite[15.6.1]{CFTbook}. Our choice of normalisation in \eqref{Boson OPE} will later ensure that the compactified free boson $\X$ and its conjugate momentum $\P$, defined in \S\ref{sec: compactified boson}, satisfy canonical commutation relation with the standard normalisations.

Note that the derivatives of the free bosons \eqref{Chiral compact bosons decomposition} have mode decompositions
\begin{equation} \label{free boson derivatives}
\partial_z \chi[z] = - \frac{i}{\sqrt{4\pi}} \sum_{n \in \ZZ} \b_n z^{-n-1}, \qquad
\partial_{\bar z} \bar \chi[\bar z] = - \frac{i}{\sqrt{4\pi}} \sum_{n \in \ZZ} \bar\b_n \bar z^{-n-1}.
\end{equation}
Up to overall factors, these are of the same form as \eqref{Mode expansion}. Moreover, the relations in \eqref{Compactified boson commutation relations a} of the Lie algebra $\Heis$, take the same form as \eqref{Current mode commutation relations} for $\hg \oplus \hbg$ with $\g = \langle \b \rangle$ a $1$-dimensional abelian Lie algebra, which can be identified with a Cartan subalgebra $\h = \langle J^3 \rangle$ of $\sl_2$. In this sense, the derivatives \eqref{free boson derivatives} of the chiral and anti-chiral free bosons correspond to untwisted affine Kac--Moody algebras associated with the abelian Lie algebra $\h$. The basic representation of $\widehat{\sl}_2$ at level $1$ extends this to a realisation of the untwisted affine Kac--Moody algebra for the whole of $\sl_2$ by realising the Kac--Moody currents $J^\pm[z]$ in terms of normal ordered exponentials of the same chiral free boson $\chi[z]$, and similarly for the anti-chiral currents $\bar J^\pm[\bar z]$. Explicitly, for any $\alpha \in \RR$ we define the vertex operators
\begin{equation} \label{normal ordered exponential}
\nord{e^{i \alpha \, \chi[z]}} = e^{i \alpha \, \chi_+[z]} e^{i \alpha \, \chi_-[z]}, \qquad
\nord{e^{i \alpha \bar \chi[\bar z]}} = e^{i \alpha \, \bar \chi_+[\bar z]} e^{i \alpha \, \bar \chi_-[\bar z]}.
\end{equation}
The basic representation of the chiral and anti-chiral $\sl_2$-currents at level $1$ is then given by
\begin{subequations} \label{su2 1 currents}
\begin{alignat}{2}
J^3[z] &\coloneqq i \sqrt{8 \pi} \partial_z \chi[z], &\qquad J^{\pm}[z] &\coloneqq \nord{e^{\pm i \sqrt{8 \pi} \chi[z]}},\\
\bar J^3[\bar z] &\coloneqq - i \sqrt{8 \pi} \partial_{\bar z} \bar \chi[\bar z], &\qquad \bar J^{\pm}[\bar z] &\coloneqq \nord{e^{\mp i \sqrt{8 \pi} \bar \chi[\bar z]}}.
\end{alignat}
\end{subequations}
The factors of $\sqrt{8 \pi}$ appearing here relate back to the normalisation by $\frac{1}{4 \pi}$ in the operator product expansions \eqref{Boson OPE}, cf. \cite[15.6.1]{CFTbook}. It is important to note here that the signs in the exponents for chiral currents $J^{\pm}[z]$ and their anti-chiral counterparts $\bar J^{\pm}[\bar z]$ are opposite, as is the overall sign in the expression for $J^3[z]$ compared to $\bar J^3[\bar z]$.

\paragraph{Basic representation on the cylinder.}

Under the inverse of the conformal transformation $u \mapsto z = e^{2 \pi i u / L}$, the chiral and anti-chiral free bosons \eqref{Chiral compact bosons decomposition} transform simply as scalars, i.e.
\begin{equation} \label{chi change of coord}
\chi(u) = \chi[z], \qquad \bar\chi(\bar u) = \bar\chi[\bar z].
\end{equation}
However, for reasons that will be explained shortly, we introduce a finer decomposition of these fields by separating out the zero-mode part and defining
\begin{equation} \label{chi finer decomposition}
\chi(u) = \chi_+(u) + \chi_0(u) + \chi_-(u), \qquad \bar\chi(\bar u) = \bar\chi_+(\bar u) + \bar\chi_0(\bar u) + \bar\chi_-(\bar u)    
\end{equation}
where each summand is given explicitly in the $u$ coordinate by
\begin{subequations} \label{chi u decomp pm 0}
\begin{alignat}{2}
\chi_\pm(u) &\coloneqq \frac{i}{\sqrt{4\pi}} \sum_{\mp n > 0} \frac{1}{n} \b_n e^{-2 \pi i n u/L}, &\qquad
\chi_0(u) &\coloneqq \x_0 + \sqrt{\pi} \b_0 \frac{u}{L},\\
\bar \chi_\pm(\bar u) &\coloneqq \frac{i}{\sqrt{4\pi}} \sum_{\mp n > 0} \frac{1}{n} \bar \b_n e^{2 \pi i n \bar u/L}, &\qquad
\bar \chi_0(\bar u) &\coloneqq \bar\x_0 - \sqrt{\pi} \bar \b_0 \frac{\bar u}{L}.
\end{alignat}
\end{subequations}
Recall that $\chi_-[z]$ and $\bar\chi_-[\bar z]$ in \eqref{Chiral compact bosons definitions} were only defined on the cut $z$-plane. Correspondingly, the fields $\chi(u)$ and $\bar \chi(\bar u)$ are a priori only defined on a vertical strip $\{ u = x + i t \,|\, x \in [0, L) \}$ in the complex $u$-plane. However, it is clear that the definitions \eqref{chi u decomp pm 0} of these fields extend naturally to the whole complex $u$-plane. In particular, $\chi_\pm(u)$ and $\bar\chi_\pm(\bar u)$ both extend periodically with period $L$ while, due to the zero-mode parts, we have the periodicity properties
\begin{equation} \label{chi bchi periodicity}
\chi(u+L) = \chi(u) + \sqrt{\pi} \b_0, \qquad
\bar\chi(\bar u+L) = \bar \chi(\bar u) - \sqrt{\pi} \bar \b_0
\end{equation}
for the chiral and anti-chiral free boson.

In terms of the $u$ coordinate, the reality conditions \eqref{reality boson modes} on the modes of the chiral and anti-chiral free bosons translate to the simple reality condition
\begin{equation} \label{chi reality conditions}
\chi(u)^\dagger = \chi(\bar u), \qquad \bar \chi(u)^\dagger = \bar \chi(\bar u)
\end{equation}
on the fields themselves.

Using the finer decomposition \eqref{chi finer decomposition} of the chiral and anti-chiral free bosons $\chi(u) = \chi[z]$ and $\bar\chi(\bar u) = \bar\chi[\bar z]$ which separates out the zero-modes, we use a slightly different notion of normal ordered exponential to define \textbf{chiral and anti-chiral vertex operators}
\begin{equation} \label{normal ordered exponential 0mode}
\nord{e^{i \alpha \, \chi(u)}} = e^{i \alpha \, \chi_+(u)} e^{i \alpha \, \chi_0(u)} e^{i \alpha \, \chi_-(u)}, \qquad
\nord{e^{i \alpha \, \bar \chi(\bar u)}} = e^{i \alpha \, \bar \chi_+(\bar u)} e^{i \alpha \, \bar \chi_0(\bar u)} e^{i \alpha \, \bar \chi_-(\bar u)},
\end{equation}
which is to be compared with \eqref{normal ordered exponential}. It is important to note, in particular, that the normal ordered exponentials \eqref{normal ordered exponential 0mode} are still defined relative to the coordinate $z$ since we are still using the notion of creation and annihilation operators relative to this coordinate in \eqref{chi u decomp pm 0}.
We can then rewrite the basic representation \eqref{su2 1 currents} in the $u$ coordinate as
\begin{subequations} \label{su2 1 currents cyl}
\begin{alignat}{2}
J^3(u) &= \sqrt{8 \pi} \partial_u \chi(u), &\qquad
J^{\pm}(u) &= \frac{2 \pi}{L} \nord{e^{\pm i \sqrt{8 \pi} \chi(u)}},\\
\bar J^3(\bar u) &= \sqrt{8 \pi} \partial_{\bar u} \bar \chi(\bar u), &\qquad
\bar J^{\pm}(\bar u) &= \frac{2 \pi}{L} \nord{e^{\mp i \sqrt{8 \pi} \bar \chi(\bar u)}}.
\end{alignat}
\end{subequations}
These relations can be obtained by combining the change of coordinate formulae in \eqref{Fourier mode expansion} and \eqref{chi change of coord} with the free field realisations \eqref{su2 1 currents} in the $z$-coordinate. In particular, when deriving the expression for the currents $J^3(u)$ and $\bar J^3(\bar u)$ we note that under the change of coordinate $z = e^{2 \pi i u / L}$ we have $\partial_u = \frac{2 \pi i}{L} z \partial_z$ and $\partial_{\bar u} = - \frac{2 \pi i}{L} \bar z \partial_{\bar z}$. To see the expressions of the currents $J^\pm(u)$ and $\bar J^\pm(\bar u)$, we note that the two notions of normal ordered exponentials in \eqref{normal ordered exponential} and \eqref{normal ordered exponential 0mode} are just related by a factor of $z$ since by the Baker-Campbell-Hausdorff formula we have
\begin{equation} \label{absorb factor of z}
z \, e^{\pm i \sqrt{8 \pi} \x_0} e^{\pm i \sqrt{8 \pi} \left( - \frac{i}{\sqrt{4 \pi}}\b_0 \log z \right)} = e^{\pm i \sqrt{8 \pi} \left( \x_0 - \frac{i}{\sqrt{4 \pi}}\b_0 \log z \right)} = e^{\pm i \sqrt{8 \pi} \chi_0(u)}.
\end{equation}

It is now also immediate that the realisations of the $\sl_2$-currents in \eqref{su2 1 currents cyl} satisfy the reality conditions \eqref{reality currents sl2} as a consequence of the reality condition \eqref{chi reality conditions} on the (anti-)chiral free bosons. Specifically, to see that the normal ordered exponentials \eqref{normal ordered exponential 0mode} satisfy the reality conditions
\begin{equation} \label{real exponentials}
\big( \nord{e^{i \alpha \, \chi(u)}} \big)^\dag = \nord{e^{- i \alpha \, \chi(\bar u)}}, \qquad
\big( \nord{e^{i \alpha \, \bar\chi(\bar u)}} \big)^\dag = \nord{e^{- i \alpha \, \bar \chi(u)}}
\end{equation}
one uses the fact that $\chi_0(u)^\dag = \chi_0(\bar u)$, $\bar \chi_0(\bar u)^\dag = \bar \chi_0(u)$ and $\chi_\pm(u)^\dag = \chi_\mp(\bar u)$, $\bar \chi_\pm(\bar u)^\dag = \bar \chi_\mp(u)$. In particular, we observe that the simple reality condition \eqref{real exponentials} is a consequence of the zero-mode pieces having been combined into a single exponential. By contrast, the normal ordered exponentials \eqref{normal ordered exponential} satisfy a slightly more involved reality condition. From now on we shall refer to \eqref{su2 1 currents cyl} as $\su_2$-currents to emphasise that they satisfy the reality conditions \eqref{reality currents sl2}.

\paragraph{Energy-momentum tensor.}

Applying the basic representation \eqref{su2 1 currents} to the holomorphic and anti-holomorphic components of the energy-momentum tensor \eqref{T on plane} on the plane we find, cf. \cite[(15.233)]{CFTbook},
\begin{equation}
T[z] = - 2 \pi \nord{\partial_z \chi[z] \partial_z \chi[z]}, \qquad
\bar T[\bar z] = - 2 \pi \nord{\partial_{\bar z} \bar \chi[\bar z] \partial_{\bar z} \bar \chi[\bar z]}.
\end{equation}
Moving to the cylinder using the transformation property \eqref{T in circular coordinates}, together with the fact that $\chi(u) = \chi[z]$, $\bar\chi(\bar u) = \bar\chi[\bar z]$ and $\partial_u = \frac{2 \pi i}{L} z \partial_z$, $\partial_{\bar u} = - \frac{2 \pi i}{L} \bar z \partial_{\bar z}$ we find
\begin{equation}
T(u) = 2 \pi \nord{\partial_u \chi(u) \partial_u \chi(u)} - \frac{\pi^2}{6 L^2}, \qquad
\bar T(\bar u) = 2 \pi \nord{\partial_{\bar u} \bar \chi(\bar u) \partial_{\bar u} \bar \chi(\bar u)} - \frac{\pi^2}{6 L^2}.
\end{equation}
Subsituting this into the Hamiltonian \eqref{WZW Hamiltonian} of the WZW model on the cylinder we find
\begin{subequations} \label{WZW Hamiltonian free realisation}
\begin{equation} \label{free Ham with Casimir term}
\H_0 = \int_0^L \dd x \big( \nord{\partial_x \chi(x) \partial_x \chi(x)} + \nord{\partial_x \bar\chi(x) \partial_x \bar\chi(x)} \big) - \frac{\pi}{6 L}.
\end{equation}
In the first term we recognise the Hamiltonian of the free boson on the circle of circumference $L$, with our normalisation conventions for the chiral and anti-chiral free bosons $\chi$. Or in other words, in the expression on the right hand side of \eqref{WZW Hamiltonian} the operators $\L_0$ and $\bar\L_0$ are those of the free boson given in terms of the osscilators $\b_n$ and $\bar\b_n$ for $n \in \ZZ$ by
\begin{equation} \label{L0 bL0 free boson}
\L_0 = \frac 12 \b_0^2 + \sum_{n>0} \b_{-n} \b_n, \qquad
\bar\L_0 = \frac 12 \bar\b_0^2 + \sum_{n>0} \bar\b_{-n} \bar\b_n.
\end{equation}
\end{subequations}
The constant term $- \frac{\pi}{6 L}$ in \eqref{WZW Hamiltonian free realisation}, i.e. $- \frac{2\pi}{L} \frac{c}{12}$ with $c=1$ in \eqref{WZW Hamiltonian}, is exactly the Casimir energy, or vacuum energy, of the free boson on the cylinder with periodic boundary conditions, see for instance \cite[(6.88)--(6.89)]{CFTbook}.

\subsubsection{Compactified boson and dual boson} \label{sec: compactified boson}

There is an obvious problem with the realisation \eqref{su2 1 currents cyl} of the $\su_2$-currents $J^3(u)$, $J^\pm(u)$ and $\bar J^3(\bar u)$, $\bar J^\pm(\bar u)$. The currents were defined in \S\ref{sec: current_algebras} by their Fourier mode decompositions \eqref{Fourier mode expansion} and so are manifestly periodic under $u \mapsto u+L$.
By contrast, the definitions of the chiral and anti-chiral free bosons $\chi(u)$ and $\bar\chi(\bar u)$ in \eqref{chi u decomp pm 0} extended to the whole complex $u$-plane with the non-trivial periodicity property \eqref{chi bchi periodicity} under $u \mapsto u + L$. We therefore need to ensure that the expressions on the right hand sides of \eqref{su2 1 currents cyl} are themselves periodic under $u \mapsto u + L$.

The derivatives $\partial_u \chi(u)$ and $\partial_{\bar u} \bar\chi(\bar u)$ are both periodic under $u \mapsto u + L$, as can be seen from differentiating the relations \eqref{chi bchi periodicity}. On the other hand, the periodicity of the vertex operators $\nord{e^{\pm i \sqrt{8 \pi} \chi(u)}}$ and $\nord{e^{\pm i \sqrt{8 \pi} \bar\chi(\bar u)}}$ is not immediate. Since $\chi_\pm(u)$ and $\bar \chi_\pm(\bar u)$ are both periodic, one needs only to ensure that the zero-mode exponentials $e^{\pm i \sqrt{8 \pi} \chi_0(u)}$ and $e^{\pm i \sqrt{8 \pi} \bar \chi_0(\bar u)}$ are periodic under $u \mapsto u + L$.
This, in turn, requires that the operators $\sqrt{2} \b_0$ and $\sqrt{2} \bar\b_0$ take integer values since it will then follow that $\sqrt{8 \pi} \chi_0(u)$ and $\sqrt{8 \pi} \bar \chi_0(\bar u)$ are defined up to integer multiples of $2 \pi$.
In this case, the family of vertex operators \eqref{normal ordered exponential 0mode} will be well defined on the cylinder only for $\alpha \in \sqrt{8 \pi} \ZZ$, i.e. we will have a discrete family of \textbf{full vertex operators}
\begin{equation} \label{vertex operators V rs}
\mathcal V_{r,s}(u, \bar u) \coloneqq \nord{e^{i r \sqrt{8 \pi} \chi(u)}} \nord{e^{i s \sqrt{8 \pi} \bar \chi(\bar u)}}
\end{equation}
labeled by pairs of integers $(r, s) \in \ZZ^2$. By a slight abuse of notation we will just write these full vertex operators with a single argument as $\mathcal V_{r,s}(u)$.
Our immediate goal is thus to introduce a representation $\Fock$ of the chiral and anti-chiral free bosons on which the zero-modes $\sqrt{2} \b_0$ and $\sqrt{2} \bar\b_0$ both act as integers. The expressions of the basic representation \eqref{su2 1 currents cyl} will then produce the desired periodic $\su_2$-currents when acting on $\Fock$.

\paragraph{Fock spaces.}
For any $p, w \in \ZZ$ we let $\Fock_{p,w}$ denote the Fock space over the Lie algebra $\Heis$, with defining relations \eqref{Compactified boson commutation relations a}, whose highest weight state $|p,w\rangle$ is defined by the properties
\begin{equation}\label{Fock m n action}
\b_n | p, w \rangle = \frac{p+w}{\sqrt{2}} |p, w \rangle \delta_{n, 0}, \qquad
\bar \b_n | p, w \rangle = \frac{p-w}{\sqrt{2}} |p, w \rangle \delta_{n,0}
\end{equation}
for all $n \in \ZZ_{\geq 0}$. Specifically, $\Fock_{p,w} \coloneqq \text{Ind}^{\Heis}_{\Heis_+} \CC |p,w\rangle$ is the module over $\Heis$ induced from the trivial $1$-dimensional module $\CC |p,w\rangle$ over the Lie subalgebra $\Heis_+$ defined by the relations \eqref{Fock m n action}. For reasons to be explained shortly, we say that states in the Fock space $\Fock_{p,w}$ carry \textbf{momentum} $p \in \ZZ$ and \textbf{winding number} $w \in \ZZ$.
It follows from the relations \eqref{Compactified boson commutation relations b} that the formal exponential operators of the zero-modes $\x_0$ and $\bar\x_0$ given by
\begin{equation}\label{Intertwining}
\mathsf p \coloneqq e^{i \sqrt{2 \pi} (\x_0 + \bar\x_0)} : \Fock_{p,w} \longrightarrow \Fock_{p + 1,w}, \qquad
\mathsf w \coloneqq e^{i \sqrt{2 \pi} (\x_0 - \bar\x_0)} : \Fock_{p,w} \longrightarrow \Fock_{p,w + 1}
\end{equation}
define intertwining operators between Fock modules $\Fock_{p,w}$ for different values of $(p,w) \in \ZZ^2$.
These operators create one unit of momentum and winding number, respectively, while their inverses destroy the corresponding units. It will be convenient to define the \textbf{shift operators} as the following simpler exponential operators
\begin{subequations}\label{Zero-mode exp}
\begin{align}
\Shi &\coloneqq e^{i \sqrt{8 \pi} \x_0} : \Fock_{p,w} \longrightarrow \Fock_{p + 1, w + 1}, \\
\bar\Shi &\coloneqq e^{i \sqrt{8 \pi} \bar\x_0} : \Fock_{p,w} \longrightarrow \Fock_{p + 1, w - 1},
\end{align}
\end{subequations}
since these operators and their inverses, $\Shi^{\pm 1}$ and $\bar\Shi^{\pm 1}$, appear as a factor in the vertex operators $\nord{e^{\pm i \sqrt{8 \pi} \chi(u)}}$ and $\nord{e^{\pm i \sqrt{8 \pi} \bar \chi(\bar u)}}$, respectively. Their only non-trivial commutation relations with the generators of $\Heis$ are
\begin{equation}
\b_0 \Shi^{\pm 1} = \Shi^{\pm 1} (\b_0 \pm \sqrt{2}), \qquad
\bar\b_0 \bar\Shi^{\pm 1} = \bar\Shi^{\pm 1} (\bar\b_0 \pm \sqrt{2}).
\end{equation}
Note also that $\Shi = \mathsf p \, \mathsf w$ and $\bar\Shi = \mathsf p \, \mathsf w^{-1}$.

We introduce the direct sum of Fock spaces with \emph{even} total momentum and winding number
\begin{equation}\label{Fock}
\Fock \coloneqq \bigoplus_{\substack{p,w \in \ZZ\\ p+w \in 2 \ZZ}} \Fock_{p,w}
\end{equation}
which we refer to as the \textbf{Fock space of the compact boson}. It can be thought of as having a unique vacuum $|0\rangle \coloneqq |0,0\rangle \in \Fock_{0,0}$ since all other highest weight states $|p,w\rangle \in \Fock_{p,w}$ with $p + w\in 2\ZZ$ for the Lie algebra $\Heis$ in \eqref{Compactified boson commutation relations a} can be created by applying a combination of the intertwining operators in \eqref{Zero-mode exp}, which define endomorphisms $\Shi^{\pm 1}, \bar\Shi^{\pm 1} : \Fock \to \Fock$. Specifically, for any $p,w\in \ZZ$ with $p+w\in 2\ZZ$ we have
\begin{equation} \label{state mn}
|p, w \rangle = \mathsf p^p \mathsf w^w |0\rangle = \Shi^{\frac{p+w}2} \bar\Shi^{\frac{p-w}2} |0\rangle
= \mathcal V_{\frac{p+w}2, \frac{p-w}2}(u, \bar u) |0\rangle \Big|_{u \to i \infty}.
\end{equation}
In other words, any state $|p,w\rangle$ with $p + w\in 2\ZZ$ is created from $|0\rangle$ by the vertex operator \eqref{vertex operators V rs} on the cylinder, with labels $(r,s)=(\frac{p+w}2, \frac{p-w}2) \in \ZZ^2$, inserted at infinity.

\paragraph{Completed Fock spaces.}
By construction, the zero-mode exponentials $e^{\pm i \sqrt{8 \pi} \chi_0(u)}$ and $e^{\pm i \sqrt{8 \pi} \bar \chi_0(\bar u)}$ act on the Fock space $\Fock_{p,w}$ as $\Shi^{\pm 1} e^{2 \pi i (p+w) u / L}$ and $\bar\Shi^{\pm 1} e^{2 \pi i (w-p) \bar u / L}$, respectively. These are manifestly periodic under $u \mapsto u + L$ so that, when viewed as endomorphisms of $\Fock$, the operators $e^{\pm i \sqrt{8 \pi} \chi_0(u)}$ and $e^{\pm i \sqrt{8 \pi} \bar \chi_0(\bar u)}$ are well defined on the cylinder, as required.

Strictly speaking, however, the formulae entering the basic representation \eqref{su2 1 currents cyl} still do not have a well defined action on the representation $\Fock$. This point will be crucial in \S\ref{sec: regularisation} so it is useful to already expand on it here.

The exponentials $e^{\pm i \sqrt{8 \pi} \chi_-(u)}$ and $e^{\pm i \sqrt{8 \pi} \bar\chi_-(\bar u)}$, involving the annihilation parts of the chiral and anti-chiral free bosons, act on $\Fock$ as finite sums, as do the expressions $\partial_u \chi_-(u)$ and $\partial_{\bar u} \bar\chi_-(\bar u)$.
By contrast, the exponentials $e^{\pm i \sqrt{8 \pi} \chi_+(u)}$ and $e^{\pm i \sqrt{8 \pi} \bar\chi_+(\bar u)}$ are formal infinite sums of creation operators which do not truncate when acting on $\Fock$. The same goes for the expressions $\partial_u \chi_+(u)$ and $\partial_{\bar u} \bar\chi_+(\bar u)$. This can be remedied by introducing a formal completion $\hFock$ of $\Fock$ as follows.

Each Fock space $\Fock_{p,w}$, for any $p, w \in \ZZ$, has a natural $\ZZ_{\geq 0}$-grading
\begin{equation} \label{Fock grading}
\Fock_{p,w} = \bigoplus_{d \geq 0} \Fock_{p,w}^{(d)}
\end{equation}
defined by letting the highest weight state $|p,w\rangle$ have grade $0$ and by assigning grade $n$ to the modes $\b_{-n}$ and $\bar\b_{-n}$ for any $n \in \ZZ$. A general state in the direct sum \eqref{Fock grading} is a finite sum of states of definite grade. We define the formal completion of \eqref{Fock grading} as the direct \emph{product}
\begin{equation} \label{completed Fock pw}
\hFock_{p,w} = \prod_{d \geq 0} \Fock_{p,w}^{(d)}
\end{equation}
whose elements are formal infinite sums of states of all non-negative grades. One can equally define this as the completion of the vector space $\Fock_{p,w}$ with respect to the linear topology whose neighbourhoods of $0$ are $\bigoplus_{d \geq N} \Fock_{p,w}^{(d)}$ for all $N \geq 0$. By analogy with \eqref{Fock} we then also define the completion of $\Fock$ as the direct sum
\begin{equation}\label{hFock}
\hFock \coloneqq \bigoplus_{\substack{p,w \in \ZZ\\ p+w \in 2 \ZZ}} \hFock_{p,w}.
\end{equation}
The expressions in the basic representation \eqref{su2 1 currents cyl} then give well-defined operators on $\hFock$.

\paragraph{Canonically conjugate fields.}
Having constructed a suitable representation of the algebra $\Bos$ on which the formulae of the basic representation \eqref{su2 1 currents cyl} produce $\su_2$-currents which are periodic under $u \mapsto u + L$, we may now introduce the main field of interest, the \textbf{compact boson} $\X$. We do this by combining the chiral and anti-chiral compact bosons as
\begin{equation} \label{Compact boson}
\X(u, \bar u) \coloneqq \chi(u) + \bar \chi(\bar u).
\end{equation}
By abuse of notation we will often write this field with a single argument as $\X(u)$ even though it depends on both chiralities. In fact, since we are interested in Hamiltonian field theory we will mostly deal with the restriction of $\X$ to the real axis, with explicit mode decomposition
\begin{equation} \label{X decomposition}
\X(x) = \x_0 + \bar\x_0 + \frac{\sqrt{\pi}}{L} (\b_0 - \bar \b_0) x + \frac{i}{\sqrt{4\pi}} \sum_{n \neq 0} \frac{1}{n} ( \b_n - \bar \b_{-n} ) e^{-2 \pi i n x/L}
\end{equation}
for $x \in \RR$, which satisfies the reality condition $\X(x)^\dag = \X(x)$ as a consequence of \eqref{chi reality conditions}. The periodicity properties \eqref{chi bchi periodicity} of the chiral and anti-chiral free bosons imply that
\begin{equation} \label{X periodicity}
\X(u+L) = \X(u) + \sqrt{\pi} (\b_0 - \bar \b_0).
\end{equation}
On the Fock space $\Fock_{p,w}$ this takes the form $\X(u+L) = \X(u) + 2 \pi w R_\circ$ where $R_\circ \coloneqq 1/\sqrt{2\pi}$ is called the self-dual radius; see \S\ref{sec: changing radius}. It is in this sense that the field $\X$ is compact: we should view it as being valued in the circle of radius $R_\circ$ so that $\X \sim \X + 2 \pi R_\circ$ and hence $\X(u)$ is well defined on the cylinder since $\X(u+L) \sim \X(u)$. In the sector $\Fock_{p,w}$ it winds $w$ times around the circle of radius $R_\circ$ as we go once around the cylinder. This is also the reason why the label $w \in \ZZ$ in the Fock space $\Fock_{p,w}$ is referred to as the winding number.

Next, we define the \textbf{dual compact boson} $\tilde\X$ as
\begin{equation} \label{Dual compact boson}
\tilde \X(u, \bar u) \coloneqq \chi(u) - \bar \chi(\bar u).
\end{equation}
We will also often denote this field with a single argument as $\tilde\X(u)$ for simplicity since we will be mostly concerned with its restriction to the real axis, with explicit mode decomposition
\begin{equation} \label{tX decomposition}
\tilde\X(x) = \x_0 - \bar\x_0 + \frac{\sqrt{\pi}}{L} (\b_0 + \bar \b_0) x + \frac{i}{\sqrt{4\pi}} \sum_{n \neq 0} \frac{1}{n} ( \b_n + \bar \b_{-n} ) e^{-2 \pi i n x/L}
\end{equation}
for $x \in \RR$. It too satisfies the reality condition $\tilde \X(x)^\dag = \tilde \X(x)$.
The properties \eqref{chi bchi periodicity} now imply
\begin{equation} \label{tX periodicity}
\tilde\X(u+L) = \tilde\X(u) + \sqrt{\pi} (\b_0 + \bar \b_0),
\end{equation}
which on the Fock space $\Fock_{p,w}$ takes the form $\tilde\X(x+L) = \tilde\X(x) + 2 \pi p R_\circ$. The field $\tilde\X$ is thus also compact: we should view it as taking values in a circle of the same radius $R_\circ$ as $\X$ so that $\tilde\X \sim \tilde\X + 2 \pi R_\circ$, and hence it is well defined on the cylinder since we have $\tilde \X(u + L) \sim \tilde \X(u)$. In the sector $\Fock_{p,w}$, the dual compact boson $\tilde \X$ winds $p$ times around the circle of radius $R_\circ$.

To explain why this `dual winding number' $p$ is referred to as the momentum of the Fock space $\Fock_{p,w}$, we recall how the dual compact boson $\tilde \X$ is related to the conjugate momentum of the compact boson $\X$. Introduce the step-function of width $L$ by
\begin{equation} \label{step-function}
\varepsilon(x-y) \coloneqq \frac{1}{L}(x-y) - \frac{i}{2 \pi} \sum_{n \neq 0} \frac{1}{n} e^{2 \pi i n (x-y)/L},
\end{equation}
whose derivative is the Dirac comb with period $L$, namely
\begin{equation} \label{Dirac comb def}
\delta(x-y) \coloneqq \partial_x \varepsilon(x-y) = \frac{1}{L} \sum_{n \in \ZZ} e^{2 \pi i n (x-y)/L}.    
\end{equation}
It then follows from the Fourier mode decompositions \eqref{chi u decomp pm 0} of the chiral and anti-chiral free bosons $\chi(x)$ and $\bar\chi(x)$ for $x \in \RR$ and the relations \eqref{Compactified boson commutation relations a} that $[\chi(x), \chi(y)] = - \frac{i}{2} \varepsilon(x-y)$, $[\bar\chi(x), \bar\chi(y)] = \frac{i}{2} \varepsilon(x-y)$ and of course $[\chi(x), \bar\chi(y)] = 0$. By definitions \eqref{Compact boson} and \eqref{Dual compact boson} of the compact and dual compact bosons, we then immediately deduce that
\begin{equation} \label{commutation relations X tX}
[\X(x), \X(y)] = 0, \qquad [\X(x), \tilde\X(y)] = - i \varepsilon(x-y), \qquad [\tilde\X(x), \tilde\X(y)] = 0
\end{equation}
for any $x, y \in \RR$.
Defining the \textbf{conjugate momentum} as $\P(x) \coloneqq \partial_x \tilde \X(x)$ we then obtain the canonical commutation relations
\begin{equation} \label{X P comm rel}
[\X(x), \X(y)] = 0, \qquad [\X(x), \P(y)] = i \delta(x-y), \qquad [\P(x), \P(y)] = 0.
\end{equation}
In other words, while the conjugate momentum $\P$ can be realized as a time-derivative of the compact boson $\X$, it is also obtained as a spatial derivative of the dual boson $\tilde\X$.

Now the momentum has the following mode expansion
\begin{equation} \label{P mode expansion}
\P(x) = \frac{\sqrt{\pi}}{L} \sum_{n \in \ZZ} ( \b_n + \bar \b_{-n} ) e^{-2 \pi i n x/L}.
\end{equation}
Its integral is the zero-mode momentum $\int_0^L \dd x \, \P(x) = \sqrt{\pi} (\b_0 + \bar \b_0)$ which on the Fock space $\Fock_{p,w}$ is quantised as $p/R_\circ$, justifying the name `momentum' for the label $p$. In particular, the fact that the zero-mode momentum is quantised in integer multiples of $1/R_\circ$ is consistent with the compactification of the boson $\X$ on a circle of circumference $2 \pi R_\circ$.

\paragraph{Causal propagator.}

The real-time evolution of the free compact boson \eqref{X decomposition} on $S^1$ is given by the quantum operator in the Heisenberg picture
\begin{equation} \label{Compact boson Lorentz}
\X(x, t) \coloneqq e^{i t \H_0} \X(x) e^{-i t \H_0} = \chi(x+t) + \bar \chi(x-t)
\end{equation}
for $t \in \RR$. The two argument notation here should not be confused with the one in \eqref{Compact boson}, but since we denote the latter simply by $\X(u)$ this should not lead to confusion. Note that by using \eqref{X periodicity} we have the property $\X(x + L, t) = \X(x, t) + \sqrt{\pi} (\b_0 - \bar \b_0)$, so that \eqref{Compact boson Lorentz} is well defined on the Lorentzian cylinder in the zero-winding sector $\Fock_{p,0}$ for any $p\in \ZZ$. More generally, it is only defined on $2$-dimensional Minkowski space.
The commutator of the quantum field \eqref{Compact boson Lorentz} at two different space-time points $(x,t)$ and $(x',t')$ is given by
\begin{equation} \label{commutator causal propagator}
[ \X(x,t), \X(x',t') ] = i \Delta^{\rm cyl}(x,t;x',t'),
\end{equation}
in terms of the \textbf{causal propagator} on the Lorentzian cylinder which reads
\begin{equation} \label{causal propagator cyl}
\Delta^{\rm cyl}(x,t;x',t') \coloneqq \frac 12 \big( \varepsilon(x-x'-t+t') - \varepsilon(x-x'+t-t') \big).
\end{equation}
Finally, we note using $\P(x) = \partial_t \X(x,t)|_{t=0}$ that the canonical commutation relations \eqref{X P comm rel} all follow as a consequence of \eqref{commutator causal propagator}. In particular, the first relation follows from the fact that the causal propagator \eqref{causal propagator cyl} vanishes at space-like separated points and the second and third relations follow using $\partial_{t'} \Delta^{\rm cyl}(x,t;x',t')|_{t=t'=0} = \delta(x-x')$ and $\partial_t \partial_{t'} \Delta^{\rm cyl}(x,t;x',t')|_{t=t'=0} = 0$.
Note that \eqref{commutator causal propagator} solves the $2$-dimensional wave equation $(-\partial_t^2 + \partial_x^2) \Delta^{\rm cyl}(x,t;x',t') = 0$. It can be written as a difference
\begin{equation*}
\Delta^{\rm cyl}(x,t;x',t') = \Delta^{\rm cyl}_{\rm R}(x,t;x',t') - \Delta^{\rm cyl}_{\rm A}(x,t;x',t')
\end{equation*}
of the retarded and advanced causal Green's functions defined by
\begin{equation} \label{retarded advanced Greens}
\Delta^{\rm cyl}_{\rm R/{\rm A}}(x,t;x',t') \coloneqq \mp \theta\big(\!\pm (t-t')\big) \Delta^{\rm cyl}(x,t;x',t'),
\end{equation}
where $\theta(x-y) \coloneqq \varepsilon(x-y) + \frac 12$. Using $\partial_t \theta(t-t') = \delta(t-t')$ and the fact that we can rewrite the causal propagator \eqref{commutator causal propagator} as $\Delta^{\rm cyl}(x,t;x',t') = \frac 12 \big( \theta(x-x'-t+t') - \theta(x-x'+t-t') \big)$, we find that \eqref{retarded advanced Greens} are indeed both Green's functions for the wave operator, namely
\begin{equation}
\big( \! -\partial_t^2 + \partial_x^2 \big) \Delta^{\rm cyl}_{\rm R/{\rm A}}(x,t;x',t') = \delta(t-t') \delta(x-x').
\end{equation}

\paragraph{Free imaginary-time evolution.}
Although we are really interested in Hamiltonian field theory on the circle $S^1$, it is useful to consider our fields as living on the extended Euclidean cylinder $i \RR \times S^1$ as in \eqref{Fourier mode expansion} for the chiral and anti-chiral currents $J^a(u)$ and $\bar J^a(\bar u)$, or \eqref{chi finer decomposition} and \eqref{chi u decomp pm 0} for the chiral and anti-chiral free bosons $\chi(u)$ and $\bar \chi(\bar u)$.

This extension of fields from $S^1$ to $i \RR \times S^1$ respectively turns chiral and anti-chiral fields on $S^1$ to holomorphic and anti-holomorphic fields on $i \RR \times S^1$ equipped with the complex coordinate $u = x - i \tau$ where $x \in [0,L]$ is the coordinate along $S^1$ and $\tau \in \RR$ the coordinate along the cylinder. Indeed, it follows from the definition of the WZW Hamiltonian in \eqref{WZW Hamiltonian} that the imaginary-time evolution of the chiral and anti-chiral currents $J^a(x)$ and $\bar J^a(x)$ on $S^1$ are given by holomorphic and anti-holomorphic currents
\begin{subequations} \label{free evolutions}
\begin{equation} \label{currents free evolution}
J^a(x - i \tau) = e^{\tau \H_0} J^a(x) e^{-\tau \H_0}, \qquad
\bar J^a(x + i \tau) = e^{\tau \H_0} \bar J^a(x) e^{-\tau \H_0}.
\end{equation}
Equivalently, working in the free field realisation where $\H_0$ is the free boson Hamiltonian given by \eqref{WZW Hamiltonian free realisation}, the free imaginary-time evolution turns the chiral and anti-chiral free bosons into holomorphic and anti-holomorphic fields, respectively,
\begin{equation} \label{bosons free evolution}
\chi(x - i \tau) = e^{\tau \H_0} \chi(x) e^{-\tau \H_0}, \qquad
\bar\chi(x + i \tau) = e^{\tau \H_0} \bar\chi(x) e^{-\tau \H_0}.
\end{equation}
\end{subequations}
In particular, the compact free boson \eqref{Compact boson} on the cylinder can be recovered from its Fourier mode decomposition \eqref{X decomposition} on the circle by imaginary-time evolution, and likewise for the dual compact boson \eqref{Dual compact boson} from \eqref{tX decomposition}.
Note that fields on the cylinder $i \RR \times S^1$ can be viewed as operators in a Euclidean version of the Heisenberg picture while fields of interest on the circle $S^1$ represent the same operators but in a Euclidean version of the Schr\"odinger picture.

It is important to note, however, that we are \emph{not} working with Euclidean field theories. The imaginary time $\tau$ introduced by the extension from $S^1$ to $i \RR \times S^1$ is a purely computational tool, allowing us to use methods from conformal field theory such as operator product expansions in \eqref{Current OPE} and \eqref{Boson OPE}.
We saw another example of the use of imaginary-time evolution in \eqref{state mn} where the highest-weight state $\ket{r+s, r-s} \in \Fock$ for arbitrary $(r,s) \in \ZZ^2$ could be created from the vacuum state $\ket{0} \in \Fock$ by inserting the vertex operator $\mathcal V_{r,s}(u,\bar u)$ in \eqref{vertex operators V rs} at infinity on the cylinder. This can be depicted pictorially as:
\begin{equation*}
\raisebox{-17mm}
{\begin{tikzpicture}
\def\R{1}
\def\x{5.5}
\def\y{-1.5}
  \fill[left color   = gray!10!black,
        right color  = gray!10!black,
        middle color = gray!10,
        shading      = axis,
        opacity      = 0.15]
    (\x+\R,\y+.5*\R) -- (\x+\R,\y+2.5*\R)  arc (360:180:\R cm and 0.3cm)
          -- (\x-\R,\y+.5*\R) arc (180:360:\R cm and 0.3cm);
  \fill[top color    = gray!90!,
        bottom color = gray!2,
        middle color = gray!30,
        shading      = axis,
        opacity      = 0.03]
    (\x,\y+2.5*\R) circle (\R cm and 0.3cm);
  \draw (\x-\R,\y+2.5*\R) -- (\x-\R,\y+.5*\R) arc (180:360:\R cm and 0.3cm)
               -- (\x+\R,\y+2.5*\R) ++ (-\R,0) circle (\R cm and 0.3cm);
  \draw[dashed] (\x-\R,\y+.5*\R) arc (180:0:\R cm and 0.3cm);

  \draw[thick, dashed, darkgray] (\x-\R,\y+1.5*\R) arc (180:0:\R cm and 0.3cm);
  \draw[thick, darkgray] (\x-\R,\y+1.5*\R) arc (180:360:\R cm and 0.3cm) node[right=.5mm]{\tiny \color{darkgray} $S^1$};

  \draw (\x,\y+.2*\R) node[below=1mm]{\scriptsize \blue $\ket{r+s,r-s}$};

\end{tikzpicture}
}
=\quad\;\;\;
\raisebox{-17mm}
{\begin{tikzpicture}
\def\R{1}
\def\x{5.5}
\def\y{-1.5}
  \fill[left color   = gray!10!black,
        right color  = gray!10!black,
        middle color = gray!10,
        shading      = axis,
        opacity      = 0.15]
    (\x+\R,\y+.5*\R) -- (\x+\R,\y+2.5*\R)  arc (360:180:\R cm and 0.3cm)
          -- (\x-\R,\y+.5*\R) arc (180:360:\R cm and 0.3cm);
  \fill[top color    = gray!90!,
        bottom color = gray!2,
        middle color = gray!30,
        shading      = axis,
        opacity      = 0.03]
    (\x,\y+2.5*\R) circle (\R cm and 0.3cm);
  \draw (\x-\R,\y+2.5*\R) -- (\x-\R,\y+.5*\R) arc (180:360:\R cm and 0.3cm)
               -- (\x+\R,\y+2.5*\R) ++ (-\R,0) circle (\R cm and 0.3cm);
  \draw[dashed] (\x-\R,\y+.5*\R) arc (180:0:\R cm and 0.3cm);

  \draw[thick, dashed, darkgray] (\x-\R,\y+1.5*\R) arc (180:0:\R cm and 0.3cm);
  \draw[thick, darkgray] (\x-\R,\y+1.5*\R) arc (180:360:\R cm and 0.3cm) node[right=.5mm]{\tiny \color{darkgray} $S^1$};

  \draw (\x,\y+.2*\R) node[below=1mm]{\scriptsize $\ket{0}$};
  
  \draw[fill, blue] (\x-0.6\R,\y+\R) node[right=.5mm]{\tiny $\mathcal V_{r,s}(u, \bar u)$} circle (0.5pt);
  \draw[thick, -stealth] (\x-0.6\R,\y+.9*\R) -- (\x-0.6\R,\y+.4*\R) node[right]{\tiny $u \to i \infty$};
\end{tikzpicture}
}
\end{equation*}
The Hamiltonian fields and states live on the circle $S^1$ at $\tau = 0$ and the physical time-derivative of fields in the Heisenberg picture is obtained by taking commutators with $i \H_0$, cf. \eqref{Compact boson Lorentz}.

\paragraph{Local operators.}
Coming back to the crucial point raised at the end of \S\ref{sec: Fourier decomp}, we can use the basic representation \eqref{su2 1 currents cyl} to express local operators of interest, given by integrals over the circle of composite operators \eqref{composite operator}, directly in terms of modes of the chiral and anti-chiral free bosons in $\Heis$. Recall that there are two classes of local operators to consider.

Integrals of purely chiral (or anti-chiral) composite operators can be expressed as infinite sums of normal ordered monomials $\nord{\b_{n_1} \ldots \b_{n_r}}$, with $n_1 + \ldots + n_r$ bounded, multiplied by a finite number of shift operators \eqref{Zero-mode exp}. For example, when $\g = \sl_2$ the chiral local operator \eqref{chiral local op example} can be expressed as
\begin{equation} \label{chiral local op example Bos}
\frac{1}{2 \pi} \int_0^L \dd x \, \nord{J^3(x) J^3(x)} = 4 \int_0^L \dd x \, \nord{\partial_x \chi(x) \partial_x \chi(x)} = \frac{4 \pi}{L} \sum_{n > 0} \b_{-n} \b_n.
\end{equation}
Since $\Fock$ is a direct sum \eqref{Fock} of Fock spaces $\Fock_{p,w}$ which are highest weight representations of $\Heis$, such normal ordered infinite sums are a well defined endomorphism of $\Fock$.

On the other hand, integrals of a composite operator \eqref{composite operator} which contains both chiral and anti-chiral pieces lead to infinite sums of products of chiral creation opeators ($\b_n$ with $n<0$) with anti-chiral creation operators ($\bar\b_n$ with $n<0$). For instance, in the case $\g=\sl_2$ the local operator \eqref{non-chiral local op example} can be expressed in terms of chiral and anti-chiral free boson modes as
\begin{equation} \label{non-chiral local op example Bos}
\frac{1}{2 \pi} \int_0^L \dd x \, J^3(x) \bar J^3(x) = 4 \int_0^L \dd x \, \nord{\partial_x \chi(x) \partial_x \bar\chi(x)} = \frac{4 \pi}{L} \sum_{n \in \ZZ} \b_n \bar \b_n.
\end{equation}
Although the infinite sum over positive modes $n \geq 0$ has a well defined action on the direct sum of Fock spaces $\Fock$, the infinite sum over negative modes $n <0$ does not. On the other hand, the infinite sum over negative modes in \eqref{non-chiral local op example Bos} has a well defined action on the direct sum $\hFock$ of completed Fock spaces, by definition of the latter, but the infinite sum over positive modes does not. For instance, $\sum_{m > 0} \b_{-m} \bar\b_{-m} |p,w\rangle \in \hFock$ is a well defined state in the completion but
\begin{equation} \label{infinite sums example}
\sum_{n > 0} \b_n \bar\b_n \sum_{m > 0} \b_{-m} \bar\b_{-m} |p,w\rangle = \sum_{n > 0} n^2 |p,w\rangle
\end{equation}
produces a divergent infinite sum times the highest weight state $|p,w\rangle$, which is ill-defined.

\subsubsection{Changing the compactification radius} \label{sec: changing radius}

We saw in \S\ref{sec: compactified boson} that for the basic representation \eqref{su2 1 currents cyl} to correctly produce $\su_2$-currents at level $1$ living on the cylinder, we had to work in a particular representation $\Fock$ of the algebra $\Bos$ ensuring that the boson $\X$ and dual boson $\tilde\X$ could be compactified on a circle of radius $R_\circ = 1/\sqrt{2\pi}$. Recall that the compactifications
\begin{equation} \label{compactifications}
\X \sim \X + 2 \pi R_\circ, \qquad \tilde\X \sim \tilde\X + \frac{1}{R_\circ}
\end{equation}
(where in the second relation we wrote $2 \pi R_\circ = 1/R_\circ$ for later convenience) were enforced by the fact that these fields appear as exponents in the two parameter family \eqref{vertex operators V rs} of full vertex operators
\begin{equation} \label{vertex operators V rs X tX}
\mathcal V_{r,s}(x) = \nord{e^{i \sqrt{8 \pi} ( r \chi(x) + s \bar \chi(x) )}} = \nord{e^{\frac{i}{R_\circ} (r+s) \X(x) + 2 \pi i R_\circ (r-s) \tilde\X(x)}},
\end{equation}
labeled by pairs of integers $(r,s) \in \ZZ^2$.

\paragraph{Rescaled canonically conjugate fields.}
It is clear from the compactification rules, written in the form \eqref{compactifications}, how to change the compactification radius from the self-dual radius $R_\circ$ to a generic radius $R > 0$. One should simply rescale the boson $\X$ by $R/R_\circ$ and correspondingly rescale the dual boson $\tilde \X$ by $R_\circ/R$. The latter also ensures that the canonical commutation relations \eqref{commutation relations X tX} are preserved.
Instead of dealing with the radius $R$ directly, it will be more convenient to work with the parameter $\beta = 2/R$ so that the self-dual radius $R_\circ$ corresponds to the value $\beta_\circ = \sqrt{8 \pi}$. Thus, for generic $\beta > 0$, the \textbf{rescaled compact boson and dual boson} are defined as
\begin{equation}\label{Rescaled Boson Dual Boson}
\Phi(u, \bar u) \coloneqq \frac{\sqrt{8 \pi}}{\beta} \X(u, \bar u), \qquad \tilde \Phi(u, \bar u) \coloneqq \frac{\beta}{\sqrt{8 \pi}}\tilde \X(u, \bar u)
\end{equation}
for any $u \in \mathbb{C}$. These have the corresponding compactification rules
\begin{equation} \label{compactifications rescaled}
\Phi \sim \Phi + \frac{4 \pi}{\beta}, \qquad \tilde\Phi \sim \tilde\Phi + \frac{\beta}{2}.
\end{equation}
The rescaled fields \eqref{Rescaled Boson Dual Boson} should, more precisely, be denoted by $\Phi_\beta$ and $\widetilde{\Phi}_\beta$ to emphasise the dependence on the compactification radius $R = 2/\beta$. However, to ease notation we will usually suppress the subscript $\beta$ if the value of $\beta$ is clear from the context. We will always denote the compact boson and dual boson at the self-dual radius by $\X = \Phi_{\sqrt{8 \pi}}$ and $\tilde\X = \tilde\Phi_{\sqrt{8 \pi}}$.

\medskip

The rescaling \eqref{Rescaled Boson Dual Boson} would be a benign transformation at the classical level. However, as we will now explain, it is more subtle at the quantum level since it corresponds to a Bogoliubov transformation on the modes of the chiral and anti-chiral free bosons $\chi$ and $\bar\chi$ which has the effect of mixing the chiralities as well as the notion of creation/annihilation operators.

By design, the rescaled fields \eqref{Rescaled Boson Dual Boson} satisfy the same canonical commutation relations as the original boson and dual boson, cf. \eqref{commutation relations X tX},
\begin{equation} \label{commutation relations Phi tPhi}
[\Phi(x), \Phi(y)] = 0, \qquad [\Phi(x), \tilde\Phi(y)] = - i \varepsilon(x-y), \qquad [\tilde\Phi(x), \tilde\Phi(y)] = 0
\end{equation}
for any $x, y \in \RR$. It follows that the rescaled fields \eqref{Rescaled Boson Dual Boson} admit a mode decomposition of the same form as the original boson and dual boson. That is, we can decompose them into chiral and anti-chiral free bosons $\phi$ and $\bar\phi$ as, cf. \eqref{Compact boson} and \eqref{Dual compact boson},
\begin{equation} \label{Phi tPhi decomp}
\Phi(u, \bar u) = \phi(u) + \bar\phi(\bar u), \qquad
\tilde\Phi(u, \bar u) = \phi(u) - \bar\phi(\bar u).
\end{equation}
In turn, we can decompose $\phi$ and $\bar\phi$ themselves into creation, zero-mode and annihilation parts, cf. \eqref{chi finer decomposition} and \eqref{chi u decomp pm 0},
\begin{equation} \label{phi finer decomposition}
\phi(u) = \phi_+(u) + \phi_0(u) + \phi_-(u), \qquad \bar\phi(u) = \bar\phi_+(u) + \bar\phi_0(u) + \bar\phi_-(u)    
\end{equation}
where each summand is given explicitly in the $u$ coordinate by
\begin{subequations} \label{phi u decomp pm 0}
\begin{alignat}{2}
\phi_\pm(u) &\coloneqq \frac{i}{\sqrt{4\pi}} \sum_{\mp n > 0} \frac{1}{n} \a_n e^{-2 \pi i n u/L}, &\qquad
\phi_0(u) &\coloneqq \q_0 + \sqrt{\pi} \a_0 \frac{u}{L},\\
\bar \phi_\pm(\bar u) &\coloneqq \frac{i}{\sqrt{4\pi}} \sum_{\mp n > 0} \frac{1}{n} \bar \a_n e^{2 \pi i n \bar u/L}, &\qquad
\bar \phi_0(\bar u) &\coloneqq \bar\q_0 - \sqrt{\pi} \bar \a_0 \frac{\bar u}{L}.
\end{alignat}
\end{subequations}
Here the modes $\q_0, \bar \q_0, \a_n$ and $\bar \a_n$ for $n \in \mathbb{Z}$ generate an infinite-dimensional Lie algebra with the same defining relations as in \eqref{Compactified boson commutation relations}, namely
\begin{subequations} \label{Compactified rescaled boson commutation relations}
\begin{alignat}{2}
\label{Compactified rescaled boson commutation relations a} [\a_n, \a_m] &= n\delta_{n+m,0}, &\qquad [\bar\a_n, \bar\a_m] &= n\delta_{n+m,0}, \\
\label{Compactified rescaled boson commutation relations b} [\q_0, \a_n] &= \frac{i}{\sqrt{4 \pi}} \delta_{n,0}, &\qquad [\bar \q_0, \bar \a_n] &= \frac{i}{\sqrt{4 \pi}} \delta_{n,0}.
\end{alignat}
\end{subequations}

\paragraph{Bogolyubov transformation.} Restricting the relations \eqref{Rescaled Boson Dual Boson} to the real axis and using the expressions \eqref{Compact boson}, \eqref{Dual compact boson} and \eqref{Phi tPhi decomp} we find that the chiral and anti-chiral bosons $\phi(x)$ and $\bar\phi(x)$ associated with a general $\beta > 0$ are related to original ones $\chi(x)$ and $\bar\chi(x)$ associated with the self-dual value of $\sqrt{8\pi}$ as
\begin{equation} \label{phi vs chi}
\phi(x) = \alpha_+ \chi(x) - \alpha_-\bar \chi(x),\qquad
\bar \phi(x) = \alpha_+\bar \chi(x) - \alpha_-\chi(x)
\end{equation}
for all $x \in \RR$, where we have introduced the parameters
\begin{equation}\label{Alpha PM}
\alpha_\pm \coloneqq \frac{1}{2}\bigg(\frac{\beta}{\sqrt{8 \pi}} \pm \frac{\sqrt{8 \pi}}{\beta} \bigg).
\end{equation}
Comparing the mode expansions of both sides of \eqref{phi vs chi}, and noting that the chiral modes $\b_n$ in \eqref{chi u decomp pm 0} come multiplied by the exponential $e^{- 2\pi i n x/L}$ while the anti-chiral modes $\bar \b_n$ come with an oppositely signed exponential $e^{2\pi i n x/L}$, and likewise in the expansions \eqref{phi u decomp pm 0}, we obtain the Bogolyubov type transformation
\begin{subequations} \label{modes a vs b}
\begin{alignat}{2}
\label{modes a vs b 1} \q_0 &= \alpha_+ \x_0 - \alpha_-\bar \x_0, &\qquad \a_n &= \alpha_+ \b_n + \alpha_-\bar \b_{-n}, \\
\label{modes a vs b 2} \bar\q_0 &= \alpha_+ \bar\x_0 - \alpha_- \x_0, &\qquad \bar \a_n &= \alpha_+\bar \b_n + \alpha_-\b_{-n}
\end{alignat}
\end{subequations}
for all $n \in \ZZ$. We thus see that the modes of the rescaled chiral boson $\Phi$ not only mix the chiral and anti-chiral modes of $\X$, but also its creation and annihilation modes.
An immediate consequence of this is that the notion of normal ordering depends on the compactification radius $R$, or equivalently on the parameter $\beta = 2/R$. We will come back to how this affects the family of vertex operators introduced in \eqref{vertex operators V rs X tX}, in \S\ref{sec: Regularised vertex operators} below.

It follows from the reality conditions \eqref{chi reality conditions} that the chiral and anti-chiral parts \eqref{phi vs chi} of the rescaled compact boson and dual boson \eqref{Phi tPhi decomp} satisfy the same reality conditions
\begin{equation} \label{phi reality conditions}
\phi(x)^\dagger = \phi(x), \qquad \bar \phi(x)^\dagger = \bar \phi(x)
\end{equation}
for $x \in \RR$. Equivalently, their modes \eqref{modes a vs b} satisfy the same reality conditions $\q_0^\dagger = \q_0$, $\bar\q_0^\dagger = \bar\q_0$,  $\a_n^\dagger = \a_{-n}$ and $\bar \a_n^\dagger = \bar \a_{-n}$ as in \eqref{reality boson modes}.

To close this section, we check that the Bogolyubov transformation \eqref{modes a vs b} reproduces the expected periodicity property on the rescaled field $\Phi$. In terms of the zero-modes $\q_0$ and $\bar \q_0$ the state \eqref{state mn} reads
\begin{equation*}
| p, w \rangle = e^{\ii \big( \frac{p\beta}{2} + \frac{4 \pi w}{\beta} \big) \q_0} e^{\ii \big( \frac{p\beta}{2} - \frac{4 \pi w}{\beta} \big)\bar \q_0} |0\rangle.
\end{equation*}
Crucially, by virtue of the second relations in \eqref{modes a vs b} it is clear that the modes $\a_n$ and $\bar\a_n$ for $n > 0$ do \emph{not} annihilate the states $|p, w \rangle$. However, the zero modes act as
\begin{equation} \label{a0 ba0 on hw states}
\a_0 | p, w \rangle = \bigg( \frac{p \beta}{4 \sqrt{\pi}} + \frac{2 w \sqrt{\pi}}{\beta} \bigg) |p, w \rangle, \qquad
\bar \a_0 | p, w \rangle = \bigg( \frac{p \beta}{4 \sqrt{\pi}} - \frac{2 w \sqrt{\pi}}{\beta} \bigg) |p, w \rangle.
\end{equation}
It follows from the decomposition \eqref{Phi tPhi decomp} of $\Phi(u)$ and the mode expansions of $\phi(u)$ and $\bar\phi(\bar u)$ in \eqref{phi u decomp pm 0} that $\Phi(u+L) = \Phi(u) + \sqrt{\pi} (\a_0 - \bar \a_0)$ and therefore, on the Fock space $\Fock_{p, w}$, we indeed have the expected periodicity relation
$\Phi(u+L) = \Phi(u) + 2 \pi n R$ with $R \coloneqq 2/\beta$.

\section{Renormalisation of Hamiltonians} \label{sec: Renormalisation}

In this section we will develop a Wilsonian approach to the renormalisation of Hamiltonian field theories on the circle $S^1$.
We begin in \S\ref{sec: regularisation} by introducing a short-distance cutoff in the conformal field theory of the compactified free boson discussed in \S\ref{sec: compactified boson}. This regularisation serves to control ultraviolet divergences, such as the one exhibited in \eqref{infinite sums example}.
In \S\ref{sec: effective Hamiltonians} we then implement the Wilsonian renormalisation programme, whose guiding principle is that long-distance physics should be insensitive to the details of the short-distance cutoff. When applied to perturbations of the compactified free boson, this framework yields renormalisation group equations governing the flow of all coupling constants.

\subsection{Regularisation} \label{sec: regularisation}

The first step in Wilsonian renormalisation is to introduce a high energy (or short distance) cutoff $\Lambda$ into the theory so as to keep ultraviolet divergences under control. And the standard approach is to truncate the mode expansions of all the fields $\phi$ of the theory by working instead with truncated fields $\phi_{\leq \Lambda}(x)$ whose mode expansions only involve momenta of magnitude below the Wilsonian cutoff $\Lambda$. As emphasised in \S\ref{sec: Free field realisations}, we cannot apply such a truncation procedure directly to the (anti-)chiral Kac--Moody currents $J^a(u)$ and $\bar J^a(u)$ of the WZW model since these are non-abelian. So our approach consists in using free field realisations of these currents, which in our $\su_2$ case at level $1$ is given by the basic representation \eqref{su2 1 currents cyl} in terms of chiral and anti-chiral free bosons $\chi(u)$ and $\bar\chi(u)$, and regularise the free fields instead.

However, since we are working with a (Hamiltonian) field theory on a compact space, the circle $S^1$, the mode decomposition of our free bosons $\chi(u)$ and $\bar\chi(u)$ is a sum over the \emph{discrete} set of modes $\b_n$ and $\bar\b_n$ with $n \in \ZZ$, given in \eqref{chi finer decomposition} and \eqref{chi u decomp pm 0}. In this setting, a high energy Wilsonian cutoff is an \emph{integer} $\Lambda \in \ZZ_{> 0}$ at which we choose to truncate these sum of modes, since momentum is measured in integer multiples of $\frac{2 \pi}{L}$. Yet our main goal is to describe the renormalisation group equations for the couplings of perturbations of the WZW model, which are conventionally described as differential equations with respect to a continuum length or mass scale. The purpose of the present section is therefore to replace the above naive sharp truncation of the chiral and anti-chiral bosons $\chi$ and $\bar\chi$ by a more general \emph{smooth} regularisation depending on a smooth cutoff function $\eta : \RR_{\geq 0} \to \RR$ and a continuous length scale $\epsilon > 0$.

\medskip

In order to motivate such smooth regularisations, there is a mathematical analogy for why smoothing the cutoff is a meaningful thing to do, coming from the regularisation of divergent series \cite[\S3.7]{Tao-blog}. We note that this analogy has already been explored recently in the context of renormalisation in $4$-dimensional quantum field theories \cite{PadillaSmith1, PadillaSmith2, Smith:2025kvh}. In our $2$-dimensional Hamiltonian setting on the circle $S^1$, however, the connection to regularisation of divergent series is more than an analogy. Indeed, the divergent series in question, of the type $\sum_{n \geq 1} n^k$ for some $k \geq -1$, are precisely the kind we encounter when computing the action of certain non-chiral local operators such as \eqref{non-chiral local op example Bos} on the completed Fock space $\hFock$, as in \eqref{infinite sums example}.

One way to obtain a finite result for a divergent series such as $\sum_{n \geq 1} n$ would be to use \textbf{zeta function regularisation}: the Riemann zeta function is defined as $\zeta(s) = \sum_{n = 1}^\infty \frac{1}{n^s}$ for $\mathrm{Re}(s)>1$, and can be analytically continued to $\CC \setminus \{ 1 \}$. In particular, we have $\zeta(-1) = - \frac{1}{12}$ which suggests the finite value of $- \frac{1}{12}$ for the series. This result is obtained through complex analytic methods, but in fact there is a way to find this result through real analytic methods and regularising in the right way. If we sharply cut off the sum, we have
\begin{equation} \label{sharply cut off sum}
    \sum_{n = 1}^N n = \frac{1}{2}N^2 + \frac{1}{2}N,
\end{equation}
and each term is divergent as $N \rightarrow \infty$. To introduce the smooth regularisation we start by viewing the sum as $\sum_{n = 1}^\infty n \, \eta(\frac nN)$ where $\eta(x) = \mathbf{1}_{x \leq 1}$ denotes the indicator function for the interval $[0,1] \subset \RR_{\geq 0}$. The function $\eta$ suffers a discontinuity at 1, and so we smoothly regularise by allowing $\eta: \mathbb{R}_{\geq 0} \rightarrow \mathbb{R}$ to be any smooth function equal to $1$ at the origin and tending to zero sufficiently fast at infinity. For any such $\eta$, one finds the result
\begin{equation} \label{regularised series intro}
\sum_{n = 1}^\infty n \, \eta\big( \tfrac nN \big) = N^2 \int_0^\infty x \, \eta(x)\dd x - \frac{1}{12} + O\big( \tfrac 1N \big).
\end{equation}
While there is a large amount of choice for $\eta$, and a divergence as $N \rightarrow \infty$ generically remains, the constant term $- \frac{1}{12}$ is the same for each. It is also interesting to observe that for certain well chosen functions $\eta$, referred to as enhanced regulators in \cite{PadillaSmith1}, the regularised series \eqref{regularised series intro} actually converges to the constant term in the limit $N \to \infty$; see \S\ref{sec: Mellin transform}. The divergent term will be important in our calculations, but analogously the results will not depend on $\eta$, as expected for an arbitrarily chosen regulator.

\medskip

We will now discuss how to implement these ideas to regularising our theory. We begin in \S\ref{sec: trunc to reg} by explaining how to view the usual truncation of fields as a sharp regularisation of the mode algebra \eqref{Compactified boson commutation relations}. This then naturally extends to the notion of smooth regularisation of the chiral and anti-chiral bosons in \S\ref{sec: smooth regularisations}. In \S\ref{sec: Mellin transform} we recall how the theory of the Mellin transform can be used to efficiently compute the leading asymptotics of regularised divergent series that we will encounter in the smoothly regularised theory. Finally, in \S\ref{sec: Regularised vertex operators} we close the discussion initiated in \S\ref{sec: changing radius} for changing the compactification radius by discussing the effect of a Bogolyubov transformation \eqref{phi vs chi} on smoothly regularised vertex operators.

\subsubsection{Truncation vs regularisation} \label{sec: trunc to reg}

\paragraph{Truncated fields.}

A natural way to regularise our theory is to truncate all the free fields in the basic realisation \eqref{su2 1 currents cyl} to have energy below some fixed cutoff $\Lambda \in \ZZ_{>0}$. More precisely, we introduce the Lie subalgebra $\Heis^\Lambda_{\rm trc} \subset \Heis$ spanned by modes $\b_n$, $\bar\b_n$ of the chiral and anti-chiral bosons for which $|n| \leq \Lambda$, and denote its universal enveloping algebra by $\Bos^\Lambda_{\rm trc} \coloneqq U(\Heis^\Lambda_{\rm trc}) \subset \Bos$.

Following the exact same steps as in \S\ref{sec: chiral free boson}, we can then introduce the \textbf{truncated chiral and anti-chiral bosons} $\chi^\Lambda_{\rm trc}$ and $\bar\chi^\Lambda_{\rm trc}$, directly in the cylinder coordinate $u$, as
\begin{subequations} \label{truncated fields def}
\begin{equation}
\chi^\Lambda_{\rm trc}(u) = \chi^\Lambda_{{\rm trc},+}(u) + \chi_0(u) + \chi^\Lambda_{{\rm trc},-}(u), \qquad \bar\chi^\Lambda_{\rm trc}(\bar u) = \bar\chi^\Lambda_{{\rm trc},+}(\bar u) + \bar\chi_0(\bar u) + \bar\chi^\Lambda_{{\rm trc},-}(\bar u),
\end{equation}
where the zero-mode parts are the same as in \eqref{chi finer decomposition} but the creation and annihilation parts are replaced with finite mode decompositions, namely
\begin{equation} \label{truncated fields def b}
\chi^\Lambda_{{\rm trc},\pm}(u) \coloneqq \frac{i}{\sqrt{4\pi}} \sum_{0 < \mp n \leq \Lambda} \frac{1}{n} \b_n e^{-2 \pi i n u/L},\qquad
\bar \chi^\Lambda_{{\rm trc},\pm}(\bar u) \coloneqq \frac{i}{\sqrt{4\pi}} \sum_{0 < \mp n \leq \Lambda} \frac{1}{n} \bar \b_n e^{2 \pi i n \bar u/L}.
\end{equation}
\end{subequations}

We can then \emph{define} the \textbf{truncated $\su_2$-currents} at level $1$ by the exact same formulae as in \eqref{su2 1 currents cyl} but in terms of the truncated chiral and anti-chiral bosons, namely we set
\begin{subequations} \label{basic representation trunc}
\begin{alignat}{2}
J^{3,\Lambda}_{\rm trc}(u) &\coloneqq \sqrt{8 \pi} \partial_u \chi^\Lambda_{\rm trc}(u), &\qquad
J^{\pm,\Lambda}_{\rm trc}(u) &\coloneqq \frac{2 \pi}{L} \nord{e^{\pm i \sqrt{8 \pi} \chi^\Lambda_{\rm trc}(u)}},\\
\bar J^{3,\Lambda}_{\rm trc}(\bar u) &\coloneqq \sqrt{8 \pi} \partial_{\bar u} \bar \chi^\Lambda_{\rm trc}(\bar u), &\qquad
\bar J^{\pm,\Lambda}_{\rm trc}(\bar u) &\coloneqq \frac{2 \pi}{L} \nord{e^{\mp i \sqrt{8 \pi} \bar \chi^\Lambda_{\rm trc}(\bar u)}}.
\end{alignat}
\end{subequations}
Note that the truncated chiral and anti-chiral bosons have the same periodicity property as in \eqref{chi bchi periodicity} so that the discussion from \S\ref{sec: compactified boson} carries over to the truncated setting. In particular, we can define the Fock spaces $\Fock_{p,w}^\Lambda$ for $p,w \in \ZZ$ over the truncated Heisenberg Lie algebra $\Heis^\Lambda_{\rm trc}$ using the same relations as in \eqref{Fock m n action} but for $|n| \leq \Lambda$, and then set
\begin{equation} \label{Fock truncated}
\Fock^\Lambda \coloneqq \bigoplus_{\substack{p,w \in \ZZ\\ p+w \in 2 \ZZ}} \Fock_{p,w}^\Lambda.
\end{equation}
Since the algebra of modes $\Heis^\Lambda_{\rm trc}$ is truncated, there is no need for completion here.

The \textbf{truncated compact boson} is then defined as $\X^\Lambda_{\rm trc}(u, \bar u) \coloneqq \chi^\Lambda_{\rm trc}(u) + \bar \chi^\Lambda_{\rm trc}(\bar u)$ for $u \in \CC$ and the corresponding conjugate momentum is truncated to $\P^\Lambda_{\rm trc}(x) \coloneqq \partial_x \chi^\Lambda_{\rm trc}(x) - \partial_x \bar \chi^\Lambda_{\rm trc}(x)$ for $x \in \RR$.
These now satisfy the regularised canonical commutation relations, cf. \eqref{X P comm rel},
\begin{equation} \label{trunc comm relations}
[\X^\Lambda_{\rm trc}(x), \X^\Lambda_{\rm trc}(y)] = 0, \qquad
[\X^\Lambda_{\rm trc}(x), \P^\Lambda_{\rm trc}(y)] = i \delta_\Lambda(x - y), \qquad
[\P^\Lambda_{\rm trc}(x), \P^\Lambda_{\rm trc}(y)] = 0
\end{equation}
where we have introduced the \textbf{regularised Dirac comb} $\delta_\Lambda(x-y) \coloneqq \frac{1}{L} \sum_{n = -\Lambda}^{\Lambda} e^{2\pi i n(x-y)/L}$, cf. \eqref{Dirac comb def}, such that $\delta_\Lambda(x-y) \to \delta(x-y)$ in the limit $\Lambda \to \infty$ when the cutoff is removed.

Finally, we can introduce the truncated Hamiltonian of the WZW model \eqref{WZW Hamiltonian free realisation} as
\begin{subequations} \label{WZW Hamiltonian free realisation trunc}
\begin{equation}
\H_{0, {\rm trc}}^{\Lambda} \coloneqq \int_0^L \dd x \big( \nord{\partial_x \chi^\Lambda_{\rm trc}(x) \partial_x \chi^\Lambda_{\rm trc}(x)} + \nord{\partial_x \bar\chi^\Lambda_{\rm trc}(x) \partial_x \bar\chi^\Lambda_{\rm trc}(x)} \big) = \frac{2 \pi}{L} \big( \L_{0,{\rm trc}}^\Lambda + \bar \L_{0,{\rm trc}}^\Lambda \big).
\end{equation}
Note that we have not included the constant shift by $- \frac{\pi}{6 L}$ as this will not play a role in when we come to discuss the renormalisation of the sine-Gordon model in \S\ref{sec: sine-Gordon}. In the last expression we have defined the truncated Virasoro zero-modes as
\begin{equation} \label{L0 bL0 free boson trunc}
\L_{0,{\rm trc}}^\Lambda = \frac 12 \b_0^2 + \sum_{n=1}^\Lambda \b_{-n} \b_{n}, \qquad
\bar\L_{0,{\rm trc}}^\Lambda = \frac 12 \bar\b_0^2 + \sum_{n=1}^\Lambda \bar\b_{-n} \bar\b_n.
\end{equation}
\end{subequations}
Notice that the truncated Hamiltonian \eqref{WZW Hamiltonian free realisation trunc} still generates the free imaginary-time evolution on the truncated chiral and anti-chiral bosons \eqref{truncated fields def} exactly as in \eqref{bosons free evolution}, namely we have
\begin{subequations} \label{free evolutions trunc}
\begin{equation} \label{bosons free evolution trunc}
\chi^\Lambda_{\rm trc}(x - i \tau) = e^{\tau \H_{0, {\rm trc}}^\Lambda} \chi^\Lambda_{\rm trc}(x) e^{-\tau \H_{0, {\rm trc}}^\Lambda}, \qquad
\bar\chi^\Lambda_{\rm trc}(x + i \tau) = e^{\tau \H_{0, {\rm trc}}^\Lambda} \bar\chi^\Lambda_{\rm trc}(x) e^{-\tau \H_{0, {\rm trc}}^\Lambda}.
\end{equation}
The same free imaginary-time evolution then also holds for the truncated $\su_2$-currents \eqref{basic representation trunc}, exactly as in \eqref{currents free evolution}, i.e.
\begin{equation} \label{currents free evolution trunc}
J^{a, \Lambda}_{\rm trc}(x - i \tau) = e^{\tau \H_{0, {\rm trc}}^\Lambda} J^{a, \Lambda}_{\rm trc}(x) e^{-\tau \H_{0, {\rm trc}}^\Lambda}, \qquad
\bar J^{a, \Lambda}_{\rm trc}(x + i \tau) = e^{\tau \H_{0, {\rm trc}}^\Lambda} \bar J^{a, \Lambda}_{\rm trc}(x) e^{-\tau \H_{0, {\rm trc}}^\Lambda}.
\end{equation}
\end{subequations}

\medskip

With these definitions in place, we can now justify the truncation by revisiting the crucial observation at the end of \S\ref{sec: Fourier decomp}. A local operator such as \eqref{non-chiral local op example Bos}, which in the original theory was expressed as a problematic infinite sum of modes, is replaced in the truncated theory by
\begin{equation} \label{non-chiral local op example Bos trunc}
\frac{1}{2 \pi} \int_0^L \dd x \, J^{3, \Lambda}_{\rm trc}(x) \bar J^{3, \Lambda}_{\rm trc}(x) = 4 \int_0^L \dd x \, \partial_x \chi^\Lambda_{\rm trc}(x) \partial_x \bar \chi^\Lambda_{\rm trc}(x) = \frac{4 \pi}{L} \sum_{n = - \Lambda}^\Lambda \b_n \bar \b_n.
\end{equation}
More generally, any integral over $S^1$ of a composite operator \eqref{composite operator} built from chiral and anti-chiral pieces is replaced in the truncated theory by a finite sum of modes. As a consequence, such truncated local operators have well-defined actions on the Fock representation \eqref{Fock truncated}. For example, the analogue of the problematic computation \eqref{infinite sums example} in the truncated setting would be to act with the positive mode part of \eqref{non-chiral local op example Bos trunc} on the state $\sum_{m=1}^\Lambda \b_{-m} \bar\b_{-m} |p,w\rangle \in \Fock^\Lambda_{\rm trc}$, i.e.
\begin{equation} \label{infinite sums example trunc}
\sum_{n = 1}^\Lambda \b_n \bar\b_n \sum_{m = 1}^\Lambda \b_{-m} \bar\b_{-m} |p,w\rangle = \sum_{n =1}^\Lambda n^2 |p,w\rangle.
\end{equation}
In other words, the ill-defined divergent infinite sum in \eqref{infinite sums example} is now replaced by a finite sum.
However, as explained at the start of this section, we would like to replace such sharply cut off sums by smoothly regularised ones.
As a first step in this direction, we begin by observing that the above regularisation method can be implemented directly at the level of the algebra by suitably modifying the Heisenberg Lie algebra relations \eqref{Compactified boson commutation relations a}.

\paragraph{Regularised fields.}

Let $\Heis^\Lambda_{\rm log}$ be the \textbf{sharply regularised} Lie algebra of modes of the chiral and anti-chiral free bosons, generated by $\b_n^\Lambda, \bar \b_n^\Lambda$ for $n \in \ZZ$ and $\x_0^\Lambda$, $\bar \x_0^\Lambda$ subject to the relations, cf. \eqref{Compactified boson commutation relations},
\begin{subequations} \label{Heisenberg log relations Lambda}
\begin{alignat}{2}
\label{Heisenberg Fourier Lambda comm a} [\b_n^\Lambda, \b_m^\Lambda] &= n {\bf 1}_{|n| \leq \Lambda} \delta_{n + m, 0}, &\qquad
[\bar\b_n^\Lambda, \bar\b_m^\Lambda] &= n {\bf 1}_{|n| \leq \Lambda} \delta_{n + m, 0},\\
\label{Heisenberg Fourier Lambda comm b} [\x_0^\Lambda, \b_n^\Lambda] &= \frac{i}{\sqrt{4 \pi}} \delta_{n,0}, &\qquad
[\bar\x_0^\Lambda, \bar\b_n^\Lambda] &= \frac{i}{\sqrt{4 \pi}} \delta_{n,0}
\end{alignat}
\end{subequations}
for all $m,n \in \ZZ$, with all other Lie brackets between generators being zero as before. In other words, rather than discarding the high-energy modes $\b_n$ and $\bar \b_n$ with $|n| > \Lambda$ altogether, we simply declare that they are central elements of the Lie algebra. We also let $\Heis^\Lambda \subset \Heis^\Lambda_{\rm log}$ denote the Lie subalgebra spanned by the modes $\b_n^\Lambda$ and $\bar\b_n^\Lambda$ for $n \in \ZZ$ and let $\Bos^\Lambda \coloneqq U(\Heis^\Lambda)$ be its universal enveloping algebra.

We introduce the \textbf{sharply regularised chiral and anti-chiral bosons} $\chi^\Lambda$ and $\bar\chi^\Lambda$ on $S^1$ by the exact same formulae as in \eqref{chi finer decomposition} and \eqref{chi u decomp pm 0} with $u = x \in [0,L]$, but where all the modes $\b_n$, $\bar \b_n$ for $n \in \ZZ$ and $\x_0$, $\bar \x_0$ are replaced by their regularised counterparts $\b_n^\Lambda$, $\bar \b_n^\Lambda$ for $n \in \ZZ$ and $\x_0^\Lambda$, $\bar \x_0^\Lambda$, respectively.

Using these we now \emph{define} the \textbf{sharply regularised $\su_2$-current algebra} at level $1$ by the same formulae as in \eqref{su2 1 currents cyl} but in terms of the sharply regularised chiral and anti-chiral bosons, namely for $x \in [0, L]$ we set
\begin{subequations}
\begin{alignat}{2}
J^{3,\Lambda}(x) &\coloneqq \sqrt{8 \pi} \partial_x \chi^\Lambda(x), &\qquad
J^{\pm,\Lambda}(x) &\coloneqq \frac{2 \pi}{L} \nord{e^{\pm i \sqrt{8 \pi} \chi^\Lambda(x)}},\\
\bar J^{3,\Lambda}(x) &\coloneqq \sqrt{8 \pi} \partial_x \bar \chi^\Lambda(x), &\qquad
\bar J^{\pm,\Lambda}(x) &\coloneqq \frac{2 \pi}{L} \nord{e^{\mp i \sqrt{8 \pi} \bar \chi^\Lambda(x)}}.
\end{alignat}
\end{subequations}
The discussion from \S\ref{sec: compactified boson} carries over verbatim to this sharply regularised setting. However, by contrast with the truncated case above, since we now have modes $\b_n^\Lambda$ and $\bar\b_n^\Lambda$ for all $n\in \ZZ$ we do need to consider the completed Fock spaces $\hFock_{p,w}^\Lambda$ for $p,w \in \ZZ$ over the sharply regularised Heisenberg Lie algebra $\Heis^\Lambda$, which are defined using the same relations as in \eqref{Fock m n action} but now for the regularised modes $\b_n^\Lambda$ and $\bar\b_n^\Lambda$ for all $n \in \ZZ$. We then set
\begin{equation} \label{Fock regularised}
\hFock^\Lambda \coloneqq \bigoplus_{\substack{p,w \in \ZZ\\ p+w \in 2 \ZZ}} \hFock_{p,w}^\Lambda.
\end{equation}

The \textbf{sharply regularised compact boson} $\X^\Lambda$ and the regularised conjugate momentum $P^\Lambda$ on $S^1$ are defined exactly as in \S\ref{sec: compactified boson} but now in terms of the regularised modes $\b_n^\Lambda$, $\bar\b_n^\Lambda$ for $n \in\ZZ$ and $\x_0^\Lambda$, $\bar\x_0^\Lambda$. The Lie algebra relations \eqref{Heisenberg log relations Lambda} then lead to a regularised version of the canonical commutation relations
\begin{equation*}
[\X^\Lambda(x), \X^\Lambda(y)] = 0, \qquad
[\X^\Lambda(x), \P^\Lambda(y)] = i \delta_\Lambda(x - y), \qquad
[\P^\Lambda(x), \P^\Lambda(y)] = 0,
\end{equation*}
where $\delta_\Lambda(x-y)$ is the regularised Dirac comb introduced after \eqref{trunc comm relations}.

Following the definition of the free Hamiltonian \eqref{WZW Hamiltonian free realisation trunc} in the truncated case, we could now similarly introduce the sharply regularised Hamiltonian of the WZW model \eqref{WZW Hamiltonian free realisation} as
\begin{subequations} \label{WZW Hamiltonian free realisation shrp reg}
\begin{equation}
\H_0^{\Lambda} \coloneqq \int_0^L \dd x \big( \nord{\partial_x \chi^\Lambda(x) \partial_x \chi^\Lambda(x)} + \nord{\partial_x \bar\chi^\Lambda(x) \partial_x \bar\chi^\Lambda(x)} \big) = \frac{2 \pi}{L} \big( \L_0^\Lambda + \bar \L_0^\Lambda \big).
\end{equation}
As in the truncation case we have not included the vacuum energy term as this will not be needed in \S\ref{sec: sine-Gordon} and we have introduced the sharply regularised Virasoro zero-modes as
\begin{equation} \label{L0 bL0 free boson shrp reg}
\L_0^\Lambda = \frac 12 \big( \b_0^\Lambda \big)^2 + \sum_{n > 0} \b_{-n}^\Lambda \b_n^\Lambda, \qquad
\bar\L_0^\Lambda = \frac 12 \big( \bar\b_0^\Lambda \big)^2 + \sum_{n > 0} \bar\b_{-n}^\Lambda \bar\b_n^\Lambda.
\end{equation}
\end{subequations}
However, the crucial difference with $\H_0^\Lambda$ in the truncated setting is that the Hamiltonian \eqref{WZW Hamiltonian free realisation shrp reg} does \emph{not} generate free imaginary-time evolution \eqref{free evolutions trunc}. Indeed, this is because the commutation relations \eqref{Heisenberg Fourier Lambda comm a} for the modes $\b_n^\Lambda$ and $\bar\b_n^\Lambda$ with $|n| > \Lambda$ are frozen and hence $\H_0^{\Lambda}$ induces trivial dynamics on these modes. But since the sharply regularised bosons $\chi^\Lambda$ and $\bar\chi^\Lambda$ are given by the same formula as in \eqref{chi finer decomposition} with Fourier expansions \eqref{chi u decomp pm 0}, the terms with mode numbers $|n|\leq \Lambda$ evolve freely while those with $|n| > \Lambda$ do not evolve. This leads to a complicated time evolution for $\chi^\Lambda$ and $\bar\chi^\Lambda$, which is no longer free.

To circumvent this problem, note that since the generators $\b_n^\Lambda$ for $|n| > \Lambda$ span a \emph{central} ideal $\mathfrak I^\Lambda \coloneqq \langle \b_n^\Lambda, \bar\b_n^\Lambda \rangle_{|n| > \Lambda}$ in the sharply regularised Heisenberg Lie algebra \eqref{Heisenberg log relations Lambda}, we should instead be working in the quotient $\Heis^\Lambda / \mathfrak I^\Lambda$ by this ideal. Of course, this just recovers the earlier truncated setting since we have a canonical isomorphism of Lie algebras
\begin{equation} \label{trunc vs quotiented sharp reg}
\Heis^\Lambda / \mathfrak I^\Lambda \cong \Heis^\Lambda_{\rm trc}.
\end{equation}
In particular, in the quotiented theory the Hamiltonian \eqref{WZW Hamiltonian free realisation shrp reg} does generate free evolution as in the truncated theory. The advantage of rephrasing truncation in terms of sharp regularisation is that it will allows to vastly generalise the concept of truncation, see \S\ref{sec: smooth regularisations} below.

\medskip

In the sharply regularised theory, a local operator such as \eqref{non-chiral local op example Bos} would take the exact same form as in the original theory without cutoff, namely
\begin{equation} \label{non-chiral local op example Bos sharp reg}
\frac{1}{2 \pi} \int_0^L \dd x \, J^{3, \Lambda}(x) \bar J^{3, \Lambda}(x) = 4 \int_0^L \dd x \, \nord{\partial_x \chi^\Lambda(x) \partial_x \bar\chi^\Lambda(x)} = \frac{4 \pi}{L} \sum_{n \in \ZZ} \b_n^\Lambda \bar \b_n^\Lambda.
\end{equation}
In particular, this is still an infinite sum but of the corresponding sharply regularised modes.
However, the key point is that the high energy modes $\b_n^\Lambda$ and $\bar\b_n^\Lambda$ for $|n| > \Lambda$ span a \emph{central} ideal in $\Heis^\Lambda$ and hence completely decouple from any computation.
In other words, since $\mathfrak{H}^\Lambda_{\rm trc}$ is a quotient of $\mathfrak{H}^\Lambda$ by this central ideal, any computation in the sharply regularised theory agrees with the corresponding computation in the truncated theory after quotienting by this central ideal.
For example, the analogue of the problematic computation \eqref{infinite sums example} in the present sharply regularised setting would be to act with the positive mode part of \eqref{non-chiral local op example Bos sharp reg} on the state $\sum_{m > 0} \b_{-m} \bar\b_{-m} |p,w\rangle \in \hFock^\Lambda$, namely
\begin{equation} \label{infinite sums example sharp reg}
\sum_{n > 0} \b_n^\Lambda \bar\b_n^\Lambda \sum_{m > 0} \b_{-m}^\Lambda \bar\b_{-m}^\Lambda |p,w\rangle = \sum_{n > 0} n^2 {\bf 1}_{|n| \leq \Lambda} |p,w\rangle = \sum_{n =1}^\Lambda n^2 |p,w\rangle,
\end{equation}
yielding exactly the same finite sum as in the truncated case \eqref{infinite sums example trunc}.

\subsubsection{Smooth regularisations} \label{sec: smooth regularisations}

The advantage of having reformulated truncation in terms of sharp regularisation is that this allows us to generalise the truncation scheme of \S\ref{sec: trunc to reg} by replacing the sharp cutoff function ${\bf 1}_{|n| \leq \Lambda}$ appearing in the regularised Heisenberg Lie algebra \eqref{Heisenberg log relations Lambda} by a smooth one.

\paragraph{Smooth cutoffs.}

We call $\eta: \mathbb{R}_{\geq 0} \rightarrow \mathbb{R}$ a \textbf{cutoff function} if it is smooth, analytic at zero with $\eta(0) = 1$ and decays faster than any inverse power of the argument $x$ as $x \rightarrow \infty$, that is,
\begin{equation} \label{cutoff function condition}
\eta(x) = O(x^{-r}) \quad \text{as} \quad x\to \infty \quad \text{for any} \quad r \in \mathbb{Z}_{\geq 0}.
\end{equation}
For any $\epsilon > 0$, let $\Heis^\epsilon_{{\rm log},\eta}$ be the \textbf{smoothly regularised Heisenberg Lie algebra} with cutoff function $\eta$ generated by $\b_n^\epsilon, \bar \b_n^\epsilon$ for $n \in \ZZ$ and $\x_0^\epsilon$, $\bar \x_0^\epsilon$ with commutation relations
\begin{subequations} \label{Heisenberg log relations general}
\begin{alignat}{2}
\label{Heisenberg Fourier comm general a} [\b_n^\epsilon, \b_m^\epsilon] &= n \, \eta\big( \tfrac{2 \pi |n| \epsilon}{L} \big) \delta_{n+m, 0}, &\qquad
[\bar\b_n^\epsilon, \bar\b_m^\epsilon] &= n \, \eta\big( \tfrac{2 \pi |n| \epsilon} L \big) \delta_{n+m, 0},\\
\label{Heisenberg Fourier comm general b} [\x_0^\epsilon, \b_n^\epsilon] &= \frac{i}{\sqrt{4 \pi}} \delta_{n,0}, &\qquad
[\bar\x_0^\epsilon, \bar\b_n^\epsilon] &= \frac{i}{\sqrt{4 \pi}} \delta_{n,0}
\end{alignat}
\end{subequations}
for all $m,n \in \ZZ$, with all other Lie brackets between generators being zero as before.
Note that we have switched to using a short-distance cutoff $\epsilon$ rather than the high-energy cutoff $\Lambda$ used until now.
In particular, $\epsilon$ has dimensions of length so that the argument of the cutoff function $\eta$ in \eqref{Heisenberg Fourier comm general a} is dimensionless. Let $\Heis^\epsilon_\eta \subset \Heis^\epsilon_{{\rm log}, \eta}$ be the Lie subalgebra spanned by the modes $\b_n^\epsilon$ and $\bar\b_n^\epsilon$ for $n \in \ZZ$ and let $\Bos^\epsilon_\eta \coloneqq U(\Heis^\epsilon_\eta)$ be its universal enveloping algebra.

Recall from the discussion at the start of \S\ref{sec: Free field realisations} that sharp truncations cannot be applied to the untwisted affine Kac--Moody algebras \eqref{Current mode commutation relations}. Let us briefly comment here on why smooth regularisations, in the above sense but applied directly to the algebra \eqref{Current mode commutation relations}, are not possible either. In order to produce smoothly regularised series, see \eqref{infinite sums example smooth reg} below, we would need such a smooth regularisation to modify the central term, i.e. the second term on the right hand side of \eqref{Current mode commutation relations}, multiplying it by $\eta(\frac{2\pi |n|\epsilon}{L})$. The Jacobi identity for such a smoothly regularised algebra would then enforce $(n +m)\eta(\frac{2\pi (|n + m|)\epsilon}{L}) = n\eta(\frac{2\pi |n|\epsilon}{L}) + m\eta(\frac{2\pi |m|\epsilon}{L})$ for any $n,m\in \ZZ$, which would be incompatible with the conditions for $\eta$ to be a cutoff function.

We can introduce smoothly regularised fields following the exact same steps as for sharply regularised fields in \S\ref{sec: trunc to reg}. Specifically, we introduce the \textbf{smoothly regularised chiral and anti-chiral bosons} $\chi^\epsilon$ and $\bar\chi^\epsilon$ on $S^1$ exactly as in \eqref{chi finer decomposition} and \eqref{chi u decomp pm 0} with $u = x \in [0,L]$ but using smoothly regularised modes $\b_n^\epsilon$, $\bar \b_n^\epsilon$ for $n \in \ZZ$ and $\x_0^\epsilon$, $\bar \x_0^\epsilon$. Using these we then define the \textbf{smoothly regularised $\su_2$-current algebra} at level $1$ as, cf. \eqref{su2 1 currents cyl},
\begin{subequations} \label{smooth reg Fock}
\begin{alignat}{2}
J^{3,\epsilon}(x) &\coloneqq \sqrt{8 \pi} \partial_x \chi^\epsilon(x), &\qquad
J^{\pm,\epsilon}(x) &\coloneqq \frac{2 \pi}{L} \nord{e^{\pm i \sqrt{8 \pi} \chi^\epsilon(x)}},\\
\bar J^{3,\epsilon}(x) &\coloneqq \sqrt{8 \pi} \partial_x \bar \chi^\epsilon(x), &\qquad
\bar J^{\pm,\epsilon}(x) &\coloneqq \frac{2 \pi}{L} \nord{e^{\mp i \sqrt{8 \pi} \bar \chi^\epsilon(x)}}.
\end{alignat}
\end{subequations}
We then introduce the completed Fock spaces $\hFock_{p,w}^\epsilon$ for $p,w \in \ZZ$ over the smoothly regularised Heisenberg Lie algebra $\Heis^\epsilon$, which are defined using the same relations as in \eqref{Fock m n action} but now for the smoothly regularised modes $\b_n^\epsilon$ and $\bar\b_n^\epsilon$ for all $n \in \ZZ$, and we also set
\begin{equation} \label{Fock regularised smooth}
\hFock^\epsilon \coloneqq \bigoplus_{\substack{p,w \in \ZZ\\ p+w \in 2 \ZZ}} \hFock_{p,w}^\epsilon.
\end{equation}
The \textbf{smoothly regularised compact boson} $\X^\epsilon$ and conjugate momentum $P^\epsilon$ are defined in terms of the smoothly regularised modes $\b_n^\epsilon$, $\bar\b_n^\epsilon$ for $n \in\ZZ$ and $\x_0^\epsilon$, $\bar\x_0^\epsilon$. They satisfy
\begin{equation*}
[\X^\epsilon(x), \X^\epsilon(y)] = 0, \qquad
[\X^\epsilon(x), \P^\epsilon(y)] = i \delta_\epsilon(x - y), \qquad
[\P^\epsilon(x), \P^\epsilon(y)] = 0,
\end{equation*}
where we have introduced the \textbf{smoothly regularised Dirac comb}
\begin{equation} \label{smooth regularised Dirac comb}
\delta_\epsilon(x-y) \coloneqq \frac{1}{L} \sum_{n \in \ZZ} \eta\bigg( \frac{2 \pi |n| \epsilon}L \bigg) e^{2\pi i n(x-y)/L}
\end{equation}
which has the property that $\delta_\epsilon(x-y) \to \delta(x-y)$ as the cutoff $\epsilon \to 0$ is removed.

Finally, we could proceed as in the sharply regularised case \eqref{WZW Hamiltonian free realisation shrp reg} and define the smoothly regularised Hamiltonian of the WZW model \eqref{WZW Hamiltonian free realisation} as
\begin{equation*}
\H_0^\epsilon \overset{?}= \int_0^L \dd x \big( \nord{\partial_x \chi^\epsilon(x) \partial_x \chi^\epsilon(x)} + \nord{\partial_x \bar\chi^\epsilon(x) \partial_x \bar\chi^\epsilon(x)} \big),
\end{equation*}
where again we have omitted the vacuum energy term as it will not be needed in \S\ref{sec: sine-Gordon}. However, we would face the same problem as noted in the sharply regularised case, namely that this definition of $\H_0^\epsilon$ would not induce free imaginary-time evolution of the smoothly regularised fields $\chi^\epsilon$ and $\bar\chi^\epsilon$, or indeed of the associated $\su_2$-currents \eqref{smooth reg Fock}. To preserve the free imaginary-time dynamics we will assume that the smooth cutoff function $\eta$ is \emph{positive}, namely $\eta(x) > 0$ for all $x \in \RR_{\geq 0}$, and define the free Hamiltonian instead as
\begin{subequations} \label{WZW Hamiltonian free realisation smth reg}
\begin{equation}
\H_0^\epsilon \coloneqq \frac{2 \pi}{L} \big( \L_0^\epsilon + \bar \L_0^\epsilon \big),
\end{equation}
in terms of the smoothly regularised Virasoro zero-modes which are now defined as, cf. \eqref{L0 bL0 free boson},
\begin{equation} \label{L0 bL0 free boson smth reg}
\L_0^\epsilon \coloneqq \frac 12 \big( \b_0^\epsilon \big)^2 + \sum_{n > 0} \frac{1}{\eta\big( \frac{2 \pi n \epsilon}{L} \big)} \b_{-n}^\epsilon \b_n^\epsilon, \qquad
\bar\L_0^\epsilon \coloneqq \frac 12 \big( \bar\b_0^\epsilon \big)^2 + \sum_{n > 0} \frac{1}{\eta\big( \frac{2 \pi n \epsilon}{L} \big)} \bar\b_{-n}^\epsilon \bar\b_n^\epsilon.
\end{equation}
\end{subequations}
In particular, the non-trivial prefactors in the sum over $n > 0$ are introduced to compensate for the fact that the oscillators $\b_n^\epsilon$ and $\bar\b_n^\epsilon$ are no longer canonically normalised in the smoothly regularised relations \eqref{Heisenberg Fourier comm general a}. As a result, we find that the free imaginary-time evolution of the smoothly regularised chiral and anti-chiral bosons $\chi^\epsilon$ and $\bar\chi^\epsilon$ is restored, namely
\begin{subequations} \label{free evolutions reg}
\begin{equation} \label{bosons free evolution reg}
\chi^\epsilon(x - i \tau) = e^{\tau \H_0^\epsilon} \chi^\epsilon(x) e^{-\tau \H_0^\epsilon}, \qquad
\bar\chi^\epsilon(x + i \tau) = e^{\tau \H_0^\epsilon} \bar\chi^\epsilon(x) e^{-\tau \H_0^\epsilon},
\end{equation}
and similarly for the associated $\su_2$-currents \eqref{smooth reg Fock}, exactly as in \eqref{currents free evolution}, i.e.
\begin{equation} \label{currents free evolution reg}
J^{a, \epsilon}(x - i \tau) = e^{\tau \H_0^\epsilon} J^{a, \epsilon}(x) e^{-\tau \H_0^\epsilon}, \qquad
\bar J^{a, \epsilon}(x + i \tau) = e^{\tau \H_0^\epsilon} \bar J^{a, \epsilon}(x) e^{-\tau \H_0^\epsilon}.
\end{equation}
\end{subequations}
In the limit $\epsilon \to 0$ when the cutoff is removed, we find using the property $\eta(0) = 1$ that the action of $\H_0^\epsilon$ on states in the regularised Fock space \eqref{Fock regularised smooth} generated by creation operators $\b_{-n}^\epsilon$ and $\bar\b_{-n}^\epsilon$ with mode numbers $n$ well below the cutoff, in the sense that $n \ll \frac{L}{2 \pi \epsilon}$, tends to the action of the original WZW Hamiltonian $\H_0$ in the basic representation \eqref{WZW Hamiltonian free realisation}, so that \eqref{WZW Hamiltonian free realisation smth reg} is indeed a regularisation of the latter. In other words, we have
\begin{equation} \label{H0 up to epsilon}
\H_0^\epsilon = \int_0^L \dd x \big( \nord{\partial_x \chi^\epsilon(x) \partial_x \chi^\epsilon(x)} + \nord{\partial_x \bar\chi^\epsilon(x) \partial_x \bar\chi^\epsilon(x)} \big) + O(\epsilon)
\end{equation}
when acting  on states if $\hFock^\epsilon$ well below the cutoff in the above sense.

More generally, we could consider also cases where $\eta(x) \geq 0$ for all $x \in \RR_{\geq 0}$ (rather than $\eta(x) > 0$), for example in the case when $\eta$ is compactly supported. In such cases one should proceed as in the sharply regularised setting by first quotienting by the ideal $\mathfrak I^\epsilon_\eta$ generated by all $\b_n^\epsilon$ and $\bar\b_n^\epsilon$ for which $\eta( 2 \pi |n| \epsilon/L) = 0$. Working in the quotient $\Heis^\epsilon_\eta / \mathfrak I^\epsilon_\eta$, we can then define the free Hamiltonian $\H_0^\epsilon$ exactly as in \eqref{WZW Hamiltonian free realisation smth reg} but where the sum over $n > 0$ in \eqref{L0 bL0 free boson smth reg} is only over modes $\b_n^\epsilon$ and $\bar\b_n^\epsilon$ that have not been quotiented out, i.e. for which $\eta( 2 \pi |n| \epsilon/L) \neq 0$. In the quotiented theory we then maintain the free time evolution \eqref{bosons free evolution reg} of the chiral and anti-chiral smoothly regularised bosons, and of the corresponding currents in \eqref{currents free evolution reg}.

\medskip

In the smoothly regularised theory, divergent infinite sums appearing in any computation will automatically be smoothly regularised by the cutoff function $\eta$.
For example, considering once again the problematic computation \eqref{infinite sums example}, in the present smoothly regularised setting we would find the regularised sum
\begin{equation} \label{infinite sums example smooth reg}
\sum_{n > 0} \b_n^\epsilon \bar\b_n^\epsilon \sum_{m > 0} \b_{-m}^\epsilon \bar\b_{-m}^\epsilon |p,w\rangle = \sum_{n > 0} n^2 \eta\bigg( \frac{2 \pi n \epsilon}{L} \bigg) |p,w\rangle.
\end{equation}

\paragraph{Examples of cutoffs.} There is a huge freedom in the choice of cutoff function $\eta : \RR_{\geq 0} \to \RR$. We see from \eqref{smooth regularised Dirac comb} that $\eta$ determines the Fourier coefficients of the smoothly regularised Dirac comb $\delta_\epsilon(x-y)$. We will generally keep the cutoff function $\eta$ arbitrary, but for completeness we list here some examples related to various standard ways of regularising the Dirac comb.

The heat kernel $K_L(t, x)$ on the circle of circumference $L$ is the fundamental solution to the $1$-dimensional heat equation $\partial_t K_L = \partial_x^2 K_L$ with periodicity $K_L(t, x+L) = K_L(t,x)$ and initial condition the Dirac comb $K_L(0, x) = \delta(x)$. It is given explicitly in terms of the Jacobi theta function $\theta_3(z,q) = \sum_{n \in \ZZ} q^{n^2} e^{2 i nz}$ by
\begin{equation}
K_L(t,x) = \frac{1}{L} \theta_3\bigg( \frac{\pi x}{L}, e^{- 4 \pi^2 t / L^2} \bigg).
\end{equation}
Since by definition $K_L(t,x-y) \to \delta(x - y)$ as $t \to 0$, the heat kernel provides a regularisation of the Dirac comb with time $t$ acting as the regularisation parameter. So if we define
\begin{equation} \label{heat kernel reg}
\delta_{\epsilon, {\rm hk}}(x - y) \coloneqq \frac{1}{L} K_L(\epsilon^2, x-y) = \frac{1}{L} \sum_{n \in \ZZ} e^{- 4 \pi^2 n^2 \epsilon^2 / L^2} e^{2 \pi i n (x-y) / L}
\end{equation}
then $\delta_{\epsilon, {\rm hk}}(x-y) \to \delta(x-y)$ as $\epsilon \to 0$. Comparing \eqref{heat kernel reg} with the general form of the smoothly regularised Dirac comb in \eqref{smooth regularised Dirac comb}, we see that the heat kernel regularisation corresponds to the choice of smooth cutoff function
\begin{equation}
\eta_{\rm hk}(x) = e^{-x^2}.
\end{equation}

Another natural way to regularise the Dirac comb $\delta(x-y)$ is to use point splitting, or the $i\epsilon$-prescription. To do so, we split the sum over $n \in \ZZ$ in \eqref{Dirac comb def} into two separate sums over $n \geq 0$ and $n < 0$, i.e. write $\delta(x-y) = \delta_+(x-y) + \delta_-(x-y)$ with $\delta_+(x-y) = \frac 1L \sum_{n \geq 0} e^{2 \pi i n (x-y)/L}$ and $\delta_-(x-y) = \frac 1L \sum_{n > 0} e^{- 2 \pi i n (x-y)/L}$. Both sums can be made to converge by shifting their arguments by $\pm i \epsilon$, respectively. We then define
\begin{equation} \label{point split reg}
\delta_{\epsilon, {\rm ps}}(x-y) \coloneqq \delta_+(x-y+i\epsilon) + \delta_-(x - y - i \epsilon) = \frac 1L \sum_{n \in \ZZ} e^{- 2 \pi |n| \epsilon / L} e^{2 \pi i n (x-y)/L}.
\end{equation}
Comparing \eqref{point split reg} with the general form of the regularised Dirac comb in \eqref{smooth regularised Dirac comb}, we see that the point splitting regularisation corresponds to the choice of smooth cutoff function
\begin{equation} \label{point splitting cutoff}
\eta_{\rm ps}(x) = e^{-x}.
\end{equation}

To produce another example of a smoothly regularised Dirac comb, consider the sharply regularised Dirac comb from \S\ref{sec: trunc to reg}, which we write here as
\begin{equation}
\delta_{\epsilon, {\rm sh}}(x-y) = \frac 1L \sum_{n = - \lfloor \frac{L}{2 \pi \epsilon} \rfloor}^{\lfloor \frac{L}{2 \pi \epsilon} \rfloor} e^{2 \pi i n (x-y) / L} = \frac 1L \sum_{n \in \ZZ} \eta_{\rm sh}\bigg( \frac{2 \pi |n| \epsilon}{L} \bigg) e^{2 \pi i n (x-y) / L}
\end{equation}
where $\lfloor \frac{L}{2 \pi \epsilon} \rfloor$ denotes the integer part of $\frac{L}{2 \pi \epsilon}$
and $\eta_{\rm sh}(x) = {\bf 1}_{x \leq 1}$ is the sharp cutoff function. We then approximate this sharp cutoff function by a smooth bump function, e.g.
\begin{equation} \label{bump function}
\eta_{\rm bf}(x) = \left\{
\begin{array}{ll}
e^{-\frac{x^2}{1-x^2}} & \text{if} \; x \leq 1\\
0 & \text{if} \; x > 1.
\end{array}
\right.
\end{equation}
to obtain the following smooth regularisation of the Dirac comb
\begin{equation}
\delta_{\epsilon, {\rm bf}}(x-y) = \frac 1L \sum_{n = - \lfloor\frac{L}{2 \pi \epsilon} \rfloor}^{\lfloor\frac{L}{2 \pi \epsilon}\rfloor}
\exp \bigg( \frac{4 \pi^2 n^2 \epsilon^2}{4 \pi^2 n^2 \epsilon^2 - L^2} \bigg) e^{2 \pi i n (x-y) / L}.
\end{equation}
This is a finite sum as a result of the smooth cutoff function \eqref{bump function} being compactly supported.
Note, in particular, that this is an example of a cutoff function for which $\eta_{\rm bf}(x) \geq 0$.

\subsubsection{Asymptotics of harmonic sums} \label{sec: Mellin transform}

In \S\ref{sec: smooth regularisations} we set up a formalism that replaces divergent infinite sums with smoothly regularised sums. We now need a way to efficiently compute the singular part of the asymptotic behaviour of such regularised sum in the limit $\epsilon \to 0$ when the cutoff is removed. A powerful framework for this is given by Mellin transform theory which we now review, closely following \cite{MellinHarmonic}.

\paragraph{Mellin transform.}
If $f : (0, \infty) \to \RR$ is a real-valued locally Lebesgue integrable function then its \textbf{Mellin transform} is
\begin{equation}
\mathscr{M}[f(x);s] = f^\ast(s) = \int_0^\infty f(x) x^{s - 1} \dd x,
\end{equation}
where $s$ takes values on an open vertical strip $\langle\alpha, \beta\rangle \coloneqq \{s = \sigma + it \,| \, \sigma \in (\alpha, \beta)\}$ of the complex plane. The largest such strip where the integral converges for $f$ is its \textbf{fundamental strip}.

We will only consider functions $f$ which decay faster than any power of $x$ at infinity, that is, $f(x) = O(x^{-r})$ as $x \rightarrow \infty$ for all $r \in \mathbb{Z}_{\geq 0}$. The fundamental strip is then given by a half-plane open strip $\langle\alpha, \infty\rangle$ whose left boundary $\alpha \in \mathbb{R}$ is controlled by the leading order asymptotics $f(x) \sim c \, x^{-\alpha}$ as $x \rightarrow 0^+$, for some $c \neq 0$. More generally, the knowledge of lower order terms in the asymptotics of $f(x)$ as $x \rightarrow 0^+$ allows us to extend $f^\ast$ to a meromorphic function on a larger strip. There is a remarkable correspondence between terms in the asymptotic expansion of $f$ and terms in the singular expansion of $f^\ast$, where the singular expansion of a meromorphic function refers to the formal sum of the principal parts at each of its poles following \cite{MellinHarmonic}.

Suppose that as $x \rightarrow 0^+$ we have the asymptotics
\begin{equation*}
    f(x) = \sum _{(\xi, k) \in A} c_{\xi, k} x^{\xi} (\log x)^k + O(x^\gamma)
\end{equation*}
where the sum is over a finite subset $A \subset \big\{ (\xi, k) \in \RR \times \ZZ_{\geq 0} \,|\, -\gamma < -\xi \leq \alpha \big\}$. Then the Mellin transform $f^\ast(s)$ can be analytically continued to a meromorphic function on the open strip $\langle -\gamma, \infty\rangle$ with the singular expansion \cite[Theorem 3]{MellinHarmonic}
\begin{equation*}
    f^\ast(s) \asymp \sum_{(\xi, k) \in A} \frac{(-1)^k k! c_{\xi, k}}{(s + \xi)^{k + 1}}.
\end{equation*}
This result is known as the direct mapping theorem, giving the singular expansion of the Mellin transform in terms of the coefficients of the asymptotic expansion of the original function $f$. There is also a converse that holds under mild conditions, known as the converse mapping theorem, which is stated as follows.

Let $f : (0, \infty) \to \RR$ be continuous, with Mellin transform $f^\ast(s)$ having a fundamental strip $\langle \alpha, \infty\rangle$ for some $\alpha \in \RR$. Assume that $f^\ast(s)$ further admits a meromorphic continuation to the open strip $\langle\gamma, \infty\rangle$ for some $\gamma < \alpha$ with a finite number of poles, and is analytic on $\mathrm{Re}(s) = \gamma$. Assume also that there exists a real number $\eta \in (\alpha, \infty)$ such that 
\begin{equation*}
    f^\ast(s) = O(|s|^{-r}) \quad \text{with} \quad r > 1,
\end{equation*}
when $|s| \rightarrow \infty$ in $\gamma \leq \mathrm{Re}(s) \leq \eta$. If $f^\ast(s)$ admits the singular expansion
\begin{equation*}
    f^\ast(s) \asymp \sum_{(\xi, k) \in A} \frac{d_{\xi, k}}{(s - \xi)^k}
\end{equation*}
for $s \in \langle \gamma, \alpha\rangle$, where the sum is over a finite subset $A \subset \big\{ (\xi, k) \in \RR \times \ZZ_{\geq 0} \,|\, \gamma < \xi \leq \alpha \big\}$, then an asymptotic expansion of $f(x)$ at $0$ is
\begin{equation*}
    f(x) = \sum_{(\xi, k) \in A} d_{\xi, k}
    \frac{(-1)^{k - 1}}{(k - 1)!}x^{-\xi}(\log x)^{k-1} + O(x^{-\gamma}).
\end{equation*}

\begin{example}
The Mellin transform of the point splitting cutoff function \eqref{point splitting cutoff} is given by the Gamma function
\begin{equation*}
\eta_{\rm ps}^\ast(s) = \int_0^\infty e^{-x} x^{s-1} \, \dd x = \Gamma(s).
\end{equation*}
This is analytic on the open strip $\langle 0, \infty \rangle$ and extends to a meromorphic function $\eta^\ast_{\rm ps} : \CC \to \CC$ with singular expansion
\begin{equation*}
\eta_{\rm ps}^\ast(s) \asymp \sum_{n \geq 0} \frac{(-1)^n}{n!} \frac{1}{s+n},
\end{equation*}
whose coefficients coincide with those of the Taylor expansion of \eqref{point splitting cutoff} at $x = 0$.
\end{example}

\begin{example}
The Mellin transform of the sharp high energy cutoff function $\eta_{\rm sh}(x) = {\bf 1}_{x \leq 1}$ is
\begin{equation*}
\eta_{\rm sh}^\ast(s) = \int_0^\infty {\bf 1}_{x \leq 1} x^{s-1} \, \dd x = \int_0^1 x^{s-1} \, \dd x = \frac{1}{s}.
\end{equation*}
This is clearly analytic on the open strip $\langle 0, \infty \rangle$ and defines a meromorphic function on $\CC$ with a simple pole at the origin of residue $1$. The coefficients of the singular expansion $\eta_{\rm sh}^\ast(s) \asymp \frac{1}{s}$ evidently match those of the Taylor expansion $\eta_{\rm sh}(x) = 1$ at $x=0$.
\end{example}

In both of the above examples the Mellin transform could be computed exactly in terms of standard functions whose singular expansion is well known. However, the full power of the direct mapping theorem comes into play in situations when the Mellin transform cannot be computed exactly. Indeed, despite not always having a closed formula for the Mellin transform, its singular expansion can always be obtained directly from the asymptotic expansion of the original function by virtue of the direct mapping theorem.

\paragraph{Harmonic sums.}
Our interest in Mellin transform theory is that it will allow us to compute the singular behaviour of the asymptotics of smoothly regularised divergent series of the form
\begin{equation} \label{regularised sums of interest}
\sum_{n > 0} n^r \eta\bigg( \frac{2 \pi n \epsilon}L \bigg),
\end{equation}
for any $r \in \ZZ_{\geq -1}$, in the limit $\epsilon \to 0$ when the cutoff is removed.

Consider more generally a series of the form $G(x) = \sum_{n > 0} \lambda_n \, g(\mu_n x)$, called a \textbf{harmonic sum}, where $g : \RR_{>0} \to \RR$ is the \textbf{base function}, $(\mu_n)_{n > 0}$ is the sequence of \textbf{frequencies} and $(\lambda_n)_{n > 0}$ the sequence of \textbf{amplitudes}. A key property of the Mellin transform is that, under suitable conditions specified in \cite[Lemma 2]{MellinHarmonic}, when applied to a harmonic sum it separates the frequency-amplitude pair from the base function, in the sense that it factorises as
\begin{equation} \label{Harmonic sum Mellin}
G^\ast(s) = \Lambda(s) g^\ast(s)
\end{equation}
where $\Lambda(s) \coloneqq \sum_{n > 0} \lambda_n (\mu_n)^{-s}$ is the associated Dirichlet series which encodes the information about the frequencies and amplitudes.

In the case \eqref{regularised sums of interest} of interest for us, we have $\lambda_n = n^r$ and $\mu_n = n$ so that the associated Dirichlet series is the shifted Riemann $\zeta$-function $\Lambda(s) = \zeta(s - r)$. Moreover, the base function is a choice of smooth cutoff function $\eta : \RR_{\geq 0} \to \RR$ which satisfies all the assumptions of \cite[Lemma 2]{MellinHarmonic} so the Mellin transform of \eqref{regularised sums of interest} factorises as
\begin{equation} \label{regularised sums of interest Mellin}
\mathscr{M}\left[\sum_{n=1}^\infty n^r \eta(nx); s\right] = \zeta(s - r) \eta^\ast(s).
\end{equation}
Our strategy now is to recover the asymptotics of the sum \eqref{regularised sums of interest} by finding the principal parts of poles on the right-hand side of \eqref{regularised sums of interest Mellin}, then applying the converse mapping theorem.

We have the first two terms in the Laurent expansion $\zeta(s - r) = \frac{1}{s-r-1} + \gamma + O(s)$, where $\gamma$ here is the \textbf{Euler–Mascheroni constant}. As for $\eta^\ast(s)$, since we are assuming that $\eta$ is analytic at $x=0$ with $\eta(0) = 1$, see \S\ref{sec: smooth regularisations}, we have the asymptotic expansion at $x=0$ of the form $\eta(x) = 1 + O(x)$. The direct mapping theorem then tells us that $\eta^\ast$ is meromorphic on the open strip $\langle -1, \infty\rangle$ with singular expansion $\eta^\ast(s) \asymp 1/s$. In the case $r=-1$ we shall also need the constant term in the Laurent expansion of $\eta^\ast(s)$ at $s=0$, which reads
\begin{equation*}
\eta^\ast(s) = \frac{1}{s} + \log C_\eta - \gamma + O(s),
\end{equation*}
where $C_\eta$ is defined by $\log C_\eta - \gamma \coloneqq - \int_0^\infty \eta'(x)\log x\, \dd x$. The shift by the Euler–Mascheroni constant is introduced for later convenience. Indeed, the constant term in the expansion is
\begin{align*}
\frac{\dd}{\dd s} ( s \eta^\ast(s) ) |_{s=0}
= - \frac{\dd}{\dd s} \big( \mathscr M[x \eta'(x); s] \big) \big|_{s=0} = - \mathscr M[x (\log x) \eta'(x); 0] = \log C_\eta - \gamma,
\end{align*}
where in the first two steps we used basic properties of the Mellin transform listed in \cite[Fig. 1]{MellinHarmonic}, explicitly $\mathscr M[x \eta'(x); s] = - s \eta^\ast(s)$ and $\frac{\dd}{\dd s} f^\ast(s) = \mathscr M[ (\log x) f(x); s]$ for any $f$, and the last step is by definition of $C_\eta$.

Putting together the above, when $r=-1$ we find that the Mellin transform of the smoothly regularised harmonic series has the following singular expansion
\begin{equation*}
\mathscr{M}\left[\sum_{n = 1}^\infty \frac{1}{n}\eta(nx); s\right] \asymp \frac{1}{s^2} + \frac{\log C_\eta}s.
\end{equation*}
Using the converse mapping theorem then gives the asymptotic expansion
\begin{equation} \label{harmonic series regularised}
\sum_{n > 0} \frac{1}{n}\eta\bigg( \frac{2 \pi n \epsilon}L \bigg) = - \log\bigg( \frac{2 \pi \epsilon}{L} \bigg) + \log C_\eta + O(\epsilon),
\end{equation}
where we have rescaled the variable $x = 2 \pi \epsilon / L$.
On the other hand, when $r \geq 0$ we find that the right hand side of \eqref{regularised sums of interest Mellin} has the following singular expansion
\begin{equation*}
\mathscr{M}\left[\sum_{n = 1}^\infty n^r \eta(nx); s\right] \asymp \frac{\zeta(-r)}{s} + \frac{C_{\eta, r}}{s-r-1},
\end{equation*}
where $C_{\eta,r} \coloneqq \eta^\ast(r+1) = \int_0^\infty x^r \eta(x) \dd x$. We immediately deduce using the converse mapping theorem that we have the asymptotic expansion
\begin{equation} \label{regularised n^r series}
\sum_{n > 0} n^r \eta\bigg( \frac{2 \pi n \epsilon}L \bigg) = \bigg( \frac{L}{2 \pi} \bigg)^{r+1} \frac{C_{\eta, r}}{\epsilon^{r + 1}} + \zeta(-r) + O(\epsilon).
\end{equation}
where again we have rescaled the variable $x = 2 \pi \epsilon / L$. Note that the asymptotic expansion \eqref{regularised n^r series} can also be derived using the Euler-Maclaurin formula under slightly more stringent conditions on the cutoff function $\eta$.

\paragraph{Enhanced cutoff functions.}
As noticed in \cite{PadillaSmith1}, it is possible to find \textbf{enhanced cutoff functions} $\eta$ such that the coefficient $C_{\eta, r}$ of the singular term in the asymptotic expansion \eqref{regularised n^r series} vanishes for certain values of $r \in \ZZ_{\geq 0}$. In this case, the singular term is removed and the regularised sum \emph{converges} in the limit $\epsilon \rightarrow 0$ to the constant value $\zeta(-r)$. For example, with the choice of smooth cutoff function $\eta(x) = e^{-x}\cos(x)$ we have $C_{\eta, 1} = 0$ so that
\begin{equation*}
\sum_{n > 0} n \, e^{-2\pi n \epsilon / L} \cos\bigg( \frac{2 \pi n \epsilon}L \bigg) \to -\frac{1}{12} \quad \text{as} \quad \epsilon \to 0.
\end{equation*}
Such enhanced cutoff functions can be constructed for any non-negative integer $r$, as well as for any subset of non-negative integers. We refer the reader to \cite[\S II.B]{PadillaSmith1} for more details.

\medskip

In order for the coefficient $C_{\eta, r} = \int_0^\infty x^r \eta(x) \dd x$ of the singular term in \eqref{regularised n^r series} to vanish for some $r \in \ZZ_{\geq 0}$, the cutoff function $\eta$ must necessarily take negative values. Yet this would lead to several issues in our approach. Firstly, recall from \S\ref{sec: smooth regularisations} that in order to ensure that the smoothly regularised Hamiltonian $\H_0^\epsilon$ in \eqref{WZW Hamiltonian free realisation smth reg} generates free imaginary-time evolution on the smoothly regularised chiral and anti-chiral bosons $\chi^\epsilon$ and $\bar\chi^\epsilon$, we had to assume that the smooth cutoff function $\eta$ was positive.
Secondly, and relatedly, within our implementation of the regularisation at the level of the Heisenberg algebra \eqref{Heisenberg log relations general}, enhanced cutoff functions are intrinsically unphysical: for any sufficiently small cutoff $\epsilon$, there exists an $n \in \ZZ_{\geq 1}$ for which the corresponding operator $\b_{-n}^\epsilon \in \mathfrak B^\epsilon_\eta$ creates a state of negative norm.

This phenomenon is familiar in quantum field theory where unphysical regulators are routinely employed. For example, Pauli--Villars regularisation introduces heavy auxiliary fields with propagators of opposite sign to those of the physical fields, and hence with associated negative-norm excitations. Dimensional regularisation (i.e. analytic continuation in the spacetime dimension) is another widely used scheme whose lack of direct physical interpretation is outweighed by the technical simplification it provides, in particular the elimination of all but logarithmic divergences. Nevertheless, to avoid the complications mentioned in the previous paragraph, in our approach we will not consider enhanced regulators.

\medskip

By contrast, it is clear that there is no notion of enhanced cutoff function for the regularised series \eqref{harmonic series regularised} since, regardless of $\eta$, the $\log \epsilon$ divergence will always be present.
The significance of the logarithmic divergence in \eqref{harmonic series regularised} over the power-law divergences in \eqref{regularised n^r series} mirrors their significance in renormalisation, where power-law divergences are regulator-dependent, while logarithmic divergences are not and hence can be considered to be universal.

\subsubsection{Regularised vertex operators}\label{sec: Regularised vertex operators}

Recall from \S\ref{sec: changing radius} that moving from the self-dual compactification radius $R_\circ = 1/\sqrt{2 \pi}$ to a generic radius $R = 2 / \beta$ with $\beta > 0$ is achieved simply by rescaling the compact boson and dual boson as in \eqref{Rescaled Boson Dual Boson}. In particular, reintroducing the parameter $\beta$ in the notation of the compact boson $\Phi_\beta$ and dual boson $\tilde\Phi_\beta$ to specify the compactification radius $R = 2/\beta$, moving between two different radii $R = 2/\beta$ and $R' = 2/\beta$ is achieved by the simple rescaling
\begin{equation}
\beta' \Phi_{\beta'} = \beta \Phi_\beta, \qquad \beta'^{-1} \tilde\Phi_{\beta'} = \beta^{-1} \tilde\Phi_\beta.
\end{equation}
On the other hand, the behaviour of the family of full vertex operators \eqref{vertex operators V rs X tX} under a change of compactification radius is more delicate. Indeed, the definition of these operators depends on a notion of normal ordering taken with respect to the creation and annihilation modes of the chiral and anti-chiral free bosons $\chi(u)$ and $\bar\chi(\bar u)$. However, these fields mix under the rescaling \eqref{Rescaled Boson Dual Boson} via the relation \eqref{phi vs chi}. Consequently, the corresponding mode operators are related by the Bogolyubov transformation \eqref{modes a vs b}, which intertwines creation and annihilation operators and therefore complicates the transformation properties of the normal-ordered products.

In order to both define the analogue of the family of full vertex operators \eqref{vertex operators V rs X tX} at a generic radius $R = 2 / \beta$ and relate such vertex operators at two different radii $R = 2/\beta$ and $R'=2/\beta'$, it will be necessary to work with smoothly regularised bosons as introduced in \S\ref{sec: smooth regularisations}. We shall therefore work with a smoothly regularised version of \S\ref{sec: changing radius}, in particular with the rescaled versions of the smoothly regularised compact boson and dual boson, cf. \eqref{Rescaled Boson Dual Boson}, but restricted to $S^1$ so
\begin{equation} \label{Rescaled Boson Dual Boson reg}
\Phi^\epsilon(x) \coloneqq \frac{\sqrt{8\pi}}{\beta} \X^\epsilon(x), \qquad
\tilde\Phi^\epsilon(x) \coloneqq \frac{\beta}{\sqrt{8\pi}} \tilde\X^\epsilon(x).
\end{equation}
In what follows we fix a value of $\beta$ so we will drop the subscript $\beta$ from all fields for now.

Let $\cnord{\--}_\beta$ denote the normal ordering with respect to the smoothly regularised chiral and anti-chiral free bosons $\phi^\epsilon(x)$ and $\bar \phi^\epsilon(x)$, i.e. using the analogue of the decomposition \eqref{phi finer decomposition} for the smoothly regularised fields. Consider the family of full vertex operators
\begin{align} \label{vertex operators V rs Phi tPhi}
\mathcal V^{\beta, \epsilon}_{r,s}(x) &\coloneqq \cnord{e^{\frac{i \beta}{2} (r+s) \Phi^\epsilon(x) + \frac{4 \pi i}{\beta} (r-s) \tilde\Phi^\epsilon(x)}}_\beta\\
&\, = \cnord{e^{i \sqrt{8\pi} ( \alpha_+ r + \alpha_- s) \phi^\epsilon(x)}}_\beta \; \cnord{e^{i \sqrt{8\pi} ( \alpha_- r + \alpha_+ s) \bar\phi^\epsilon(x)}}_\beta, \notag
\end{align}
for $(r,s) \in \ZZ^2$. In other words, the exponent in the first line is just the smoothly regularised version of \eqref{vertex operators V rs X tX} rewritten using the rescaling \eqref{Rescaled Boson Dual Boson reg}, but the normal ordering used is $\cnord{\--}_\beta$ rather than $\nord{\--} = \cnord{\--}_{\sqrt{8 \pi}}$. In the second line we have split the vertex operator into its chiral and anti-chiral parts using the decomposition \eqref{Phi tPhi decomp} and the definitions \eqref{Alpha PM}. Explicitly, by analogy with the definition \eqref{normal ordered exponential 0mode}, or rather its smoothly regularised version, and the definition of the normal ordering $\cnord{\--}_\beta$ we have
\begin{equation} \label{normal ordered exponential phi}
\cnord{e^{i \alpha \phi^\epsilon(x)}}_\beta = e^{i \alpha \phi^\epsilon_+(x)} e^{i \alpha \phi^\epsilon_0(x)} e^{i \alpha \phi^\epsilon_-(x)}, \qquad
\cnord{e^{i \alpha \bar \phi^\epsilon(x)}}_\beta = e^{i \alpha \bar \phi^\epsilon_+(x)} e^{i \alpha \bar\phi^\epsilon_0(x)} e^{i \alpha \bar \phi^\epsilon_-(x)}
\end{equation}
for $\alpha \in \RR$. Following the discussion at the start of \S\ref{sec: compactified boson}, we note that the zero-mode parts of these operators are not periodic under $x \mapsto x + L$ for general $\alpha \in \RR$, but one checks that the zero-mode part of the particular combination in the second line of \eqref{vertex operators V rs Phi tPhi} is periodic under $x \mapsto x + L$ when acting on the representation \eqref{smooth reg Fock}. Therefore \eqref{vertex operators V rs Phi tPhi} is a well-defined operator on the direct sum of completed smoothly regularised Fock spaces \eqref{smooth reg Fock}.

The easiest way to relate the smoothly regularised full vertex operator \eqref{vertex operators V rs Phi tPhi} for generic $\beta$ to the smoothly regularised version of \eqref{vertex operators V rs X tX}, defined at the self-dual radius, is to relate both to the exponential of field
\begin{equation} \label{exponent Phi vs X}
\frac{i \beta}{2} (r+s) \Phi^\epsilon(x) + \frac{4 \pi i}{\beta} (r-s) \tilde\Phi^\epsilon(x) = \frac{i}{R_\circ} (r+s) \X^\epsilon(x) + 2 \pi i R_\circ (r-s) \tilde\X^\epsilon(x)
\end{equation}
without normal ordering. Of course, working with un-normal-ordered exponentials is only possible since we are using regularized fields.
On the right hand sides of the expressions \eqref{normal ordered exponential phi} we can form the exponential without normal ordering by combining the different exponentials into a single exponential using the Baker-Campbell-Hausdorff formula, which gives
\begin{align}
\cnord{e^{i \alpha \phi^\epsilon(x)}}_\beta &= e^{i \alpha \phi^\epsilon(x)}\exp\left({\frac{\alpha^2}{8\pi} \sum_{n > 0} \frac{1}{n} \eta\left(\frac{2\pi n\epsilon}{L}\right)}\right),\\
\cnord{e^{i \alpha \bar \phi^\epsilon(x)}}_\beta &= e^{i \alpha \bar \phi^\epsilon(x)}\exp\left({\frac{\alpha^2}{8\pi} \sum_{n > 0} \frac{1}{n} \eta\left(\frac{2\pi n\epsilon}{L}\right)}\right).
\end{align}
Combining these results we obtain the desired expression for the family of full vertex operators \eqref{vertex operators V rs Phi tPhi} in terms of a single exponential without normal ordering, namely
\begin{align*}
\mathcal V^{\beta, \epsilon}_{r,s}(x) = e^{\frac{i}{R_\circ} (r+s) \X^\epsilon(x) + 2 \pi i R_\circ (r-s) \tilde\X^\epsilon(x)} \exp\left( \bigg( (r+s)^2 \frac{\beta^2}{16\pi} + (r-s)^2 \frac{4 \pi}{\beta^2} \bigg) \sum_{n > 0} \frac{1}{n} \eta\left(\frac{2\pi n \epsilon}{L}\right)\right),
\end{align*}
where in the first exponential we have used the identity \eqref{exponent Phi vs X}. Since this first exponential is independent of $\beta$, we see that the full vertex operators \eqref{vertex operators V rs Phi tPhi} are related for different values of $\beta$ by the exponential of a multiple of the smoothly regularised divergent sum \eqref{harmonic series regularised}.
Using the asymptotic expansion for the latter as $\epsilon \to 0$, from the results of \S\ref{sec: Mellin transform}, for any $\beta, \beta' > 0$ we obtain the asymptotics
\begin{align*}
\mathcal V^{\beta', \epsilon}_{r,s}(x) \underset{\epsilon \to 0}\sim \bigg( \frac{2 \pi \epsilon}{C_\eta L} \bigg)^{(r+s)^2 \big( \frac{\beta^2}{16\pi} - \frac{\beta'^2}{16\pi} \big) + (r-s)^2 \big( \frac{4 \pi}{\beta^2} - \frac{4 \pi}{\beta'^2} \big)} \mathcal V^{\beta, \epsilon}_{r,s}(x).
\end{align*}
Later we will be particularly interested in the special case when $r=s$, for which the full vertex operator \eqref{vertex operators V rs Phi tPhi} simplifies to $\mathcal V^{\beta, \epsilon}_{r,r}(x) = \cnord {e^{ir\beta \Phi^\epsilon_{\beta}(x)}}_{\beta}$ and only involves the smoothly regularised compact boson $\Phi^\epsilon_\beta$, not the dual boson. Here we have reintroduced the subscript $\beta$ for clarity. Under a change of compactification radius from $R= 2/\beta$ to $R' = 2/\beta'$ we then obtain the simplified asymptotics
\begin{equation}
\cnord {e^{ir\beta' \Phi^\epsilon_{\beta'}(x)}}_{\beta'} \underset{\epsilon \to 0}\sim 
\left( \frac{2\pi \epsilon}{C_\eta L} \right)^{\frac{r^2}{4\pi}(\beta^2 - \beta'^2)}
\cnord {e^{ir\beta \Phi^\epsilon_{\beta}(x)}}_{\beta}.
\end{equation}
In the particular case with $r = \pm 1$ and $\beta' = \sqrt{8 \pi}$, which corresponds to the self-dual radius $R_\circ = 1/ \sqrt{2\pi}$, we find
\begin{align} \label{normal order change exp}
\nord{e^{\pm i \sqrt{8\pi}\X^\epsilon(x)}} &= \exp\left( \bigg( 2 - \frac{\beta^2}{4\pi} \bigg) \sum_{n > 0} \frac{1}{n} \eta\left(\frac{2\pi n \epsilon}{L}\right)\right) \cnord {e^{\pm i \beta \Phi^\epsilon_{\beta}(x)}}_{\beta} \notag\\
&\underset{\epsilon \to 0}\sim 
\left( \frac{2\pi \epsilon}{C_\eta L} \right)^{\frac{\beta^2}{4\pi} - 2} \cnord {e^{\pm i \beta \Phi^\epsilon_{\beta}(x)}}_{\beta}
\end{align}
where for later purposes we have also included the exact expression in $\epsilon$.

\subsection{Effective Hamiltonians} \label{sec: effective Hamiltonians}

In \S\ref{sec: regularisation} we introduced a general procedure for regularising the ultraviolet divergences that arise in the Hamiltonian/operator formulation of the $\su_2$ WZW model at level $1$ on the circle $S^1$ (see \S\ref{sec: background}). This was achieved by imposing a high-energy cutoff $\Lambda$, equivalently a short-distance cutoff $\epsilon$, on the chiral and anti-chiral free bosons $\chi$ and $\bar\chi$ on $S^1$ in terms of which the currents of the $\su_2$ WZW model are realised (see \S\ref{sec: Free field realisations}).

\medskip

The next step in the Wilsonian approach to renormalisation is to isolate a `shell' of higher-energy modes from the regularised free fields $\chi^\Lambda$ and $\bar\chi^\Lambda$, namely those with energies between the original cutoff $\Lambda$ and a lower cutoff $\Lambda' < \Lambda$. Or in terms of short-distance cutoffs, this corresponds to isolating a `shell' of shorter-distance degrees of freedom from the regularised fields $\chi^\epsilon$ and $\bar\chi^\epsilon$, lying between the original cutoff $\epsilon$ and a longer cutoff $\epsilon' > \epsilon$.

Given two short distance cutoffs $\epsilon' > \epsilon > 0$, in \S\ref{sec: short long split} we describe how to split the smoothly regularised Heisenberg Lie algebra $\Heis^\epsilon_\eta$ associated with the shorter cutoff $\epsilon$ into the same Lie algebra $\Heis^{\epsilon'}_\eta$ associated with the longer cutoff $\epsilon'$ and a new `shell' Lie algebra $\Heis^{\epsilon'\bsl\epsilon}_\eta$ representing short distance modes between the two cutoffs $\epsilon$ and $\epsilon'$. Specifically, this split is encoded as an embedding of Lie algebras $\Heis^\epsilon_\eta \to \Heis^{\epsilon'}_\eta \oplus \Heis^{\epsilon'\bsl\epsilon}_\eta$; more precisely, see \eqref{embedding algebras epsilon eta}. This embedding is then used to introduce a representation $\mathcal H^{\epsilon', \epsilon}$ of the smoothly regularised Heisenberg Lie algebra $\Heis^\epsilon_\eta$ which canonically splits into a direct sum \eqref{long short decomposition complete} of `long' and `short' distance subspaces denoted respectively as $\mathcal H^{\epsilon', \epsilon}_l$ and $\mathcal H^{\epsilon', \epsilon}_s$.

\medskip

The final and key step in Wilsonian renormalisation is to `integrate out' the short distance degrees of freedom. The phrase `integrate out' here comes from the action formalism where the short distance degrees of freedom between two energy scales $\Lambda '< \Lambda$ are explicitly integrated out in the path-integral to produce an effective action for the low energy degrees of freedom below the cutoff $\Lambda'$. The key feature of this low energy effective action at the lower cutoff $\Lambda'$ is that it captures the same physics as the original action at the higher energy cutoff $\Lambda$ in the sense that their corresponding partition functions agree.

The purpose of \S\ref{sec: integrating out} is to implement this `integrating out' procedure at the Hamiltonian level. Our starting point is to consider a Hamiltonian $\H^\epsilon$ in the regularised theory at some length scale cutoff $\epsilon > 0$ which is described as a perturbation $\H^\epsilon = \H_0^\epsilon + \Vop(\chi^\epsilon, \bar\chi^\epsilon)$ of the free Hamiltonian $\H_0^\epsilon$ of the smoothly regularised theory by a potential term which couples together the chiral and anti-chiral bosons $\chi^\epsilon$ and $\bar\chi^\epsilon$. This Hamiltonian acts on the space of states $\mathcal H^{\epsilon', \epsilon}$ constructed in \S\ref{sec: short long split} which splits into a direct sum $\mathcal H^{\epsilon', \epsilon} = \mathcal H^{\epsilon', \epsilon}_l \oplus \mathcal H^{\epsilon', \epsilon}_s$ of `long' and `short' distance subspaces. We then seek to construct an effective Hamiltonian $\tilde\H^{\epsilon',\epsilon}_{\rm eff}$ built in terms of the degrees of freedom of the regularised theory at a larger cutoff $\epsilon' > \epsilon$, and thus acting on the long distance subspace $\mathcal H^{\epsilon', \epsilon}_l$, which captures the same dynamics as the original Hamiltonian $\H^\epsilon$ restricted to this subspace. More precisely, our proposal is to define the effective Hamiltonian $\tilde\H^{\epsilon',\epsilon}_{\rm eff}$ by requiring that its imaginary-time evolution operator agrees with that of the original Hamiltonian $\H^\epsilon$ when restricted to states in the long distance subspace $\mathcal H^{\epsilon', \epsilon}_l$.

\medskip

By letting the length scale cutoff $\epsilon$ vary infinitesimally, in \S\ref{sec: RG flow} we derive a Hamiltonian version of Polchinski's equation \cite{Polchinski} for the effective potential $\Vop(\chi^\epsilon, \bar\chi^\epsilon)$ at the cutoff $\epsilon > 0$, see \eqref{RG eq as Polchinski}, which describes the variation of the Hamiltonian $\H^\epsilon = \H_0^\epsilon + \Vop(\chi^\epsilon, \bar\chi^\epsilon)$ with respect to the cutoff $\epsilon$ as we `integrate out' a thin shell of short distance modes. We then use this to relate the variation of the Hamiltonian under this `integrating out' procedure to the beta functions for the couplings of the interaction terms appearing in the effective potential $\Vop(\chi^\epsilon, \bar\chi^\epsilon)$.

\subsubsection{Short/long distance splitting} \label{sec: short long split}

\paragraph{Truncation and sharp regularisation.}

When dealing with truncated chiral and anti-chiral bosons $\chi^\Lambda_{\rm trc}$ and $\bar\chi^\Lambda_{\rm trc}$, separating out the higher-energy modes is straightforward. Indeed, recall that these were defined in \eqref{truncated fields def} as Fourier polynomials on $S^1$. We can single out the modes with energies between $\Lambda$ and a lower energy cutoff $\Lambda' < \Lambda$ by introducing the `shell' fields
\begin{equation*}
\chi^{\Lambda\bsl\Lambda'}_{\rm trc}(u) = \chi^{\Lambda\bsl\Lambda'}_{{\rm trc},+}(u) + \chi^{\Lambda\bsl\Lambda'}_{{\rm trc},-}(u), \qquad \bar\chi^{\Lambda\bsl\Lambda'}_{\rm trc}(\bar u) = \bar\chi^{\Lambda\bsl\Lambda'}_{{\rm trc},+}(\bar u) + \bar\chi^{\Lambda\bsl\Lambda'}_{{\rm trc},-}(\bar u),
\end{equation*}
with the creation and annihilation parts defined by
\begin{equation*}
\chi^{\Lambda\bsl\Lambda'}_{{\rm trc},\pm}(u) \coloneqq \frac{i}{\sqrt{4\pi}} \sum_{\Lambda' < \mp n \leq \Lambda} \frac{1}{n} \b_n e^{-2 \pi i n u/L},\qquad
\bar \chi^{\Lambda\bsl\Lambda'}_{{\rm trc},\pm}(\bar u) \coloneqq \frac{i}{\sqrt{4\pi}} \sum_{\Lambda' < \mp n \leq \Lambda} \frac{1}{n} \bar \b_n e^{2 \pi i n \bar u/L}.
\end{equation*}
This allows us to decompose the truncated fields as 
\begin{equation} \label{trunc field decomposition}
\chi^\Lambda_{\rm trc}(u) = \chi^{\Lambda'}_{\rm trc}(u) + \chi^{\Lambda\bsl\Lambda'}_{\rm trc}(u), \qquad \bar\chi^\Lambda_{\rm trc}(\bar u) = \bar \chi^{\Lambda'}_{\rm trc}(\bar u) + \bar \chi^{\Lambda\bsl\Lambda'}_{\rm trc}(\bar u),
\end{equation}
which is the usual decomposition of truncated fields into high and low energy parts.

In order to obtain an analogue of \eqref{trunc field decomposition} for smoothly regularised fields, following \S\ref{sec: trunc to reg} we first need to reformulate this decomposition for truncated fields in terms of sharply regularised fields.
For any $\Lambda > \Lambda' > 0$, let us therefore introduce the Lie algebra $\Heis^{\Lambda\bsl \Lambda'}$ with generators $\b_n^{\Lambda\bsl\Lambda'}$ and $\bar\b_n^{\Lambda\bsl\Lambda'}$ for $n \in \ZZ \setminus \{0\}$ subject to the relations
\begin{subequations} \label{Heisenberg log relations Lambda high}
\begin{alignat}{2}
\label{Heisenberg Fourier Lambda high comm a} \big[ \b_n^{\Lambda \bsl \Lambda'}, \b_m^{\Lambda \bsl \Lambda'} \big] &= n \big( {\bf 1}_{|n| \leq \Lambda} - {\bf 1}_{|n| \leq \Lambda'} \big) \delta_{n+m, 0}, \\
\label{Heisenberg Fourier Lambda high comm b} \big[ \bar\b_n^{\Lambda \bsl \Lambda'}, \bar\b_m^{\Lambda \bsl \Lambda'} \big] &= n \big( {\bf 1}_{|n| \leq \Lambda} - {\bf 1}_{|n| \leq \Lambda'} \big) \delta_{n+m, 0},
\end{alignat}
\end{subequations}
for $m, n \in \ZZ \setminus \{0\}$. As in the definition of the Lie algebra $\Heis^\Lambda_{\rm log}$ in \S\ref{sec: trunc to reg}, the modes $\b_n^{\Lambda\bsl\Lambda'}$ and $\bar\b_n^{\Lambda\bsl\Lambda'}$ exist for every $n\in \ZZ \setminus \{0\}$ but we enforce that they are central if $n \in \ZZ \setminus \{0\}$ lies outside of the range $\Lambda' < n \leq \Lambda$.
Let $\Bos^{\Lambda\bsl\Lambda'} \coloneqq U(\Heis^{\Lambda\bsl\Lambda'})$ denote the universal enveloping algebra.

The analogue of the high/low energy splitting \eqref{trunc field decomposition} is then implemented in the sharply regularised setting by noting that we have a natural embedding of Lie algebras
\begin{subequations} \label{embedding algebras Lambda}
\begin{equation}
\varsigma_{\Lambda \bsl \Lambda'} : \Heis^\Lambda_{\rm log} \longrightarrow \Heis^{\Lambda'}_{\rm log} \oplus \Heis^{\Lambda \bsl \Lambda'}
\end{equation}
defined on generators as mapping
\begin{alignat}{3}
\x_0^\Lambda &\longmapsto \x_0^{\Lambda'}, &\qquad
\b_0^\Lambda &\longmapsto \b_0^{\Lambda'}, &\qquad
\b_n^\Lambda &\longmapsto \b_n^{\Lambda'} + \b_n^{\Lambda \bsl \Lambda'},\\
\bar \x_0^\Lambda &\longmapsto \bar \x_0^{\Lambda'}, &\qquad
\bar \b_0^\Lambda &\longmapsto \bar \b_0^{\Lambda'}, &\qquad
\bar \b_n^\Lambda &\longmapsto \bar \b_n^{\Lambda'} + \bar \b_n^{\Lambda \bsl \Lambda'}
\end{alignat}
\end{subequations}
for $n \in \ZZ \setminus \{ 0 \}$.
If we introduce the `shell' chiral boson $\chi^{\Lambda\bsl\Lambda'}(x) \coloneqq \chi^{\Lambda\bsl\Lambda'}_+(x) + \chi^{\Lambda\bsl\Lambda'}_-(x)$ and anti-chiral boson $\bar\chi^{\Lambda\bsl\Lambda'}(x) \coloneqq \bar\chi^{\Lambda\bsl\Lambda'}_+(x) + \bar\chi^{\Lambda\bsl\Lambda'}_-(x)$ where the creation and annihilation parts are defined by the same formulae as in \eqref{chi u decomp pm 0} but using the modes $\b_n^{\Lambda\bsl\Lambda'}$ and $\bar\b_n^{\Lambda\bsl\Lambda'}$, then the decompositions in \eqref{trunc field decomposition} get replaced by the statements
\begin{equation}
\varsigma_{\Lambda \bsl \Lambda'}\big( \chi^\Lambda(x) \big) = \chi^{\Lambda'}(x) + \chi^{\Lambda\bsl\Lambda'}(x), \qquad
\varsigma_{\Lambda \bsl \Lambda'}\big( \bar\chi^\Lambda(x) \big) = \bar \chi^{\Lambda'}(x) + \bar \chi^{\Lambda\bsl\Lambda'}(x).
\end{equation}
Note that \eqref{embedding algebras Lambda} induces a morphism of algebras $\varsigma_{\Lambda \bsl \Lambda'} : \Bos^\Lambda \to \Bos^{\Lambda'} \otimes \Bos^{\Lambda \bsl \Lambda'}$.

\paragraph{Smooth regularisation.}
Recall the smoothly regularised Heisenberg Lie algebra $\Heis^\epsilon_\eta$ with cutoff function $\eta : \RR_{>0} \to \RR$ as introduced in \S\ref{sec: smooth regularisations}. Having just reformulated the separation of high and low energy modes as a morphism of sharply regularised Lie algebras \eqref{embedding algebras Lambda}, we can now similarly separate the short and long distance degrees of freedom in the smoothly regularised setting as follows. For any $\epsilon' > \epsilon > 0$ we can introduce, by direct analogy with \eqref{Heisenberg log relations Lambda high}, the Lie algebra $\Heis^{\epsilon'\bsl\epsilon}_\eta$ with generators $\b_n^{\epsilon'\bsl\epsilon}$ and $\bar\b_n^{\epsilon'\bsl\epsilon}$ for $n \in \ZZ \setminus \{ 0 \}$ satisfying the relations
\begin{subequations} \label{Heisenberg shell}
\begin{alignat}{2}
\label{Heisenberg shell a} \big[ 
\b_n^{\epsilon' \bsl \epsilon}, \b_m^{\epsilon' \bsl \epsilon} 
\big] &= 
n \big[ \eta\big( \tfrac{2 \pi |n| \epsilon}{L} \big) - \eta\big( \tfrac{2 \pi |n| \epsilon'}{L} \big) \big] \delta_{n+m, 0},\\
\label{Heisenberg shell b} \big[ \bar\b_n^{\epsilon' \bsl \epsilon}, \bar\b_m^{\epsilon' \bsl \epsilon} \big] &= 
n \big[ \eta\big( \tfrac{2 \pi |n| \epsilon}{L} \big) - \eta\big( \tfrac{2 \pi |n| \epsilon'}{L} \big) \big] \delta_{n+m, 0},
\end{alignat}
\end{subequations}
for all $m, n \in \ZZ \setminus \{ 0 \}$.
As in the sharply regularised case \eqref{embedding algebras Lambda}, the short/long distance splitting is then implemented in the smoothly regularised setting as an embedding of Lie algebras
\begin{subequations} \label{embedding algebras epsilon eta}
\begin{equation}
\varsigma_{\epsilon'\bsl\epsilon}:\Heis^\epsilon_{{\rm log}, \eta} \longrightarrow \Heis^{\epsilon'}_{{\rm log},\eta} \oplus \Heis^{\epsilon'\bsl\epsilon}_\eta
\end{equation}
which is defined on generators as
\begin{alignat}{3}
\x_0^\epsilon &\longmapsto \x_0^{\epsilon'}, &\qquad
\b_0^\epsilon &\longmapsto \b_0^{\epsilon'}, &\qquad
\b_n^\epsilon &\longmapsto \b_n^{\epsilon'} + \b_n^{\epsilon'\bsl\epsilon}, \\
\bar \x_0^\epsilon &\longmapsto \bar \x_0^{\epsilon'}, &\qquad
\bar \b_0^\epsilon &\longmapsto \bar \b_0^{\epsilon'}, &\qquad
\bar \b_n^\epsilon &\longmapsto \bar \b_n^{\epsilon'} + \bar \b_n^{\epsilon'\bsl\epsilon}.
\end{alignat}
\end{subequations}
for every $n \in \ZZ \setminus \{0\}$.
Introducing the `shell' chiral boson $\chi^{\epsilon'\bsl\epsilon}(x) \coloneqq \chi^{\epsilon'\bsl\epsilon}_+(x) + \chi^{\epsilon'\bsl\epsilon}_-(x)$ and the `shell' anti-chiral boson $\bar\chi^{\epsilon'\bsl\epsilon}(x) \coloneqq \bar\chi^{\epsilon'\bsl\epsilon}_+(x) + \bar\chi^{\epsilon'\bsl\epsilon}_-(x)$ where the creation and annihilation parts are defined by the same formulae as in \eqref{chi u decomp pm 0} but using the modes $\b_n^{\epsilon'\bsl\epsilon}$ and $\bar\b_n^{\epsilon'\bsl\epsilon}$, we have the decompositions
\begin{equation} \label{splitting of chi bar chi}
\varsigma_{\epsilon'\bsl\epsilon}\big( \chi^\epsilon(x) \big) = \chi^{\epsilon'}(x) + \chi^{\epsilon'\bsl\epsilon}(x), \qquad
\varsigma_{\epsilon'\bsl\epsilon}\big( \bar\chi^\epsilon(x) \big) = \bar \chi^{\epsilon'}(x) + \bar \chi^{\epsilon'\bsl\epsilon}(x).
\end{equation}
Introducing the `shell' compact boson $\X^{\epsilon'\bsl\epsilon}(x) = \chi^{\epsilon'\bsl\epsilon}(x) + \bar\chi^{\epsilon'\bsl\epsilon}(x)$ we have the corresponding decomposition $\varsigma_{\epsilon'\bsl\epsilon}\big( \X^\epsilon(x) \big) = \X^{\epsilon'}(x) + \X^{\epsilon'\bsl\epsilon}(x)$.
Letting $\Bos^{\epsilon'\bsl\epsilon}_\eta \coloneqq U(\Heis^{\epsilon'\bsl\epsilon}_\eta)$ denote the universal enveloping algebra of $\Heis^{\epsilon'\bsl\epsilon}_\eta$, as usual, we note that \eqref{embedding algebras epsilon eta} induces a morphism of algebras
\begin{equation} \label{Bos embedding algebras epsilon eta}
\varsigma_{\epsilon'\bsl\epsilon} : \Bos^\epsilon_\eta \longrightarrow \Bos^{\epsilon'}_\eta \otimes \Bos^{\epsilon'\bsl\epsilon}_\eta.
\end{equation}

\paragraph{Fock space of shell.}
Recall from \S\ref{sec: compactified boson} the family of Fock spaces $\Fock_{p,w}$ over the Heisenberg Lie algebra $\Heis$ with highest weight state $\ket{p,w}$, for any $p,w \in \ZZ$. The exponentials of the zero-modes $\x_0$ and $\bar\x_0$ of the Lie algebra $\Heis_{\rm log}$ introduced in \eqref{Intertwining} provided intertwining operators between these different Fock spaces. The completed Fock spaces $\hFock_{p,w}$ were defined in \eqref{completed Fock pw} using the natural $\ZZ_{\geq 0}$-grading on $\Fock_{p,w}$ by total mode number. In \S\ref{sec: smooth regularisations} we also introduced the analogues $\Fock^\epsilon_{p,w}$ and $\hFock^\epsilon_{p,w}$ over the smoothly regularised Heisenberg Lie algebra $\Heis^\epsilon_\eta$.

We now consider the construction of Fock spaces over the Lie algebra $\Heis^{\epsilon'\bsl\epsilon}_\eta$ with defining relations \eqref{Heisenberg shell}. The key difference is that since this Lie algebra does not involve zero-modes, we can only define a single Fock space over it, which we will denote by $\Fock^{\epsilon'\bsl\epsilon}$, whose highest weight state $\ket{0}^{\epsilon'\bsl\epsilon}$ is defined by the properties
\begin{equation}
\b_n^{\epsilon'\bsl\epsilon} \ket{0}^{\epsilon'\bsl\epsilon} = 0, \qquad
\bar\b_n^{\epsilon'\bsl\epsilon} \ket{0}^{\epsilon'\bsl\epsilon} = 0
\end{equation}
for all $n \in \ZZ_{> 0}$. There is a natural $\ZZ_{\geq 0}$-grading, cf. \eqref{Fock grading},
\begin{equation} \label{shell Fock}
\Fock^{\epsilon'\bsl\epsilon} = \bigoplus_{d \geq 0} \Fock^{\epsilon'\bsl\epsilon}_d
\end{equation}
defined by letting the highest weight state $\ket{0}^{\epsilon'\bsl\epsilon}$ have grade $0$ and by assigning grade $n$ to the modes $\b_{-n}^{\epsilon'\bsl\epsilon}$ and $\bar\b_{-n}^{\epsilon'\bsl\epsilon}$ for any $n \in \ZZ \setminus \{0\}$. It will be convenient to also introduce the subspace of strictly positive grade states
\begin{equation} \label{shell Fock >0}
\Fock_{>0}^{\epsilon'\bsl\epsilon} = \bigoplus_{d > 0} \Fock^{\epsilon'\bsl\epsilon}_d.
\end{equation}
Noting that the grade $0$ subspace is spanned by the highest weight state, i.e. $\Fock^{\epsilon'\bsl\epsilon}_0 = \CC \ket{0}^{\epsilon'\bsl\epsilon}$, we then have the following important direct sum decomposition
\begin{equation} \label{long short decomposition pre}
\Fock^{\epsilon'\bsl\epsilon} = \CC \ket{0}^{\epsilon'\bsl\epsilon} \oplus \Fock_{>0}^{\epsilon'\bsl\epsilon}.
\end{equation}

We will be interested in the tensor product
\begin{equation} \label{tensor product Fock}
\Fock^{\epsilon', \epsilon} \coloneqq \Fock^{\epsilon'} \otimes \Fock^{\epsilon'\bsl\epsilon}
\end{equation}
which is canonically a module over the direct sum Lie algebra $\Heis^{\epsilon'}_\eta \oplus \Heis^{\epsilon'\bsl\epsilon}_\eta$. Crucially, using the embedding of Lie algebras \eqref{embedding algebras epsilon eta}, or more precisely its restriction $\varsigma_{\epsilon'\bsl\epsilon} : \Heis^\epsilon_\eta \to \Heis^{\epsilon'}_\eta \oplus \Heis^{\epsilon'\bsl\epsilon}_\eta$, the tensor product \eqref{tensor product Fock} defines a representation over $\Heis^\epsilon_\eta$. Consider the subspaces
\begin{subequations} \label{long short distance subspaces}
\begin{align}
\label{long distance subspace} \Fock^{\epsilon', \epsilon}_l &\coloneqq \Fock^{\epsilon'} \otimes \CC \ket{0}^{\epsilon'\bsl\epsilon}, \\
\label{short distance subspace} \Fock^{\epsilon', \epsilon}_s &\coloneqq \Fock^{\epsilon'} \otimes \Fock^{\epsilon'\bsl\epsilon}_{>0}
\end{align}
\end{subequations}
of $\Fock^{\epsilon', \epsilon}$, which we refer to as its \textbf{long and short distance subspaces}, respectively. Indeed, the subspace $\Fock^{\epsilon', \epsilon}_l$ only contains excitations by modes $\b_n^{\epsilon'}$ and $\bar\b_n^{\epsilon'}$ of the Heisenberg Lie algebra $\Heis^{\epsilon'}_\eta$ with the longer length scale cutoff $\epsilon' > \epsilon$. By contrast, the subspace $\Fock^{\epsilon', \epsilon}_s$ contains at least one excitation by the modes $\b_n^{\epsilon'\bsl\epsilon}$ and $\bar\b_n^{\epsilon'\bsl\epsilon}$ from the short distance `shell' Lie algebra $\Heis^{\epsilon'\bsl\epsilon}_\eta$. Importantly, using \eqref{long short decomposition pre} we obtain a direct sum decomposition of vector spaces
\begin{equation} \label{long short decomposition}
\Fock^{\epsilon', \epsilon} = \Fock^{\epsilon', \epsilon}_l \,\oplus\, \Fock^{\epsilon', \epsilon}_s.
\end{equation}

Following \eqref{completed Fock pw}, we consider the completions of the tensor product \eqref{tensor product Fock} and its long and short distance subspaces \eqref{long short distance subspaces} with respect to the total $\ZZ_{\geq 0}$-grading on the tensor product. We will denote these completions by $\mathcal H^{\epsilon', \epsilon}$, $\mathcal H^{\epsilon', \epsilon}_l$ and $\mathcal H^{\epsilon', \epsilon}_s$, respectively. The decomposition \eqref{long short decomposition} extends to these completions, namely
\begin{equation} \label{long short decomposition complete}
\mathcal H^{\epsilon', \epsilon} = \mathcal H^{\epsilon', \epsilon}_l \,\oplus\, \mathcal H^{\epsilon', \epsilon}_s.
\end{equation}
Note that we have the important canonical isomorphism
\begin{equation} \label{iso Hl}
\mathcal H^{\epsilon', \epsilon}_l \cong \hFock^{\epsilon'}
\end{equation}
so that $\mathcal H^{\epsilon', \epsilon}$ includes as a direct summand the completed space of states \eqref{Fock regularised smooth} in the smoothly regularised theory at the larger length scale cutoff $\epsilon' > \epsilon$.

It will be useful to introduce the following piece of notation and terminology. Let
\begin{equation} \label{projections Pl Ps}
\P_l : \mathcal H^{\epsilon', \epsilon} \to \mathcal H^{\epsilon', \epsilon}_l, \qquad
\P_s : \mathcal H^{\epsilon', \epsilon} \to \mathcal H^{\epsilon', \epsilon}_s
\end{equation}
onto the long and short distance subspaces of $\mathcal H^{\epsilon', \epsilon}$ relative to the decomposition \eqref{long short decomposition complete}. We say that an operator $\O\in \End \mathcal H^{\epsilon', \epsilon}$ is \textbf{block diagonal} if it does not mix the short and long distance subspaces, i.e. we have $\P_l \O \P_s = \P_s \O \P_l = 0$ so that $\O = \P_l \O \P_l + \P_s \O \P_s$. On the other hand, we say that $\O$ is \textbf{pure mixing}, or \textbf{block off-diagonal}, if $\P_l \O \P_l = \P_s \O \P_s = 0$ so that $\O = \P_l \O \P_s + \P_s \O \P_l$. We can always decompose any operator $\O\in \End \mathcal H^{\epsilon', \epsilon}$ into its block diagonal and pure mixing parts as $\O = \O_{\rm bd} + \O_{\rm pm}$ where
\begin{equation} \label{bd pm decomposition}
\O_{\rm bd} \coloneqq \P_l \O \P_l + \P_s \O \P_s, \qquad
\O_{\rm pm} \coloneqq \P_l \O \P_s + \P_s \O \P_l.
\end{equation}

\subsubsection{Integrating out a thin shell} \label{sec: integrating out}

\paragraph{Infinitesimally thin shell.}
Now that we are working with an arbitrary smooth cutoff function $\eta$, it makes sense to vary the cutoff $\epsilon$ infinitesimally. Indeed, recall that in the truncation setting of \S\ref{sec: trunc to reg} the cutoff $\Lambda$ necessarily had to be an integer, since $\pm \Lambda$ represented the bounds in the truncated sums, such as in \eqref{truncated fields def b}. Likewise, in the sharply regularised setting, although $\Lambda$ could now be a real number, the sharply regularised algebra \eqref{Heisenberg log relations Lambda} only depends on the integer part $\lfloor\Lambda\rfloor$ of $\Lambda$ since for all $n \in \ZZ$ the condition that $|n| \leq \Lambda$ is equivalent to $|n| \leq \lfloor\Lambda\rfloor$. So in the sharply regularised setting the cutoff $\Lambda$ is effectively still an integer.

By contrast, the smoothly regularised algebra \eqref{Heisenberg log relations general} depends smoothly on the length scale cutoff $\epsilon$, which can be an arbitrary positive real number. Since we will be interested in varying the cutoff smoothly in \S\ref{sec: RG flow}, to derive renormalisation group flows, from this section onward we will focus on the case when the thickness of the `shell' $\delta\epsilon = \epsilon' - \epsilon$ is infinitesimally small, and will only work to first order in $\delta\epsilon$.
We will make extensive use of the fact that in this `thin shell' limit the `shell' Heisenberg Lie algebra \eqref{Heisenberg shell} expands to first order in $\delta\epsilon$ as
\begin{subequations} \label{Heisenberg thin shell}
\begin{alignat}{2}
\label{Heisenberg thin shell a} \big[ 
\b_n^{\epsilon' \bsl \epsilon}, \b_m^{\epsilon' \bsl \epsilon} 
\big] &= 
- n |n| \frac{2 \pi}{L} \eta'\big( \tfrac{2 \pi |n| \epsilon}{L} \big) \delta_{n+m, 0} \delta\epsilon + O( \delta\epsilon^2 ),\\
\label{Heisenberg thin shell b} \big[ \bar\b_n^{\epsilon' \bsl \epsilon}, \bar\b_m^{\epsilon' \bsl \epsilon} \big] &= 
- n |n| \frac{2 \pi}{L} \eta'\big( \tfrac{2 \pi |n| \epsilon}{L} \big) \delta_{n+m, 0} \delta\epsilon + O( \delta\epsilon^2 ).
\end{alignat}
\end{subequations}
for all $n,m \in \ZZ \setminus \{ 0 \}$.
Recall from \S\ref{sec: short long split} that the smoothly regularised short distance chiral and anti-chiral bosons on the circle $S^1$ have no zero-mode contribution, only creation/annihilation parts, i.e. $\chi^{\epsilon'\bsl\epsilon}(u) = \chi^{\epsilon'\bsl\epsilon}_+(u) + \chi^{\epsilon'\bsl\epsilon}_-(u)$ and $\bar\chi^{\epsilon'\bsl\epsilon}(\bar u) = \bar\chi^{\epsilon'\bsl\epsilon}_+(\bar u) + \bar\chi^{\epsilon'\bsl\epsilon}_-(\bar u)$ as in \eqref{chi finer decomposition} but with the zero-modes removed. We can write their mode expansions explicitly as
\begin{equation} \label{X shell decomposition}
\chi^{\epsilon'\bsl\epsilon}(u) = \frac{i}{\sqrt{4\pi}} \sum_{n \neq 0} \frac{1}{n} \b_n^{\epsilon'\bsl\epsilon} e^{-2 \pi i n u/L}, \qquad
\bar \chi^{\epsilon'\bsl\epsilon}(u) = \frac{i}{\sqrt{4\pi}} \sum_{n \neq 0} \frac{1}{n} \bar \b_n^{\epsilon'\bsl\epsilon} e^{2 \pi i n \bar u/L}.
\end{equation}
The $2$-point functions of these chiral and anti-chiral bosons coincide up to conjugation and can be computed in the thin shell limit using \eqref{Heisenberg thin shell} to be, for $u_1, u_2 \in \CC$,
\begin{subequations} \label{shell 2-point functions}
\begin{align}
\null^{\epsilon'\bsl\epsilon} \bra{0} \chi^{\epsilon'\bsl\epsilon}(u_1) \chi^{\epsilon'\bsl\epsilon}(u_2) \ket{0}^{\epsilon'\bsl\epsilon} &= \delta\epsilon \, \Delta^\epsilon_s(u_1, u_2) + O( \delta\epsilon^2 ),\\
\null^{\epsilon'\bsl\epsilon} \bra{0} \bar \chi^{\epsilon'\bsl\epsilon}(\bar u_1) \bar \chi^{\epsilon'\bsl\epsilon}(\bar u_2) \ket{0}^{\epsilon'\bsl\epsilon} &= \delta\epsilon \, \bar\Delta^\epsilon_s(\bar u_1, \bar u_2) + O( \delta\epsilon^2 )
\end{align}
\end{subequations}
where we have defined
\begin{subequations} \label{propagator shell modes}
\begin{align}
\label{propagator shell modes a} \Delta^\epsilon_s(u_1, u_2) &\coloneqq - \frac{1}{2L} \sum_{n > 0} \eta'\big( \tfrac{2 \pi n \epsilon}{L} \big) \exp\big( \tfrac{2 \pi i n}L (u_2 - u_1) \big),\\
\label{propagator shell modes b} \bar\Delta^\epsilon_s(\bar u_1, \bar u_2) &\coloneqq - \frac{1}{2L} \sum_{n > 0} \eta'\big( \tfrac{2 \pi n \epsilon}{L} \big) \exp\big( \tfrac{2 \pi i n}L (\bar u_1 - \bar u_2) \big).
\end{align}
\end{subequations}

\paragraph{Introducing interactions.}

Recall the free Hamiltonian $\H_0^\epsilon$ in the regularised theory with cutoff $\epsilon$ defined in \eqref{WZW Hamiltonian free realisation smth reg}. We are interested in adding to it an interaction term $\Vop(\chi^\epsilon, \bar\chi^\epsilon)$ which couples together the two chiralities of the compact boson $\X^\epsilon$. The interaction may also depend on spatial derivatives of $\chi^\epsilon$ and $\bar\chi^\epsilon$ but we will always omit those from the notation for simplicity. Consider the Hamiltonian
\begin{equation} \label{H = H_0 + V}
\H^\epsilon \coloneqq \H_0^\epsilon + \Vop(\chi^\epsilon, \bar\chi^\epsilon).
\end{equation}
This operator defines an endomorphism $\H^\epsilon \in \End \hFock^\epsilon$ of the completed space of states \eqref{Fock regularised smooth} in the smoothly regularised theory at cutoff $\epsilon$. 
However, in order to be able to `integrate out' the short distance modes we first need to let this Hamiltonian act on the space of states $\mathcal H^{\epsilon', \epsilon}$.

To begin with, the free part $\H_0^\epsilon$ of the Hamiltonian \eqref{H = H_0 + V} generates free time evolution on the fields $\chi^\epsilon$, $\bar\chi^\epsilon$ regularised at the cutoff $\epsilon > 0$, so in the effective Hamiltonian acting on the short distance subspace $\mathcal H^{\epsilon', \epsilon}_l$ we can simply replace this term by the sum of the free Hamiltonian $\H_0^{\epsilon'}$, which generates the same free time evolution on $\chi^{\epsilon'}$, $\bar\chi^{\epsilon'}$ regularised at the larger cutoff $\epsilon' > \epsilon$, and a free Hamiltonian $\H_0^{\epsilon'\bsl\epsilon}$, which generates the free time evolution on the shell fields $\chi^{\epsilon'\bsl\epsilon}$, $\bar\chi^{\epsilon'\bsl\epsilon}$. The latter has an explicit expression similar to \eqref{WZW Hamiltonian free realisation smth reg} assuming that $\eta$ is monotonically decreasing so that the difference $\eta( \frac{2 \pi |n| \epsilon}{L} ) - \eta( \frac{2 \pi |n| \epsilon'}{L} )$ appearing on the right hand side of \eqref{Heisenberg shell} is non-vanishing.

Dealing with the interaction term in \eqref{H = H_0 + V} is more complicated. To let it act on $\mathcal H^{\epsilon', \epsilon}$ we apply the splitting morphism \eqref{embedding algebras epsilon eta} which using \eqref{splitting of chi bar chi} has the effect of replacing the fields $\chi^\epsilon$ and $\bar\chi^\epsilon$ by $\chi^{\epsilon'} + \chi^{\epsilon'\bsl\epsilon}$ and $\bar\chi^{\epsilon'} + \bar\chi^{\epsilon'\bsl\epsilon}$, respectively. This leads to the Hamiltonian
\begin{equation} \label{H = H_0 + V split}
\tilde\H^\epsilon = \H_0^{\epsilon'} + \H_0^{\epsilon'\bsl\epsilon} + \Vop\big( \chi^{\epsilon'} + \chi^{\epsilon'\bsl\epsilon}, \bar\chi^{\epsilon'} + \bar\chi^{\epsilon'\bsl\epsilon} \big).
\end{equation}
The issue, of course, is that this potential term need not preserve the long distance subspace $\mathcal H^{\epsilon', \epsilon}_l$ since it can contain pure mixing terms, in the terminology introduced at the end of \S\ref{sec: short long split}, that would take us out of the long distance space $\mathcal H^{\epsilon', \epsilon}_l$ and into the short distance one $\mathcal H^{\epsilon', \epsilon}_s$. To isolate these problematic terms it is useful to start by expanding the potential in \eqref{H = H_0 + V split} in terms of the short distance bosons $\chi^{\epsilon'\bsl\epsilon}(x)$ and $\bar \chi^{\epsilon'\bsl\epsilon}(x)$ as
\begin{align} \label{potential field expansion}
\Vop\big( \chi^{\epsilon'} + \chi^{\epsilon'\bsl\epsilon}, \bar\chi^{\epsilon'} + \bar\chi^{\epsilon'\bsl\epsilon} \big) &= \Vop ( \chi^{\epsilon'}, \bar\chi^{\epsilon'} )\\
&\quad + \int_0^L \dd x \, \Vop^{\epsilon', (1,0)}(x) \chi^{\epsilon'\bsl\epsilon}(x) + \int_0^L \dd x \, \Vop^{\epsilon', (0,1)}(x) \bar\chi^{\epsilon'\bsl\epsilon}(x) \notag\\
&\quad + \int_0^L \dd x_1 \int_0^L \dd x_2 \Vop^{\epsilon', (2,0)}(x_1,x_2) \chi^{\epsilon'\bsl\epsilon}(x_1) \chi^{\epsilon'\bsl\epsilon}(x_2) \notag\\
&\quad + \int_0^L \dd x_1 \int_0^L \dd x_2 \Vop^{\epsilon', (0,2)}(x_1,x_2) \bar\chi^{\epsilon'\bsl\epsilon}(x_1) \bar\chi^{\epsilon'\bsl\epsilon}(x_2) \notag\\
&\quad + \int_0^L \dd x_1 \int_0^L \dd x_2 \Vop^{\epsilon', (1,1)}(x_1,x_2) \chi^{\epsilon'\bsl\epsilon}(x_1) \bar\chi^{\epsilon'\bsl\epsilon}(x_2) + \ldots \notag
\end{align}
where the coefficient operators $\Vop^{\epsilon', (i,j)}(x)$ depend only on the long distance (anti-)chiral bosons $\chi^{\epsilon'}(x)$, $\bar\chi^{\epsilon'}(x)$ and their $\partial_x$-derivatives.  By using \eqref{X shell decomposition} we can rewrite \eqref{potential field expansion} explicitly as an expansion in terms of chiral and anti-chiral short distance modes as
\begin{align} \label{potential expansion}
\Vop\big( \chi^{\epsilon'} + \chi^{\epsilon'\bsl\epsilon}, \bar\chi^{\epsilon'} + \bar\chi^{\epsilon'\bsl\epsilon} \big) &=\Vop ( \chi^{\epsilon'}, \bar\chi^{\epsilon'} ) + \frac{i}{\sqrt{4 \pi}} \sum_{n \neq 0} \frac{1}{n} \Vop^{\epsilon', (1,0)}_n \b_n^{\epsilon'\bsl\epsilon} + \frac{i}{\sqrt{4 \pi}} \sum_{n \neq 0} \frac{1}{n} \Vop^{\epsilon', (0,1)}_n \bar\b_n^{\epsilon'\bsl\epsilon} \notag\\
&\quad - \frac{1}{4 \pi} \sum_{n, m \neq 0} \frac{1}{n m} \Vop^{\epsilon', (2,0)}_{n,m} \b_n^{\epsilon'\bsl\epsilon} \b_m^{\epsilon'\bsl\epsilon} - \frac{1}{4 \pi} \sum_{n, m \neq 0} \frac{1}{n m} \Vop^{\epsilon', (0,2)}_{n,m} \bar\b_n^{\epsilon'\bsl\epsilon} \bar\b_m^{\epsilon'\bsl\epsilon} \notag\\
&\quad - \frac{1}{4 \pi} \sum_{n, m \neq 0} \frac{1}{n m} \Vop^{\epsilon', (1,1)}_{n,m} \b_n^{\epsilon'\bsl\epsilon} \bar\b_m^{\epsilon'\bsl\epsilon} + \ldots
\end{align}
where for example the coefficients of the purely chiral terms shown are given by
\begin{align}
\Vop^{\epsilon', (1,0)}_n &\coloneqq \int_0^L \dd x \, \Vop^{\epsilon', (1,0)}(x) e^{- 2 \pi i n x / L}, \\
\Vop^{\epsilon', (2,0)}_{n,m} &\coloneqq \int_0^L \dd x_1 \int_0^L \dd x_2 \, \Vop^{\epsilon', (2,0)}(x_1,x_2) e^{- 2 \pi i n x_1 / L} e^{- 2 \pi i m x_2 / L}.
\end{align}
Each sum on the right hand side of \eqref{potential expansion} is a mixture of block diagonal and pure mixing parts. For instance, in the first sum over $n \neq 0$, the chiral modes with $n < 0$ send $\mathcal H^{\epsilon'\bsl\epsilon}_l$ to $\mathcal H^{\epsilon'\bsl\epsilon}_s$ and hence contribute to the pure mixing part, but they also send $\mathcal H^{\epsilon'\bsl\epsilon}_s$ to itself, therefore also contributing to the block diagonal part. The chiral terms with $n > 0$ also contribute to both the block diagonal and pure mixing parts.
Explicitly, for $n > 0$ we have
\begin{alignat}{4}
\big( \b_{-n}^{\epsilon'\bsl\epsilon} \big)_{\rm bd} &= \b_{-n}^{\epsilon'\bsl\epsilon} \P_s, &\qquad
\big( \b_{-n}^{\epsilon'\bsl\epsilon} \big)_{\rm pm} &= \b_{-n}^{\epsilon'\bsl\epsilon} \P_l, &\qquad
\big( \b_n^{\epsilon'\bsl\epsilon} \big)_{\rm bd} &= \P_s \b_n^{\epsilon'\bsl\epsilon}, &\qquad
\big( \b_n^{\epsilon'\bsl\epsilon} \big)_{\rm pm} &= \P_l \b_n^{\epsilon'\bsl\epsilon} \notag\\
\big( \bar\b_{-n}^{\epsilon'\bsl\epsilon} \big)_{\rm bd} &= \bar\b_{-n}^{\epsilon'\bsl\epsilon} \P_s, &\qquad
\big( \bar\b_{-n}^{\epsilon'\bsl\epsilon} \big)_{\rm pm} &= \bar\b_{-n}^{\epsilon'\bsl\epsilon} \P_l, &\qquad
\big( \bar\b_n^{\epsilon'\bsl\epsilon} \big)_{\rm bd} &= \P_s \bar\b_n^{\epsilon'\bsl\epsilon}, &\qquad
\big( \bar\b_n^{\epsilon'\bsl\epsilon} \big)_{\rm pm} &= \P_l \bar\b_n^{\epsilon'\bsl\epsilon}.
\end{alignat}
However, notice that the purely (anti-)chiral double sums on the second line of \eqref{potential expansion} with $n > 0$ and $m = -n$ only contribute to the block diagonal part since they create and then annihilate the same short distance (anti-)chiral excitation.

\paragraph{Effective Hamiltonian.}

Consider the decomposition of \eqref{H = H_0 + V split} into its block diagonal and pure mixing parts
\begin{equation} \label{H epsilon split}
\tilde\H^\epsilon = \tilde\H^\epsilon_{\rm bd} + \tilde\H^\epsilon_{\rm pm}.
\end{equation}
Our goal is to describe an effective Hamiltonian $\tilde\H^{\epsilon',\epsilon}_{\rm eff}$ on the long distance subspace $\mathcal H^{\epsilon', \epsilon}_l$ which captures the same dynamics as the original Hamiltonian \eqref{H epsilon split} restricted to this subspace.

Naively, one could try to define the effective Hamiltonian on the long distance subspace $\mathcal H^{\epsilon', \epsilon}_l$ by simply projecting $\tilde\H^\epsilon$ onto this subspace, namely
\begin{equation} \label{crude approximation Heff}
\tilde\H^{\epsilon',\epsilon}_{\rm eff} \overset{?}= \P_l \tilde\H^\epsilon \P_l \in \End \mathcal H^{\epsilon', \epsilon}_l.
\end{equation}
The problem is that the latter depends only on the block diagonal piece $\tilde\H^\epsilon_{\rm bd}$ and is completely independent of the pure mixing term $\tilde\H^\epsilon_{\rm pm}$. In other words, the approximation in \eqref{crude approximation Heff} would be exact only if the original Hamiltonian had no pure mixing term, i.e. if $\tilde\H^\epsilon_{\rm pm} = 0$.

\medskip

It is important at this point to emphasise that we are \emph{not}, at least in this section, treating the potential term in \eqref{H = H_0 + V split} as a small perturbation. Instead, as previously mentioned, we are working perturbatively in the shell thickness $\delta\epsilon$, specifically to first order. In fact, as we shall see, since the right hand side of the `shell' Heisenberg Lie algebra \eqref{Heisenberg thin shell} is of order $\delta\epsilon$, virtual excursion into the shell subspace $\mathcal H^{\epsilon',\epsilon}_s$ and back will cost a factor of $\delta\epsilon$. But $\tilde \H^\epsilon_{\rm pm}$ is pure mixing so its effect is precisely to move between the short and long distance subspaces. This means that we will effectively be working perturbatively in $\tilde\H^\epsilon_{\rm pm}$, and more specifically to second order. We are therefore treating $\tilde\H^\epsilon_{\rm pm}$ in the decomposition \eqref{H epsilon split} as a perturbation and seeking corrections to the naive effective Hamiltonian \eqref{crude approximation Heff} of second order in $\tilde\H^\epsilon_{\rm pm}$.

\medskip

In order to construct a long distance effective Hamiltonian that reproduces the \emph{dynamics} of the full Hamiltonian $\tilde\H^\epsilon$ on the long distance subspace, instead of focusing on the Hamiltonians themselves, as in \eqref{crude approximation Heff}, we will consider directly their associated evolution operators. Just as in the free theory, see \S\ref{sec: compactified boson}, it will be convenient to consider imaginary-time evolution to construct the effective Hamiltonian.
So consider the exponential operator
\begin{equation} \label{exp T H}
e^{- T \tilde\H^\epsilon} = e^{- T(\tilde\H^\epsilon_{\rm bd} + \tilde\H^\epsilon_{\rm pm})} = e^{- T \tilde\H^\epsilon_{\rm bd}} \mathcal T \, \overleftarrow{\exp} \bigg[ - \int_0^T \dd \tau \tilde\H^\epsilon_{\rm pm}(\tau) \bigg]
\end{equation}
for any $T \in \RR_{>0}$. In the last expression here we have introduced the imaginary-time ordering symbol $\mathcal T$ and the notation
\begin{equation} \label{imaginary time evolve def}
\O(\tau) \coloneqq e^{\tau \tilde\H^\epsilon_{\rm bd}} \O e^{- \tau \tilde\H^\epsilon_{\rm bd}}.
\end{equation}
for the imaginary-time evolution of an operator $\mathsf O \in \End \mathcal H^{\epsilon', \epsilon}$ by the block diagonal part $\tilde\H^\epsilon_{\rm bd}$ of the full interacting Hamiltonian $\tilde\H^\epsilon$. In order to derive the last expression in \eqref{exp T H}, it is useful to consider the operator $U(T) \coloneqq e^{T \tilde\H^\epsilon_{\rm bd}} e^{- T \tilde\H^\epsilon}$. 
This satisfies the first order linear differential equation $\frac{\dd}{\dd T} U(T) = - e^{T \tilde\H^\epsilon_{\rm bd}} \tilde\H^\epsilon_{\rm pm} e^{- T \tilde\H^\epsilon} = - \tilde\H^\epsilon_{\rm pm}(T) U(T)$ and initial condition $U(0) = {\bf 1}$, which specifies it uniquely. But another solution of this differential equation and initial condition is given by the imaginary-time ordered exponential on the right hand side of \eqref{exp T H}.

We can now introduce the effective Hamiltonian $\tilde\H^{\epsilon',\epsilon}_{\rm eff} \in \End \mathcal H^{\epsilon',\epsilon}_l$ by the condition that the associated imaginary-time evolution operator under a time $T > 0$ coincides with the restriction to the long distance subspace $\mathcal H^{\epsilon',\epsilon}_l$ of the imaginary-time evolution operator \eqref{exp T H} in the full theory, i.e.
\begin{equation} \label{Heff def}
e^{- T \tilde\H^{\epsilon',\epsilon}_{\rm eff}} \coloneqq \P_l e^{- T \tilde\H^\epsilon} \P_l.
\end{equation}
It is important to note that such a relation cannot hold for every $T \in \RR_{> 0}$. Indeed, although the left hand side of \eqref{Heff def} is a representation of the semi-group $(\RR_{> 0}, +)$ under composition, i.e. $e^{- T \tilde\H^{\epsilon',\epsilon}_{\rm eff}} e^{- T' \tilde\H^{\epsilon',\epsilon}_{\rm eff}} = e^{- (T+T') \tilde\H^{\epsilon',\epsilon}_{\rm eff}}$ for any $T, T' \in \RR_{> 0}$, this is not the case of the right hand side since between two successive imaginary-time evolutions by times $T'$ and $T$ under the full Hamiltonian $\tilde\H^\epsilon$ we are artificially projecting back to the long distance subspace. Thus $T > 0$ in \eqref{Heff def} will be a \emph{fixed} imaginary time on which the effective Hamiltonian explicitly depends.

Since formula \eqref{Heff def} is defining an effective Hamiltonian in the regularised theory at the length scale cutoff $\epsilon'$, a natural choice for the imaginary time $T$ would be the cutoff $\epsilon'$ itself, i.e. $T = \epsilon'$ the smallest available distance in the regularised theory. However, as we shall see in the main example of \S\ref{sec: sine-Gordon}, the choice of $T$ will not have any impact on the renormalisation group flow of the theory, at least not to the perturbative order we will be working at. So from now on we will keep the imaginary time $T$ in \eqref{Heff def} fixed but arbitrary.

\medskip

Recall that we only wish to work up to second order in the pure mixing part $\tilde\H^\epsilon_{\rm pm}$ of \eqref{H epsilon split} since this will be sufficient to determine the effective Hamiltonian $\tilde\H^{\epsilon',\epsilon}_{\rm eff}$ up to the first order in the shell thickness $\delta\epsilon$. In what follows we begin to unpack the definition \eqref{Heff def} to obtain the desired expression for the effective Hamiltonian $\tilde\H^{\epsilon',\epsilon}_{\rm eff}$ at order $\delta\epsilon$.

Restricting both sides of \eqref{exp T H} to the long distance subspace $\mathcal H^{\epsilon',\epsilon}_l$ and using the definition \eqref{Heff def} of the effective Hamiltonian on the left hand side, we obtain
\begin{equation} \label{P exp T H P}
e^{- T \tilde\H^{\epsilon',\epsilon}_{\rm eff}} = e^{-T \P_l \tilde\H^\epsilon \P_l} \; \P_l \bigg( \mathcal T \, \overleftarrow{\exp} \bigg[ - \int_0^T \dd \tau \tilde\H^\epsilon_{\rm pm}(\tau) \bigg] \bigg) \P_l.
\end{equation}
On the right hand side we have used the fact that
\begin{align*}
\P_l e^{-T \tilde\H^\epsilon_{\rm bd}} &= \P_l e^{-T \P_l \tilde\H^\epsilon \P_l - T \P_s \tilde\H^\epsilon \P_s} = \P_l e^{-T \P_l \tilde\H^\epsilon \P_l} e^{- T \P_s \tilde\H^\epsilon \P_s} = e^{-T \P_l \tilde\H^\epsilon \P_l} \P_l,
\end{align*}
which follows from repeatedly using the identities $\P_l \P_s = \P_s \P_l = 0$, in particular in the second equality to show that $\P_l \tilde\H^\epsilon \P_l$ and $\P_s \tilde\H^\epsilon \P_s$ commute since their product in either order vanishes.

It is convenient to introduce the long distance effective potential $\Vop^{\epsilon',\epsilon}_{\rm eff}(T)$, for an interaction of imaginary-time $T > 0$, as the operator logarithm of the second factor on the right hand side of \eqref{P exp T H P}. Specifically, we set
\begin{equation} \label{effective potential def}
e^{- T \Vop^{\epsilon',\epsilon}_{\rm eff}(T)} \coloneqq \P_l \bigg( \mathcal T \, \overleftarrow{\exp} \bigg[ - \int_0^T \dd \tau \tilde\H^\epsilon_{\rm pm}(\tau) \bigg] \bigg) \P_l.
\end{equation}
This allows us to rewrite the definition of the effective Hamiltonian \eqref{P exp T H P} as
\begin{equation} \label{Heff Veff}
e^{- T \tilde\H^{\epsilon',\epsilon}_{\rm eff}} = e^{-T \P_l \tilde\H^\epsilon \P_l} \; e^{- T \Vop^{\epsilon',\epsilon}_{\rm eff}(T)}.
\end{equation}
We will see shortly that the effective potential is second order in $\tilde\H^\epsilon_{\rm pm}$, so in order to work to the desired first order in $\delta\epsilon$ we will only need an expression for the effective Hamiltonian up to first order in $\Vop^{\epsilon',\epsilon}_{\rm eff}(T)$. Using the expansion of the Baker-Campbell-Hausdorff formula in the form
\begin{equation*}
\log(e^X e^Y) = X + \frac{\ad_X}{1 - e^{- \ad_X}}Y + O(Y^2),
\end{equation*}
which follows from a well-known integral expression for the Baker-Campbell-Hausdorff formula, see for instance \cite{Hall:2015xtd}, and applying it to the right hand side of \eqref{Heff Veff} we obtain
\begin{equation} \label{Heff Veff BCH}
\tilde\H^{\epsilon',\epsilon}_{\rm eff} = \P_l \tilde\H^\epsilon \P_l - \frac{T \ad_{\tilde\H^\epsilon_{\rm bd}}}{1 - e^{T \ad_{\tilde\H^\epsilon_{\rm bd}}}} \Vop^{\epsilon',\epsilon}_{\rm eff}(T) + O\big( \Vop^{\epsilon',\epsilon}_{\rm eff}(T)^2 \big).
\end{equation}
In the second term on the right hand side we have implicitly used the fact that the adjoint action of $\tilde\H^\epsilon_{\rm bd}$ on an operator in $\End \mathcal H^{\epsilon', \epsilon}_l$ coincides with the adjoint action of $\P_l \tilde\H^\epsilon \P_l$.

\paragraph{Block diagonal contribution.}

Since we are working to first order in $\delta\epsilon$, we can obtain an explicit expression for the block diagonal contribution $\P_l \tilde\H^\epsilon \P_l$ to the effective Hamiltonian \eqref{Heff Veff BCH}.
The free Hamiltonian $\H^{\epsilon'}_0 + \H_0^{\epsilon'\bsl\epsilon}$ in \eqref{H = H_0 + V split} satisfies $\P_l (\H^{\epsilon'}_0 + \H_0^{\epsilon'\bsl\epsilon}) \P_l = \H^{\epsilon'}_0$, so we focus on the potential.

In the expansion \eqref{potential expansion} of the potential, any term containing an odd number of chiral or anti-chiral shell oscillators will necessarily have a different number of creation and annihilation operators, and thus such terms cannot contribute to $\P_l \tilde\H^\epsilon \P_l$. In the same vein, the terms in the expansion \eqref{potential expansion} which contain an even number of chiral and anti-chiral shell oscillators will only contribute to $\P_l \tilde\H^\epsilon \P_l$ if they consist of normal ordered creation/annihilation pairs of the same chirality, i.e. if they are of the form
\begin{equation} \label{multiple excursions}
\prod_{i=1}^N \b_{n_i}^{\epsilon'\bsl\epsilon} \b_{-n_i}^{\epsilon'\bsl\epsilon} \prod_{j=1}^M \bar\b_{m_j}^{\epsilon'\bsl\epsilon} \bar\b_{-m_j}^{\epsilon'\bsl\epsilon}
\end{equation}
for some $n_i > 0$ for $i = 1, \ldots, N$ and $m_j > 0$ for $j = 1, \ldots, M$ with $N, M \in \ZZ_{\geq 1}$. Using the `thin shell' Heisenberg Lie algebra relations \eqref{Heisenberg thin shell} we see that such an operator contributes a term to $\P_l \tilde\H^\epsilon \P_l$ which is of order $\delta\epsilon^{N+M}$. Since we are working to first order in $\delta\epsilon$, the only terms contributing to this order are the ones with $(N,M) = (0,0)$, $(1,0)$ and $(0,1)$, i.e. the leading term $\Vop(\chi^{\epsilon'}, \bar\chi^{\epsilon'})$ and the purely chiral/anti-chiral terms in the second line on the right hand side of \eqref{potential expansion}. In other words, we have
\begin{align} \label{potential expansion PlPl}
\P_l \tilde\H^\epsilon \P_l &= \H^{\epsilon'}_0 + \Vop ( \chi^{\epsilon'}, \bar\chi^{\epsilon'} ) + \delta\epsilon \int_0^L \dd x_1 \int_0^L \dd x_2 \Big( \Vop^{\epsilon', (2,0)}(x_1,x_2) \Delta^\epsilon_s(x_1, x_2) \notag\\
&\qquad\qquad\qquad\qquad\qquad\qquad\qquad\qquad\qquad + \Vop^{\epsilon', (0,2)}(x_1,x_2) \bar\Delta^\epsilon_s(x_1, x_2) \Big) + O(\delta\epsilon^2) \notag\\
&= \H^{\epsilon'}_0 + \Vop ( \chi^{\epsilon'}, \bar\chi^{\epsilon'} ) - \frac{\delta\epsilon}{2L} \sum_{n > 0} \eta'\big( \tfrac{2 \pi n \epsilon}{L} \big)  \Big( \Vop^{\epsilon', (2,0)}_{n,-n} + \Vop^{\epsilon', (0,2)}_{n,-n} \Big) + O(\delta\epsilon^2).
\end{align}
Notice that if the potential in \eqref{H = H_0 + V split} was originally normal ordered then we would in fact have $\Vop^{\epsilon', (2,0)}_{n,-n} = \Vop^{\epsilon', (0,2)}_{n,-n} = 0$ and so $\P_l \tilde\H^\epsilon \P_l = \H^{\epsilon'}_0 + \Vop ( \chi^{\epsilon'}, \bar\chi^{\epsilon'} ) + O(\delta\epsilon^2)$. In this case, to first order in $\delta\epsilon$ the block diagonal piece $\P_l \tilde\H^\epsilon \P_l$ of the effective Hamiltonian \eqref{Heff Veff BCH} at the larger cutoff $\epsilon' > \epsilon$ takes the exact same form as the original Hamiltonian \eqref{H = H_0 + V} at the smaller cutoff $\epsilon$.

One can think of each individual chiral and anti-chiral creation/annihilation pair in \eqref{multiple excursions}, namely the individual products $\b_{n_i}^{\epsilon'\bsl\epsilon} \b_{-n_i}^{\epsilon'\bsl\epsilon}$ and $\bar\b_{m_j}^{\epsilon'\bsl\epsilon} \bar\b_{-m_j}^{\epsilon'\bsl\epsilon}$, as virtual excursions into the short distance sector $\mathcal H^{\epsilon'\bsl\epsilon}_s$ of the space of states $\mathcal H^{\epsilon'\bsl\epsilon}$. In other words, even though $\P_l \tilde\H^\epsilon \P_l$ is an operator in $\End \mathcal H^{\epsilon'\bsl\epsilon}_l$, it receives contributions at order $\delta\epsilon$ from processes which involve the temporary excitation of a mode of the short distance (anti-)chiral bosons $\chi^{\epsilon'\bsl\epsilon}$ and $\bar\chi^{\epsilon'\bsl\epsilon}$.

\paragraph{Effective potential contribution.}

We now turn to the evaluation of the $O(\delta\epsilon)$ contribution of the effective potential $\Vop^{\epsilon',\epsilon}_{\rm eff}(T)$ to the effective Hamiltonian $\tilde\H^{\epsilon',\epsilon}_{\rm eff}$, i.e. the second term on the right hand side of \eqref{Heff Veff BCH}.

Expanding the right hand side of \eqref{effective potential def} to second order in $\tilde\H^\epsilon_{\rm pm}$ gives
\begin{align} \label{T exp expand}
e^{- T \Vop^{\epsilon',\epsilon}_{\rm eff}(T)} = \P_l + \frac{1}{2} \int_0^T \dd \tau_1 \int_0^T \dd \tau_2 \, \P_l \Big( \mathcal T \big( \tilde\H^\epsilon_{\rm pm}(\tau_1) \tilde\H^\epsilon_{\rm pm}(\tau_2) \big) \Big) \P_l + O\big( (\tilde\H^\epsilon_{\rm pm})^4 \big).
\end{align}
The absence of the linear term on the right hand side follows using the fact that
\begin{align} \label{Pl V Pl is 0}
\P_l \tilde\H^\epsilon_{\rm pm}(\tau) \P_l &= \P_l e^{\tau \tilde\H^\epsilon_{\rm bd}} (\P_l \tilde\H^\epsilon_{\rm pm} \P_s + \P_s \tilde\H^\epsilon_{\rm pm} \P_l) e^{-\tau \tilde\H^\epsilon_{\rm bd}} \P_l \notag\\
&= \P_l e^{\tau \tilde\H^\epsilon_{\rm bd}} \P_l \tilde\H^\epsilon_{\rm pm} \P_s e^{-\tau \tilde\H^\epsilon_{\rm bd}} \P_l + \P_l e^{\tau \tilde\H^\epsilon_{\rm bd}} \P_s \tilde\H^\epsilon_{\rm pm} \P_l e^{-\tau \tilde\H^\epsilon_{\rm bd}} \P_l = 0
\end{align}
where in the first step we used the definitions \eqref{imaginary time evolve def} and \eqref{bd pm decomposition}. Then the last equality follows from the fact that $\tilde\H^\epsilon_{\rm bd}$ is block diagonal so that $\P_l e^{\tau \tilde\H^\epsilon_{\rm bd}} \P_s = 0$ using $\P_l \P_s = \P_s \P_l = 0$, and similarly we have that $\P_s e^{-\tau \tilde\H^\epsilon_{\rm bd}} \P_l = 0$. A similar argument shows that all the terms in the expansion with an odd power of $\tilde\H^\epsilon_{\rm pm}$ vanish, hence why \eqref{T exp expand} holds to $O \big( (\tilde\H^\epsilon_{\rm pm})^4 \big)$.

Combining this with the expansion of the operator logarithm $\log({\bf 1}+\mathsf A) = \mathsf A + O(\mathsf A^2)$ for some operator $\mathsf A \in \End \mathcal H^{\epsilon', \epsilon}_l$, and noting that $\P_l$ acts as the identity on $\mathcal H^{\epsilon', \epsilon}_l$, we can expand the effective potential \eqref{effective potential def} as
\begin{align} \label{log expansion}
\Vop^{\epsilon',\epsilon}_{\rm eff}(T) &= - \frac{1}{T} \; \log \Bigg( \P_l \bigg( \mathcal T \, \overleftarrow{\exp} \bigg[ - \int_0^T \dd \tau \tilde\H^\epsilon_{\rm pm}(\tau) \bigg] \bigg) \P_l \Bigg) \notag\\
&= - \frac{1}{2T} \int_0^T \dd \tau_1 \int_0^T \dd \tau_2 \, \P_l \Big( \mathcal T \big( \tilde\H^\epsilon_{\rm pm}(\tau_1) \tilde\H^\epsilon_{\rm pm}(\tau_2) \big) \Big) \P_l + O\big( (\tilde\H^\epsilon_{\rm pm})^4 \big).
\end{align}
Note that the $N^{\rm th}$ term in this expansion, i.e. the term of order $(\tilde \H^\epsilon_{\rm pm})^{2N}$, represents $N$ virtual excursions from the long distance subspace into the shell subspace and back; schematically
\begin{equation*}
\begin{tikzcd}[row sep=1mm, column sep=7mm]
& \mathcal H^{\epsilon', \epsilon}_s \arrow[ld, "\P_l \tilde \H^\epsilon \P_s"' near start] &
& \cdots \arrow[ld] &
& \mathcal H^{\epsilon', \epsilon}_s \arrow[ld, "\P_l \tilde \H^\epsilon \P_s"' near start] &\\
\mathcal H^{\epsilon', \epsilon}_l & &
\mathcal H^{\epsilon', \epsilon}_l \arrow[lu, "\P_s \tilde \H^\epsilon \P_l"' near end] & &
\mathcal H^{\epsilon', \epsilon}_l \arrow[lu] & &
\mathcal H^{\epsilon', \epsilon}_l \arrow[lu, "\P_s \tilde \H^\epsilon \P_l"' near end].
\end{tikzcd}
\end{equation*}
In particular, when the operator $\tilde \H^\epsilon_{\rm pm}(\tau)$ for $\tau \in [0,T]$ is acting on the long distance subspace it can be replaced by $\P_s \tilde \H^\epsilon(\tau) \P_l$ and when it is acting on the shell subspace it can be replaced instead by $\P_l \tilde \H^\epsilon(\tau) \P_s$.

It will be convenient to split the double integral over the square $(\tau_1,\tau_2) \in [0, T]^2$ in \eqref{log expansion} into the two sub-regions where the difference $\tau \coloneqq \tau_1 - \tau_2$ is such that $\tau \in [0,T]$ and $\tau \in [-T,0]$. Explicitly, we can rewrite the effective potential \eqref{log expansion} as
\begin{align} \label{Veff split}
\Vop^{\epsilon',\epsilon}_{\rm eff}(T) &= - \frac{1}{2T} \int_0^T \dd \tau \int_0^{T-\tau} \dd \tau_2 \, e^{\tau_2 \ad_{\tilde\H^\epsilon_{\rm bd}}} \P_l \tilde\H^\epsilon_{\rm pm}(\tau) \tilde\H^\epsilon_{\rm pm} \P_l \notag\\
&\qquad\qquad - \frac{1}{2T} \int_{-T}^0 \dd \tau \int_{-\tau}^T \dd \tau_2 \, e^{\tau_2 \ad_{\tilde\H^\epsilon_{\rm bd}}} \P_l \tilde\H^\epsilon_{\rm pm} \tilde\H^\epsilon_{\rm pm}(\tau) \P_l + O\big( (\tilde\H^\epsilon_{\rm pm})^4 \big).
\end{align}
Applying the operator acting on the effective potential in \eqref{Heff Veff BCH} then allows us to perform the integrals over $\tau_2$ explicitly to find
\begin{equation} \label{Veff term in Heff}
- \frac{T \ad_{\tilde\H^\epsilon_{\rm bd}}}{1 - e^{T \ad_{\tilde\H^\epsilon_{\rm bd}}}} \Vop^{\epsilon',\epsilon}_{\rm eff}(T)
= - \frac{1}{2} \int_{-T}^T \dd \tau \, \Wop^\epsilon(\tau) \, \P_l \mathcal T \big( \tilde\H^\epsilon_{\rm pm}(\tau) \tilde\H^\epsilon_{\rm pm} \big) \P_l + O\big( (\tilde\H^\epsilon_{\rm pm})^4 \big)
\end{equation}
where $\Wop^\epsilon(\tau)$ is a weighting operator defined by
\begin{equation} \label{weighting operator}
\Wop^\epsilon(\tau) = \frac{1}{1 - e^{T \ad_{\tilde\H^\epsilon_{\rm bd}}}} \left\{
\begin{array}{ll}
\Big( 1 - e^{(T-\tau) \ad_{\tilde\H^\epsilon_{\rm bd}}} \Big), & \quad \text{if} \; 0 \leq \tau \leq T,\\
\Big( e^{-\tau \ad_{\tilde\H^\epsilon_{\rm bd}}} - e^{T \ad_{\tilde\H^\epsilon_{\rm bd}}} \Big), & \quad \text{if} \; -T \leq \tau < 0.
\end{array}
\right.
\end{equation}
Note, in particular, that $\Wop^\epsilon(\tau)$ is continuous at $\tau = 0$ where it is just the identity operator $\Wop^\epsilon(0) = \id$ and it vanishes at the boundaries of the integration domain, namely $\Wop^\epsilon(\pm T) = 0$.

Finally, by the same argument that led to \eqref{potential expansion PlPl}, the only contribution to the effective potential \eqref{log expansion} at order $\delta\epsilon$ is
\begin{align} \label{effective potential expand detail}
- \frac{T \ad_{\tilde\H^\epsilon_{\rm bd}}}{1 - e^{T \ad_{\tilde\H^\epsilon_{\rm bd}}}} \Vop^{\epsilon',\epsilon}_{\rm eff}(T) &= - \frac{\delta\epsilon}{2} \int_0^T \dd \tau \int_0^L \dd x_1 \int_0^L \dd x_2 \, \Wop^\epsilon(\tau)\\
&\qquad\qquad \times \Big( \Vop^{\epsilon', (1,0)}(x_1; \tau) \Vop^{\epsilon', (1,0)}(x_2) \Delta^\epsilon_s(x_1 - i \tau, x_2)\notag\\
&\qquad\qquad\qquad + \Vop^{\epsilon', (0,1)}(x_1; \tau) \Vop^{\epsilon', (0,1)}(x_2) \bar\Delta^\epsilon_s(x_1 + i \tau, x_2) \Big) \notag\\
&\quad - \frac{\delta\epsilon}{2} \int_{-T}^0 \dd \tau \int_0^L \dd x_1 \int_0^L \dd x_2 \, \Wop^\epsilon(\tau) \notag\\
&\qquad\qquad \times \Big( \Vop^{\epsilon', (1,0)}(x_2) \Vop^{\epsilon', (1,0)}(x_1; \tau) \Delta^\epsilon_s(x_2, x_1 - i \tau)\notag\\
&\qquad\qquad\qquad + \Vop^{\epsilon', (0,1)}(x_2) \Vop^{\epsilon', (0,1)}(x_1; \tau) \bar\Delta^\epsilon_s(x_2, x_1 + i \tau) \Big) + O( \delta\epsilon^2), \notag
\end{align}
where the semicolon notation for the imaginary-time dependence, such as $\Vop^{\epsilon', (1,0)}(x_1; \tau)$, is defined by conjugating by $e^{\tau \tilde\H^\epsilon_{\rm bd}}$ as in \eqref{imaginary time evolve def}, e.g. $\Vop^{\epsilon', (1,0)}(x_1; \tau) \coloneqq e^{\tau \tilde\H^\epsilon_{\rm bd}} \Vop^{\epsilon', (1,0)}(x_1) e^{- \tau \tilde\H^\epsilon_{\rm bd}}$.

It will be convenient to abuse the notation of the imaginary-time ordering symbol $\mathcal T$ here and write the above expression more simply as
\begin{align} \label{effective potential expand}
- \frac{T \ad_{\tilde\H^\epsilon_{\rm bd}}}{1 - e^{T \ad_{\tilde\H^\epsilon_{\rm bd}}}} \Vop^{\epsilon',\epsilon}_{\rm eff}(T) &= - \frac{\delta\epsilon}{2} \int_{-T}^T \dd \tau \int_0^L \dd x_1 \int_0^L \dd x_2 \, \Wop^\epsilon(\tau)\\
&\qquad\qquad \times \mathcal T \Big( \Vop^{\epsilon', (1,0)}(x_1; \tau) \Vop^{\epsilon', (1,0)}(x_2) \Delta^\epsilon_s(x_1 - i \tau, x_2)\notag\\
&\qquad\qquad\qquad + \Vop^{\epsilon', (0,1)}(x_1; \tau) \Vop^{\epsilon', (0,1)}(x_2) \bar\Delta^\epsilon_s(x_1 + i \tau, x_2) \Big) + O( \delta\epsilon^2). \notag
\end{align}
In other words, when the $2$-point functions $\Delta^\epsilon_s(x_1 - i \tau, x_2)$ and $\bar\Delta^\epsilon_s(x_1 + i \tau, x_2)$ of the chiral and anti-chiral shell bosons, defined in \eqref{propagator shell modes}, appear inside an imaginary-time ordered product $\mathcal T$ of two operators, the order of their arguments depends on which of the two operators comes first, as can be seen in the full expression \eqref{effective potential expand detail}.

In order to see why \eqref{effective potential expand detail} holds, we first observe that since the free part $\H_0^{\epsilon'} + \H_0^{\epsilon'\bsl\epsilon}$ in \eqref{H = H_0 + V split} is block diagonal it does not contribute to $\tilde\H^\epsilon_{\rm pm}(\tau)$. The contribution from the potential part is found using the expansion \eqref{potential field expansion} to be
\begin{align} \label{potential field expansion evolution}
\tilde\H^\epsilon_{\rm pm}(\tau) &= \int_0^L \dd x \, \Vop^{\epsilon', (1,0)}(x; \tau) \chi^{\epsilon'\bsl\epsilon}(x - i \tau)_{\rm pm} + \int_0^L \dd x \, \Vop^{\epsilon', (0,1)}(x;\tau) \bar\chi^{\epsilon'\bsl\epsilon}(x + i \tau)_{\rm pm}\\
&\quad + \int_0^L \dd x_1 \int_0^L \dd x_2 \Vop^{\epsilon', (2,0)}(x_1,x_2; \tau) \big( \chi^{\epsilon'\bsl\epsilon}(x_1 - i \tau) \chi^{\epsilon'\bsl\epsilon}(x_2 - i \tau) \big)_{\rm pm} \notag\\
&\quad\quad + \int_0^L \dd x_1 \int_0^L \dd x_2 \Vop^{\epsilon', (0,2)}(x_1,x_2; \tau) \big( \bar\chi^{\epsilon'\bsl\epsilon}(x_1 + i \tau) \bar\chi^{\epsilon'\bsl\epsilon}(x_2 + i \tau) \big)_{\rm pm} \notag\\
&\quad\quad\quad + \int_0^L \dd x_1 \int_0^L \dd x_2 \Vop^{\epsilon', (1,1)}(x_1,x_2; \tau) \big( \chi^{\epsilon'\bsl\epsilon}(x_1 - i \tau) \bar\chi^{\epsilon'\bsl\epsilon}(x_2 + i \tau) \big)_{\rm pm} + \ldots \notag
\end{align}
where the dependence on the imaginary time $\tau$ is defined by conjugating by $e^{\tau \tilde\H^\epsilon_{\rm bd}}$ as in \eqref{imaginary time evolve def}. In particular, it follows using the Baker--Campbell--Hausdorff formula that the imaginary time evolution of $\chi^{\epsilon'\bsl\epsilon}(x)_{\rm pm}$ and $\bar\chi^{\epsilon'\bsl\epsilon}(x)_{\rm pm}$ by $e^{\tau \tilde\H^\epsilon_{\rm bd}}$ coincide with their free evolution by $e^{\tau \H^{\epsilon'\bsl\epsilon}_0}$, up to terms that are either at least quadratic in the short distance modes $\b^{\epsilon'\bsl\epsilon}_n$ or of order at least $\delta\epsilon$.
Similarly, every other term in the expansion \eqref{potential field expansion evolution}, from the second line onwards on the right hand side, is at least of quadratic order in the short distance modes $\b^{\epsilon'\bsl\epsilon}_n$ and of order at least $\delta\epsilon$. Hence, since we are working to first order in $\delta\epsilon$, the only contribution to the right hand side of \eqref{Veff term in Heff} comes from the term written, of order $(\tilde\H_{\rm pm}^\epsilon)^2$, with $\tilde\H^\epsilon_{\rm pm}(\tau)$ replaced simply by $\int_0^L \dd x \, \Vop^{\epsilon', (1,0)}(x; \tau) \chi^{\epsilon'\bsl\epsilon}(x - i\tau)_{\rm pm} + \int_0^L \dd x \, \Vop^{\epsilon', (0,1)}(x;\tau) \bar\chi^{\epsilon'\bsl\epsilon}(x + i\tau)_{\rm pm}$.
The desired result \eqref{effective potential expand detail} now follows since $\P_l \chi^{\epsilon'\bsl\epsilon}(u_1) \chi^{\epsilon'\bsl\epsilon}(u_2) \P_l$ and $\P_l \bar\chi^{\epsilon'\bsl\epsilon}(\bar u_1) \bar\chi^{\epsilon'\bsl\epsilon}(\bar u_2) \P_l$ are just given by multiplication by the shell propagators $\delta\epsilon \, \Delta^\epsilon_s(u_1, u_2)$ and $\delta\epsilon \, \bar\Delta^\epsilon_s(\bar u_1, \bar u_2)$ defined in \eqref{propagator shell modes}.

\subsubsection{Renormalisation group flow} \label{sec: RG flow}

It is useful at this point to summarise the result of \S\ref{sec: integrating out}. We considered in \eqref{H = H_0 + V} a family of interacting Hamiltonians $\H^\epsilon$, labelled by the cutoff scale $\epsilon > 0$, defined by adding to the free Hamiltonian $\H_0^\epsilon$ of the smoothly regularised theory at cutoff $\epsilon$, see \eqref{WZW Hamiltonian free realisation smth reg}, a potential term $\Vop(\chi^\epsilon, \bar\chi^\epsilon)$ of the smoothly regularised chiral and anti-chiral bosons $\chi^\epsilon$ and $\bar\chi^\epsilon$. Starting from a given cutoff $\epsilon > 0$, we introduced an effective Hamiltonian $\tilde\H^{\epsilon',\epsilon}_{\rm eff}$ at a larger cutoff $\epsilon' > \epsilon$ by the requirement in \eqref{Heff def} that its imaginary-time evolution for a fixed time $T > 0$ coincides with that of the original Hamiltonian $\H^\epsilon$ when restricted to states in the long distance subspace \eqref{iso Hl}.
When the cutoff is varied infinitesimally, we worked out this effective Hamiltonian to first order in $\delta\epsilon$. Combining \eqref{Heff Veff BCH} with \eqref{potential expansion PlPl} and \eqref{effective potential expand}, the final expression takes the form
\begin{align} \label{effective Hamiltonian final}
\tilde\H^{\epsilon',\epsilon}_{\rm eff} &= \H^{\epsilon'} + \delta\epsilon \int_0^L \dd x_1 \int_0^L \dd x_2 \Big( \Vop^{\epsilon', (2,0)}(x_1,x_2) \Delta^\epsilon_s(x_1, x_2) + \Vop^{\epsilon', (0,2)}(x_1,x_2) \bar\Delta^\epsilon_s(x_1, x_2) \Big) \notag\\
&\quad - \frac{\delta\epsilon}{2} \int_{-T}^T \dd \tau \int_0^L \dd x_1 \int_0^L \dd x_2 \, \Wop^\epsilon(\tau) \, \mathcal T \Big( \Vop^{\epsilon', (1,0)}(x_1, \tau) \Vop^{\epsilon', (1,0)}(x_2) \Delta^\epsilon_s(x_1-i\tau, x_2) \notag\\
&\qquad\qquad\qquad\qquad\qquad + \Vop^{\epsilon', (0,1)}(x_1, \tau) \Vop^{\epsilon', (0,1)}(x_2) \bar\Delta^\epsilon_s(x_1+i\tau, x_2) \Big) + O( \delta\epsilon^2),
\end{align}
where we recall definition of the propagators $\delta\epsilon \, \Delta^\epsilon_s(u_1, u_2)$ and $\delta\epsilon \, \bar\Delta^\epsilon_s(\bar u_1, \bar u_2)$ for both the shell chiral and anti-chiral bosons $\chi^{\epsilon'\bsl\epsilon}(u)$ and $\bar\chi^{\epsilon'\bsl\epsilon}(u)$ in \eqref{propagator shell modes}.

\paragraph{Cutoff independence.}

Given two distinct cutoffs $\epsilon' > \epsilon$, we cannot directly compare the two Hamiltonians $\H^\epsilon$ and $\H^{\epsilon'}$ since they are defined in the smoothly regularised theories using the same cutoff function $\eta$ but at the different cutoff scales $\epsilon$ and $\epsilon'$, respectively. The effective Hamiltonian $\tilde\H^{\epsilon',\epsilon}_{\rm eff}$ is what enables such a comparison. Indeed, by definition, it encodes the same (imaginary-time $T$) evolution at the longer cutoff scale $\epsilon' > \epsilon$ as the Hamiltonian $\H^\epsilon$ does at the smaller cutoff scale $\epsilon$. We can thus compare $\H^\epsilon$ and $\H^{\epsilon'}$ by considering the difference
\begin{equation} \label{Hamiltonian variation}
\delta \H^{\epsilon'} \coloneqq \H^{\epsilon'} - \tilde\H^{\epsilon',\epsilon}_{\rm eff}
\end{equation}
in the smoothly regularised theory at the cutoff $\epsilon'$. If this difference were zero, i.e. $\delta \H^{\epsilon'} = 0$, then the Hamiltonian $\H^\epsilon$ would be independent of the cutoff $\epsilon$ in the sense that the (imaginary-time $T$) evolution of the Hamiltonians $\H^\epsilon$ and $\H^{\epsilon'}$ for two different cutoffs $\epsilon' > \epsilon$ would coincide when projected onto the long distance subspace $\mathcal H^{\epsilon',\epsilon}_l$. However, it follows from \eqref{effective Hamiltonian final} that in the limit $\delta\epsilon = \epsilon' - \epsilon \to 0$ the difference \eqref{Hamiltonian variation} can be expressed to leading order in $\delta\epsilon$ as
\begin{align} \label{Hamiltonian variation explicit}
\delta \H^{\epsilon'} &= \frac{\delta\epsilon}{2} \int_{-T}^T \dd \tau \int_0^L \dd x_1 \int_0^L \dd x_2 \, \Wop^\epsilon(\tau) \, \mathcal T \Big( \Vop^{\epsilon, (1,0)}(x_1; \tau) \Vop^{\epsilon, (1,0)}(x_2) \Delta^\epsilon_s(x_1-i\tau, x_2)\\
&\qquad\qquad\qquad\qquad\qquad\qquad\qquad\qquad\qquad + \Vop^{\epsilon, (0,1)}(x_1; \tau) \Vop^{\epsilon, (0,1)}(x_2) \bar\Delta^\epsilon_s(x_1+i\tau, x_2) \Big) \notag\\
&\quad - \delta\epsilon \int_0^L \dd x_1 \int_0^L \dd x_2 \Big( \Vop^{\epsilon, (2,0)}(x_1,x_2) \Delta^\epsilon_s(x_1, x_2) + \Vop^{\epsilon, (0,2)}(x_1,x_2) \bar\Delta^\epsilon_s(x_1, x_2) \Big) + O( \delta\epsilon^2). \notag
\end{align}
Although the second term on the right hand side will vanish if the potential $\Vop(\chi^\epsilon, \bar\chi^\epsilon)$ is normal ordered, as explained in \S\ref{sec: integrating out}, there is no reason for the first term to vanish. Indeed, as we will see in the main example in \S\ref{sec: sine-Gordon}, this term typically produces an infinite series in $\epsilon$.

\medskip

The key idea behind Wilsonian renormalisation, in the present Hamiltonian setting, is to allow various parameters in the Hamiltonian $\H^\epsilon$ of the smoothly regularised theory at cutoff $\epsilon$ to depend themselves on the cutoff in such a way as to compensate for the variation \eqref{Hamiltonian variation} of the Hamiltonian as we integrate out short distance degrees of freedom.

In other words, if the family of smoothly regularised interacting Hamiltonians $\H^\epsilon$ depends on a (possibly infinite) collection of parameters $(g_j)$, namely $\H^\epsilon = \H^\epsilon[(g_j)]$, then we let these parameters depend on $\epsilon$ and require that
\begin{equation} \label{RG equation}
\sum_i \epsilon \partial_\epsilon g_i(\epsilon) \frac{\partial \H^\epsilon}{\partial g_i}\big[(g_j(\epsilon))\big] + \lim_{\epsilon' \to \epsilon} \epsilon \frac{\delta\H^{\epsilon'}}{\delta\epsilon}\big[(g_j(\epsilon))\big] = 0.
\end{equation}
Here, the second term encodes the variation \eqref{Hamiltonian variation} of $\H^\epsilon[(g_j(\epsilon))]$ arising from the process of integrating out the short distance degrees of freedom $\chi^{\epsilon'\bsl\epsilon}(u)$ and $\bar\chi^{\epsilon'\bsl\epsilon}(u)$ in a thin shell of thickness $\delta\epsilon$.
The first sum on the left hand side describes the compensatory variation of $\H^\epsilon[(g_j(\epsilon))]$ coming from the explicit dependence of the parameters $g_j(\epsilon)$ on the cutoff.
The condition \eqref{RG equation} is known as the \textbf{renormalisation group equation}. It ensures that the full Hamiltonian $\H^\epsilon[(g_j(\epsilon))]$, with its parameters flowing with the cutoff $\epsilon$, is actually independent of the cutoff $\epsilon$, as it should be since the cutoff is purely artificial and therefore unphysical.

\paragraph{Beta functions.}

To be more explicit, we consider the usual situation where the potential $\Vop(\chi^\epsilon, \bar\chi^\epsilon)$ in \eqref{H = H_0 + V} is given by a linear combination of local operators
\begin{equation} \label{couplings}
\Vop(\chi^\epsilon, \bar\chi^\epsilon) = \sum_j \epsilon^{\Delta_j - 2} g_j(\epsilon) \int_0^L \dd x \, \O^\epsilon_j(x),
\end{equation}
where the densities $\O^\epsilon_j(x)$ are operators on $S^1$ built out of the smoothly regularised (anti-)chiral bosons $\chi^\epsilon(x)$, $\bar\chi^\epsilon(x)$ and their derivatives, with conformal dimensions $\Delta_j$. The parameters $g_j(\epsilon)$ are the \textbf{coupling constants} for each of the local operators and in view of the above discussion they are given an explicit cutoff dependence. We assume that the coupling constants are all dimensionless and so we include an explicit factor of $\epsilon^{\Delta_j - 2}$ to ensure that the potential itself has inverse length dimension, matching the dimension of the Hamiltonian.

As usual, the local operators appearing in the potential $\Vop(\chi^\epsilon, \bar\chi^\epsilon)$ as in \eqref{couplings} fall into three categories. The operator $\int_0^L \dd x \, \O^\epsilon_j(x)$ is said to be
\begin{center}
\textbf{relevant} if $\Delta_j < 2$, \qquad \textbf{marginal} if $\Delta_j = 2$ \qquad and \qquad \textbf{irrelevant} if $\Delta_j > 2$.
\end{center}
Crucially, relevant/irrelevant operators are multiplied by negative/positive powers of the cutoff $\epsilon$, while marginal ones come without any additional cutoff dependent factors.

One should include in the potential \eqref{couplings} of the effective theory at the cutoff $\epsilon$, \emph{all} of the possible local operators compatible with the symmetries of the theory under consideration. Indeed, when performing the integration over the thin shell of thickness $\delta\epsilon$, the complicated $O(\delta\epsilon)$ expression on the right hand side of \eqref{Hamiltonian variation explicit} will typically produce an infinite sum over all possible local operators compatible with the symmetries of the theory. In other words, if we assume that the collection of local operators in \eqref{couplings} is a complete set of all local operators compatible with the symmetries of our theory then we can write
\begin{equation} \label{beta functions}
\lim_{\epsilon' \to \epsilon} \epsilon \frac{\delta\H^{\epsilon'}}{\delta\epsilon}\big[(g_j(\epsilon))\big] = - \sum_j \epsilon^{\Delta_j - 2} \beta_j^{\rm qu}\big[ (g_k(\epsilon) ) \big] \int_0^L \dd x \, \O^\epsilon_j(x)
\end{equation}
for some coefficients $\beta_j^{\rm qu}\big[ (g_k(\epsilon) ) \big]$ depending on the original couplings in \eqref{couplings}.
Substituting \eqref{couplings} and \eqref{beta functions} into the left hand side of the renormalisation group equation \eqref{RG equation}, and setting to zero the coefficients of each local operator $\int_0^L \dd x \, \O_j^\epsilon(x)$, we obtain the beta equation for the corresponding coupling $g_j(\epsilon)$, namely
\begin{equation} \label{beta equations}
\epsilon \partial_\epsilon g_j = (2 - \Delta_j) g_j + \beta_j^{\rm qu}[ (g_k)] \eqqcolon \beta_j[ (g_k)].
\end{equation}
The first term on the right hand side is the `classical' scaling dimension which stems from the dimensionful prefactor $\epsilon^{\Delta_j - 2}$ in front of the coupling in \eqref{couplings}, while the second term is the `quantum' correction coming from integrating out the thin shell degrees of freedom. The sum of both terms defines the beta function $\beta_j[ (g_k)]$ of the coupling $g_j$.

\paragraph{Comparison to Wilson-Polchinski.}
Our renormalisation group equation \eqref{RG equation} for the running of the couplings $g_j(\epsilon)$ in the Hamiltonian $\H^\epsilon[(g_j(\epsilon))]$ bears a very close resemblance to Polchinski's equation \cite{Polchinski} for the running of the couplings in the effective interaction $S^{\rm int}_\Lambda$ with a smooth high energy cutoff $\Lambda$.

To see this, we can use the explicit form \eqref{Hamiltonian variation explicit} of the variation $\delta \H^{\epsilon'}$ to first order in $\delta\epsilon$ to compute the limit $\epsilon' \to \epsilon$ on the left hand side of \eqref{RG equation}. Specifically, dividing \eqref{Hamiltonian variation explicit} by $\delta\epsilon$ and taking the limit $\delta\epsilon \to 0$, all higher order terms which are not explicitly written in \eqref{Hamiltonian variation explicit} disappear. On the other hand, since the couplings are all contained in the potential part of the interacting Hamiltonian \eqref{H = H_0 + V}, we can write the first term on the left hand side of \eqref{RG equation} in terms of the potential as
\begin{equation}
\sum_i \epsilon \partial_\epsilon g_i(\epsilon) \frac{\partial \H^\epsilon}{\partial g_i}\big[(g_j(\epsilon))\big] = \sum_i \epsilon \partial_\epsilon g_i(\epsilon) \frac{\partial \Vop(\chi^\epsilon,\bar\chi^\epsilon)}{\partial g_i}\big[(g_j(\epsilon))\big] \eqqcolon \epsilon \partial_\epsilon \Vop(\chi^\epsilon, \bar\chi^\epsilon).
\end{equation}
Here the right hand side is just a shorthand for the middle expression, namely the derivative $\epsilon \partial_\epsilon$ is to be understood as acting only on the couplings in $\Vop(\chi^\epsilon, \bar\chi^\epsilon)$.
Putting all this together we can then rewrite the renormalisation group equation \eqref{RG equation} as
\begin{align} \label{RG eq as Polchinski}
\epsilon \partial_\epsilon \Vop(\chi^\epsilon, \bar\chi^\epsilon) &= \epsilon \int_0^L \dd x_1 \int_0^L \dd x_2 \Big( \Vop^{\epsilon, (2,0)}(x_1,x_2) \Delta^\epsilon_s(x_1, x_2) + \Vop^{\epsilon, (0,2)}(x_1,x_2) \bar\Delta^\epsilon_s(x_1, x_2) \Big) \notag\\
&\quad - \frac{\epsilon}{2} \int_{-T}^T \dd \tau \int_0^L \dd x_1 \int_0^L \dd x_2 \, \Wop^\epsilon(\tau) \, \mathcal T \Big( \Vop^{\epsilon, (1,0)}(x_1; \tau) \Vop^{\epsilon, (1,0)}(x_2) \Delta^\epsilon_s(x_1-i\tau, x_2) \notag\\
&\qquad\qquad\qquad\qquad\qquad + \Vop^{\epsilon, (0,1)}(x_1; \tau) \Vop^{\epsilon, (0,1)}(x_2) \bar\Delta^\epsilon_s(x_1+i\tau, x_2) \Big).
\end{align}
This equation is structurally of the same form as Polchinski's equation \cite{Polchinski}. We can depict it schematically as follows:
\begin{equation*}
\epsilon \partial_\epsilon \;\;
\raisebox{-12mm}
{\begin{tikzpicture}
\def\R{1}
\def\x{5.5}
\def\y{-1.5}
  \fill[left color   = gray!10!black,
        right color  = gray!10!black,
        middle color = gray!10,
        shading      = axis,
        opacity      = 0.15]
    (\x+\R,\y+.5*\R) -- (\x+\R,\y+2.5*\R)  arc (360:180:\R cm and 0.3cm)
          -- (\x-\R,\y+.5*\R) arc (180:360:\R cm and 0.3cm);
  \fill[top color    = gray!90!,
        bottom color = gray!2,
        middle color = gray!30,
        shading      = axis,
        opacity      = 0.03]
    (\x,\y+2.5*\R) circle (\R cm and 0.3cm);
  \draw (\x-\R,\y+2.5*\R) -- (\x-\R,\y+.5*\R) arc (180:360:\R cm and 0.3cm)
               -- (\x+\R,\y+2.5*\R) ++ (-\R,0) circle (\R cm and 0.3cm);
  \draw[dashed] (\x-\R,\y+.5*\R) arc (180:0:\R cm and 0.3cm);

  \draw[thick, dashed, darkgray] (\x-\R,\y+1.5*\R) arc (180:0:\R cm and 0.3cm);
  \draw[thick, darkgray] (\x-\R,\y+1.5*\R) arc (180:360:\R cm and 0.3cm) node[right=.5mm]{\tiny \color{darkgray} $S^1$};

  \draw[fill, red] (\x,\y+1.2*\R)
                                circle (1.2pt);
  \draw[thick, red] (\x,\y+1.2*\R) -- (\x-0.2\R,\y+.95*\R);
  \draw[thick, red] (\x,\y+1.2*\R) -- (\x+0.05\R,\y+.85*\R);
  \draw[thick, red] (\x,\y+1.2*\R) -- (\x+0.35\R,\y+1.1*\R);
  \draw[thick, red] (\x,\y+1.2*\R) -- (\x+0.3\R,\y+1.45*\R);
  \draw[thick, red] (\x,\y+1.2*\R) -- (\x-0.15\R,\y+1.5*\R);
  \draw[thick, red] (\x,\y+1.2*\R) -- (\x-0.3\R,\y+1.3*\R);
\end{tikzpicture}
}
\;\; = \qquad
\raisebox{-12mm}
{\begin{tikzpicture}
\def\R{1}
\def\x{5.5}
\def\y{-1.5}
  \fill[left color   = gray!10!black,
        right color  = gray!10!black,
        middle color = gray!10,
        shading      = axis,
        opacity      = 0.15]
    (\x+\R,\y+.5*\R) -- (\x+\R,\y+2.5*\R)  arc (360:180:\R cm and 0.3cm)
          -- (\x-\R,\y+.5*\R) arc (180:360:\R cm and 0.3cm);
  \fill[top color    = gray!90!,
        bottom color = gray!2,
        middle color = gray!30,
        shading      = axis,
        opacity      = 0.03]
    (\x,\y+2.5*\R) circle (\R cm and 0.3cm);
  \draw (\x-\R,\y+2.5*\R) -- (\x-\R,\y+.5*\R) arc (180:360:\R cm and 0.3cm)
               -- (\x+\R,\y+2.5*\R) ++ (-\R,0) circle (\R cm and 0.3cm);
  \draw[dashed] (\x-\R,\y+.5*\R) arc (180:0:\R cm and 0.3cm);

  \draw[thick, dashed, darkgray] (\x-\R,\y+1.5*\R) arc (180:0:\R cm and 0.3cm);
  \draw[thick, darkgray] (\x-\R,\y+1.5*\R) arc (180:360:\R cm and 0.3cm) node[right=.5mm]{\tiny \color{darkgray} $S^1$};

  \draw[thick, dotted, black]
    (\x,\y+1.2*\R) .. controls +(.1,.07) and +(-.05,-.3) ..
    (\x+.3*\R,\y+1.5*\R)
      .. controls +(-.3,.2) and +(.05,.2) ..
    (\x,\y+1.2*\R);

  \draw[fill, red] (\x,\y+1.2*\R)
                                circle (1.2pt);
  \draw[thick, red] (\x,\y+1.2*\R) -- (\x-0.25\R,\y+1*\R);
  \draw[thick, red] (\x,\y+1.2*\R) -- (\x-0.1\R,\y+.9*\R);
  \draw[thick, red] (\x,\y+1.2*\R) -- (\x+0.1\R,\y+.9*\R);
  \draw[thick, red] (\x,\y+1.2*\R) -- (\x-0.15\R,\y+1.5*\R);
  \draw[thick, red] (\x,\y+1.2*\R) -- (\x+0.35\R,\y+1.1*\R);
  \draw[thick, red] (\x,\y+1.2*\R) -- (\x-0.3\R,\y+1.3*\R);
\end{tikzpicture}
}
\;\; + \quad \sum \;\;
\raisebox{-12mm}
{\begin{tikzpicture}
\def\R{1}
\def\x{5.5}
\def\y{-1.5}
  \fill[left color   = gray!10!black,
        right color  = gray!10!black,
        middle color = gray!10,
        shading      = axis,
        opacity      = 0.15]
    (\x+\R,\y+.5*\R) -- (\x+\R,\y+2.5*\R)  arc (360:180:\R cm and 0.3cm)
          -- (\x-\R,\y+.5*\R) arc (180:360:\R cm and 0.3cm);
  \fill[top color    = gray!90!,
        bottom color = gray!2,
        middle color = gray!30,
        shading      = axis,
        opacity      = 0.03]
    (\x,\y+2.5*\R) circle (\R cm and 0.3cm);
  \draw (\x-\R,\y+2.5*\R) -- (\x-\R,\y+.5*\R) arc (180:360:\R cm and 0.3cm)
               -- (\x+\R,\y+2.5*\R) ++ (-\R,0) circle (\R cm and 0.3cm);
  \draw[dashed] (\x-\R,\y+.5*\R) arc (180:0:\R cm and 0.3cm);
  \fill[blue!90,
        opacity      = 0.15]
    (\x+\R,\y+1.08*\R) -- (\x+\R,\y+1.92*\R)  arc (0:180:\R cm and 0.3cm)
          -- (\x-\R,\y+1.08*\R) arc (180:0:\R cm and 0.3cm);
  \fill[blue!90,
        opacity      = 0.3]
    (\x+\R,\y+1.08*\R) -- (\x+\R,\y+1.92*\R)  arc (360:180:\R cm and 0.3cm)
          -- (\x-\R,\y+1.08*\R) arc (180:360:\R cm and 0.3cm);

  \draw[thick, dashed, darkgray] (\x-\R,\y+1.5*\R) arc (180:0:\R cm and 0.3cm);
  \draw[thick, darkgray] (\x-\R,\y+1.5*\R) arc (180:360:\R cm and 0.3cm) node[right=.5mm]{\tiny \color{darkgray} $S^1$};

  \draw[thick, dotted, black]
    (\x,\y+1.2*\R) to (\x-0.5*\R,\y+1.5*\R);

  \draw[fill, red] (\x,\y+1.2*\R)
                                circle (1.2pt);
  \draw[thick, red] (\x,\y+1.2*\R) -- (\x-0.1\R,\y+.9*\R);
  \draw[thick, red] (\x,\y+1.2*\R) -- (\x+0.1\R,\y+.9*\R);
  \draw[thick, red] (\x,\y+1.2*\R) -- (\x+0.35\R,\y+1.1*\R);
  \draw[thick, red] (\x,\y+1.2*\R) -- (\x+0.3\R,\y+1.4*\R);

  \draw[fill, blue] (\x-0.5*\R,\y+1.5*\R)
                                circle (1.2pt);
  \draw[thick, blue] (\x-0.5*\R,\y+1.5*\R) -- (\x-0.7*\R,\y+1.65*\R);
  \draw[thick, blue] (\x-0.5*\R,\y+1.5*\R) -- (\x-0.8*\R,\y+1.45*\R);
\end{tikzpicture}
}
\end{equation*}
The left hand side represents the flow of the coupling in front of one of the local operators in the potential \eqref{couplings}. Specifically, the red vertices in the diagrams represent densities $\O_j^\epsilon(x)$, with each solid red line representing a constituent long distance (anti-)chiral boson $\chi^\epsilon(x)$, $\bar\chi^\epsilon(x)$ or their $\partial_x$-derivatives. (We are simplifying the story here for illustration purposes. In the main example of \S\ref{sec: sine-Gordon} below we will be interested in densities $\O_j^\epsilon(x)$ given by products of chiral and anti-chiral vertex operators as in \eqref{smooth reg Fock}, which diagrammatically would be represented by vertices with arbitrarily many legs.) The blue vertex in the last diagram represents a density $\O_j^\epsilon(x-i \tau)$ that is shifted in the imaginary time direction, with each solid blue line representing a constituent long distance (anti-)chiral boson $\chi^\epsilon(x-i\tau)$, $\bar\chi^\epsilon(x+i\tau)$ or their $\partial_x$-derivatives.
The dotted lines represent short distance propagators $\Delta^\epsilon_s$ for the (anti-)chiral shell bosons $\chi^{\epsilon'\bsl\epsilon}$, $\bar\chi^{\epsilon'\bsl\epsilon}$ as defined in \eqref{propagator shell modes}. Each red vertex is integrated over $S^1$ while the blue vertex in the last diagram is integrated over the shaded blue region $[-T, T] \times S^1$ on the cylinder $\RR \times S^1$. The sum in the last diagram is over all ways of building the given density $\O_j^\epsilon(x)$ on the left hand side from two other densities in the potential \eqref{couplings}.

\paragraph{Renormalised trajectories.}

Each solution of the renormalisation group equations \eqref{beta equations} represents a particular quantum field theory built from the smoothly regularised (anti-)chiral bosons $\chi^\epsilon$ and $\bar\chi^\epsilon$. In particular, the Hamiltonian $\H^\epsilon[(g_j(\epsilon))] = \H_0^\epsilon + \Vop(\chi^\epsilon, \bar\chi^\epsilon)$ in \eqref{H = H_0 + V}, with potential given by \eqref{couplings}, describes the same theory at different values of the cutoff $\epsilon > 0$ in the sense described in \S\ref{sec: integrating out}.

\medskip

A fundamentally important class of solutions to the renormalisation group equation \eqref{beta equations} are the constant ones, which occur at the zeros of the beta function. These fixed points typically represent conformal field theories which are intrinsically scale-invariant, allowing the cutoff to be removed. Indeed, because the couplings do not run, the Hamiltonian $\H^\epsilon$ is invariant under the flow and thus independent of the short distance cutoff $\epsilon$.

Any Hamiltonian that is normal-ordered, has only marginal couplings and is block diagonal in the sense introduced at the end of \S\ref{sec: short long split}, defines such a constant solution. Indeed, as we saw in \S\ref{sec: integrating out}, given any block diagonal Hamiltonian $\tilde \H^\epsilon$ at the cutoff $\epsilon$, the effective Hamiltonian at the longer cutoff $\epsilon' > \epsilon$ is given simply by the naive projection \eqref{crude approximation Heff} of $\tilde \H^\epsilon$ onto the long distance subspace $\mathcal H_l^{\epsilon', \epsilon}$, i.e. $\tilde\H^{\epsilon',\epsilon}_{\rm eff} = \P_l \tilde\H^\epsilon \P_l$. In particular, there is no correction coming from the effective potential as in \eqref{Heff Veff BCH}. Moreover, if $\tilde \H^\epsilon$ is normal-ordered then by the discussion after \eqref{potential expansion PlPl} we have $\tilde\H^{\epsilon',\epsilon}_{\rm eff} = \H^{\epsilon'} + O(\delta\epsilon^2)$. In other words, the variation of the Hamiltonian \eqref{Hamiltonian variation} resulting from integrating out a thin shell of short distance modes vanishes to first order in $\delta\epsilon$. The quantum correction to the beta function, introduced in \eqref{beta functions}, then vanishes when evaluated on the set of couplings $g_j^\ast$ of our Hamiltonian $\H^\epsilon$. And if all these couplings are marginal, i.e. dimensionless, then the full beta function vanishes $\beta_j[ (g_k^\ast)] = 0$, implying by \eqref{beta equations} that the couplings $g_j^\ast$ are in fact constant.

The free Hamiltonian $\H_0^\epsilon$ itself, introduced in \eqref{WZW Hamiltonian free realisation smth reg}, corresponds to a constant solution of the renormalisation group equation \eqref{beta equations}, namely the trivial fixed point $g_j^\ast = 0$. This is a smoothly regularised version of the WZW model associated with $\su_2$ at level $1$, or rather of its free field realisation \eqref{WZW Hamiltonian free realisation} in terms of the free boson $\X$ and dual boson $\tilde \X$ compactified at the self-dual radius $R_\circ = 1/\sqrt{2\pi}$ from \S\ref{sec: compactified boson}. The $1$-parameter family of free bosons $\Phi_\beta$ and dual bosons $\tilde\Phi_\beta$ compactified at a generic radius $R = 2/\beta$, introduced in \S\ref{sec: changing radius}, also describe conformal field theories. As we will see in \S\ref{sec: sine-Gordon} below, these correspond to a line of fixed points in the renormalisation group flow parametrised by $\beta$ and connected by a marginal deformation.

\medskip

We are mainly interested in non-trivial, i.e. non-constant, solutions to the renormalisation group equation \eqref{beta equations} which represent massive quantum field theories. The general solution is specified by an initial condition at a fixed scale $\epsilon_0$, namely it can be written as
\begin{equation} \label{RG flow general sol}
g_j(\epsilon) = g_j \big( g_k(\epsilon_0), \epsilon/\epsilon_0 \big)
\end{equation}
for some function $g_j$ of all the initial couplings $g_k(\epsilon_0)$ and the parameter $\epsilon/\epsilon_0 < 1$.

Since we are interested in quantum field theories whose behaviour in the UV is described by marginally relevant deformations of a $2$-dimensional conformal field theory, we are particularly interested in solutions \eqref{RG flow general sol} for which the initial couplings $g_k(\epsilon_0)$ in the UV limit $\epsilon_0 \to 0$ approach those of a $2$-dimensional conformal field theory. In other words, we are interested in solutions $g_j(\epsilon)$ to the renormalisation group equation \eqref{beta equations} which are defined all the way to the UV limit $\epsilon \to 0$ and such that:
\begin{subequations} \label{renormalised trajectory}
\begin{itemize}
  \item[a)] all the irrelevant couplings are switched off in the UV, i.e.
\begin{equation} \label{renormalised trajectory a}
\lim_{\epsilon \to 0} g_j(\epsilon) = 0 \quad \text{if} \quad \Delta_j > 2,
\end{equation}
  \item[b)] the marginal couplings approach the specific dimensionless constants $g_{j\ast}$ parameterising the $2$-dimensional conformal field theory in the UV, i.e.
\begin{equation} \label{renormalised trajectory b}
\lim_{\epsilon \to 0} g_j(\epsilon) = g_{j \ast} \quad \text{if} \quad \Delta_j = 2,
\end{equation}
  \item[c)] the relevant couplings tend to zero sufficiently fast in the UV, namely
\begin{equation} \label{renormalised trajectory c}
g_j(\epsilon) \underset{\epsilon \to 0}\sim (m_j \epsilon)^{2 - \Delta_j} \quad \text{if} \quad \Delta_j < 2
\end{equation}
for some mass scales $m_j$.
\end{itemize}
\end{subequations}
Such a solution describes the \textbf{renormalised trajectory} emanating from the conformal field theory in the UV described by the dimensionless parameters $g_j^\ast$ for $\Delta_j = 2$.

\section{The quantum sine-Gordon model on \texorpdfstring{$S^1$}{S1}} \label{sec: sine-Gordon}

We now apply the general method of Wilsonian renormalisation in the Hamiltonian framework as developed in \S\ref{sec: Renormalisation} to our main example, the quantum sine-Gordon model. In \S\ref{sec: quantum sG as def WZW} we begin by relating this model to the anisotropic deformation of the $\su_2$ WZW at level $1$. In \S\ref{sec: sG thin shell} we then compute the quantum beta functions of the marginal couplings of the anisotropic deformation to derive the renormalisation group flow of the quantum sine-Gordon model. We summarise the result obtained for this flow in \S\ref{sec: RG flow summary} and make some further comments.

\subsection{The Sine-Gordon Hamiltonian} \label{sec: quantum sG as def WZW}

In \S\ref{sec: Anisotropic def} we introduce the anisotropic deformation of the $\su_2$ WZW model at level $1$, where to ensure that the perturbation is well defined we work with the smoothly regularised version, at some cutoff $\epsilon > 0$, of the compactified boson $\X = \chi + \bar\chi$ at the self-dual radius $R_\circ = 1/\sqrt{2 \pi}$.
In \S\ref{sec: usual sG Ham} we relate this theory to the conventional description of the sine-Gordon Hamiltonian, written in terms of the compactified boson $\Phi^\epsilon$ with cutoff dependent radius $R(\epsilon) = 2/\beta(\epsilon)$ and its conjugate momentum $\Pi^\epsilon$, where $\beta(\epsilon)$ is related to one of the couplings of the anisotropic deformation of the $\su_2$ WZW model at level $1$.

\subsubsection{Anisotropic deformation of the WZW model} \label{sec: Anisotropic def}

Recall the Hamiltonian $\H_0$ of the WZW model \eqref{WZW Hamiltonian}. As already emphasised in \S\ref{sec: background}, our goal is to describe quantum field theories on the compact space $S^1$, in the Hamiltonian/operator formalism, whose behaviour in the ultraviolet is given by a marginally relevant deformation of $\H_0$.
And for simplicity we have been focusing throughout this paper on the Lie algebra $\su_2$. 

In what follows we will consider the so called \textbf{anisotropic deformation} of the $\su_2$ WZW model, which breaks the $\su_2$ symmetry of the model down to its Cartan subalgebra $\mathfrak{u}_1$, and whose Hamiltonian is given formally by
\begin{equation}\label{H deformed su(2) WZW}
\H = \H_0 + \frac{g_1}{4\pi} \int_0^L \dd x \big( J^+(x) \bar J^-(x) + J^-(x) \bar J^+(x) \big) + \frac{g_2}{8\pi} \int_0^L \dd x \, J^3(x) \bar J^3(x)
\end{equation}
for some coupling constants $g_1$ and $g_2$, where the factors of $4\pi$ and $8 \pi$ are introduced for later convenience. Recall that $J^a(x)$ and $\bar J^a(x)$ denote the chiral and anti-chiral currents introduced in \S\ref{sec: currents} but expressed in the periodic complex coordinate $u \sim u + L$ on the cylinder $i \RR \times S^1$, as in \eqref{Fourier mode expansion}. Here we are restricting these currents to the real slice $S^1$, whose periodic coordinate we denote by $x \sim x + L$, following the conventions of \S\ref{sec: background}.

The Hamiltonian \eqref{H deformed su(2) WZW} is only `formal' in the sense that, as explained at the end of \S\ref{sec: compactified boson}, the two operators defining the perturbation are integrals of local operators coupling the two chiralities together and are therefore not well defined. Indeed, formal expressions as in \eqref{non-chiral local op example Bos} produce divergent series when acting on the Fock space $\hFock$, see \eqref{infinite sums example}.

\medskip

In order to make sense of the deformed Hamiltonian \eqref{H deformed su(2) WZW} we therefore need to pass over to the smoothly regularised theory at some cutoff $\epsilon > 0$. As explained in \S\ref{sec: smooth regularisations}, the smoothly regularised $\su_2$-currents at level $1$ are then defined by the same expressions \eqref{smooth reg Fock} as the basic representation \eqref{su2 1 currents cyl} but written in terms of the smoothly regularised chiral and anti-chiral bosons $\chi^\epsilon$ and $\bar\chi^\epsilon$.
We can now make sense of the formal expressions for the two perturbing local operators in \eqref{H deformed su(2) WZW} by introducing the regularised counterparts of their densities as
\begin{subequations} \label{O1 O2 definition}
\begin{align}
\label{O1 definition} \mathsf O_1^\epsilon(x) &\coloneqq \frac{1}{4\pi} \Big( J^{+,\epsilon}(x) \bar J^{-, \epsilon}(x) + J^{-,\epsilon}(x) \bar J^{+, \epsilon}(x) \Big) \notag\\
&\,= \frac{\pi}{L^2} \Big( \nord{e^{i\sqrt{8\pi}\chi^\epsilon(x)}e^{i\sqrt{8\pi} \bar\chi^\epsilon(x)}} + 
\nord{e^{-i\sqrt{8\pi}\chi^\epsilon(x)}e^{-i\sqrt{8\pi} \bar\chi^\epsilon(x)}} \Big),\\
\label{O2 definition} \mathsf O_2^\epsilon(x) &\coloneqq \frac{1}{8\pi} \, J^{3,\epsilon}(x) \bar J^{3,\epsilon}(x) = \partial_{x}\chi^\epsilon(x) \partial_{x}\bar \chi^\epsilon(x).
\end{align}
\end{subequations}
With these definitions in place, the desired Hamiltonian of the anisotropic deformation of the $\su_2$ WZW model at level $1$ is given in the smoothly regularised theory at scale $\epsilon > 0$ by
\begin{equation}\label{sine-Gordon Hamiltonian}
\H^\epsilon = \H_0^\epsilon + g_1(\epsilon) \int_0^L \dd x \, \mathsf O_1^\epsilon(x) + g_2(\epsilon) \int_0^L \dd x \, \mathsf O_2^\epsilon(x),
\end{equation}
using the same notation convention as in \eqref{couplings} for the potential.

Since the expressions \eqref{O1 O2 definition} are both smoothly regularised versions of operators of conformal dimension $2$ in the free compactified boson theory at the self-dual radius $R_\circ = 1/\sqrt{2 \pi}$, they come multiplied by a factor of $\epsilon^0 = 1$, in accordance with \eqref{couplings}.
Technically, by the general discussion in \S\ref{sec: RG flow}, we should also include in the interaction term of the Hamiltonian \eqref{sine-Gordon Hamiltonian} all possible local operators that are compatible with the symmetry of the two terms already written. In particular, this will include an infinite number of irrelevant operators $\int_0^L \dd x\, \mathsf O_j^\epsilon(x)$, say labeled by $j > 3$, with their own independent couplings $g_j(\epsilon)$. Indeed, we shall see below in \S\ref{sec: sG thin shell} that an infinite number of irrelevant terms is generated in the variation of the Hamiltonian \eqref{sine-Gordon Hamiltonian} after integrating out a thin shell. Since we will only be concerned with deriving the flows of the couplings $g_1$ and $g_2$ in \eqref{sine-Gordon Hamiltonian}, from now on we will always suppress all the irrelevant operators.
However, for illustration purposes we will compute in \S\ref{sec: H22 var} a contribution to the flow of the constant term in the Hamiltonian, so we explicitly include this coupling in our Hamiltonian and write
\begin{align}\label{Regularized bare Hamiltonian}
\H^\epsilon &\coloneqq \H_0^\epsilon +\frac{\pi g_1(\epsilon)}{L^2}\int_0^L \dd x \Big(\nord{e^{i\sqrt{8\pi}\chi^\epsilon(x)}e^{i\sqrt{8\pi} \bar\chi^\epsilon(x)}} + \nord{e^{-i\sqrt{8\pi}\chi^\epsilon(x)}e^{-i\sqrt{8\pi} \bar\chi^\epsilon(x)}} \Big) \notag\\
& \qquad\qquad\qquad + g_2(\epsilon)\int_0^L \dd x\, \partial_x\chi^\epsilon(x) \partial_x\bar \chi^\epsilon(x) + \frac{L}{\epsilon^2}g_3(\epsilon).
\end{align}
The factor of $\frac{L}{\epsilon^2}$ comes from the fact that we should view the constant term as arising from integrating the identity operator $\mathsf O_3^\epsilon(x) = 1$ around the spatial direction, giving the factor of $L = \int_0^L \dd x \, \mathsf O_3^\epsilon(x)$. And since the identity has conformal dimension $0$, in order to have $g_3(\epsilon)$ non-dimensional we include an explicit factor of $\epsilon^{-2}$, in line with the general structure \eqref{couplings}.

It will also be convenient later to split \eqref{O1 definition} into its positively and negatively `charged' parts as $\mathsf{O}_1^\epsilon(x) = \mathsf{O}_{1+}^\epsilon(x) + \mathsf{O}_{1-}^\epsilon(x)$ with
\begin{equation} \label{O1 pm defintion}
\mathsf{O}_{1\pm}^\epsilon(x) \coloneqq \frac{1}{4\pi} J^{\pm,\epsilon}(x) \bar J^{\mp, \epsilon}(x) = \frac{\pi}{L^2} \nord{e^{\pm i\sqrt{8\pi}\chi^\epsilon(x)}e^{\pm i\sqrt{8\pi} \bar\chi^\epsilon(x)}}.
\end{equation}

\subsubsection{Rewriting as the sine-Gordon Hamiltonian} \label{sec: usual sG Ham}

Recall from \eqref{H0 up to epsilon} that $\H_0^\epsilon$, defined in \eqref{WZW Hamiltonian free realisation smth reg}, is the usual expression for the free Hamiltonian in terms of the smoothly regularised chiral and anti-chiral bosons $\chi^\epsilon$ and $\bar\chi^\epsilon$, up to $O(\epsilon)$. By adding to this the perturbation by the marginal operator $\int_0^L \dd x\, \mathsf O_2^\epsilon(x)$ given by the first term on the second line of \eqref{Regularized bare Hamiltonian}, we obtain the quadratic Hamiltonian
\begin{align*}
\H_0^\epsilon + g_2(\epsilon) \int_0^L \dd x \, \mathsf O_2^\epsilon(x) = \int_0^L \dd x \bigg( \frac{1 + \frac 12 g_2(\epsilon)}{2} \nord{(\partial_x \X^\epsilon(x))^2} + \, \frac{1 - \frac 12 g_2(\epsilon)}{2} \nord{(\P^\epsilon(x))^2} \bigg) + O(\epsilon),
\end{align*}
written in terms of the smoothly regularised version $\X^\epsilon(x)$ of the compactified boson \eqref{X decomposition} at the self-dual radius $R_\circ = 1/\sqrt{2 \pi}$ and the regularised version $\P^\epsilon(x)$ of its conjugate momentum \eqref{P mode expansion}. Recall that $\P(x) = \partial_x \tilde\X(x)$ where $\tilde\X(x)$ is the dual compactified boson \eqref{tX decomposition}.

Now observe that by suitably rescaling the compactified boson and dual boson as in \eqref{Rescaled Boson Dual Boson}, for some $\beta = \beta(\epsilon)$ depending on $g_2(\epsilon)$, we can bring the above expression back into a canonically normalised kinetic term for the rescaled compactified boson $\Phi^\epsilon(x)$, with radius $R(\epsilon) = 2/\beta(\epsilon)$, and its conjugate momentum $\Pi^\epsilon(x) = \partial_x \tilde\Phi^\epsilon(x)$. Specifically, we can write
\begin{align} \label{canonical kinetic term}
\H_0^\epsilon + g_2(\epsilon) \int_0^L \dd x \, \mathsf O_2^\epsilon(x) &= \frac{\nu(\epsilon)}{2} \int_0^L \dd x \big(\cnord{(\partial_x\Phi^\epsilon(x))^2}_{\beta(\epsilon)} + \cnord{(\Pi^\epsilon(x))^2}_{\beta(\epsilon)} \big)\\
&\qquad\qquad - \frac{\nu(\epsilon)}{2} \bigg( \frac{\beta(\epsilon)}{\sqrt{8\pi}} - \frac{\sqrt{8\pi}}{\beta(\epsilon)} \bigg)^2 \frac{2 \pi}{L} \sum_{n > 0} n \eta\bigg( \frac{2 \pi n \epsilon}{L} \bigg) + O(\epsilon), \notag
\end{align}
where the parameter $\beta(\epsilon)$ determining the new radius of compactification $R(\epsilon) = 2/\beta(\epsilon)$ and the overall normalisation factor $\nu(\epsilon)$ are given by
\begin{equation} \label{beta nu coupling def}
\frac{\beta(\epsilon)^2}{8\pi} = \sqrt{\frac{1 - \frac 12 g_2(\epsilon)}{1 + \frac 12 g_2(\epsilon)}}, \qquad
\nu(\epsilon) \coloneqq \sqrt{1 - \tfrac 14 g_2(\epsilon)^2} = \frac{16 \pi \beta(\epsilon)^2}{64 \pi^2 + \beta(\epsilon)^4}.
\end{equation}
The term in the second line of \eqref{canonical kinetic term} is a normal ordering constant which comes from the fact that we have changed the notion of normal ordering from $\nord{\--}$ to $\cnord{\--}_{\beta(\epsilon)}$. The former denotes normal ordering with respect to the raising and lowering operators of the chiral and anti-chiral bosons $\chi^\epsilon$ and $\bar\chi^\epsilon$ while the latter denotes normal ordering with respect to the raising and lowering operators of the chiral and anti-chiral bosons $\phi^\epsilon$ and $\bar\phi^\epsilon$. Specifically, we find
\begin{align*}
\frac{8 \pi}{\beta(\epsilon)^2} \nord{(\partial_x \X^\epsilon(x))^2} &= \cnord{(\partial_x\Phi^\epsilon(x))^2}_{\beta(\epsilon)} + \bigg( 1 - \frac{8 \pi}{\beta(\epsilon)^2} \bigg) \frac{2 \pi}{L^2} \sum_{n > 0} n \eta\bigg( \frac{2 \pi n \epsilon}{L} \bigg),\\
\frac{\beta(\epsilon)^2}{8 \pi} \nord{(\P^\epsilon(x))^2} &= \cnord{(\Pi^\epsilon(x))^2}_{\beta(\epsilon)} + \bigg( 1 - \frac{\beta(\epsilon)^2}{8 \pi} \bigg) \frac{2 \pi}{L^2} \sum_{n > 0} n \eta\bigg( \frac{2 \pi n \epsilon}{L} \bigg)
\end{align*}
and adding these together leads to \eqref{canonical kinetic term}.

It remains to rewrite the second term on the right hand side of \eqref{Regularized bare Hamiltonian}, namely the operator $\int_0^L \dd x \, \mathsf O_1^\epsilon(x) = \frac{4\pi}{L^2} \int_0^L \dd x \, \nord{\cos\big( \sqrt{8\pi} \X^\epsilon(x) \big)}$, in terms of the rescaled compactified boson $\Phi^\epsilon(x)$ and re-normal-order it at the new radius $R(\epsilon) = 2/\beta(\epsilon)$. Doing so, we find
\begin{equation} \label{cosine potential Bogo}
g_1(\epsilon) \int_0^L \dd x \, \mathsf O_1^\epsilon(x) =
\epsilon^{\frac{\beta(\epsilon)^2}{4 \pi} - 2} \nu(\epsilon) \frac{2\pi \lambda(\epsilon)}{L^{\frac{\beta(\epsilon)^2}{4 \pi}}} \int^L_0 \dd x \, \cnord{\cos\big( \beta(\epsilon) \Phi^\epsilon(x) \big)}_{\beta(\epsilon)}
\end{equation}
where we have included an overall factor of $\nu(\epsilon)$ matching the one of the kinetic term in \eqref{canonical kinetic term} and the coupling $\lambda(\epsilon)$ is given using the exact expression on the right hand side of \eqref{normal order change exp} by
\begin{equation} \label{lambda coupling def}
\lambda(\epsilon) = \bigg( \frac{\epsilon}{L} \bigg)^{2 - \frac{\beta(\epsilon)^2}{4\pi}} \exp\left( \bigg( 2 - \frac{\beta(\epsilon)^2}{4\pi} \bigg) \sum_{n > 0} \frac{1}{n} \eta\left(\frac{2\pi n \epsilon}{L}\right)\right) \frac{g_1(\epsilon)}{\sqrt{1 - \frac 14 g_2(\epsilon)^2}}.
\end{equation}
This is a dimensionless coupling since $g_1(\epsilon)$ and $g_2(\epsilon)$ are marginal and hence dimensionless, and the prefactor in \eqref{lambda coupling def} is a function of the dimensionless ratio $\frac{\epsilon}{L}$. Moreover, let us denote the limiting value of the marginal coupling $g_2(\epsilon)$ in the UV limit by $g_{2\ast}$ and suppose that the marginal coupling $g_1(\epsilon)$ in this limit tends to $0$.
More explicitly, let us suppose that
\begin{equation*}
g_1(\epsilon) \underset{\epsilon \to 0}\sim (\mu_\ast \epsilon)^{2 - \frac{\beta_\ast^2}{4 \pi}}, \qquad g_{2\ast} \coloneqq \lim_{\epsilon \to 0} g_2(\epsilon),
\end{equation*}
for some mass scale $\mu_\ast$. Let $\beta_\ast \coloneqq \lim_{\epsilon \to 0} \beta(\epsilon)$ denote the corresponding limiting value of the marginal coupling $\beta(\epsilon)$ defined in \eqref{beta nu coupling def}. We then have
\begin{equation} \label{lambda UV limit}
\lambda(\epsilon) \underset{\epsilon \to 0}\sim ( M_\ast \epsilon )^{2 - \frac{\beta_\ast^2}{4 \pi}}, \qquad M_\ast \coloneqq \frac{\mu_\ast C_\eta}{2 \pi} \bigg( \frac{4}{\beta_\ast^2} + \frac{\beta_\ast^2}{16 \pi^2} \bigg)^{\frac{4 \pi}{8 \pi - \beta_\ast^2}},
\end{equation}
where $M_\ast$ is a scheme-dependent mass scale (i.e. depending on the choice of $\eta$ through $C_\eta$).

By comparing the right hand side of \eqref{cosine potential Bogo} with the general form \eqref{couplings} of an interaction term in the Hamiltonian, and using the fact that $\lambda(\epsilon)$ is dimensionless, we read off from the power of $\epsilon$ the well-known conformal dimension $\Delta_\ast = \beta_\ast^2/ 4 \pi$ of the full vertex operators
\begin{equation} \label{full VO at beta}
\cnord{e^{\pm i \beta_\ast \Phi(x)}}_{\beta_\ast} = \cnord{e^{\pm i \beta_\ast \phi(x)}}_{\beta_\ast} \cnord{e^{\pm i \beta_\ast \bar\phi(x)}}_{\beta_\ast},
\end{equation}
in the conformal field theory of the compactified boson $\Phi(x) = \phi(x) + \bar\phi(x)$, defined in \S\ref{sec: changing radius}, with compactification radius $R_\ast = 2/\beta_\ast$. Specifically, \eqref{full VO at beta} is the product of a chiral and an anti-chiral vertex operator, both of which have conformal dimension $\beta_\ast^2/8 \pi$ \cite[(6.60)]{CFTbook}.

Note also that the origin of the factor of $L^{- \beta_\ast^2/4\pi}$ appearing in \eqref{cosine potential Bogo}, in the limit $\epsilon \to 0$, is the same as that of the factor of $L^{-1}$ in the free field realisations of the currents $J^\pm(u)$ and $\bar J^\pm(u)$ in \eqref{su2 1 currents cyl}, which led to the factor of $L^{-2}$ in the cosine potential in the Hamiltonian \eqref{Regularized bare Hamiltonian}. Indeed, since $\cnord{e^{\pm i \beta_\ast \phi[z]}}_{\beta_\ast}$ is a conformal primary of conformal dimension $\beta_\ast^2/8 \pi$ it picks up a factor of $(2 \pi z/L)^{\beta_\ast^2/8 \pi}$ under a coordinate transformation from the $z$-coordinate on the plane to the $u$-coordinate on the cylinder. The power of $z^{\beta_\ast^2/8 \pi}$ is then absorbed into the definition of the vertex operator $\cnord{e^{\pm i \beta_\ast \phi(u)}}_{\beta_\ast}$ where the zero-mode are combined into a single exponential as in \eqref{normal ordered exponential 0mode}. Specifically, the analogue of the computation \eqref{absorb factor of z} reads
\begin{equation} \label{z factor absorbed}
z^{\frac{\beta_\ast^2}{8 \pi}} \, e^{\pm i \beta_\ast \q_0} e^{\pm i \beta_\ast \left( - \frac{i}{\sqrt{4 \pi}}\a_0 \log z \right)} = e^{\pm i \beta_\ast \left( \q_0 - \frac{i}{\sqrt{4 \pi}}\a_0 \log z \right)} = e^{\pm i \beta_\ast \phi_0(u)}
\end{equation}
where in the first step we have used the commutation relations \eqref{Compactified rescaled boson commutation relations} and in the last step we used the explicit change of coordinate $z = e^{2 \pi i u / L}$ and the definition \eqref{phi u decomp pm 0} of the zero-mode $\phi_0(u)$ in the cylinder coordinate $u$. The same holds for the anti-chiral vertex operator.

In summary, the anisotropic deformation of the $\su_2$ WZW model at level $1$ in the smoothly regularised theory at cutoff $\epsilon > 0$, defined in \eqref{Regularized bare Hamiltonian}, takes the form
\begin{align} \label{sG Hamiltonian Bogo}
\H^\epsilon &= \nu(\epsilon) \int_0^L \dd x \bigg( \tfrac 12 \cnord{(\partial_x\Phi^\epsilon(x))^2}_{\beta(\epsilon)} + \tfrac 12 \cnord{(\Pi^\epsilon(x))^2}_{\beta(\epsilon)} \notag\\
&\qquad\qquad\qquad\qquad\qquad\qquad\qquad\qquad\qquad + \epsilon^{\frac{\beta(\epsilon)^2}{4 \pi} - 2} \frac{2\pi \lambda(\epsilon)}{L^{\frac{\beta(\epsilon)^2}{4 \pi}}} \cnord{\cos\big( \beta(\epsilon) \Phi^\epsilon(x) \big)}_{\beta(\epsilon)} \bigg) \notag\\
&\qquad\qquad\qquad\qquad + \frac{L}{\epsilon^2} \Bigg( g_3(\epsilon) - \frac{\nu(\epsilon)}{4 \pi} \bigg( \frac{\beta(\epsilon)}{\sqrt{8\pi}} - \frac{\sqrt{8\pi}}{\beta(\epsilon)} \bigg)^2 C_{\eta, 1} \Bigg) - \frac{\pi \delta c_\ast}{6 L} + O(\epsilon).
\end{align}
In the first two lines on the right hand side we recognise the usual sine-Gordon Hamiltonian, up to an overall normalisation factor $\nu(\epsilon)$, written in terms of the smoothly regularised boson $\Phi^\epsilon(x)$ with compactification radius $R(\epsilon) = 2 / \beta(\epsilon)$ and its conjugate momentum $\Pi^\epsilon(x)$. The first term in the last line on the right hand side of \eqref{sG Hamiltonian Bogo} is a scheme-dependent correction (i.e. depending on the choice of smooth cutoff function $\eta$) to the divergent $O(\epsilon^{-2})$ term introduced by hand in \eqref{Regularized bare Hamiltonian}. The last term on the right hand side of \eqref{sG Hamiltonian Bogo} is a scheme-independent shift to the finite Casimir energy, namely the constant term $- \frac{\pi c}{6 L}$ with $c=1$ in \eqref{free Ham with Casimir term} which we have been omitting, where
\begin{equation}
\delta c_\ast \coloneqq - \frac{\nu_\ast}{2} \bigg( \frac{\beta_\ast}{\sqrt{8\pi}} - \frac{\sqrt{8\pi}}{\beta_\ast} \bigg)^2
\end{equation}
and we have introduced the UV limit $\nu_\ast = \lim_{\epsilon \to 0} \nu(\epsilon)$.

Applying the renormalisation procedure from \S\ref{sec: Renormalisation} to the Hamiltonian $\H^\epsilon$ in \S\ref{sec: sG thin shell} below will lead to the renormalisation group flows of all the parameters $g_1$, $g_2$ and $g_3$ of the anisotropic deformation of the $\su_2$ WZW at level $1$ introduced in \eqref{Regularized bare Hamiltonian}. We will then be able to map this to the renormalisation group flow of the couplings $\lambda$, $\beta$ and $\nu$ of the sine-Gordon Hamiltonian \eqref{sG Hamiltonian Bogo} using the explicit transformations \eqref{beta nu coupling def} and \eqref{lambda coupling def}.

\subsection{Integrating out a thin shell} \label{sec: sG thin shell}

We are now in a position to apply the general renormalisation procedure described in \S\ref{sec: Renormalisation} to the regularised Hamiltonian \eqref{Regularized bare Hamiltonian} at hand. The majority of this subsection will be dedicated to calculating the variation $\delta \H^{\epsilon'}$ of the Hamiltonian \eqref{Regularized bare Hamiltonian} from integrating out a thin shell as in \S\ref{sec: integrating out}. We will use the explicit form of this variation given by \eqref{Hamiltonian variation explicit}. In fact, since \eqref{Regularized bare Hamiltonian} is normal-ordered, the second term on the right hand side of that expression vanishes; see the discussion after \eqref{potential expansion PlPl}. The expression we shall use for the variation is then
\begin{align} \label{delta H normal-ordered}
\delta \H^{\epsilon'} &= \frac{\delta\epsilon}{2} \int_{-T}^T \dd \tau \int_0^L \dd x_1 \int_0^L \dd x_2 \, \Wop^\epsilon(\tau) \, \mathcal T \Big( \Vop^{\epsilon', (1,0)}(x_1; \tau) \Vop^{\epsilon', (1,0)}(x_2)
\Delta^{\epsilon}_s(x_1-i\tau, x_2)\\
&\qquad\qquad\qquad\qquad\qquad\qquad\qquad + \Vop^{\epsilon', (0,1)}(x_1; \tau) \Vop^{\epsilon', (0,1)}(x_2)\bar\Delta^{\epsilon}_s(x_1+i\tau, x_2) \Big) + O(\delta\epsilon^2). \notag
\end{align}

From now on we will be working perturbatively in the couplings $g_1$ and $g_2$, see \eqref{Regularized bare Hamiltonian}, in order to simplify the time evolution of operators by $\tilde \H_{\rm bd}^\epsilon = \H_0^{\epsilon'} + \Vop(\chi^{\epsilon'} + \chi^{\short}, \bar \chi^{\epsilon'} + \bar \chi^{\short})_{\rm bd}$. The expression \eqref{delta H normal-ordered} for the variation $\delta \H^{\epsilon'}$ of a normal-ordered potential is quadratic in the potential and therefore already quadratic in these couplings $g_1$ and $g_2$. The imaginary-time evolutions $\Vop^{\epsilon', (1,0)}(x_1; \tau)$ and $\Vop^{\epsilon', (0,1)}(x_1; \tau)$ of these potentials defined by conjugation by the exponential of $\tau \tilde \H_{\rm bd}^\epsilon$ rather than the exponential of $\tau \, \H_0^{\epsilon'}$, will only introduce further powers of $g_1$ and $g_2$. Thus the imaginary-time evolution by $\tilde \H_{\rm bd}^\epsilon$ can simply be replaced by the free imaginary-time evolution by $\H_0^{\epsilon'}$ to leading order in the couplings $g_1$ and $g_2$. 
In other words, we shall work with the variation
\begin{align} \label{delta H normal-ordered g0 out}
\delta \H^{\epsilon'} &= \frac{\delta\epsilon}{2} \int_{-T}^T \dd \tau \int_0^L \dd x_1 \int_0^L \dd x_2 \, \Wop^\epsilon(\tau) \, \mathcal T \Big( \Vop^{\epsilon', (1,0)}(x_1, \tau) \Vop^{\epsilon', (1,0)}(x_2)
\Delta^{\epsilon}_s(x_1-i\tau, x_2) \notag\\
&\qquad\qquad\qquad\qquad\qquad + \Vop^{\epsilon', (0,1)}(x_1, \tau) \Vop^{\epsilon', (0,1)}(x_2)\bar\Delta^{\epsilon}_s(x_1+i\tau, x_2) \Big) + O(\delta\epsilon^2),
\end{align}
where we use the simple comma notation for the dependence on the imaginary-time $\tau$, by contrast with the semicolon notation in \eqref{delta H normal-ordered} denoting imaginary-time evolution by the full Hamiltonian $\tilde \H_{\rm bd}^\epsilon$. It will be important to recall from \eqref{bosons free evolution reg} that the free imaginary-time $\tau$ evolution of the smoothly regularised chiral and anti-chiral bosons $\chi^\epsilon(x)$ and $\bar\chi^\epsilon(x)$ is given simply by a shift of their arguments by $\mp i \tau$, respectively.

The regularised potential is
\begin{align}\label{Regularised potential}
    \Vop(\chi^\epsilon, \bar \chi^\epsilon) &= \frac{\pi g_1(\epsilon)}{L^2}\int_0^L
        \dd x \Big(\nord{e^{i\sqrt{8\pi}\chi^\epsilon(x)}e^{i\sqrt{8\pi} \bar\chi^\epsilon(x)}} + 
            \nord{e^{-i\sqrt{8\pi}\chi^\epsilon(x)}e^{-i\sqrt{8\pi} \bar\chi^\epsilon(x)}} \Big) \notag\\
    &\qquad + g_2(\epsilon)\int_0^L \dd x \, \partial_x\chi^\epsilon(x)\partial_x \bar \chi^\epsilon(x) + \frac{L}{\epsilon^2}g_3(\epsilon).
\end{align}
To compute $\Vop^{\epsilon', (1, 0)}$ and $\Vop^{\epsilon', (0, 1)}$, as defined in \eqref{potential field expansion}, we apply the splitting morphism \eqref{embedding algebras epsilon eta} to the expression \eqref{Regularised potential}, which amounts to replacing $\chi^\epsilon$ by $\chi^{\epsilon'} + \chi^{\epsilon'\bsl\epsilon}$ and $\bar\chi^\epsilon$ by $\bar\chi^{\epsilon'} + \bar\chi^{\epsilon'\bsl\epsilon}$, and then expand to first order in the short distance fields $\chi^{\epsilon'\bsl\epsilon}$ and $\bar\chi^{\epsilon'\bsl\epsilon}$.
Bringing the $g_2$ term in this first order expansion to the same form as in \eqref{potential field expansion} requires performing an integration by parts. Since $\chi^{\short}(x)$ has no zero-mode it is periodic and therefore there is no boundary term from the integration by parts. The resulting expression for $\Vop^{\epsilon', (1, 0)}$ can be written as
\begin{align} \label{V 10 explicit}
\Vop^{\epsilon', (1, 0)}(x) = \Vop^{\epsilon', (1, 0)}_1(x) + \Vop^{\epsilon', (1, 0)}_2(x),
\end{align}
where to break up the calculation of $\delta \H^{\epsilon'}$ later we have split this into two contributions
\begin{subequations} \label{potential two parts}
\begin{align}
\label{potential two parts a} \Vop^{\epsilon', (1, 0)}_1(x) &= \frac{\pi}{L^2} i\sqrt{8\pi} \Big( \nord{e^{i\sqrt{8\pi} \chi^{\epsilon'}(x)}e^{i\sqrt{8\pi} \bar \chi^{\epsilon'}(x)}} - \nord{e^{-i\sqrt{8\pi} \chi^{\epsilon'}(x)}e^{-i\sqrt{8\pi} \bar \chi^{\epsilon'}(x)}} \Big), \\
\label{potential two parts b} \Vop^{\epsilon', (1, 0)}_2(x) &= - \partial^2_x \bar \chi^{\epsilon'}(x).
\end{align}
\end{subequations}
The expression for $\Vop^{\epsilon', (0, 1)}(x) = \Vop^{\epsilon', (0, 1)}_1(x) + \Vop^{\epsilon', (0, 1)}_2(x)$ is identical, but with the roles of the fields $\chi^{\epsilon'}$ and $\bar\chi^{\epsilon'}$ interchanged. In particular, we note that
\begin{equation} \label{V1 01 = V1 10}
\Vop^{\epsilon', (0, 1)}_1(x) = \Vop^{\epsilon', (1, 0)}_1(x).
\end{equation}

Splitting the potential into two pieces allows us to break up the calculation of $\delta \H^{\epsilon'}$ into four parts, namely $\delta \H^{\epsilon'} = \sum_{i,j=1}^2 \delta \H^{\epsilon'}_{ij}$ with
\begin{align}\label{Delta H ij}
    \delta \H^{\epsilon'}_{ij} &= \frac{\delta\epsilon}{2} g_i(\epsilon) g_j(\epsilon) \int_{-T}^T \dd \tau \int_0^L \dd x_1 \int_0^L \dd x_2 \, \Wop^\epsilon(\tau) \, \mathcal T \Big( \Vop_i^{\epsilon', (1,0)}(x_1, \tau) \Vop_j^{\epsilon', (1,0)}(x_2)\Delta^{\epsilon}_s(x_1-i\tau, x_2) \notag\\
&\qquad\qquad\qquad\qquad\qquad + \Vop_i^{\epsilon', (0,1)}(x_1, \tau) \Vop_j^{\epsilon', (0,1)}(x_2)\bar\Delta^{\epsilon}_s(x_1+i\tau, x_2) \Big) + O(\delta\epsilon^2).
\end{align}
To remove the time-ordering symbol it will be convenient to further split each piece into two, by breaking up the integral over $\tau \in [-T,T]$ into the forward-time part $\tau \in [0,T]$, namely
\begin{align}\label{Delta H ij plus}
    \delta \H^{\epsilon'}_{ij \uparrow} &= \frac{\delta\epsilon}{2} g_i(\epsilon) g_j(\epsilon) \int_0^T \dd \tau \int_0^L \dd x_1 \int_0^L \dd x_2 \, \Wop^\epsilon(\tau) \, \Big( \Vop_i^{\epsilon', (1,0)}(x_1, \tau) \Vop_j^{\epsilon', (1,0)}(x_2) \Delta^{\epsilon}_s(x_1-i\tau, x_2)\notag\\
&\qquad\qquad\qquad\qquad\qquad + \Vop_i^{\epsilon', (0,1)}(x_1, \tau) \Vop_j^{\epsilon', (0,1)}(x_2) \bar\Delta^{\epsilon}_s(x_1+i\tau, x_2)\Big) + O(\delta\epsilon^2),
\end{align}
and the backward-time part $\tau \in [-T, 0]$ which we denote by $\delta \H^{\epsilon'}_{ij \downarrow}$.
We will also split these into their chiral parts, with a superscript `$(1,0)$' for the $\Vop^{\epsilon,(1, 0)}(x)$-dependent term, i.e.
\begin{align} \label{Delta H ij plus plus}
\delta \H^{\epsilon'(1,0)}_{ij\uparrow} &= \frac{\delta\epsilon}{2} g_i(\epsilon) g_j(\epsilon) \int_0^T \dd \tau \int_0^L \dd x_1 \int_0^L \dd x_2 \, \Wop^\epsilon(\tau)\\
&\qquad\qquad\qquad\qquad \times \Vop_i^{\epsilon', (1,0)}(x_1, \tau) \Vop_j^{\epsilon', (1,0)}(x_2) \Delta^{\epsilon}_s(x_1-i\tau, x_2) + O(\delta\epsilon^2), \notag
\end{align}
and a superscript `$(0,1)$' for the corresponding $\Vop^{\epsilon,(0, 1)}(x)$-dependent term.
Finally, it will also be convenient to split $\Vop^{\epsilon', (1, 0)}_1(x)$ in \eqref{potential two parts a} into its two different charges, introducing
\begin{equation} \label{V pm split}
\Vop^{\epsilon', (1, 0)}_{1\pm}(x) = \pm\frac{\pi}{L^2} i \sqrt{8\pi}\nord{e^{\pm i\sqrt{8\pi} \chi^{\epsilon'}(x)}e^{\pm i\sqrt{8\pi} \bar \chi^{\epsilon'}(x)}}
\end{equation}
so that $\Vop^{\epsilon', (1, 0)}_1(x) = \Vop^{\epsilon', (1, 0)}_{1+}(x) + \Vop^{\epsilon', (1, 0)}_{1-}(x)$ and similarly $\Vop^{\epsilon', (0, 1)}_1(x) = \Vop^{\epsilon', (0, 1)}_{1+}(x) + \Vop^{\epsilon', (0, 1)}_{1-}(x)$.

\subsubsection{\texorpdfstring{$\delta\H^{\epsilon'}_{22}$}{H22} variation} \label{sec: H22 var}

We will calculate the contribution $\delta \H^{\epsilon' (1,0)}_{22\uparrow}$ to $\delta\H^{\epsilon'}_{22} = \delta\H^{\epsilon'(1,0)}_{22\uparrow} + \delta\H^{\epsilon'(1,0)}_{22\downarrow} + \delta\H^{\epsilon'(0,1)}_{22\uparrow} + \delta\H^{\epsilon'(0,1)}_{22\downarrow}$, the calculations for the other three terms being almost identical.

Written out in full, $\delta \H^{\epsilon' (1,0)}_{22\uparrow}$ is
\begin{align}\label{H22 variation def}
\delta \H_{22\uparrow}^{\epsilon' (1,0)} &= \frac 12 \delta\epsilon g_2(\epsilon)^2 \int_0^T \dd \tau \int_0^L \dd x_1 \int_0^L \dd x_2 \Wop^\epsilon(\tau)\\
&\qquad\qquad\qquad\qquad \times \Delta^\epsilon_s(x_1-i\tau, x_2) \partial_{x_1}^2 \bar \chi^{\epsilon'}(x_1 + i\tau) \partial^2_{x_2} \bar \chi^{\epsilon'}(x_2) + O(\delta\epsilon^2), \notag
\end{align}
where recall that we are working to leading order in the couplings so $\bar\chi^{\epsilon'}(x_1 + i \tau)$ represents the \emph{free} imaginary-time evolution of the anti-chiral boson $\bar\chi^{\epsilon'}(x_1)$, as defined in \eqref{bosons free evolution reg}.
Writing out the mode expansions for the $2$-point function $\Delta^{\epsilon}_s(x_1 - i\tau, x_2)$ in \eqref{propagator shell modes a} and for the anti-chiral boson $\bar\chi^{\epsilon'}$, c.f. \eqref{chi finer decomposition} and \eqref{chi u decomp pm 0}, and evaluating the integrals over $x_1$ and $x_2$ gives
\begin{align}\label{H22 variation non-normal ordered}
\delta \H^{\epsilon'(1,0)}_{22\uparrow} = \frac{\pi^3}{L^3}\delta\epsilon g_2(\epsilon)^2 \int_0^T\dd \tau \Wop^\epsilon(\tau) \sum_{n > 0} n^2 \eta'\left(\frac{2\pi n \epsilon'}{L}\right) \bar \b_n^{\epsilon'} \bar \b_{-n}^{\epsilon'} e^{-4 \pi n \tau/L} + O(\delta\epsilon^2).
\end{align}
Recall from discussion in \S\ref{sec: RG flow} that the Wilsonian renormalisation procedure computes the flows of the couplings for a complete basis of local operators that can appear in the potential, as in \eqref{couplings}. Since any regularised operator can always be rewritten as a linear combination of normal-ordered operators, it is natural to work with a basis of normal-ordered local operators in the expansion \eqref{couplings}.
We therefore commute the modes on the right hand side of \eqref{H22 variation non-normal ordered} to get the normal-ordered expression
\begin{align}\label{H22 variation normal ordered}
    \delta \H^{\epsilon'(1,0)}_{22\uparrow} &= \frac{\pi^3}{L^3}\delta\epsilon g_2(\epsilon)^2
    \int_0^T\dd \tau \Wop^\epsilon(\tau) \sum_{n > 0} 
    n^2 \eta'\left(\frac{2\pi n \epsilon'}{L}\right)
    \bar \b_{-n}^{\epsilon'} \bar \b_{n}^{\epsilon'} e^{-4 \pi n \tau/L}\\
    &\qquad + \frac{\pi^3}{L^3}\delta\epsilon g_2(\epsilon)^2
    \int_0^T\dd \tau \Wop^\epsilon(\tau) \sum_{n > 0} 
    n^3 \eta'\left(\frac{2\pi n \epsilon'}{L}\right)
    \eta\left(\frac{2\pi n \epsilon'}{L}\right) e^{-4 \pi n \tau/L}
+ O(\delta\epsilon^2). \notag
\end{align}
To determine the beta function of the couplings in front of each local operator, as defined in \eqref{beta functions}, we compute the following limit
\begin{align}\label{H22 variational flow}
    \lim_{\epsilon' \to \epsilon}\epsilon\frac{\delta \H^{\epsilon'(1,0)}_{22\uparrow}}{\delta\epsilon} &= \frac{\pi^3 \epsilon}{L^3} g_2(\epsilon)^2
    \int_0^T\dd \tau \Wop^\epsilon(\tau) \sum_{n > 0} 
    n^2 \eta'\left(\frac{2\pi n \epsilon}{L}\right)
    \bar \b_{-n}^{\epsilon} \bar \b_{n}^{\epsilon} e^{-4 \pi n \tau/L}\\
    &\qquad + \frac{\pi^3 \epsilon}{L^3} g_2(\epsilon)^2
    \int_0^T\dd \tau \Wop^\epsilon(\tau) \sum_{n > 0} 
    n^3 \eta'\left(\frac{2\pi n \epsilon}{L}\right)
    \eta\left(\frac{2\pi n \epsilon}{L}\right) e^{-4 \pi n \tau/L}. \notag
\end{align}
The first term on the right hand side represents an infinite sum of irrelevant operators. This can be seen by expanding the operator $\Wop^\epsilon(\tau)$ in non-negative powers of $\tau$, as we will explain in detail below, and expanding the smooth function $\eta'\big(\frac{2\pi n \epsilon}{L}\big)$ in non-negative powers of $\epsilon$. Note that the latter expansion is justified since each term in the sum over $n > 0$ corresponds to a different operator $\bar\b_{-n}^\epsilon \bar\b_n^\epsilon$. So, for instance, at leading order in this $\epsilon$-expansion we obtain the formal infinite sum $\frac{\epsilon}{L^3} \sum_{n > 0} n^2 \bar\b_{-n}^\epsilon \bar\b_n^\epsilon e^{-4 \pi n \tau/L}$ which is a well-defined operator in the smoothly regularised theory. Expanding also the operator $\Wop^\epsilon(\tau)$ in $\tau$, to zeroth order it can be replaced by the identity operator (see below). Therefore after performing the integral over $\tau \in [0, T]$ we obtain the operator $\frac{1}{L^2} \sum_{n > 0} n \bar\b_{-n}^\epsilon \bar\b_n^\epsilon$ which is indeed proportional to the irrelevant operator $\int_0^L \nord{\partial_x \bar\chi^\epsilon(x) \partial_x^2 \bar\chi^\epsilon(x)}$ of conformal dimension $\Delta = 3$ and comes multiplied by $\epsilon = \epsilon^{\Delta - 2}$, in agreement with the general structure of local operators in \eqref{couplings}.

The second term on the right hand side of \eqref{H22 variational flow} is a regularised divergent sum which will contribute to the running of the constant term $g_0(\epsilon)$ of the Hamiltonian \eqref{Regularized bare Hamiltonian}. Its asymptotics can be obtained using Mellin transform theory as outlined in \S\ref{sec: Mellin transform}.

Before calculating these asymptotics, let us first discuss the effect of $\Wop^\epsilon(\tau)$ which we can expand as $\Wop^{\epsilon}(\tau) = \id + \sum_{k > 0} \tau^k \Wop^{\epsilon(k)}$, recalling from its definition \eqref{weighting operator} that $\Wop^\epsilon(0) = \id$. The $\tau$ integral in the second line of \eqref{H22 variational flow} can then be evaluated term by term, and due to the exponential factor $e^{-4 \pi n \tau/L}$, at order $\tau^k$ for $k > 0$ the $\tau$ integral will contribute an additional negative power $n^{-k}$ to the sum over $n > 0$. Importantly, the more negative the power of $n$ in the sum is, the less singular the resulting asymptotics in $\epsilon$ will be. Therefore, if we focus on the leading singular behaviour for the sum, we can replace $\Wop^\epsilon(\tau)$ with the identity operator.
Making this simplification and evaluating the $\tau$ integral we get
\begin{align} \label{delta H22 computation}
\lim_{\epsilon' \to \epsilon}\epsilon\frac{\delta \H^{\epsilon'(1,0)}_{22\uparrow}}{\delta\epsilon} = \frac{\pi^2 \epsilon}{4 L^2} g_2(\epsilon)^2
\sum_{n > 0} n^2 \eta'\left(\frac{2\pi n \epsilon}{L}\right) \eta\left(\frac{2\pi n \epsilon}{L}\right) +\text{subleading in }\epsilon,
\end{align}
and using the asymptotic expansion \eqref{regularised n^r series} for the sum over $n > 0$, with $r = 2$ and the choice of cutoff function $\eta$ given here by $\eta' \eta$, we find
\begin{align} \label{var H22 final pre}
\lim_{\epsilon' \to \epsilon}\epsilon\frac{\delta \H^{\epsilon'(1,0)}_{22\uparrow}}{\delta\epsilon} = \frac{C_{\eta'\eta,2}}{32 \pi}\frac{L}{\epsilon^2} g_2(\epsilon)^2 +\text{subleading in }\epsilon.
\end{align}

In fact, for this particular calculation we can say something stronger about the subleading terms, although the above argument still holds and will be used again later. In the case at hand, $\Wop^\epsilon(\tau)$ is being applied to the identity operator. From the definition of $\Wop^\epsilon(\tau)$ given in \eqref{weighting operator}, and since we are considering here the case $0 \leq \tau \leq T$, we can write its expansion in powers of $\tau$ explicitly as
\begin{equation} \label{Wop explicit pre}
\Wop^\epsilon(\tau) = 1 + \sum_{n > 0}e^{nT\rm ad_{\tilde \H^\epsilon_{\rm bd}}} (1 - e^{-\tau\rm ad_{\tilde \H^\epsilon_{\rm bd}}})
= 1 + \sum_{k > 0} \tau^k \frac{(-1)^{k+1}}{k!} \sum_{n > 0} e^{nT\rm ad_{\tilde \H^\epsilon_{\rm bd}}} \ad_{\tilde \H^\epsilon_{\rm bd}}^k,
\end{equation}
which provides the explicit form of the operators $\Wop^{\epsilon(k)}$ for $k > 0$ introduced earlier. In fact, recall from the paragraph following \eqref{delta H normal-ordered} that since we are working only perturbatively to second order in the couplings, to this order we can replace all instances of the block-diagonal part of the interacting Hamiltonian $\tilde \H^\epsilon_{\rm bd}$ by the free Hamiltonian $\H^{\epsilon'}_0$. In other words, we can rewrite \eqref{Wop explicit pre} simply as
\begin{equation} \label{Wop explicit}
\Wop^\epsilon(\tau) = 1 + \sum_{k > 0} \tau^k \frac{(-1)^{k+1}}{k!} \sum_{n > 0} e^{nT\rm ad_{\tilde \H^{\epsilon'}_0}} \ad_{\tilde \H^{\epsilon'}_0}^k.
\end{equation}
Now observe that since ${\rm ad}_{\tilde \H^{\epsilon'}_0}$ (or indeed also ${\rm ad}_{\tilde \H^\epsilon_{\rm bd}}$) annihilates the identity operator, so do all the operators $\Wop^{\epsilon(k)}$ for $k > 0$. And so, in the present case, it turns out that the approximation we made above of replacing $\Wop^\epsilon(\tau)$ with the identity operator is in fact exact. There are therefore no subleading terms in $\epsilon$ in \eqref{delta H22 computation} so that \eqref{var H22 final pre} becomes, more precisely,
\begin{align} \label{var H22 final}
\lim_{\epsilon' \to \epsilon}\epsilon\frac{\delta \H^{\epsilon'(1,0)}_{22\uparrow}}{\delta\epsilon} = \frac{C_{\eta'\eta,2}}{32 \pi}\frac{L}{\epsilon^2} g_2(\epsilon)^2 + O(\epsilon).
\end{align}

Using \eqref{var H22 final} and the analogous results from the other pieces $\delta\H^{\epsilon'(1,0)}_{22\downarrow}, \delta \H^{\epsilon'(0,1)}_{22\uparrow}$ and $\delta \H^{\epsilon'(0,1)}_{22\downarrow}$, and recalling the definition of the coupling $g_3(\epsilon)$ in \eqref{Regularized bare Hamiltonian}, we obtain the contribution from the variation $\delta \H^{\epsilon}_{22}$ to the quantum beta function for $g_3(\epsilon)$ as defined by \eqref{beta functions}. Specifically, recalling the minus sign on the right hand side of \eqref{beta functions} we have 
\begin{equation} \label{qu beta for g3}
\beta^{\rm qu}_3\big[ (g_k(\epsilon)) \big] = - \frac{C_{\eta'\eta, 2}}{8\pi} g_2(\epsilon)^2 + \ldots,
\end{equation}
with `$\ldots$' denoting possible corrections coming from the other variations.

\subsubsection{\texorpdfstring{$\delta \H^{\epsilon'}_{21}$ and $\delta \H^{\epsilon'}_{12}$}{H21 and H12} variations} \label{sec: H21 and H12 var}

We begin by evaluating the $\delta \H^{\epsilon' (1,0)}_{21\uparrow}$ component, defined in \eqref{Delta H ij plus plus}, of the variation $\delta \H^{\epsilon'}_{21}$. In fact, recalling the notation \eqref{V pm split} of the two summands in \eqref{potential two parts a}, it will be convenient to further split $\delta \H^{\epsilon' (1,0)}_{21\uparrow}$ into two pieces
\begin{align*}
\delta \H^{\epsilon'(1,0)}_{21\uparrow,\pm} &= \frac{\delta\epsilon}{2} g_1(\epsilon) g_2(\epsilon) \int_0^T \dd \tau \int_0^L \dd x_1 \int_0^L \dd x_2 \, \Wop^\epsilon(\tau) \\
&\qquad\qquad\qquad\qquad\qquad\qquad \times \Vop_2^{\epsilon', (1,0)}(x_1, \tau) \Vop_{1\pm}^{\epsilon', (1,0)}(x_2) \Delta^{\epsilon}_s(x_1 - i\tau, x_2) + O(\delta\epsilon^2)
\end{align*}
where the positively/negatively charged component \eqref{V pm split} of $\Vop_1^{\epsilon', (1, 0)}(x)$ is taken. Written out in full using \eqref{potential two parts b} and \eqref{V pm split}, the above explicitly reads
\begin{align}\label{H21 variation def}
    \delta \H_{21\uparrow,\pm}^{\epsilon'(1,0)} &= \mp\frac{\pi \delta\epsilon}{2L^2} g_1(\epsilon)g_2(\epsilon)
    \int_0^T \dd \tau \int_0^L \dd x_1 \int_0^L \dd x_2 \Wop^\epsilon(\tau) \Delta^\epsilon_s(x_1 - i\tau, x_2)\\
    &\qquad\qquad\qquad\qquad\qquad \times \partial_{x_1}^2 \bar \chi^{\epsilon'}(x_1 + i\tau)i\sqrt{8\pi
    }\nord{e^{\pm i\sqrt{8\pi} \chi^{\epsilon'}(x_2)}e^{\pm i\sqrt{8\pi} \bar \chi^{\epsilon'}(x_2)}} + O(\delta\epsilon^2). \notag
\end{align}
Using the explicit expression \eqref{propagator shell modes a} for the $2$-point function $\Delta^\epsilon_s(x_1 - i\tau, x_2)$ and expanding the operator $\partial_{x_1}^2 \bar \chi^{\epsilon'}(x_1 + i\tau)$ in terms of modes, we can evaluate the $x_1$ integral to obtain
\begin{align}\label{H21 variation x_1 int}
    \delta \H_{21\uparrow,\pm}^{\epsilon' (1,0)} &= \pm \frac{\pi \delta\epsilon}{2L^2} g_1(\epsilon)g_2(\epsilon)
    \int_0^T \dd \tau \int_0^L \dd x_2 \Wop^\epsilon(\tau) \sum_{n > 0} \bar \b_n^{\epsilon'}
    \nord{e^{\pm i\sqrt{8\pi} \chi^{\epsilon'}(x_2)}e^{\pm i\sqrt{8\pi} \bar \chi^{\epsilon'}(x_2)}}\\
    &\qquad\qquad\qquad\qquad \times \frac{1}{\sqrt{2}}\left(\frac{2\pi}{L}\right)^2 n\eta'\left(\frac{2\pi n \epsilon'}{L}\right)e^{-4 \pi n \tau/L}e^{2\pi i n x_2/L} + O(\delta\epsilon^2). \notag
\end{align}
Next, as in \S\ref{sec: H22 var}, we bring this expression to normal-order form by commuting the mode $\bar \b_n^{\epsilon'}$ past the vertex operator, using its commutator with the anti-chiral vertex operator
\begin{equation} \label{bn commute e chi}
\big[ \bar \b_n^{\epsilon'}, \nord{e^{\pm i\sqrt{8\pi}\bar \chi^{\epsilon'}(x)}} \big] = \pm \sqrt{2}e^{-2\pi i n x/L}\eta\left(\frac{2\pi n \epsilon'}{L}\right)\nord {e^{\pm i\sqrt{8\pi}\bar \chi^{\epsilon'}(x)}}.
\end{equation}
The resulting normal-ordered version of \eqref{H21 variation x_1 int} then takes the form
\begin{align}\label{H21 variation commute mode}
    \delta \H_{21\uparrow,\pm}^{\epsilon'(1,0)} &= \pm \frac{\pi \delta\epsilon}{2 L^2} g_1(\epsilon)g_2(\epsilon)
    \int_0^T \dd \tau \int_0^L \dd x_2 \Wop^\epsilon(\tau) \sum_{n > 0} \nord{e^{\pm i\sqrt{8\pi} \chi^{\epsilon'}(x_2)}e^{\pm i\sqrt{8\pi} \bar \chi^{\epsilon'}(x_2)}}\bar \b_n^{\epsilon'} \\
    &\qquad\qquad\qquad\qquad \times \frac{1}{\sqrt{2}}\left(\frac{2\pi}{L}\right)^2 n \eta'\left(\frac{2\pi n \epsilon'}{L}\right)e^{-4\pi n \tau/L}e^{2\pi i n x_2/L}\notag\\
    &\quad  +\frac{\pi \delta\epsilon}{2 L^2} g_1(\epsilon)g_2(\epsilon)
    \int_0^T \dd \tau \int_0^L \dd x_2 \Wop^\epsilon(\tau) \nord{e^{\pm i\sqrt{8\pi} \chi^{\epsilon'}(x_2)}e^{\pm i\sqrt{8\pi} \bar \chi^{\epsilon'}(x_2)}} \notag\\
    &\qquad\qquad\qquad\qquad \times \left(\frac{2\pi}{L}\right)^2\sum_{n > 0}n\eta'\left(\frac{2\pi n \epsilon'}{L}\right)\eta\left(\frac{2\pi n \epsilon'}{L}\right)e^{-4\pi n \tau/L} + O(\delta\epsilon^2). \notag
\end{align}
For similar reasons to the discussion after \eqref{H22 variational flow} in \S\ref{sec: H22 var}, the first of these terms, which takes up the first two lines, represents an infinite sum of irrelevant operators. Indeed, it consists of operators of conformal dimension at least $3$ and after dividing by $\delta\epsilon$ and multiplying through by $\epsilon$ to compute the limit \eqref{beta functions}, these operators will all come multiplied by strictly positive powers of $\epsilon$, in accordance with the general pattern \eqref{couplings}.

Consider now the second term on the right hand side of \eqref{H21 variation commute mode}, which takes up the last two lines. Using the argument from \S\ref{sec: H22 var} we can replace $\Wop^\epsilon(\tau)$ by the identity operator up to subleading terms in $\epsilon$, and in fact such terms will be non-singular. We can then evaluate the $\tau$ integral in the second term on the right hand side of \eqref{H21 variation commute mode} to obtain
\begin{align}\label{H21 variation tau integral computed}
    \delta \H_{21\uparrow,\pm}^{\epsilon'(1,0)} &= \frac 14 \delta\epsilon g_1(\epsilon)g_2(\epsilon) \frac{\pi}{L^2} \int_0^L \dd x_2 \nord{e^{\pm i\sqrt{8\pi} \chi^{\epsilon'}(x_2)}e^{\pm i\sqrt{8\pi} \bar \chi^{\epsilon'}(x_2)}}\\
    &\qquad\qquad\qquad \times \frac{2 \pi}{L}\sum_{n > 0}(1 - e^{-4\pi n T/L})\eta'\left(\frac{2\pi n \epsilon'}{L}\right)\eta\left(\frac{2\pi n \epsilon'}{L}\right) + O(\log \epsilon', \delta\epsilon^2), \notag
\end{align}
where the $O(\log \epsilon')$ terms account for the infinite sum of irrelevant operators coming from the $O(\tau)$ terms in the expansion of $\Wop^{\epsilon}(\tau)$. Indeed, as explained in \S\ref{sec: H22 var}, due to the presence of the exponential $e^{-4 \pi n \tau / L}$, a power $\tau^k$ with $k > 0$ will introduce an additional factor of $n^{-k}$ upon integrating over $\tau$, rendering the sum over $n > 0$ in the second term on the right hand side of \eqref{H21 variation commute mode} more convergent. The term which is linear in $\tau$ produces a regularised harmonic sum whose leading asymptotics is a term $O(\log \epsilon')$ by \eqref{harmonic series regularised}.
The factor of $1 - e^{-4\pi n T/L}$ in \eqref{H21 variation tau integral computed} can be replaced with $1$ as the sum involving the exponential $e^{-4\pi n T/L}$ produces a non-singular expression in $\epsilon'$. Dividing the expression \eqref{H21 variation tau integral computed} by $\delta\epsilon$ and multiplying through by $\epsilon$ to compute the limit as in \eqref{beta functions} we find
\begin{align}\label{H21 variational derivative}
\lim_{\epsilon' \to \epsilon}\epsilon'\frac{\delta \H_{21\uparrow,\pm}^{\epsilon'(1,0)}}{\delta\epsilon} = \frac 14 \epsilon g_1(\epsilon)g_2(\epsilon)
\frac{2\pi}{L}\sum_{n > 0}\eta'\left(\frac{2\pi n \epsilon}{L}\right)\eta\left(\frac{2\pi n \epsilon}{L}\right) \int_0^L \dd x \, \mathsf O_{1\pm}^{\epsilon}(x) + O(\epsilon \log \epsilon)
\end{align}
where we have used the defintion \eqref{O1 pm defintion}.
The asymptotics of the sum is computed using Mellin transform theory from \S\ref{sec: Mellin transform}, specifically the regularised series \eqref{regularised n^r series} with $r = 0$, yielding
\begin{align}\label{H21 variational derivative 2}
\lim_{\epsilon' \to \epsilon}\epsilon\frac{\delta \H_{21\uparrow,\pm}^{\epsilon'(1,0)}}{\delta\epsilon} = \frac 14 g_1(\epsilon)g_2(\epsilon) C_{\eta'\eta, 0} \int_0^L \dd x \, \mathsf O_{1\pm}^{\epsilon}(x) + O(\epsilon).
\end{align}
It is important to note here that the constant $C_{\eta'\eta, 0} = \int_0^\infty\eta'(x)\eta(x) \dd x = \frac{1}{2}[\eta(x)^2]_0^\infty = -\frac{1}{2}$ is, in fact, independent of the choice of cutoff function $\eta : \RR_{\geq 0} \to \RR_{>0}$. This is to be expected since we are computing here the running of the coupling of a marginal operator $\int_0^L \dd x \, \mathsf O_{1\pm}^{\epsilon'}(x)$ which is tied to logarithmic divergences and, unlike polynomial divergences, these are generally scheme independent. Indeed, the result \eqref{H21 variational derivative 2} is to be compared with the result \eqref{var H22 final} that controlled the running of the relevant coupling $g_3(\epsilon)$ of the identity operator, i.e. the constant term in the Hamiltonian \eqref{Regularized bare Hamiltonian}, and which is tied to a polynomial divergence whose scheme-dependence is manifest in the fact that the coefficient $C_{\eta'\eta,2}$ appearing in \eqref{var H22 final} is $\eta$-dependent.

Finally, adding together the two contributions obtained in \eqref{H21 variational derivative 2} and using the fact that $\mathsf{O}_1^\epsilon(x) = \mathsf{O}_{1+}^\epsilon(x) + \mathsf{O}_{1-}^\epsilon(x)$ from \S\ref{sec: Anisotropic def} gives
\begin{align}\label{H21 variational derivative 4}
\lim_{\epsilon' \to \epsilon}\epsilon\frac{\delta \H_{21\uparrow}^{\epsilon'(1,0)}}{\delta\epsilon} = - \frac 18 g_1(\epsilon)g_2(\epsilon) \int_0^L \dd x\, \mathsf O_1^{\epsilon}(x) + O(\epsilon).
\end{align}
Similar calculations for the variations $\delta \H_{21\uparrow}^{\epsilon'(0,1)}$, $\delta \H_{21\downarrow}^{\epsilon'(1,0)}$ and $\delta \H_{21\downarrow}^{\epsilon'(0,1)}$ lead to the same result as the right hand side of \eqref{H21 variational derivative 4}, so that altogether we obtain
\begin{align}\label{H21 variational derivative 5}
\lim_{\epsilon' \to \epsilon}\epsilon\frac{\delta \H_{21}^{\epsilon'}}{\delta\epsilon} = - \frac 12 g_1(\epsilon)g_2(\epsilon) \int_0^L \dd x \, \mathsf O_1^{\epsilon}(x) + O(\epsilon).
\end{align}

Consider now the variation $\delta \H^{\epsilon'}_{12}$ in which the vertex operators $\Vop^{\epsilon',(1,0)}_1(x_1)$ and $\Vop^{\epsilon',(0,1)}_1(x_1)$ are imaginary-time evolved. The result in this case turns out to be exactly the same as \eqref{H21 variational derivative 5} but the intermediate steps are slightly different. For illustration purposes we will focus here on the backward-time evolved part $\delta \H^{\epsilon'}_{12\downarrow}$ of the variation, recall the definition \eqref{Delta H ij plus}, since in our previous computation we looked at the forward-time evolved one. We will briefly sketch the computation of the variation $\delta \H^{\epsilon'(1,0)}_{12\downarrow,\pm}$, the other cases forming $\delta \H^{\epsilon'}_{12}$ being very similar. So our starting point is
\begin{align}\label{H12 variation def}
    \delta \H_{12\downarrow,\pm}^{\epsilon'(1,0)} &= \mp \frac{\pi \delta\epsilon}{2 L^2} g_1(\epsilon)g_2(\epsilon)
    \int_{-T}^0 \dd \tau \int_0^L \dd x_1 \int_0^L \dd x_2  \Wop^\epsilon(\tau) \Delta^\epsilon_s(x_2, x_1 - i \tau)\\
    &\qquad\qquad\qquad\qquad \times \partial_{x_2}^2 \bar \chi^{\epsilon'}(x_2) i\sqrt{8\pi} \nord{e^{\pm i\sqrt{8\pi} \chi^{\epsilon'}(x_1 - i\tau)}e^{\pm i\sqrt{8\pi} \bar \chi^{\epsilon'}(x_1 + i\tau)}} + O(\delta\epsilon^2), \notag
\end{align}
which is to be compared with the starting point \eqref{H21 variation def} of our previous computation. Note, in particular, that it is now the vertex operator which is being imaginary-time evolved and, since we are considering the range of integration $\tau \in [-T, 0]$, it is sitting to the right of $\partial_{x_2}^2 \bar \chi^{\epsilon'}(x_2)$. Recall, moreover, the abuse of notation used in \eqref{effective potential expand} which explains the order of the arguments in the $2$-point function $\Delta^\epsilon_s(x_2, x_1 - i\tau)$.

Carrying out the $x_2$ integral and commuting the mode $\bar \b_n^{\epsilon'}$ past the vertex operator using the identity in \eqref{bn commute e chi} gives
\begin{align}\label{H12 variation commute mode}
    \delta \H_{12\downarrow,\pm}^{\epsilon'(1,0)} &= \text{irrelevant operators}\\
    &\quad + \frac{\pi \delta\epsilon}{2 L^2} g_1(\epsilon)g_2(\epsilon) \int_{-T}^0 \dd \tau \int_0^L \dd x_1
    \Wop^\epsilon(\tau) \nord{e^{\pm i\sqrt{8\pi} \chi^{\epsilon'}(x_1 -i\tau)}e^{\pm i\sqrt{8\pi} \bar \chi^{\epsilon'}(x_1 + i\tau)}} \notag\\
    &\qquad\qquad\qquad\qquad \times \left(\frac{2\pi}{L}\right)^2\sum_{n > 0}n\eta'\left(\frac{2\pi n \epsilon'}{L}\right)\eta\left(\frac{2\pi n \epsilon'}{L}\right)e^{4\pi n \tau/L}
    + O(\delta\epsilon^2), \notag
\end{align}
which is to be compared with \eqref{H21 variation commute mode} from the above computation. The `irrelevant operators' here refers to the analogue of the first two lines of \eqref{H21 variation commute mode} which consists of irrelevant operators, again by the same arguments as in the discussion after \eqref{H22 variational flow} in \S\ref{sec: H22 var}.
The term written in \eqref{H12 variation commute mode} and the second term on the right hand side of \eqref{H21 variation commute mode} are almost identical. Aside from the fact that \eqref{H12 variation commute mode} involves an integral over $\tau \in [-T, 0]$ and a positive exponential $e^{4\pi n \tau/L}$, as we are considering the backward-time evolved case, the main difference is that the vertex operators appearing in \eqref{H12 variation commute mode} are imaginary-time evolved, while the ones in \eqref{H21 variation commute mode} were not. 

To deal with the additional time-evolution of the vertex operators, we can expand them in $\tau$ and use the same argument as given before \eqref{delta H22 computation} to show that the $\tau$-dependent terms in this expansion will correspond to irrelevant operators. We can thus effectively remove the imaginary-time dependence of the vertex operators in \eqref{H12 variation commute mode}, and the rest of the computation then follows through exactly as above and leads to the same result as in \eqref{H21 variational derivative 5}, namely
\begin{align}\label{H12 variational derivative 5}
\lim_{\epsilon' \to \epsilon}\epsilon\frac{\delta \H_{12}^{\epsilon'}}{\delta\epsilon} = - \frac 12 g_1(\epsilon)g_2(\epsilon) \int_0^L \dd x \, \mathsf O_1^{\epsilon}(x) + O(\epsilon).
\end{align}
Combining the results \eqref{H21 variational derivative 5} and \eqref{H12 variational derivative 5} we can finally read off the quantum beta function, defined in \eqref{beta functions}, for the coupling $g_1(\epsilon)$ of the operator $\int_0^L \dd x \,\mathsf O_1^\epsilon(x)$ to be
\begin{equation} \label{qu beta for g1}
    \beta^{\rm qu}_1[ (g_k(\epsilon)) ] = g_1(\epsilon)g_2(\epsilon) + \ldots,
\end{equation}
where as in \eqref{qu beta for g3}, `$\ldots$' denotes possible corrections coming from the other variations.

\subsubsection{\texorpdfstring{$\delta \H^{\epsilon'}_{11}$}{H11} variation} \label{sec: H11 var}

As with the previous calculations, to compute $\delta \H^{\epsilon'}_{11}$ we will start with the $\delta \H^{\epsilon'(1,0)}_{11\uparrow}$ piece. In fact, it follows from the definition \eqref{Delta H ij plus plus} and using the relation \eqref{V1 01 = V1 10} that $\delta \H^{\epsilon'(0,1)}_{11\uparrow}$ is given by the exact same expression as $\delta \H^{\epsilon'(1,0)}_{11\uparrow}$ but with $\Delta_s^\epsilon(x_1 - i\tau, x_2)$ replaced by $\bar\Delta_s^\epsilon(x_1 + i\tau, x_2)$.
Recalling from the definition of the shell $2$-point functions \eqref{propagator shell modes} that $\bar\Delta_s^\epsilon(x_1 + i\tau, x_2)$ is the complex conjugate of $\Delta_s^\epsilon(x_1 - i\tau, x_2)$, we introduce their sum
\begin{align} \label{hat Delta def}
\widehat{\Delta}^\epsilon(x_1 - x_2, \tau) &\coloneqq \Delta^\epsilon_s(x_1-i\tau, x_2) + \bar\Delta^\epsilon_s(x_1+i\tau, x_2) = 2 \text{Re} \big(\Delta^\epsilon_s(x_1-i\tau, x_2) \big)\\
&\, = - \frac{1}{L} \sum_{n > 0} \eta'\big( \tfrac{2 \pi n \epsilon}{L} \big) \cos\big( \tfrac{2 \pi n}L (x_2 - x_1) \big) e^{-2\pi n \tau/L}, \notag
\end{align}
where we have dropped the `$s$'-subscript, standing for `shell', for simplicity.
The forward-time contribution \eqref{Delta H ij plus} to the variation $\delta \H^{\epsilon'}_{11}$ is then given simply by
\begin{align}\label{Delta H 11 plus}
\delta \H^{\epsilon'}_{11\uparrow} &= \frac{\delta\epsilon}{2} g_1(\epsilon)^2 \int_0^T \dd \tau \int_0^L \dd x_1 \int_0^L \dd x_2 \, \Wop^\epsilon(\tau) \widehat{\Delta}^{\epsilon}(x_1 - x_2, \tau) \\
&\qquad\qquad\qquad\qquad\qquad\qquad\qquad\qquad \times \Vop_1^{\epsilon', (1,0)}(x_1, \tau) \Vop_1^{\epsilon', (1,0)}(x_2) + O(\delta\epsilon^2), \notag
\end{align}
and the corresponding backward-time contribution takes the same form with the integral over $\tau \in [-T, 0]$, the two vertex operators swapped and $\widehat{\Delta}^{\epsilon}(x_1 - x_2, \tau)$ replaced by $\widehat{\Delta}^{\epsilon}(x_1 - x_2, -\tau)$.
We further split the computation of \eqref{Delta H 11 plus} into four parts by defining
\begin{align}\label{Delta H 11 rs}
\delta \H^{\epsilon'}_{11\uparrow, rs} &\coloneqq \frac{\delta\epsilon}{2} g_1(\epsilon)^2 \int_{0}^T \dd \tau \int_0^L \dd x_1 \int_0^L \dd x_2 \, \Wop^\epsilon(\tau) \widehat{\Delta}^{\epsilon}(x_1 - x_2, \tau) \\
&\qquad\qquad\qquad\qquad\qquad\qquad\qquad\qquad \times \Vop_{1r}^{\epsilon', (1,0)}(x_1, \tau) \Vop_{1s}^{\epsilon', (1,0)}(x_2) + O(\delta\epsilon^2) \notag\\
&\, = \frac{\delta\epsilon}{2} \left(\frac{\pi}{L^2}\right)^2 (-8\pi rs) g_1(\epsilon)^2 \int_0^T \dd \tau \int_0^L \dd x_1 \int_0^L \dd x_2 \Wop^\epsilon(\tau) \widehat{\Delta}^\epsilon(x_1 - x_2, \tau) \notag\\
&\qquad \times \nord{e^{ir\sqrt{8\pi} \chi^{\epsilon'}(x_1 - i\tau)}e^{ir\sqrt{8\pi} \bar \chi^{\epsilon'}(x_1 + i\tau)}}\nord{e^{i s \sqrt{8\pi} \chi^{\epsilon'}(x_2)}e^{is\sqrt{8\pi} \bar \chi^{\epsilon'}(x_2)}} + O(\delta\epsilon^2), \notag
\end{align}
where $r, s \in \{ \pm 1 \}$ and it is understood that when these appear as subscripts we just write the corresponding sign $\pm$ rather than $\pm 1$, e.g. $\delta \H^{\epsilon'}_{11\uparrow, rs} = \delta \H^{\epsilon'}_{11\uparrow, ++}$ when $r=s=1$.

To evaluate the right hand side of \eqref{Delta H 11 rs} it will be useful to first compute the operator product of two regularised vertex operators, of charge $r$ and $s$, with one located at $x_1 - i\tau$ and the other located at $x_2$. Recalling that normal ordering of vertex operators in the coordinate on the cylinder is defined as in \eqref{normal ordered exponential 0mode}, and likewise for the regularised versions, we find for the chiral part
\begin{align}
    \nord{e^{ir\sqrt{8\pi}\chi^{\epsilon'}(x_1 - i\tau)}}
    \nord{e^{is\sqrt{8\pi}\chi^{\epsilon'}(x_2)}} &=
    \nord{e^{i\sqrt{8\pi}(r \chi^{\epsilon'}(x_1 - i\tau) + s\chi^{\epsilon'}(x_2))}}\\
    &\qquad\qquad \times e^{-8\pi rs [\chi^{\epsilon'}_-(x_1 - i\tau), \chi_+^{\epsilon'}(x_2)]}e^{2rs\pi i (x_1 - x_2 - i\tau)/L} \notag
\end{align}
using the Baker–Campbell–Hausdorff formula.
In particular, the final exponential factor comes from combining the two zero mode exponentials into a single exponential. The commutator can be evaluated to give
\begin{equation}
    [\chi^{\epsilon'}_-(x_1 - i\tau), \chi^{\epsilon'}_+(x_2)] = 
    \frac{1}{4\pi} \sum_{n > 0} \frac{1}{n} \eta\left(\frac{2\pi n \epsilon'}{L}\right)
    e^{-2\pi i n (x_1 - x_2 - i\tau)/L}.
\end{equation}
Combining with the anti-chiral regularised vertex operators, we have
\begin{align}\label{Full vertex operator OPE}
&\nord{e^{ir\sqrt{8\pi}\chi^{\epsilon'}(x_1 - i\tau)}
e^{ir\sqrt{8\pi}\bar\chi^{\epsilon'}(x_1 + i\tau)}}
\nord{e^{is\sqrt{8\pi}\chi^{\epsilon'}(x_2)}
e^{is\sqrt{8\pi}\bar\chi^{\epsilon'}(x_2)}} \\
&\quad =
\nord{e^{i \sqrt{8\pi} ( r \chi^{\epsilon'}(x_1 - i\tau) + s \chi^{\epsilon'}(x_2) )}}
\nord{e^{i\sqrt{8\pi} ( r \bar\chi^{\epsilon'}(x_1 + i\tau) + s \bar\chi^{\epsilon'}(x_2) )}}
\; e^{-4rs \textup{Re} (G(x_1 - x_2 - i\tau; \epsilon'))}
e^{4rs\pi\tau/L},\notag
\end{align}
where we have introduced the smoothly regularised series depending on a complex parameter $u \in \CC$,
\begin{equation} \label{G def}
G(u; \epsilon) \coloneqq \sum_{n > 0} \frac{e^{-2\pi i n u / L}}{n} \eta\big( \tfrac{2 \pi n \epsilon}L \big).
\end{equation}
Substituting \eqref{Full vertex operator OPE} into the right hand side of \eqref{Delta H 11 rs} we obtain
\begin{align}\label{Delta H 11 long}
\delta \H^{\epsilon'}_{11\uparrow, rs} &= \frac{\delta\epsilon}{2} \left(\frac{\pi}{L^2}\right)^2 (-8\pi rs) g_1(\epsilon)^2 \int_0^T \dd \tau \int_0^L \dd x_1 \int_0^L \dd x_2 \Wop^\epsilon(\tau)\\
&\qquad\qquad\qquad\qquad \times \nord{e^{i\sqrt{8\pi} (r \chi^{\epsilon}(x_1 - i\tau) + s \chi^{\epsilon}(x_2))}}
\nord{e^{i \sqrt{8\pi} (r \bar\chi^{\epsilon}(x_1 + i\tau) + s \bar\chi^{\epsilon}(x_2))}}\notag\\
&\qquad\qquad\qquad\qquad\qquad\qquad \times e^{-4 rs \textup{Re}(G(x_1 - x_2 - i\tau; \epsilon))}e^{4rs\pi\tau/L} \; \widehat{\Delta}^\epsilon(x_1 - x_2, \tau) + O(\delta\epsilon^2). \notag
\end{align}
Since we are working only up to $O(\delta\epsilon^2)$ and the leading term is already linear in $\delta\epsilon = \epsilon' - \epsilon$, we have replaced all $\epsilon'$ on the right hand side by $\epsilon$.

As in \S\ref{sec: H22 var} and \S\ref{sec: H21 and H12 var}, we will use the variation \eqref{Delta H 11 long} to determine the renormalisation group flows of the couplings written explicitly in \eqref{Regularized bare Hamiltonian}, according to \eqref{beta functions}. It then remains to work out the singular behaviour of \eqref{Delta H 11 long} as $\epsilon \to 0$. Specifically, we will focus on the marginal couplings $g_1$ and $g_2$ so we are primarily interested in $O(\epsilon^{-1})$ terms since when multiplied by $\epsilon$ and divided by $\delta \epsilon$, as in \eqref{beta functions}, such singularities will contribute to the flow of $g_1$ and $g_2$.

Note that the series \eqref{G def} appears in \eqref{Delta H 11 long} evaluated at $u = x_1 - x_2 - i \tau \in \CC$, for which $\text{Im}(u) \leq 0$ since $\tau \geq 0$. In this region, the series \eqref{G def} is convergent even when the regularising factor $\eta(2\pi n \epsilon/L)$ is not present, except at the point $u=0$.
In fact, we need the behaviour of the exponential $e^{- 4rs \text{Re}(G(u; \epsilon))}$ which is worked out in Appendix \ref{sec: Asymptotic of G}. When $\tau > 0$ we can apply the identity \eqref{identity for exp G} to $u = x_1 - x_2 - i \tau$, which satisfies $\text{Im}(u) < 0$, to obtain the relation
\begin{equation}\label{exp correlator}
e^{-4rs\textup{Re}(G(x_1 - x_2 - i\tau; \epsilon))} = \big| 1 - e^{-2\pi i(x_1 - x_2)/L}e^{-2\pi (\tau - a_1\epsilon)/L} \big|^{4rs} + O(\epsilon^2),
\end{equation}
where $a_1 \coloneqq \eta'(0)$ denotes the first coefficient in the Taylor expansion of $\eta$ at the origin. We will assume $a_1 < 0$ so that $\tau - a_1\epsilon > 0$ for $\tau > 0$.
Note also from the definitions \eqref{hat Delta def} and \eqref{G def} that we have the simple relation
\begin{equation} \label{Delta G relation}
\widehat{\Delta}^\epsilon(x, \tau) = - \frac{1}{2 \pi} \partial_\epsilon \text{Re}\big( G(x - i\tau; \epsilon) \big),
\end{equation}
for $x \in \RR$ and $\tau \geq 0$.
Given the expression \eqref{exp correlator}, there are now two distinct cases to consider, depending on the sign of the product $rs$ in the exponent; see in particular \eqref{limit identity for exp G}.

When $rs > 0$, the expression on the right hand side of \eqref{exp correlator} is regular in $\epsilon$ and of order at least $O(\epsilon^2)$ for all $x_1, x_2 \in [0,L]$ and $\tau \geq 0$, covering the entire integration region of \eqref{Delta H 11 long}. Moreover, the series $\widehat{\Delta}^\epsilon(x_1 - x_2, \tau)$ in \eqref{hat Delta def} will produce a divergence of order at most $O(\epsilon^{-1})$ by the theory of Mellin transforms from \S\ref{sec: Mellin transform}. So overall, the variation \eqref{Delta H 11 long} will produce only regular terms in $\epsilon$. Indeed, expanding the normal-ordered operator in \eqref{Delta H 11 long} for $u = x_1 - x_2 - i \tau$ near the origin we obtain operators of the form
\begin{equation} \label{irrelevant VO example}
\nord{\partial_{x_2}^m\chi^{\epsilon}(x_2)\partial_{x_2}^n\chi^{\epsilon}(x_2)e^{2ir\sqrt{8\pi}\chi^{\epsilon}(x_2)} e^{2ir\sqrt{8\pi}\bar\chi^{\epsilon}(x_2)}}
\end{equation}
for $m,n \in \ZZ_{\geq 0}$, which have conformal dimension $m + n + 8 > 2$ and are therefore irrelevant.

\medskip

The other case is when $rs < 0$, i.e. $r = \pm 1 = - s$, and without loss of generality we will focus on the case $r = 1$. In this case, \eqref{exp correlator} has a singularity of order $O(\epsilon^{-4})$ at the coinciding point limit $|x_1 - x_2 - i\tau| \to 0$, see \eqref{limit identity for exp G}. This is the familiar $|z-w|^{-4}$ singularity coming from the product of the leading order singularities in the operator product expansions \eqref{Current OPE} of the chiral and anti-chiral Kac--Moody currents, namely $J^+[z]$ with $J^-[w]$ and $\bar J^+[\bar z]$ with $\bar J^-[\bar w]$.
To obtain also the subleading orders in the singularity we expand the normal-ordered operators appearing in \eqref{Delta H 11 long} in the limit $|x_1 - i\tau - x_2| \to 0$. We find
\begin{align}\label{Chiral boson expansion}
&\nord{e^{i\sqrt{8\pi} (\chi^{\epsilon}(x_1 - i\tau) - \chi^{\epsilon}(x_2))}}
\nord{e^{i\sqrt{8\pi} (\bar \chi^{\epsilon}(x_1 + i\tau) - \bar\chi^{\epsilon}(x_2))}}\\
&\qquad = 1 + i\sqrt{8\pi}\frac{L}{2\pi} \Big( (e^{2\pi i(x_1 - i\tau - x_2)/L} - 1)\partial_{x_2}\chi^{\epsilon}(x_2) +
(e^{2\pi i(x_1 + i\tau - x_2)/L} - 1)\partial_{x_2}\bar \chi^{\epsilon}(x_2) \Big) \notag\\
&\qquad\qquad -8\pi \left(\frac{L}{2\pi}\right)^2 \Big( \big| e^{2\pi i(x_1 - x_2)/L}e^{-2\pi \tau/L} - 1 \big|^2 \partial_{x_2}\chi^{\epsilon}(x_2)\partial_{x_2}\bar\chi^{\epsilon}(x_2) \notag\\
&\qquad\qquad\qquad\qquad\qquad\qquad + \tfrac 12 \big( e^{2\pi i(x_1 - x_2)/L}e^{-2\pi \tau/L} - 1 \big)^2 \nord{(\partial_{x_2}\chi^{\epsilon}(x_2))^2} \notag\\
&\qquad\qquad\qquad\qquad\qquad\qquad\qquad + \tfrac 12 \big( e^{-2\pi i(x_1 - x_2)/L}e^{-2\pi \tau/L} - 1 \big)^2 \nord{(\partial_{x_2}\bar\chi^{\epsilon}(x_2))^2} \Big) + \ldots \notag
\end{align}
where for later convenience we performed the expansion in the coordinate $z = e^{iu}$ on the plane, but still wrote coefficient fields in this expansion using the cylinder coordinate $u$.

The leading $1$ on the right hand side of \eqref{Chiral boson expansion} will produce a correction of order $O(g_1^2)$ to the quantum beta function \eqref{qu beta for g3} of the coupling $g_3$ of the identity operator, i.e. the constant term in the Hamiltonian \eqref{Regularized bare Hamiltonian}. However, the detailed computation of $\delta \H^{\epsilon'}_{22}$ in \S\ref{sec: H22 var} was for illustration purposes since we are focusing on deriving the flows of the marginal couplings $g_1$ and $g_2$, and from now on we will therefore ignore the contribution from $1$ in \eqref{Chiral boson expansion}.

Next, let us consider the second term on the right hand side of \eqref{Chiral boson expansion}, involving the linear combination of the operators $\partial_{x_2}\chi^{\epsilon}(x_2)$ and $\partial_{x_2}\bar\chi^{\epsilon}(x_2)$. Upon substituting it into \eqref{Delta H 11 long} with $r=-s=1$ and performing the integral over $x_2$ (after changing variables in the double integral over $x_1$ and $x_2$ to the variables $x = x_1 - x_2$ and $x_2$), we are left simply with a linear combination of the zero-modes $\b_0^{\epsilon}$ and $\bar\b_0^{\epsilon}$. These are acted on by $\Wop^\epsilon(\tau)$ but using the explicit form \eqref{Wop explicit} of this operator and of the regularised free Hamiltonian in \eqref{WZW Hamiltonian free realisation smth reg}, we see that $\Wop^\epsilon(\tau) \b_0^{\epsilon} = \b_0^{\epsilon}$ and $\Wop^\epsilon(\tau) \bar\b_0^{\epsilon} = \bar\b_0^{\epsilon}$. This particular piece of the variation $\delta \H^{\epsilon'}_{11+, +-}$ would therefore produce a flow in the coupling of the zero mode of the chiral and anti-chiral bosons $\int_0^L \dd x \partial_x \chi^\epsilon(x)$ and $\int_0^L \dd x \partial_x \bar\chi^\epsilon(x)$, which we have not explicitly included in our original Hamiltonian \eqref{Regularized bare Hamiltonian}. We will also ignore these since we are focusing on the marginal couplings $g_1$ and $g_2$. In fact, just like the identity operator, these zero-mode operators have minimal impact on renormalisation as they are block diagonal in the sense of \S\ref{sec: short long split} and hence do not contribute to the quantum beta function of any coupling.
Note also that terms proportional to higher order derivatives of the fields, starting with $\partial_{x_2}^2\chi^{\epsilon}(x_2)$ and $\partial_{x_2}^2\bar\chi^{\epsilon}(x_2)$, which we have omitted in \eqref{Chiral boson expansion}, are total derivatives of periodic operators in $x_2$ and hence vanish upon integration over $x_2 \in [0, L]$.

\medskip

Let us then focus on the last term written on the right hand side of \eqref{Chiral boson expansion}, which takes up the last three lines. We will focus on the first of these three lines, namely the term proportional to $\partial_{x_2}\chi^{\epsilon}(x_2) \partial_{x_2} \bar \chi ^{\epsilon}(x_2)$, since the other two lines will turn out not to be singular in $\epsilon$. We will comment on this point at the end of this section. Since we are focusing on a particular term from the expansion \eqref{Chiral boson expansion}, let us denote its contribution to the variation $\delta \H^{\epsilon'}_{11\uparrow, +-}$ as
\begin{align} \label{H11 contribution to g2}
\delta \H^{\epsilon'}_{11\uparrow, +-}\big|_{\mathsf O_2^\epsilon} &= -\frac{8\pi^2}{L^2}\delta\epsilon g_1(\epsilon)^2 \int_0^T \dd \tau \int_{-L/2}^{L/2} \dd x \int_0^L \dd x_2 \Wop^\epsilon(\tau) \mathsf O_2^\epsilon(x_2)\\
&\qquad\qquad \times |1 - e^{2\pi i x/L}e^{-2\pi \tau/L}|^2 \widehat{\Delta}^\epsilon(x, \tau)e^{4\textup{Re}(G(x - i\tau; \epsilon))} e^{-4\pi\tau/L} + O(\epsilon^0, \delta\epsilon^2). \notag
\end{align}
Here we have used the fact that the integrand in \eqref{Delta H 11 long} is periodic in $x_1$ to shift the integration range from $x_1, x_2 \in [0,L]$ to $x_1 \in [-L/2+x_2,L/2+x_2]$ and $x_2 \in [0,L]$. Then after inserting the expansion \eqref{Chiral boson expansion} we have changed variable from $x_1$ to $x \coloneqq x_1 - x_2 \in [-L/2,L/2]$.

In \eqref{H11 contribution to g2} we are also anticipating that the first term on the right hand side will generate infinitely many non-singular terms as $\epsilon \to 0$, for instance the subleading terms in the expansion of $\Wop^\epsilon(\tau)$ in powers of $\tau$, and we are including a term $O(\epsilon^0)$ to cancel these off. In other words, as the notation suggests, \eqref{H11 contribution to g2} corresponds exactly to the term inducing the quantum beta function of the coupling $g_2(\epsilon)$, i.e. of the operator $\int_0^L \dd x \, \mathsf O_2^\epsilon(x)$. Recall the definition \eqref{O2 definition} of the operator $\mathsf O_2^\epsilon(x)$. Our goal is therefore to extract the singular in $\epsilon$ contribution to the first term on the right hand side of \eqref{H11 contribution to g2}.

We can use \eqref{Delta G relation} to rewrite \eqref{H11 contribution to g2} as
\begin{align} \label{H11 contribution to g2 d epsilon}
\delta \H^{\epsilon'}_{11\uparrow, +-}\big|_{\mathsf O_2^\epsilon} &= \partial_\epsilon \Bigg( \frac{\pi}{L^2} \int_0^T \dd \tau \int_{-L/2}^{L/2} \dd x \, |1 - e^{2\pi i x/L}e^{-2\pi \tau/L}|^2 e^{4\textup{Re}(G(x - i\tau; \epsilon))} e^{-4\pi\tau/L} \Bigg) \notag\\
&\qquad\qquad\qquad\qquad \times \delta\epsilon g_1(\epsilon)^2 \int_0^L \dd x_2 \Wop^\epsilon(\tau) \mathsf O_2^\epsilon(x_2) + O(\epsilon^0, \delta\epsilon^2).
\end{align}
It therefore remains to determine the singular behaviour of the double integral in brackets on the right hand side of \eqref{H11 contribution to g2 d epsilon} as $\epsilon \to 0$.
Since this comes from the singularity of $e^{4\textup{Re}(G(x - i\tau; \epsilon))}$ as $x, \tau, \epsilon \to 0$, in order to extract the most singular term we can approximate the integrand in the region where $x$, $\tau$ and $\epsilon$ are small. Explicitly, using \eqref{exp correlator} we find
\begin{equation} \label{approximation e^G and Delta}
e^{4\textup{Re}(G(x - i\tau; \epsilon))} \approx \bigg(\frac{L}{2\pi}\bigg)^4 \frac{1}{(x^2 + (\tau-a_1 \epsilon)^2)^2}.
\end{equation}
Using also the approximations $|1 - e^{2\pi i x/L}e^{-2\pi \tau/L}|^2 \approx \big( \frac{2\pi}{L} \big)^2 (x^2 + \tau^2)$ and $e^{-4\pi\tau/L} \approx 1$, and expanding the operator $\Wop^{\epsilon}(\tau) = \sum_{k \geq 0} \tau^k \Wop^{\epsilon(k)}$ in powers of $\tau$ as in \S\ref{sec: H22 var}, where we set $\Wop^{\epsilon(0)} \coloneqq \id$, we may rewrite \eqref{H11 contribution to g2 d epsilon} as
\begin{align} \label{H11 contribution to g2 expand}
\delta \H^{\epsilon'}_{11\uparrow, +-}\big|_{\mathsf O_2^\epsilon} &= \sum_{k \geq 0} \bigg( \frac{1}{4\pi} \int_0^T \dd \tau \int_{-\infty}^\infty \dd x \, \frac{\tau^k (x^2 + \tau^2)}{(x^2 + (\tau-a_1 \epsilon)^2)^2} \bigg)\\
&\qquad\qquad\qquad\qquad\qquad \times \delta\epsilon g_1(\epsilon)^2 \int_0^L \dd x_2\, \Wop^{\epsilon(k)} \mathsf O_2^\epsilon(x_2) + O(\epsilon^0, \delta\epsilon^2). \notag
\end{align}
Since the singular behaviour in $\epsilon$ comes from the region of integration near $x=0$ we have also extended the integration region from $x \in [-L/2, L/2]$ to $x \in \RR$. The double integral in the brackets in \eqref{H11 contribution to g2 expand} can now be evaluated explicitly and we find that for $k>0$ this contributes regular terms in $\epsilon$, see \S\ref{sec: asymp double int 1} for details. It follows that none of the terms $k>0$ in the above sum will contribute to the flow of marginal or relevant operators. The double integral for $k=0$ evaluates to, again see \S\ref{sec: asymp double int 1} for details,
\begin{equation} \label{Contour integral computation 1}
\frac{1}{4\pi} \int_0^T \dd \tau \int_{-\infty}^\infty \dd x \, \frac{x^2 + \tau^2}{(x^2 + (\tau-a_1 \epsilon)^2)^2} = - \frac{1}{4} \log \epsilon + O(\epsilon^0).
\end{equation}
Ignoring the regular piece as usual and substituting the singular part back into \eqref{H11 contribution to g2 expand} gives
\begin{align}\label{H11 asymptotics}
\delta \H^{\epsilon'}_{11\uparrow, +-}\big|_{\mathsf O_2^\epsilon} = -\frac{\delta\epsilon}{4 \epsilon} g_1(\epsilon)^2 \int_0^L \dd x \, \mathsf O_2^\epsilon(x) + O(\epsilon^0, \delta\epsilon^2).
\end{align}
The other contributions coming from the variations $\delta \H^{\epsilon'}_{11\downarrow, +-}$, $\delta \H^{\epsilon'}_{11\uparrow, -+}$ and $\delta \H^{\epsilon'}_{11\downarrow, -+}$ lead to the same exact expression.
By adding together these different variations, we finally read off the quantum beta function \eqref{beta functions} of the coupling $g_2$ to be
\begin{equation}\label{qu beta for g2}
\beta^{\rm qu}_2[(g_k(\epsilon))] = g_1(\epsilon)^2 + \ldots
\end{equation}
where as usual `$\ldots$' denotes possible corrections coming from the other variations.

\medskip

Finally, let us return to the comment made before \eqref{H11 contribution to g2} about the analogous computation for last two lines on the right hand side of \eqref{Chiral boson expansion}. In fact, we will focus on the chiral operator $\nord{(\partial_{x_2}\chi^{\epsilon'}(x_2))^2}$, the second last term on the right hand side of \eqref{Chiral boson expansion}, and since the computation is very similar to the one for $\mathsf O_2^\epsilon(x)$ described in detail above, we will only highlight the main differences with that computation. And the analogous computation for the anti-chiral operator $\nord{(\partial_{x_2} \bar\chi^{\epsilon'}(x_2))^2}$ in the last line on the right hand side of \eqref{Chiral boson expansion} will be almost identical. The contribution to $\delta \H^{\epsilon'}_{11\uparrow, +-}$ coming from the second last line of \eqref{Chiral boson expansion} reads, cf. \eqref{H11 contribution to g2 d epsilon},
\begin{align} \label{H11 contribution to g2 d epsilon dX}
\delta \H^{\epsilon'}_{11\uparrow, +-}\big|_{\nord{(\partial \chi^\epsilon)^2}} &= \partial_\epsilon \Bigg( \frac{\pi}{2 L^2} \int_0^T \dd \tau \int_{-L/2}^{L/2} \dd x \, \big( e^{2\pi i x/L}e^{-2\pi \tau/L} - 1 \big)^2 e^{4\textup{Re}(G(x - i\tau; \epsilon))} e^{-4\pi\tau/L} \Bigg) \notag\\
&\qquad\qquad\qquad \times \delta\epsilon g_1(\epsilon)^2 \int_0^L \dd x_2 \Wop^\epsilon(\tau) \nord{\big( \partial_{x_2} \chi^\epsilon(x_2)\big)^2} + O(\epsilon^0, \delta\epsilon^2).
\end{align}
Using the same approximation \eqref{approximation e^G and Delta} that led to \eqref{H11 contribution to g2 expand} we now have the following analoguous expansion
\begin{align} \label{H11 no g0 contribution}
\delta \H^{\epsilon'}_{11\uparrow, +-}\big|_{\nord{(\partial \chi^\epsilon)^2}} &= \sum_{k \geq 0} \bigg( \frac{1}{8\pi} \int_0^T \dd \tau \int_{-\infty}^\infty \dd x \, \frac{\tau^k (x - i \tau)^2}{(x^2 + (\tau-a_1 \epsilon)^2)^2} \bigg)\\
&\qquad\qquad\qquad \times \delta\epsilon g_1(\epsilon)^2 \int_0^L \dd x_2\, \Wop^{\epsilon(k)} \nord{(\partial_{x_2} \chi^\epsilon(x_2))^2} + O(\epsilon^0, \delta\epsilon^2). \notag
\end{align}
However, unlike the double integral with $k=0$ appearing in \eqref{H11 contribution to g2 expand}, one can show that the double integrals in \eqref{H11 no g0 contribution} are regular as $\epsilon \to 0$ for all $k \geq 0$, see \S\ref{sec: asymp double int 2}. The same is true of the variations $\delta \H^{\epsilon'}_{11\downarrow, +-}$, $\delta \H^{\epsilon'}_{11\uparrow, -+}$ and $\delta \H^{\epsilon'}_{11\downarrow, -+}$, so it follows that there is no contribution to the flow of the coefficient of the free Hamiltonian $\H_0^\epsilon$, which justifies a posteriori why we did not include an explicit $\epsilon$-dependent coefficient in front of $\H_0^\epsilon$ in \eqref{Regularized bare Hamiltonian}.

\subsection{Renormalisation group flows} \label{sec: RG flow summary}

In this section we will briefly analyse the renormalisation group flows of the couplings in the Hamiltonian \eqref{Regularized bare Hamiltonian} using their quantum beta functions derived in \S\ref{sec: sG thin shell}. Before doing so, it will be helpful to summarise the computation in \S\ref{sec: sG thin shell} and the results obtained.

Our starting point in \S\ref{sec: sG thin shell} was the formula \eqref{delta H normal-ordered g0 out} for the variation of the full interacting Hamiltonian $\H^{\epsilon'}$ after integrating out an infinitesimally thin shell of short distance degrees of freedom between the cutoffs $\epsilon' = \epsilon + \delta\epsilon$ and $\epsilon$. This formula involves the first order variations $\Vop^{\epsilon', (1,0)}(x)$ and $\Vop^{\epsilon', (0,1)}(x)$ of the potential $\Vop(\chi^\epsilon, \bar\chi^\epsilon)$ with respect to short distance chiral and anti-chiral bosons $\chi^{\epsilon'\bsl \epsilon}(x)$ and $\bar\chi^{\epsilon'\bsl\epsilon}(x)$; see \eqref{potential field expansion} for the precise definition.
The expression we took for the potential in \S\ref{sec: sG thin shell} was \eqref{Regularised potential}, which is a linear combination of the two regularised marginal operators $\mathsf O_1^\epsilon(x)$ and $\mathsf O_2^\epsilon(x)$ in \eqref{O1 O2 definition} and the relevant identity operator $\mathsf O_3^\epsilon(x) = 1$.
Since the variation $\delta\H^{\epsilon'}$ in \eqref{delta H normal-ordered g0 out} is bilinear in the variations of the potential $\Vop(\chi^\epsilon, \bar\chi^\epsilon)$ and the identity operator clearly has vanishing variations, the computation of $\delta\H^{\epsilon'}$ broke up into four parts $\delta \H^{\epsilon'} = \delta \H^{\epsilon'}_{11} + \delta \H^{\epsilon'}_{12} + \delta \H^{\epsilon'}_{21} + \delta \H^{\epsilon'}_{22}$. And by computing these four variations separately, we identified the pieces which contributed to the quantum beta functions \eqref{beta functions} of the three coupling parameters $g_1(\epsilon)$, $g_2(\epsilon)$ and $g_3(\epsilon)$ appearing in the Hamiltonian \eqref{Regularized bare Hamiltonian}. Focusing only on the marginal couplings $g_1(\epsilon)$, $g_2(\epsilon)$, we found the quantum beta functions
\begin{equation} \label{beta function summary}
\beta^{\rm qu}_1[(g_k(\epsilon))] = g_1(\epsilon)g_2(\epsilon) + \ldots, \qquad
\beta^{\rm qu}_2[(g_k(\epsilon))] = g_1(\epsilon)^2 + \ldots
\end{equation}

\subsubsection{Irrelevant coupling contributions} \label{sec: irrelevant couplings}

Recall from \S\ref{sec: Anisotropic def}, or the general discussion in \S\ref{sec: RG flow}, that we should really include into the potential $\Vop(\chi^\epsilon, \bar\chi^\epsilon)$ all possible local operators compatible with the symmetries of our theory, as in \eqref{couplings}, which will in particular include an infinite sum of irrelevant operators. Indeed, all these operators will inevitably be generated from the procedure of integrating out a thin shell of short distance modes and so should be included in the potential from the outset in order to ensure self-consistency of the renormalisation group equation \eqref{RG equation}.

The ellipses `$\ldots$' in \eqref{beta function summary} encode precisely the contributions to the quantum beta functions of the couplings $g_1(\epsilon)$ and $g_2(\epsilon)$ from the irrelevant operators we have omitted from the potential \eqref{potential field expansion}. To give an example, recall the full vertex operators \eqref{vertex operators V rs} labelled by pairs of integers $(r,s) \in \ZZ^2$, of conformal dimension $r^2 + s^2$, and consider the family of operators
\begin{equation} \label{operator O_n}
\int_0^L\dd x \, \mathsf O_{(r)}^\epsilon(x) = \frac{(2\pi)^{2 r^2 - 1}}{L^{2 r^2}} \int_0^L\dd x \Big( \nord{e^{i r \sqrt{8\pi}\chi^\epsilon(x)}e^{i r \sqrt{8\pi} \bar\chi^\epsilon(x)}} + 
\nord{e^{-i r \sqrt{8\pi}\chi^\epsilon(x)}e^{-i r \sqrt{8\pi} \bar\chi^\epsilon(x)}} \Big)
\end{equation}
for any $r \in \ZZ_{>1}$. Note that the origin of the prefactor of $(2 \pi / L)^{2r^2}$ is the same as explained in the paragraph around \eqref{z factor absorbed} since the integrand is a sum of full vertex operators of conformal dimension $2r^2$. In particular, the local operators in \eqref{operator O_n} are all irrelevant since $2r^2 > 2$ and would appear in the general potential \eqref{couplings} with a prefactor of $\epsilon^{2r^2 - 2}$.
We saw an example of such an operator being generated from the procedure of integrating the thin shell in \S\ref{sec: H11 var}, specifically the irrelevant operator \eqref{irrelevant VO example} with $m=n=0$ which corresponds to the positive charge part of \eqref{operator O_n} with $r=2$. Letting $g_{(r)}(\epsilon)$ denote the couplings of the operator \eqref{operator O_n}, its quantum beta function $\beta_{(r)}^{\rm qu}[(g_k(\epsilon))]$ would therefore receive a contribution of the form
\begin{equation} \label{beta for g(r)}
\beta_{(r)}^{\rm qu}\big[ (g_k(\epsilon)) \big] = c_{(r),1} g_1(\epsilon)^2 + \ldots
\end{equation}
for some constant $c_{(r),1}$ which could be computed following the analysis of \S\ref{sec: H11 var}.

On the other hand, the contribution of the couplings $g_{(r)}(\epsilon)$ to the quantum beta function $\beta_2^{\rm qu}[(g_k(\epsilon))]$, say, of the coupling $g_2(\epsilon)$ could be computed using a very similar calculation to that of $\delta\H^{\epsilon'}_{11}$ in \S\ref{sec: H11 var}, replacing the role of the full vertex operators $\nord{e^{\pm i\sqrt{8\pi}\chi^\epsilon(x)}e^{\pm i\sqrt{8\pi} \bar\chi^\epsilon(x)}}$ there by $\nord{e^{\pm i r \sqrt{8\pi}\chi^\epsilon(x)}e^{\pm i r \sqrt{8\pi} \bar\chi^\epsilon(x)}}$. A similar expansion as the one in \eqref{Chiral boson expansion} of these full vertex operators would also involve the quadratic operators $\partial_{x_2} \chi^\epsilon(x_2) \partial_{x_2} \bar\chi^\epsilon(x_2)$, $\nord{(\partial_{x_2} \chi^\epsilon(x_2))^2}$ and $\nord{(\partial_{x_2} \bar\chi^\epsilon(x_2))^2}$, leading to infinitely many corrections
\begin{equation} \label{beta for g2 irr corrections}
\beta^{\rm qu}_2[(g_k(\epsilon))] = g_1(\epsilon)^2 + \sum_{r > 1} c_{2, (r)} g_{(r)}(\epsilon)^2 + \ldots
\end{equation}
for some coefficients $c_{2, (r)}$ that could be computed along the same lines as in \S\ref{sec: H11 var}.
However, the corrections from the irrelevant couplings in \eqref{beta for g2 irr corrections} represent higher order corrections in the marginal couplings $g_1(\epsilon)$ and $g_2(\epsilon)$. Indeed, the general solution to the beta equation for the coupling $g_{(r)}(\epsilon)$ takes the form
\begin{equation} \label{g(r) general sol}
g_{(r)}(\epsilon) = g_{(r)}(\epsilon_0) \bigg( \frac{\epsilon_0}{\epsilon} \bigg)^{2r^2-2} + \int_{\epsilon_0}^\epsilon \frac{\dd t}t \bigg( \frac{t}{\epsilon} \bigg)^{2r^2 - 2} \beta_{(r)}^{\rm qu}\big[ (g_k(t))\big].
\end{equation}
As explained in \S\ref{sec: RG flow}, we are interested in the renormalised trajectory which is a particular solution to the renormalisation group equation for which the irrelevant couplings, such as $g_{(r)}$, are all switched off in the UV limit $\epsilon \to 0$, see \eqref{renormalised trajectory a}. In other words, for the renormalised trajectory we set the initial condition in \eqref{g(r) general sol} to $g_{(r)}(\epsilon_0) = 0$ in the continuum limit $\epsilon_0 \to 0$. Since the quantum beta function \eqref{beta for g(r)} for the coupling $g_{(r)}(\epsilon)$ is quadratic in the marginal coupling $g_1(\epsilon)$, its contribution to the quantum beta function of $g_2(\epsilon)$ in \eqref{beta for g2 irr corrections} will be subleading (of order $g_1(\epsilon)^4$). In fact, even if we are not on the renormalised trajectory, the first term on the right hand side of \eqref{g(r) general sol} is exponentially suppressed in the RG time $t = \log(\epsilon/\epsilon_0)$, so that the second term becomes dominant. This is the concept of \textbf{universality}: if we include any amount of the irrelevant couplings $g_{(r)}(\epsilon_0)$ at some small UV scale $\epsilon_0$, the trajectory will asymptotically approach the renormalised one as we increase the length scale $\epsilon$.

\subsubsection{Berezinskii--Kosterlitz--Thouless transition}

\begin{figure}[ht]
\centering
\includegraphics[width=60mm]{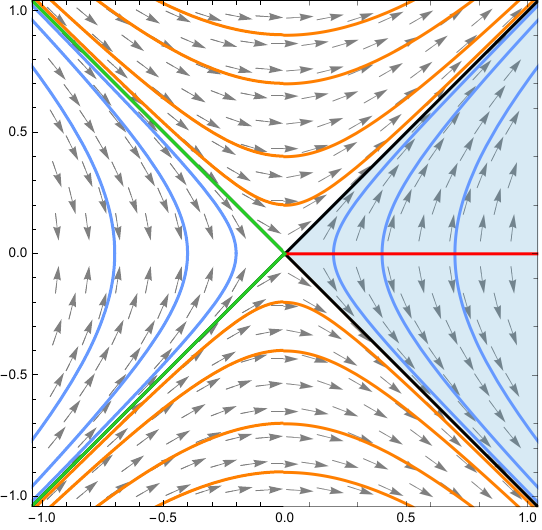}
\caption{A plot of the $2$-loop renormalisation group flow of the anisotropic deformation of the $\su_2$ WZW model at level $1$, in the marginal $(g_2, g_1)$-plane near the Kosterlitz–Thouless point $(g_2, g_1) = (0,0)$. The gray arrows indicate the direction of the flow. The blue/orange curves correspond to different negative/positive values of the $2$-loop renormalisation group invariant $C = g_1^2 - g_2^2$. The green lines are the separatrices $g_2 = - |g_1|$ corresponding to the Berezinskii--Kosterlitz--Thouless transition which separates the massless phase, where the theory flows to a massless free boson in the IR, from the massive phase where the mass grows in the IR. The shaded region where $C < 0$ and $g_2 > 0$ contains the renormalised trajectories emanating from the one parameter family of conformal field theories given by the free compactified boson with radius $R=2/\beta$, depicted by the red line along the positive real axis $g_1 = 0$ and $g_2 \geq 0$.}
\label{fig:anisotropic flow}
\end{figure}

The upshot of \S\ref{sec: irrelevant couplings} is that since we are working perturbatively to second order in the marginal couplings $g_1(\epsilon)$ and $g_2(\epsilon)$, we can ignore the ellipses `$\ldots$' in the quantum beta functions \eqref{beta function summary}. The renormalisation group equations for the two marginal couplings $g_1(\epsilon)$ and $g_2(\epsilon)$, up to second order, therefore take the simple form
\begin{equation} \label{RG equations sG}
\epsilon \partial_\epsilon g_1 = g_1 g_2, \qquad
\epsilon \partial_\epsilon g_2 = g_1^2.
\end{equation}
The integral curves of the flow are depicted in Figure \ref{fig:anisotropic flow}.

The flow in \eqref{RG equations sG} coincides with the $2$-loop renormalisation group flow of the anisotropic deformation of the $\su_2$ WZW model at level $1$ derived using conformal perturbation theory, see for instance \cite[(5.5)]{Zamolodchikov:1995xk} or also \cite[(2.3) \& (2.4)]{BernardLeClair} expanded to second order in the couplings.
To make the comparison with the literature more explicit, recall from \S\ref{sec: Anisotropic def} that the Hamiltonian we are considering is \eqref{Regularized bare Hamiltonian}, which using the definitions \eqref{smooth reg Fock} of the regularised chiral and anti-chiral currents $J^{a, \epsilon}(x)$ and $\bar J^{a, \epsilon}(x)$ can be rewritten as
\begin{equation*}
\H^\epsilon = \H_0^\epsilon +\frac{g_1(\epsilon)}{4\pi}\int_0^L \dd x \big( J^{+,\epsilon}(x) \bar J^{-,\epsilon}(x) + J^{-,\epsilon}(x) \bar J^{+,\epsilon}(x) \big) + \frac{g_2(\epsilon)}{8 \pi} \int_0^L \dd x \, J^{3,\epsilon}(x) \bar J^{3,\epsilon}(x).
\end{equation*}
Recall also from \S\ref{sec: currents} that the operator product expansion of the $\su_2$-currents $J^a[z]$ in \eqref{Current OPE}, with our convention for the normalisations of the $\sl_2$ generators in \eqref{sl2 algebra} and the bilinear form $\kappa : \sl_2 \times \sl_2 \to \CC$ defined after \eqref{sl2 algebra}, takes the explicit form
\begin{align*}
J^3[z]J^3[w] \sim \frac{2}{(z - w)^2}, \qquad
J^3[z]J^\pm[w] \sim \pm \frac{2 J^\pm[w]}{z - w}, \qquad
J^+[z] J^-[w] \sim \frac{1}{(z - w)^2} + \frac{J^3[w]}{z - w}.
\end{align*}
Comparing with \cite[(2.1)]{BernardLeClair} we see that $J^3_{\rm BL}[z] = \tfrac 12 J^3[z]$ and $J^\pm_{\rm BL}[z] = \tfrac{1}{\sqrt{2}} J^\pm[z]$, so that our definition of the marginal couplings $g_1$ and $g_2$ agrees with the one used in the perturbation \cite[(2.2)]{BernardLeClair}. Likewise, comparing with \cite[(5.2)]{Zamolodchikov:1995xk} we see that $J^3_{\rm Z}[z] = \frac 12 J^3[z]$ and $J^\pm_{\rm Z}[z] = J^\pm[z]$, so that our marginals are related to those of \cite[(5.1)]{Zamolodchikov:1995xk} by $g_1 = g_\perp$ and $g_2 = g_{||}$.

\appendix

\section{Asymptotic expansion of \texorpdfstring{$G(u;\epsilon)$}{G(u; epsilon)}} \label{sec: Asymptotic of G}

We will use Mellin transform theory from \S\ref{sec: Mellin transform} to evaluate the asymptotics with respect to $\epsilon$ of the regularised series $G(u; \epsilon)$ defined in \eqref{G def}.

The function $G(u;\epsilon)$ is a harmonic series depending on a complex parameter $u \in \CC$, with base function given by the smooth cutoff function $\eta : \RR_{\geq 0} \to \RR$, the sequence of frequencies $\mu_n = 2 \pi n/L$ and the sequence of amplitudes $\lambda_n = e^{-2\pi i n u / L}/n$. So the associated Dirichlet series, which depends on the parameter $u$, is given by
\begin{equation} \label{Dirichlet for G}
\Lambda(u; s) \coloneqq \sum_{n > 0} \lambda_n (\mu_n)^{-s} = \bigg( \frac{L}{2 \pi} \bigg)^s \sum_{n > 0} \frac{e^{-2\pi i n u / L}}{n^{s+1}},
\end{equation}
which converges absolutely for all $s \in \CC$ when $\text{Im}(u) < 0$.
Given the Dirichlet series \eqref{Dirichlet for G}, the Mellin transform of $G(u; \epsilon)$ is then the function $G^\ast(u; s)$ depending on the parameter $u$ and given by
\begin{equation} \label{G star def}
G^\ast(u; s) = \Lambda(u; s) \eta^\ast(s).
\end{equation}

Now suppose that the smooth cutoff function $\eta$ expands as $\eta(x) = 1 + \sum_{n=1}^{r-1} a_n x^n + O(x^r)$. Its Mellin transform is a holomorphic function $\eta^\ast : \langle 0, \infty \rangle \to \CC$ which by the direct mapping theorem can be analytically continued to a meromorphic function $\eta^\ast : \langle - r, \infty\rangle \to \CC$ with the singular expansion
\begin{equation*}
\eta^\ast(s) \asymp \frac{1}{s} + \sum_{n=1}^{r-1} \frac{a_n}{s + n}.
\end{equation*}
And since $\Lambda(u;s)$ is entire in $s$ if $\text{Im}(u) < 0$, it follows that \eqref{G star def} is a meromorphic function $G^\ast(u; \--) : \langle -r, \infty \rangle \to \CC$ with singular expansion, for $\text{Im}(u) < 0$, given by
\begin{equation*}
G^\ast(u; s) \asymp \frac{\Lambda(u; 0)}{s} + \sum_{n=1}^{r-1} \frac{\Lambda(u; -n) a_n}{s+n}.
\end{equation*}
To compute the coefficients $\Lambda(u; -n)$, first note that we have $\Lambda(u; 0) = - \log(1 - e^{-2\pi i u /L})$. Also, differentiating with respect to the parameter $u$ we obtain
\begin{equation*}
i \partial_u \Lambda(u; s) = \bigg( \frac{L}{2 \pi} \bigg)^{s-1} \sum_{n > 0} \frac{e^{-2\pi i n u / L}}{n^s} = \Lambda(u; s-1),
\end{equation*}
from which it follows that $\Lambda(u; -n) = (i \partial_u)^n \Lambda(u; 0)$ for every $n \geq 0$.

We therefore conclude by the converse mapping theorem that for $\text{Im}(u) < 0$ we have the asymptotic expansion
\begin{equation} \label{G epsilon general expansion}
G(u; \epsilon) = \Lambda(u; 0) + \sum_{n=1}^{r-1} a_n (i \epsilon \partial_u)^n \Lambda(u; 0) + O(\epsilon^r)
\end{equation}
as $\epsilon \to 0$.
Restricting to the case $r=2$ we may rewrite this expansion as
\begin{align} \label{G expansion epsilon}
G(u; \epsilon) &= \Lambda(u; 0) + i a_1 \epsilon \partial_u \Lambda(u; 0) + O(\epsilon^2) \notag\\
&= \Lambda\big( u + i a_1\epsilon; 0 \big) + O(\epsilon^2) = -\log\big( 1 - e^{-2\pi i u/L}e^{2\pi a_1\epsilon/L} \big) + O(\epsilon^2),
\end{align}
where the second equality follows from recognising the first two terms in the second expression as those of the Taylor expansion of $\Lambda\big( u + i a_1\epsilon; 0 \big)$ for small $\epsilon$. Notice that if we take the limit $u \to 0$ on both sides of \eqref{G expansion epsilon}, we obtain
\begin{align} \label{G expansion epsilon u 0}
\lim_{u \to 0} G(u; \epsilon) = - \log \epsilon - \log\big( \tfrac{2\pi a_1}{L} \big) + O(\epsilon),
\end{align}
which is consistent, at least at leading order, with the asymptotics of the regularised harmonic series \eqref{harmonic series regularised}. The disagreement between the subleading $O(1)$ terms in \eqref{G expansion epsilon u 0} and \eqref{harmonic series regularised} suggests that taking the limit $u \to 0$ in $G(u;\epsilon)$ does not commute with taking the $\epsilon \to 0$ asymptotics. In any case, we note that this subleading $O(1)$ term is scheme-dependent (i.e. depends on the choice of smooth cutoff function $\eta$).

Multiplying both sides of the relation \eqref{G expansion epsilon} by $-4rs$ with $r, s \in \{ \pm 1\}$ and then exponentiating we obtain
\begin{equation*}
e^{- 4rs G(u; \epsilon)} = \big( 1 - e^{-2\pi i u/L}e^{2\pi a_1\epsilon/L} \big)^{4rs} + O(\epsilon^2).
\end{equation*}
And finally, taking the modulus of both sides leads to the relation
\begin{equation} \label{identity for exp G}
e^{- 4rs \text{Re}(G(u; \epsilon))} = \big| 1 - e^{-2\pi i u/L}e^{2\pi a_1\epsilon/L} \big|^{4rs} + O(\epsilon^2),
\end{equation}
valid for $\text{Im}(u) < 0$. In particular, the right hand side of \eqref{identity for exp G} represents a regular expansion in $\epsilon$ for all $\text{Im}(u) < 0$, but in the limit $u \to 0$ it follows from \eqref{G expansion epsilon u 0} that
\begin{equation} \label{limit identity for exp G}
\lim_{u \to 0} e^{- 4rs \text{Re}(G(u; \epsilon))} = \epsilon^{4rs} + O(1),
\end{equation}
which contains a singular term if $rs < 0$.

\section{Asymptotics of a family of double integrals} \label{sec: Contour integrals}

In this appendix we compute the singular part of the asymptotic behaviour as $\epsilon \to 0$ of the family of double integrals appearing in \eqref{H11 contribution to g2 expand} and \eqref{H11 no g0 contribution}, namely
\begin{subequations} \label{I_k J_k def}
\begin{align}
\label{I_k def} I_k &\coloneqq \frac{1}{4\pi} \int_0^T \dd \tau \int_{-\infty}^\infty \dd x \frac{\tau^k(x^2 + \tau^2)}{(x^2 + (\tau - a_1\epsilon)^2)^2},\\
\label{J_k def} J_k &\coloneqq \frac{1}{8\pi} \int_0^T \dd \tau \int_{-\infty}^\infty \dd x \frac{\tau^k(x - i\tau)^2}{(x^2 + (\tau - a_1\epsilon)^2)^2}
\end{align}
\end{subequations}
labelled by $k \in \ZZ_{\geq 0}$. It will be useful to recall in what follows that $a_1 < 0$, so that $\tau - a_1\epsilon > 0$ since $\tau \geq 0$ and $\epsilon > 0$.

\subsection{Asymptotics of the double integrals \texorpdfstring{$I_k$}{Ik}} \label{sec: asymp double int 1}

The inner $x$-integral in \eqref{I_k def} can be evaluated using a contour integral. Specifically, closing the contour off using a semicircle in the upper half plane and using the residue theorem we find
\begin{equation}\label{I(tau, u) def}
\frac{1}{4\pi} \int_{-\infty}^\infty \dd x \frac{x^2 + \tau^2}{(x^2 + (\tau - a_1 \epsilon)^2)^2} = \frac{(\tau - a_1 \epsilon)^2 + \tau^2}{8 (\tau - a_1 \epsilon)^3}.
\end{equation}
Since $-a_1 \epsilon > 0$ it follows that the singularity of \eqref{I(tau, u) def} lies outside the range of integration of the $\tau$-integral in \eqref{I_k def}.
To evaluate this $\tau$-integral we break it up into two pieces $I_k = I'_k + I''_k$ where we have defined
\begin{equation}\label{I_k tau}
I'_k \coloneqq \frac{1}{8} \int_0^T \dd \tau \frac{\tau^k}{\tau - a_1\epsilon}, \qquad I''_k \coloneqq \frac{1}{8} \int_0^T \dd \tau \frac{\tau^{k+2}}{(\tau - a_1\epsilon)^3}.
\end{equation}

The first integral for $k=0$ evaluates immediately to
\begin{align}\label{I'0 contribution}
I'_0 = \frac{1}{8} \log(T - a_1\epsilon) - \frac{1}{8} \log(- a_1\epsilon) = - \frac{1}{8} \log \epsilon + O(\epsilon^0).
\end{align}
The remaining integrals $I'_k$ for $k > 0$ can be determined using the recurrence relation
\begin{align}\label{I'k formula}
I'_k &= \frac{1}{8} \int_0^T \dd \tau \frac{(\tau - a_1\epsilon)^{k}}{\tau - a_1\epsilon} - \frac{1}{8} \sum_{n = 1}^k \binom{k}{n} I'_{k - n} (-a_1\epsilon)^n \notag\\
&= \frac{(T - a_1\epsilon)^k - (-a_1\epsilon)^k}{8 k} - \frac{1}{8} \sum_{n = 1}^k \binom{k}{n} I'_{k - n} (-a_1\epsilon)^n
\end{align}
from which it follows that $I'_k = O(\epsilon^0)$ for all $k > 0$.
The second integral in \eqref{I_k tau} for $k=0$ evaluates to
\begin{align}\label{I''0 contribution}
I''_0 = \frac{1}{8} \log(T -a_1\epsilon) - \frac{1}{8} \log(-a_1\epsilon) - \frac{T}{4(T -a_1\epsilon)} + \frac{T^2 - 2 a_1\epsilon T}{16 (T -a_1\epsilon)^2} = - \frac{1}{8} \log \epsilon + O(\epsilon^0)
\end{align}
and using a similar recurrence relation as in \eqref{I'k formula} we find that $I'_k = O(\epsilon^0)$ for all $k > 0$.
Putting together all the above we deduce that, for $k \geq 0$,
\begin{equation} \label{asymp double int 1}
\frac{1}{4\pi} \int_0^T \dd \tau \int_{-\infty}^\infty \dd x \frac{\tau^k(x^2 + \tau^2)}{(x^2 + (\tau - a_1\epsilon)^2)^2} = \left\{
\begin{array}{ll}
- \frac 14 \log \epsilon + O(\epsilon^0), &\quad k = 0\\
O(\epsilon^0), &\quad k>0.
\end{array}
\right.
\end{equation}

\subsection{Asymptotics of the double integrals \texorpdfstring{$J_k$}{Jk}} \label{sec: asymp double int 2}

The inner $x$-integral in \eqref{J_k def} can similarly be evaluated using a contour integral, by closing the contour off using a semicircle in the upper half plane and using the residue theorem. This time we find
\begin{equation}\label{J(tau, u) def}
\frac{1}{8\pi} \int_{-\infty}^\infty \dd x \frac{(x - i\tau)^2}{(x^2 + (\tau - a_1 \epsilon)^2)^2} = \frac{(\tau - a_1 \epsilon)^2 - \tau^2}{16 (\tau - a_1 \epsilon)^3},
\end{equation}
so that we have $J_k = \frac{1}{2}(I'_k - I''_k)$ written in terms of the pair of integrals defined in \eqref{I_k tau}. It now follows from the above asymptotic expansions of these integrals as $\epsilon \to 0$, computed in \S\ref{sec: asymp double int 1}, that $J_k = O(\epsilon^0)$ for all $k \geq 0$, i.e.
\begin{equation} \label{asymp double int 2}
\frac{1}{8\pi} \int_0^T \dd \tau \int_{-\infty}^\infty \dd x \frac{\tau^k(x - i\tau)^2}{(x^2 + (\tau - a_1\epsilon)^2)^2} = O(\epsilon^0).
\end{equation}


\begin{thebibliography}{99}

\bibitem[AM1]{Alexanian:1998gz}
  G.~Alexanian and E.~F.~Moreno,
  \emph{Renormalization of the Hamiltonian and a geometric interpretation of asymptotic freedom},
  Phys. Rev. D \textbf{60} (1999), 105028.

\bibitem[AM2]{Alexanian:1998wu}
  G.~Alexanian and E.~F.~Moreno,
  \emph{On the renormalization of Hamiltonians},
  Phys. Lett. B \textbf{450} (1999), 149--157.

\bibitem[AGG]{Amit:1979ab}
  D.~J.~Amit, Y.~Y.~Goldschmidt and G.~Grinstein,
  \emph{Renormalisation group analysis of the phase transition in the 2D Coulomb gas, Sine-Gordon theory and XY-model},
  J. Phys. A: Math. Gen. \textbf{13} (1980), 585--620.

\bibitem[BR]{BahnsRejzner}
  D.~Bahns and K.~Rejzner, 
  \emph{The Quantum Sine-Gordon Model in Perturbative AQFT},
  Commun. Math. Phys. \textbf{357} (2018), 421--446.

\bibitem[BH]{Balog:2000qr}
  J.~Balog and A.~Hegedus,
  \emph{Two loop beta functions of the Sine-Gordon model},
  J. Phys. A \textbf{33} (2000), 6543--6548.

\bibitem[B1]{Berezinskii:1971}
  V.~Berezinskii,
  \emph{Destruction of Long-range Order in One-dimensional and 2-dimensional Systems having a Continuous Symmetry Group I. Classical Systems},
  Soviet Journal of Experimental and Theoretical Physics \textbf{32} (1971), 493.

\bibitem[B2]{Berezinskii:1972}
  V.~Berezinskii,
  \emph{Destruction of Long-range Order in One-dimensional and 2-dimensional Systems having a Continuous Symmetry Group II. Quantum Systems},
  Soviet Journal of Experimental and Theoretical Physics \textbf{34} (1972), 610.

\bibitem[BTW]{Berges:2002}
  J.~Berges, N.~Tetradis and C.~Wetterich,
  \emph{Nonperturbative renormalization flow in quantum field theory and statistical physics},
  Phys. Rept. \textbf{363} (2002), 223.

\bibitem[BL1]{Bernard:1990ys}
  D.~Bernard and A.~LeClair,
  \emph{Quantum group symmetries and nonlocal currents in 2-D QFT},
  Commun. Math. Phys. \textbf{142} (1991), 99--138.

\bibitem[BL2]{BernardLeClair}
  D.~Bernard and A.~LeClair,
  \emph{Strong-weak coupling duality in anisotropic current interactions},
  Phys. Lett. B \textbf{512} (2001), 78.

\bibitem[BP]{BogoPara}
  N.~N.~Bogolyubov and O.~S.~Parasiuk,
  \emph{On the multiplication of causal functions in the quantum theory of fields},
  Acta Math. \textbf{97} (1957), 227--266.

\bibitem[BDL]{BDL}
  S.~Brayvi, D.~DiVincenzo, D.~Loss,
  \emph{Schrieffer–Wolff transformation for quantum many-body systems},
  Ann. Phys. \textbf{326} (2011), 2793--2826.

\bibitem[BF]{BrooksFrautschi}
  E.~D.~Brooks and S.~C.~Frautschi,
  \emph{Scalars coupled to fermions in 1+1 dimensions},
  Z. Phys. C - Particles and Fields \textbf{23} (1984), 263--273.

\bibitem[CFHL]{Cohen:2021erm}
  T.~Cohen, K.~Farnsworth, R.~Houtz and M.~A.~Luty,
  \emph{Hamiltonian Truncation Effective Theory},
  SciPost Phys. \textbf{13}, no.2 (2022), 011

\bibitem[Col]{Coleman}
  S.~Coleman,
  \emph{Quantum sine-Gordon equation as the massive Thirring model},
  Phys. Rev. D \textbf{11} (1975), 2088.

\bibitem[Coll]{Collins}
  J.~C.~Collins, 
  \emph{Renormalization: An Introduction to Renormalization, the Renormalization Group and the Operator-Product Expansion},
  Cambridge University Press (1984).

\bibitem[CK1]{Connes:1999yr}
  A.~Connes and D.~Kreimer,
  \emph{Renormalization in quantum field theory and the Riemann-Hilbert problem. 1. The Hopf algebra structure of graphs and the main theorem},
  Commun. Math. Phys. \textbf{210} (2000), 249--273.

\bibitem[CK2]{Connes:2000fe}
  A.~Connes and D.~Kreimer,
  \emph{Renormalization in quantum field theory and the Riemann-Hilbert problem. 2. The beta function, diffeomorphisms and the renormalization group},
  Commun. Math. Phys. \textbf{216} (2001), 215--241.

\bibitem[Cos]{CostelloRGbook}
  K.~Costello,
  \emph{Renormalization and Effective Field Theory},
  Mathematical Surveys and Monographs, Volume 170 (2011).

\bibitem[CG2]{CGBook2}
  K.~Costello and O.~Gwilliam,
  \emph{Factorization algebras in quantum field theory, Vol. 2},
  New Mathematical Monographs 41, Cambridge University Press, Cambridge (2021).

\bibitem[DDPR]{DAngelo:2022vsh}
  E.~D'Angelo, N.~Drago, N.~Pinamonti and K.~Rejzner,
  \emph{An Algebraic QFT Approach to the Wetterich Equation on Lorentzian Manifolds},
  Annales Henri Poincar\'{e} \textbf{25}, no.4 (2024), 2295-2352.

\bibitem[DR]{DAngelo:2023tis}
  E.~D'Angelo and K.~Rejzner,
  \emph{A Lorentzian Renormalization Group Equation for Gauge Theories},
  Annales Henri Poincar\'{e} \textbf{26}, no.12 (2025), 4411-4459.

\bibitem[DD]{DavietDupuis}
  R.~Daviet and N.~Dupuis,
  \emph{Nonperturbative Functional Renormalization-Group Approach to the Sine-Gordon Model and the Lukyanov-Zamolodchikov Conjecture},
  Phys. Rev. Lett. \textbf{122} (2019), 155301.

\bibitem[Del]{Delamotte}
  B.~Delamotte,
  \emph{An Introduction to the Nonperturbative Renormalization Group}.
  In: Schwenk, A., Polonyi, J. (eds) Renormalization Group and Effective Field Theory Approaches to Many-Body Systems. Lecture Notes in Physics, vol 852. Springer, Berlin, Heidelberg (2012).

\bibitem[DFH]{Demiray:2025zqh}
  E.~Demiray, K.~Farnsworth and R.~Houtz,
  \emph{Systematic Improvement of Hamiltonian Truncation Effective Theory},
  [arXiv:2507.15941 [hep-th]].

\bibitem[DMS]{CFTbook}
  P.~di Francesco, P.~Mathieu and D.~Senechal,
  \emph{Conformal Field Theory},
  Springer (1997).

\bibitem[DH1]{Dimock:1991ig}
  J.~Dimock and T.~R.~Hurd,
  \emph{A Renormalization group analysis of the Kosterlitz-Thouless phase},
  Commun. Math. Phys. \textbf{137} (1991), 263--287.

\bibitem[DH2]{Dimock:1993qw}
  J.~Dimock and T.~R.~Hurd,
  \emph{Construction of the two-dimensional sine-Gordon model for $\beta < 8 \pi$},
  Commun. Math. Phys. \textbf{156} (1993), 547--580.

\bibitem[DH3]{Dimock:1999jt}
  J.~Dimock and T.~R.~Hurd,
  \emph{Sine-Gordon revisited},
  Annales Henri Poincare \textbf{1} (2000), 499--541.

\bibitem[Du+]{Dupuis:2020fhh}
  N.~Dupuis, L.~Canet, A.~Eichhorn, W.~Metzner, J.~M.~Pawlowski, M.~Tissier and N.~Wschebor,
  \emph{The nonperturbative functional renormalization group and its applications},
  Phys. Rept. \textbf{910} (2021), 1--114.

\bibitem[EG]{EpsteinGlaser}
  H.~Epstein and V.~Glaser
  \emph{The role of locality in perturbation theory},
  Ann. Inst. H. Poincar\'e, Section A, Physique Th\'eorique, vol 19, no. 3 (1973), 211--295.

\bibitem[FGPTW]{Feverati:2006ni}
  G.~Feverati, K.~Graham, P.~A.~Pearce, G.~Z.~Toth and G.~Watts,
  \emph{A Renormalisation group for the truncated conformal space approach},
  J. Stat. Mech. \textbf{0803} (2008), P03011.

\bibitem[FGD]{MellinHarmonic}
  P.~Flajolet, X.~Gourdon and P.~Dumas,
  \emph{Mellin transforms and asymptotics: Harmonic sums},
  Theor. Comput. Sci. \textbf{144} 1-2 (1995), 3--58.

\bibitem[Fra]{FradkinBook}
  E.~Fradkin,
  \emph{Field Theories of Condensed Matter Physics},
  2nd ed. Cambridge University Press (2013).

\bibitem[Fre]{IFrenkel}
  I.~B.~Frenkel,
  \emph{Two constructions of affine Lie algebra representations and boson-fermion correspondence in quantum field theory},
  Journal of functional analysis {\bf 44} (1981), 259--327.

\bibitem[FK]{FrenkelKac}
  I.~B.~Frenkel, V.~G.~Kac,
  \emph{Basic representations of affine Lie algebras and dual resonance models},
  Invent Math {\bf 62} (1980), 23--66.
  
\bibitem[FrS]{Frohlich:1981yn}
  J.~Frohlich and T.~Spencer,
  \emph{The Kosterlitz-thouless Transition in Two-dimensional Abelian Spin Systems and the Coulomb Gas},
  Commun. Math. Phys. \textbf{81} (1981), 527--602.

\bibitem[GLM]{GLM}
  B.~Gerganov, A.~Leclair and M.~Moriconi,
  \emph{On the Beta Function for Anisotropic Current Interactions in 2D},
  Phys. Rev. Lett. \textbf{86}, (2001), 4753--4756.

\bibitem[GiWa]{Giokas:2011ix}
  P.~Giokas and G.~Watts,
  \emph{The renormalisation group for the truncated conformal space approach on the cylinder},
  [arXiv:1106.2448 [hep-th]].

\bibitem[GWi1]{GlazekWilson}
  S.~D.~Głazek and K.~G.~Wilson,
  \emph{Renormalization of Hamiltonians},
  Phys. Rev. D \textbf{48} (1993), 5863--5872.

\bibitem[GWi2]{Glazek:1994qc}
  S.~D.~Głazek and K.~G.~Wilson,
  \emph{Perturbative renormalization group for Hamiltonians},
  Phys. Rev. D \textbf{49} (1994), 4214--4218.

\bibitem[GJ]{GlimmJaffe}
  J.~Glimm and A.~Jaffe,
  \emph{Quantum Physics: A Functional Integral Point of View},
  Springer New York, NY (1987).

\bibitem[GLPZ]{Grisaru:1990gf}
  M.~T.~Grisaru, A.~Lerda, S.~Penati and D.~Zanon,
  \emph{Renormalization Group Flows in Generalized Toda Field Theories},
  Nucl. Phys. B \textbf{346} (1990), 264--292.

\bibitem[GP]{Grisaru:1990kh}
  M.~T.~Grisaru and S.~Penati,
  \emph{Renormalization group flows in generalized Toda field theories. 2. Nonsimply laced algebras},
  Nucl. Phys. B \textbf{348} (1991), 148--177.

\bibitem[GWe]{GubankovaWeg}
  E.~L.~Gubankova and F.~Wegner,
  \emph{Flow equations for QED in light front dynamics},
  Phys. Rev. D \textbf{58} (1998), 025012.

\bibitem[Ha]{Hall:2015xtd}
  B.~C.~Hall
  \emph{Lie Groups, Lie Algebras, and Representations},
  Graduate Texts in Mathematics, Springer (2015).

\bibitem[HJMSN]{Hariharakrishnan:2024iba}
  S.~Hariharakrishnan, U.~D.~Jentschura, I.~G.~Marian, K.~Szabo and I.~N\'{a}ndori,
  \emph{Perturbative versus non-perturbative renormalization},
  J. Phys. G \textbf{51}, no.8 (2024), 085005.

\bibitem[H]{Hepp}
  K.~Hepp,
  \emph{Proof of the Bogolyubov–Parasiuk theorem on renormalization},
  Comm. Math. Phys. 2 (1966), 301--326.

\bibitem[Hol]{HollowoodBook}
  T.~J.~Hollowood,
  \emph{Renormalization Group and Fixed Points in Quantum Field Theory},
  SpringerBriefs in Physics,
  Springer Berlin, Heidelberg (2013).

\bibitem[Kac1]{KacBook}
  V.~Kac,
  \emph{Infinite Dimensional Lie Algebras: An Introduction}, Third Edition,
  Cambridge University Press (1990).

\bibitem[Kac2]{KacVA}
  V.~Kac,
  \emph{Vertex Algebras for Beginners},
  Vol. 10 of University Lecture Series, 2nd Edition, American Mathematical Society, Providence (1998).

\bibitem[K]{Kadanoff:1966wm}
  L.~P.~Kadanoff,
  \emph{Scaling laws for Ising models near $T_c$},
  Physics Physique Fizika \textbf{2} (1966), 263--272.

\bibitem[Kli]{Klimcik:2008eq}
  C.~Klimcik,
  \emph{On integrability of the Yang-Baxter sigma-model},
  J. Math. Phys. \textbf{50} (2009), 043508.

\bibitem[KW]{KogutWilson}
  J.~Kogut and K.~G.~Wilson,
  \emph{The renormalization group and the $\epsilon$ expansion},
  Physics Reports \textbf{12} (1974), 75--199.

\bibitem[Ko]{Kosterlitz:1974sm}
  J.~M.~Kosterlitz,
  \emph{The Critical properties of the two-dimensional x y model},
  J. Phys. C \textbf{7} (1974), 1046--1060.

\bibitem[KT]{KosterlitzThouless}
  J.~M.~Kosterlitz and D.~J.~Thouless,
  \emph{Ordering, metastability and phase transitions in two-dimensional systems},
  Journal of Physics C: Solid State Physics, \textbf{6} (1973), 1181-1203.

\bibitem[KLT]{Kotousov:2022azm}
  G.~A.~Kotousov, S.~Lacroix and J.~Teschner,
  \emph{Integrable Sigma Models at RG Fixed Points: Quantisation as Affine Gaudin Models},
  Annales Henri Poincar\'e \textbf{25}, no.1 (2024), 843--1006.

\bibitem[Kr1]{Kreimer:1997dp}
  D.~Kreimer,
  \emph{On the Hopf algebra structure of perturbative quantum field theories},
  Adv. Theor. Math. Phys. \textbf{2} (1998), 303--334.

\bibitem[KN]{Kulish:1976xi}
  P.~P.~Kulish and E.~R.~Nissimov,
  \emph{Conservation Laws in the Quantum Theory: cos phi in Two-Dimensions and in the Massive Thirring Model},
  JETP Lett. \textbf{24} (1976), 220--223.

\bibitem[LLT1]{Lang:2017beo}
  T.~Lang, K.~Liegener and T.~Thiemann,
  \emph{Hamiltonian renormalisation I: derivation from Osterwalder{\textendash}Schrader reconstruction},
  Class. Quant. Grav. \textbf{35}, no.24 (2018), 245011.

\bibitem[LLT2]{Lang:2017yxi}
  T.~Lang, K.~Liegener and T.~Thiemann,
  \emph{Hamiltonian Renormalisation II. Renormalisation Flow of 1+1 dimensional free scalar fields: Derivation},
  Class. Quant. Grav. \textbf{35} (2018) no.24, 245012.

\bibitem[Lo]{Lovelace:1986kr}
  C.~Lovelace,
  \emph{Stability of String Vacua. 1. A New Picture of the Renormalization Group},
  Nucl. Phys. B \textbf{273} (1986), 413--467.

\bibitem[LW]{Ludwig:2002fu}
  A.~W.~W.~Ludwig and K.~J.~Wiese,
  \emph{The Four loop beta function in the 2-D nonAbelian Thirring model, and comparison with its conjectured 'exact' form},
  Nucl. Phys. B \textbf{661} (2003), 577--607.

\bibitem[MRP]{Maestri:2026hqb}
  A.~Maestri, S.~Rodini and B.~Pasquini,
  \emph{Higher-Order Structure of Hamiltonian Truncation Effective Theory},
  [arXiv:2602.13019 [hep-ph]].

\bibitem[Ma]{Mandelstam:1975hb}
  S.~Mandelstam,
  \emph{Soliton Operators for the Quantized Sine-Gordon Equation},
  Phys. Rev. D \textbf{11} (1975), 3026.

\bibitem[MKP]{Marchetti1900}
  D.~H.~U.~Marchetti, A.~Klein and J.~F.~Perez,
  \emph{Power-law falloff in the kosterlitz-Thouless phase of a two-dimensional lattice Coulomb gas},
  J Stat Phys \textbf{60} (1990), 137--166.

\bibitem[MN]{Minic:1994ff}
  D.~Minic and V.~P.~Nair,
  \emph{Wave functionals, Hamiltonians and the renormalization group},
  Int. J. Mod. Phys. A \textbf{11} (1996), 2749--2764.

\bibitem[NNPS1]{FRG-sG}
  S.~Nagy, I.~N\'{a}ndori, J.~Polonyi, and K.~Sailer,
  \emph{Functional Renormalization Group Approach to the Sine-Gordon Model},
  Phys. Rev. Lett. \textbf{102} (2009), 241603.

\bibitem[NNPS2]{FRG-sG2}
  S.~Nagy, I.~Nandori, J.~Polonyi and K.~Sailer,
  \emph{Renormalizable parameters of the sine-Gordon model},
  Phys. Lett. B \textbf{647} (2007), 152--158.

\bibitem[NT]{Niccoli:2009jq}
  G.~Niccoli and J.~Teschner,
  \emph{The Sine-Gordon model revisited I},
  J. Stat. Mech. \textbf{1009} (2010), P09014.

\bibitem[NKOP]{RG-Thirring}
  P.~A.~Nosov, Jun-ichiro Kishine, A.~S.~Ovchinnikov and I.~Proskurin,
  \emph{Functional renormalization-group approach to the Pokrovsky-Talapov model via the modified massive Thirring fermions},
  Phys. Rev. B \textbf{96} (2017), 235126.
  
\bibitem[OaSa]{Oak:2017trw}
  P.~Oak and B.~Sathiapalan,
  \emph{Exact Renormalization Group and Sine Gordon Theory},
  JHEP \textbf{07} (2017), 103
  [erratum: JHEP \textbf{09} (2017), 077].

\bibitem[Pol]{Polchinski}
  J.~Polchinski
  \emph{Renormalization and effective lagrangians},
  Nucl. Phys. B \textbf{231} (1984), 269.

\bibitem[PS1]{PadillaSmith1}
  A.~Padilla and R.~G.~C.~Smith,
  \emph{Smoothed asymptotics: From number theory to QFT},
  Phys. Rev. D {\bf 110} (2024), 025010.

\bibitem[PS2]{PadillaSmith2}
  A.~Padilla and R.~G.~C.~Smith,
  \emph{Gauge invariance and generalized $\eta$ regularization},
  Phys. Rev. D \textbf{111}, no.12 (2025), 125013.

\bibitem[RT]{Zarate:2025qlg}
  M.~Rodriguez~Zarate and T.~Thiemann,
  \emph{Hamiltonian renormalisation VIII. P(Phi,2) quantum field theory},
  [arXiv:2505.13030 [hep-th]].

\bibitem[RV]{Rychkov:2014eea}
  S.~Rychkov and L.~G.~Vitale,
  \emph{Hamiltonian truncation study of the $\phi^4$ theory in two dimensions},
  Phys. Rev. D \textbf{91} (2015), 085011.

\bibitem[SW]{SchriefferWolff}
  J.~R.~Schrieffer and P.~A.~Wolff,
  \emph{Relation between the Anderson and Kondo Hamiltonians},
  Phys. Rev. \textbf{149} (1966), 491--492.

\bibitem[STF]{Sklyanin:1979pfu}
  E.~K.~Sklyanin, L.~A.~Takhtadzhyan and L.~D.~Faddeev,
  \emph{Quantum inverse problem method. I},
  Theor. Math. Phys. \textbf{40}, no.2 (1979), 688--706.

\bibitem[SG]{Smith:2025kvh}
  R.~G.~C.~Smith and M.~Grewar,
  \emph{$\eta$ regularisation and the functional measure},
  [arXiv:2505.01290 [hep-th]].

\bibitem[Tao]{Tao-blog}
  T.~Tao,
  \emph{Compactness and contradiction},
  American Mathematical Soc. (2013).

\bibitem[TZ]{Thiemann:2022ulb}
  T.~Thiemann and E.~A.~Zwicknagel,
  \emph{Hamiltonian renormalization. VI. Parametrized field theory on the cylinder},
  Phys. Rev. D \textbf{108}, no.12 (2023), 125006.

\bibitem[To]{Torrielli:2024bpa}
  A.~Torrielli,
  \emph{Integrability using the Sine-Gordon and Thirring Duality},
  IOP, 2024.

\bibitem[V]{Vicedo:2025vql}
  B.~Vicedo,
  \emph{Full universal enveloping vertex algebras from factorisation},
  Ann. Henri Poincar\'e (2026).

\bibitem[Wa]{Walhout1998}
  T.~S.~Walhout,
  \emph{Similarity Renormalization, Hamiltonian Flow Equations, and Dyson’s Intermediate Representation},
  Phys. Rev. D \textbf{59} (1999), 065009.

\bibitem[Weg]{Wegner:1994fdg}
  F.~Wegner,
  \emph{Flow-equations for Hamiltonians},
  Ann. Physik \textbf{3} (1994), 77--91.

\bibitem[WH]{Wegner:1972ih}
  F.~J.~Wegner and A.~Houghton,
  \emph{Renormalization group equation for critical phenomena},
  Phys. Rev. A \textbf{8} (1973), 401--412.

\bibitem[We]{Wetterich:1992yh}
  C.~Wetterich,
  \emph{Exact evolution equation for the effective potential},
  Phys. Lett. B \textbf{301} (1993), 90--94.

\bibitem[W]{Wilson1975}
  K.~G.~Wilson,
  \emph{The renormalization group: Critical phenomena and the Kondo problem},
  Rev. Mod. Phys. \textbf{47} (1975), 773.

\bibitem[YZ]{Yurov:1989yu}
  V.~P.~Yurov and A.~B.~Zamolodchikov,
  \emph{Truncated conformal space approach to scaling Lee-Yang model},
  Int. J. Mod. Phys. A \textbf{5} (1990), 3221--3246.

\bibitem[ZZ]{Zamolodchikov:1978xm}
  A.~B.~Zamolodchikov and A.~B.~Zamolodchikov,
  \emph{Factorized s Matrices in Two-Dimensions as the Exact Solutions of Certain Relativistic Quantum Field Models},
  Annals Phys. \textbf{120} (1979), 253--291.

\bibitem[Za]{Zamolodchikov:1995xk}
  A.~B.~Zamolodchikov,
  \emph{Mass scale in the sine-Gordon model and its reductions},
  Int. J. Mod. Phys. A \textbf{10} (1995), 1125--1150.

\bibitem[Zan]{Zanello}
  F.~Zanello,
  \emph{Renormalization of Higher Currents of the Sine-Gordon Model in pAQFT},
  Ann. Henri Poincar\'e \textbf{26} (2025), 1407--1442.

\bibitem[Zi]{Zimmermann:1969jj}
  W.~Zimmermann,
  \emph{Convergence of Bogolyubov's method of renormalization in momentum space},
  Commun. Math. Phys. \textbf{15} (1969), 208--234.

\end{thebibliography}
\end{document}